# Portfolio Theory and Risk Analysis Using Coefficient of Variation:

# An Alternative to the Modern Portfolio Theory

Julius O. Campeciño

Michigan State University, East Lansing, MI

## Abstract

We provided proof here that coefficient of variation (CV) is a direct measure of risk using an equation that has been derived here for the first time. We also presented a method to generate a stock CV based on return that strongly correlates with stock price performance. Consequently, we found that the price growths of stocks with low but positive CV are approximately exponential which explains our finding here that the total return of US domestic stocks within $0 \leq \text{CV} \leq 1$ between Dec 2008 – Dec 2018 averaged at around 475% and outperformed the average total return of stocks within $\text{CV} > 1$ and $\text{CV} > 4$ by 144% and 2000%, respectively. From these observations, we posit that minimizing portfolio CV does not only minimize risk but also maximizes return. Minimizing risk by minimizing the standard deviation of return (volatility) as espoused by the Modern Portfolio Theory only resulted in a meager average total return of 15%, and the low-risk (low volatility) portfolio outperformed the high-risk portfolio by only 25%. These observations suggest that CV is a more reliable measure of risk than volatility.

## I. Introduction

The 1950s saw the development of the mean-variance analysis also known as the Modern Portfolio Theory (MPT) which is still used today for optimizing security return while minimizing risk (Markowitz 1952). The method described standard deviation (and variance) of return as the measure of risk which can be mathematically defined for a portfolio with *n* securities as:

$$\sigma_P^2 = \sum_{i=1}^{n} w_i^2 \sigma_i^2 + \sum_{i=1}^{n} \sum_{j=1}^{n} 2 w_i w_j \, \rho_{i,j} \sigma_i \sigma_j \quad j \neq i \qquad Eq. 1$$

where $\sigma_i^2$ and $\sigma_P^2$ are the variances for security $i$ and portfolio $p$, respectively, while $w_i$ is the proportion of the asset allocated to each security, and $\rho_{i,j}$ is the correlation of returns between securities $i$ and $j$. While MPT is a model that produced innumerous investment successes and nurtured better investing practices, it fails to accurately model risk in the financial markets (Chen 2016). It had been argued that instead of measuring risk in terms of variance, risk should be measured only based on the probability of investment losses (Kim, Kim, and Fabozzi 2015; Keppler 1990). About 20 years later, Miller *et al.* developed a different method for risk analysis by introducing the use of coefficient of variation as a *relative measure of risk* (Miller and Karson 1977). In their method, the coefficient of variation (CV) of one security is compared to the CV of another security using a test statistic to determine if the CV of the two securities differ significantly. The test statistic is complex which may not be accessible by many ordinary investors and are, admittedly, only good when the number of securities (n) is $n \geq 10$ and $CV \leq 0.67$ for a one-sided alternative hypothesis, *i.e.*, $H_0: CV_1 = CV_2$ versus $H_a: CV_1 \neq CV_2$ (Feltz and Miller 1996), but such CV is rarely achieved in securities, especially in stock market (see Appendix). This limitation is in addition to the fact that such test is inaccurate when the data include negative values (Miller and Karson 1977) which, unfortunately, is common in financial securities' rate of return. In fact, Miller *et al.* used historical share prices instead of historical returns perhaps to avoid the inclusion of data with negative values and further assumed that the prices under consideration were taken from a normal population (Miller and Karson 1977). However, such assumption may be often incorrect as it is well known that share prices have a log-normal distribution (Antoniou et al. 2004; Rosenkrantz 2003).

This manuscript focuses on stock securities and describes an approach that generates stock CVs based on return that are strongly correlated to share price performance. CV can also be used to estimate the probability of the occurrence of negative return, herein defined as risk, using an equation that is described here for the first time. We found here that stocks with low but positive CV (low risk) have price growths that are approximately exponential and significantly outperform stocks with high or negative CV (high-risk). The method of selecting stocks using CV based on return as presented herein is mathematically simpler but more robust and reliable than the mean-variance analysis and test statistic.

## II. Deriving risk from the probability density function

The probability density function, $f_X$, is a function that describes a normal distribution and is given by:

$$f_X(x) = \frac{1}{\sigma\sqrt{2\pi}} e^{-\frac{(x-\mu)^2}{2\sigma^2}} \qquad Eq. 3$$

where $\mu$, $\sigma$, and $\sigma^2$ are the mean, standard deviation, and variance (Naghettini 2017). In finance, $\mu$ can be the average return of a security, and the distribution of returns has been theoretically and empirically proven to be normal (Fama 1965; Samuelson 2016). With $\mu$ as the average return, the mean-variance analysis (also known as Modern Porfolio Theory or MPT) then defined $\sigma$ (and $\sigma^2$) as the measure of risk. To illustrate why standard deviation is an inaccurate measure of risk, consider the theoretical normal distribution of returns of portfolios A and B shown in Figure 1.

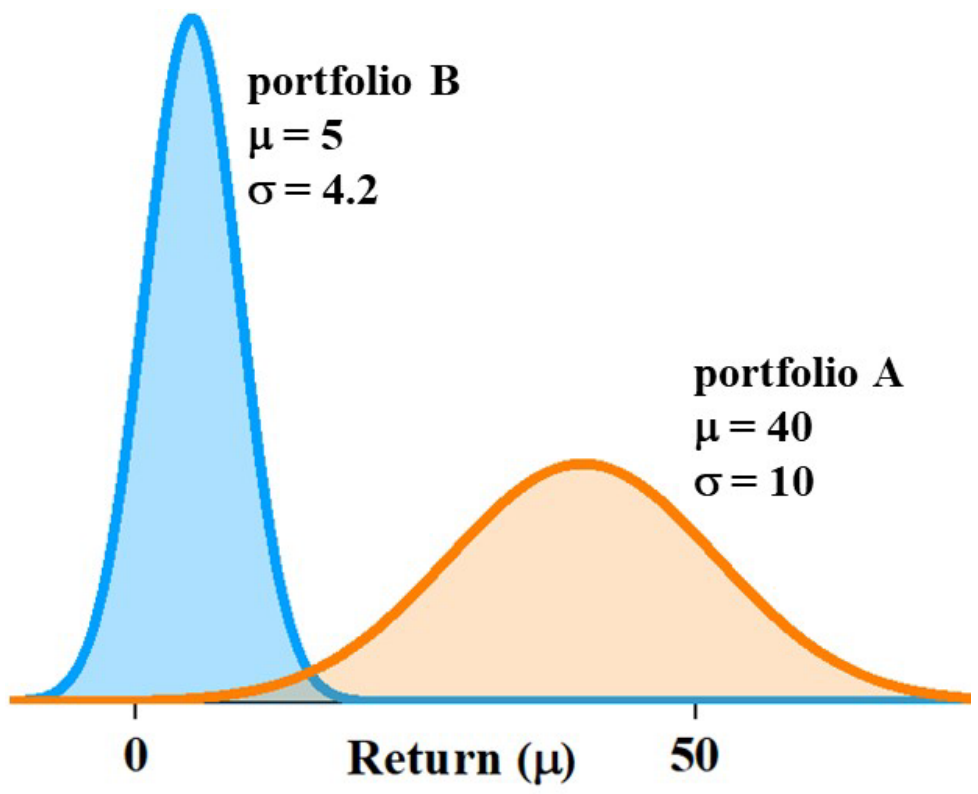

Figure 1. Theoretical normal distribution of returns of portfolios A and B.

Portfolio A has the larger standard deviation, and therefore should be the riskier portfolio according to MPT. However, the distribution of its return is well within the positive domain, suggesting that this portfolio had not incurred any losses. In contrast, portfolio B has the smaller standard deviation, and this portfolio should have the lower risk according to MPT. However, it is obvious that portfolio B incurred losses (some return fall below zero, Figure 1) and should, therefore, be the riskier portfolio. This example and the similar illustrations previously described (Kim, Kim, and Fabozzi 2015) demonstrate that risk should be defined only as the probability of incurring negative return. This probability is the area under a normal distribution curve in which the return fall below zero. This area is defined by the probability density function evaluated at $(-\infty, 0]$ shown as follows,

$$probability\ of\ negative\ return\ (risk) = \int_{-\infty}^{0} \frac{1}{\sigma\sqrt{2\pi}} e^{-\frac{(x-\mu)^2}{2\sigma^2}} dx \qquad Eq. 4$$

Equation 4 transforms into an error function (Weisstein 2002; Burkardt 2014; Marsaglia 2004) shown as follows:

$$probability\ of\ negative\ returns\ (risk) = \frac{1}{2}\left[1 + \text{erf}\left(\frac{-\mu}{\sigma\sqrt{2}}\right)\right] \qquad Eq. 5$$

which can be rewritten as:

$$\Phi_R = \frac{1}{2}\left[1 + \text{erf}\left(\frac{-1}{CV\sqrt{2}}\right)\right] x\ 100\% \qquad Eq. 6$$

where $\Phi_R$ is the % probability of occurrence of negative return (risk) and $CV = \frac{\sigma}{\mu}$. This equation reveals for the first time the direct relationship between coefficient of variation and risk. In contrast to using CV as *relative risk* as espoused by Miller *et al.*, equation 6 is straight forward and the error function (erf) can be easily accessed in most spreadsheets, calculators, and graphing softwares.

**III. $\Phi_R$ describes practical market phenomena**

Insights from equation 6 can be derived by plotting risk ($\Phi_R$) against standard deviation or volatility ($\sigma$) for various returns ($\mu$) (Figure 2). The plot suggests that as volatility ($\sigma$) increases, risk ($\Phi_R$) approaches but never cross 50%. This observation is interesting because it suggests that so long as the average return ($\mu$) is positive, the odds of profiting from an investment is better than the odds of losing. Also apparent from the plot is the inverse relationship between risk ($\Phi_R$) and return (blue arrow) which suggests that for the same volatility, assets with lower returns are riskier. Furthermore, it is apparent from the plot that for the same risk (red arrow), assets with higher returns tend to be more volatile, *e.g.*, stocks vs bonds. Overall, these observations point to the fact that equation 6 has basis from the reality of the financial market.

Interestingly, there exists a region for each return where a nonzero volatility carries zero risk. This region turns out to be within $0 \leq CV \leq 0.25$ regardless of the magnitude of the return (figure 2). The existence of this risk-free region suggests that volatility is not a risk-factor when its magnitude is no more than 0.25 of the return, highlighting the argument that volatility cannot be the true measures of risk.

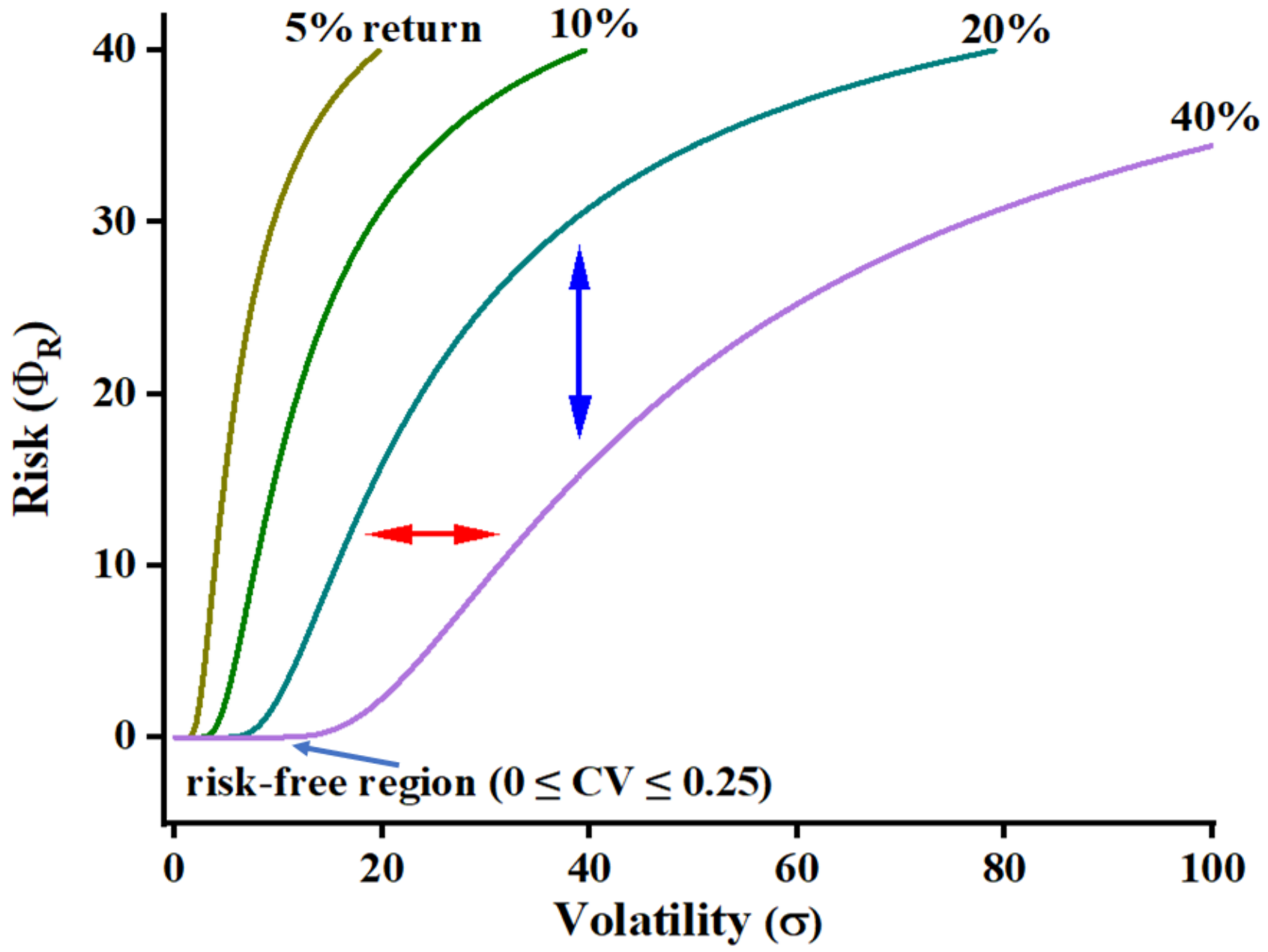

Figure 2. Plot of risk vs standard deviation for select returns.

## IV. CV derived from monthly annual returns is strongly correlated to stock performance

This section presents for the first time the strong but inverse correlation between CV and stock performance when CV is determined from annual returns calculated for each month of the period of interest. For lack of a better description, we shall call it as monthly annual return and is simply described as follows:

$$\frac{month's\ closing\ price - closing\ price\ of\ the\ same\ month\ a\ year\ ago}{closing\ price\ of\ the\ same\ month\ a\ year\ ago} x100\% \quad Eq.7$$

It can be readily shown that if CV calculation only includes annual returns based on, for example, end-of-year closing prices (Figure 3A), price fluctuations from the intervening months are unaccounted for and often lead to a weak correlation between CV and stock performance. However, when a CV is derived from monthly annual returns (Figure 3B), most of the price fluctuations are accounted for and, therefore, should lead to a strong but inverse correlation between CV and stock performance - this manner of CV calculation is akin to integral calculus. To illustrate this idea, we evaluated ~3000 US domestic stock securities and calculated their CVs based on end-of-year vs monthly annual returns between Dec 2008 – Dec 2018 (see Appendix).

Tickers (some may be valid only up to 2020) were included in the table which the reader may use for a quick verification using, for example, Yahoo!/Google Finance. For securities with no tickers, PERMNOs were provided which can be used to retrieve historical share prices from the Center for Research in Security Prices (CRSP). Indeed, we found that CVs derived from annual returns based only on end-of-the-year closing prices have weaker correlation with stock performance. For example, WRB stock (Figure 4A) appears to be riskier compared to BMRN stock (Figure 4B) by comparison of their CVs based on end-of-year closing prices. However, it is apparent from the historical share prices of these two stocks that the WRB share price growth was more consistent than that of BMRN. Reconciliation of the seemingly disconnect between CV and stock performance appears to be remediated by using monthly annual returns (compare CVs based on annual monthly returns in Figure 4). The reader is invited to verify this claim by comparing CV and performance of the stocks listed in the Appendix using Yahoo!/Google Finance as described above.

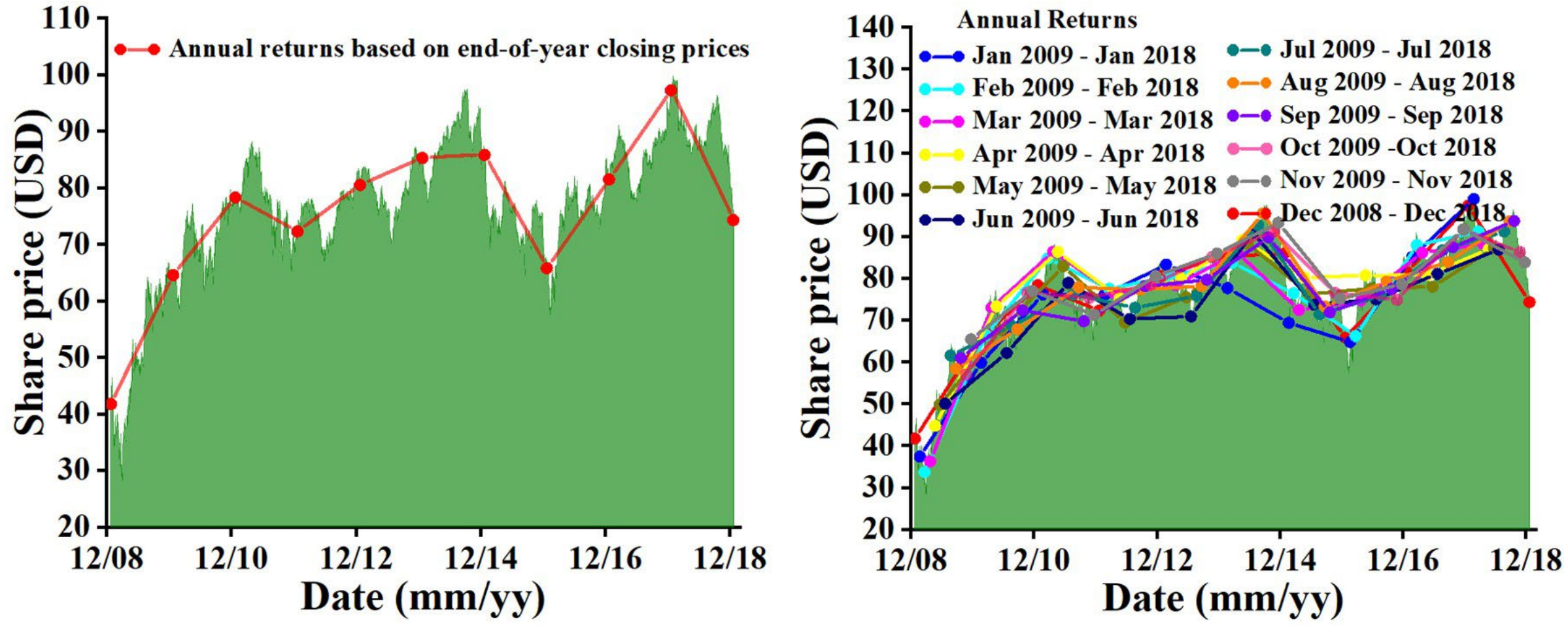

Figure 3. A. End of year annual returns. B. Monthly annual returns.

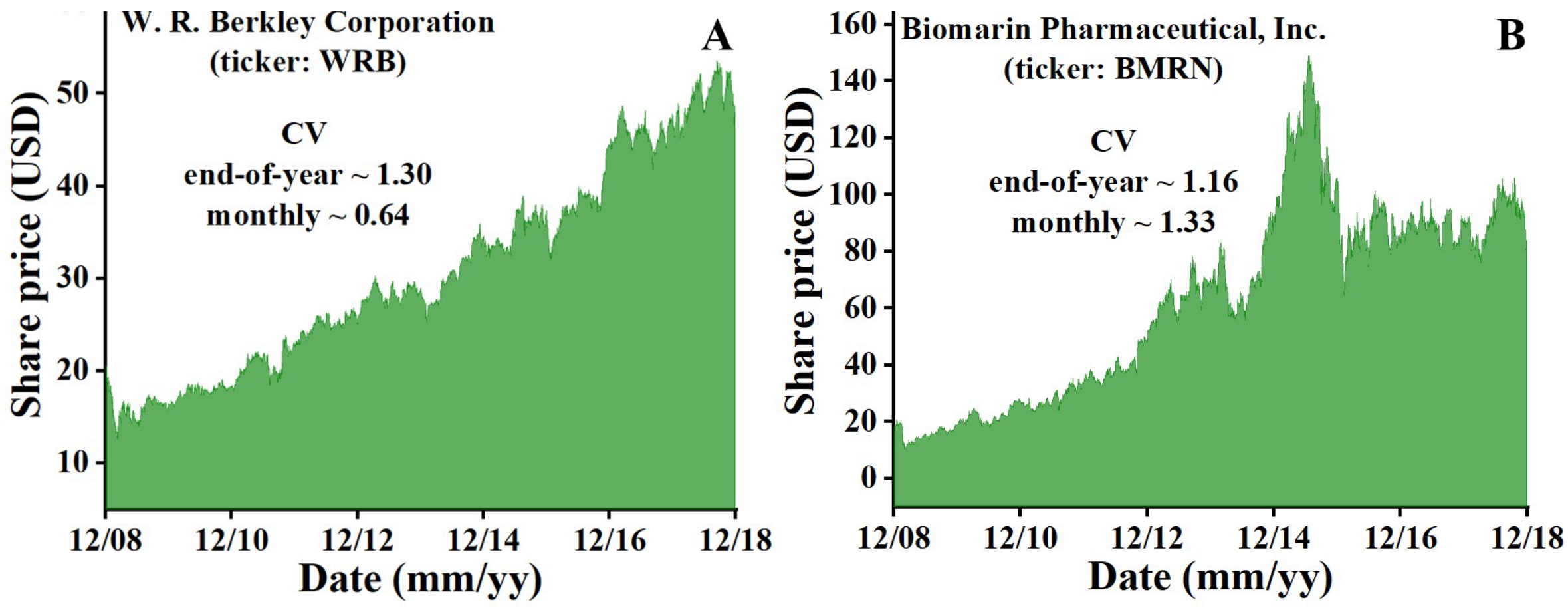

Figure 4. Comparison of WRB and BMRN end-of-year and monthly CVs and their correlation to historical share prices.

**V. Price of stocks with low but positive CV grow exponentially**

Examination of the relationship between CV (see Appendix) and price growth (via Yahoo!/Google Finance) revealed that the price growth of stocks with low but positive CVs are approximately exponential (at least between 2008 to 2018). Then, gradual flattening or appearance of large peaks and valleys occur as CV increases or a decline in price growth as CV becomes negative. This observation suggests that stocks with low but positive CVs should outperform stocks with high or negative CVs. In fact, by calculating the total return between Dec 2008 – Dec 2018 for each stock, *i.e.*,

$$total\ return = \frac{Dec\ 2018\ price - Dec\ 2008\ price}{Dec\ 2008\ price} \times 100\% \quad Eq. 8$$

we found that stocks within $0 \leq CV \leq 1$ have an average total return of 475% compared to 195% and -25% average total return for stocks within $CV > 1$ and $CV > 4$, respectively (see Appendix and figure 5A). Stocks within $0 \leq CV \leq 1$, dominated stocks within CV > 1 & CV > 4 by 144% and 2000%, respectively. It is noticeable from figure 5A that stocks within $1 \leq CV \leq 4$ appears to be still profitable. However, it needs to be pointed out that most of the stocks within this range already deviate from the exponential price growth. Previously, Baker *et al.* and many others (Baker and Haugen 2012; Haugen and Heins 1975; Haugen and Baker 1991; Jagannathan and Ma 2003) also found that low-risk (low volatility) equities/portfolios outperform high-risk (high volatility) equities/portfolios. However, when minimizing risk by minimizing volatility, the low-risk portfolio only has an average total return of no more than 15%, and dominated the high-risk portfolio by 25% only (Baker and Haugen 2012). These results are rather meager compared to the numbers generated when using CV, suggesting yet again that CV may be the better measure of risk than volatility. Also, we can show that only an exponential growth can possibly achieve zero risk. Assuming $price = ce^{a+bt}$, where a, b, c are positive constants and *t* is the number of invested years,

$$risk\text{-}free\ annual\ return = \frac{ce^{a+2b} - ce^{a+b}}{ce^{a+b}} \times 100\% = \frac{ce^{a+3b} - ce^{a+2b}}{ce^{a+2b}} \times 100$$
$$= \frac{ce^{a+b(t+1)} - ce^{a+bt}}{ce^{a+bt}} \times 100 \quad Eq. 9$$

which can be simplified to

$$risk\text{-}free\ annual\ return = (e^{b} - 1) \times 100\% \quad Eq. 10$$

and with plots of selected risk-free price growth presented in figure 5B-D. Theoretically, the risk-free return has no limit, but the question is: what distinguishes a high yield risk-free stock from a stock in a bubble? In theory, the return of a risk-free stock is constant with respect to time. In practice, the return randomly fluctuates around a mean value (figure 6 A-B). For a stock in a bubble, however, the return is either trending up during bubble formation or trending down during bubble rupture (figure 6 C-D).

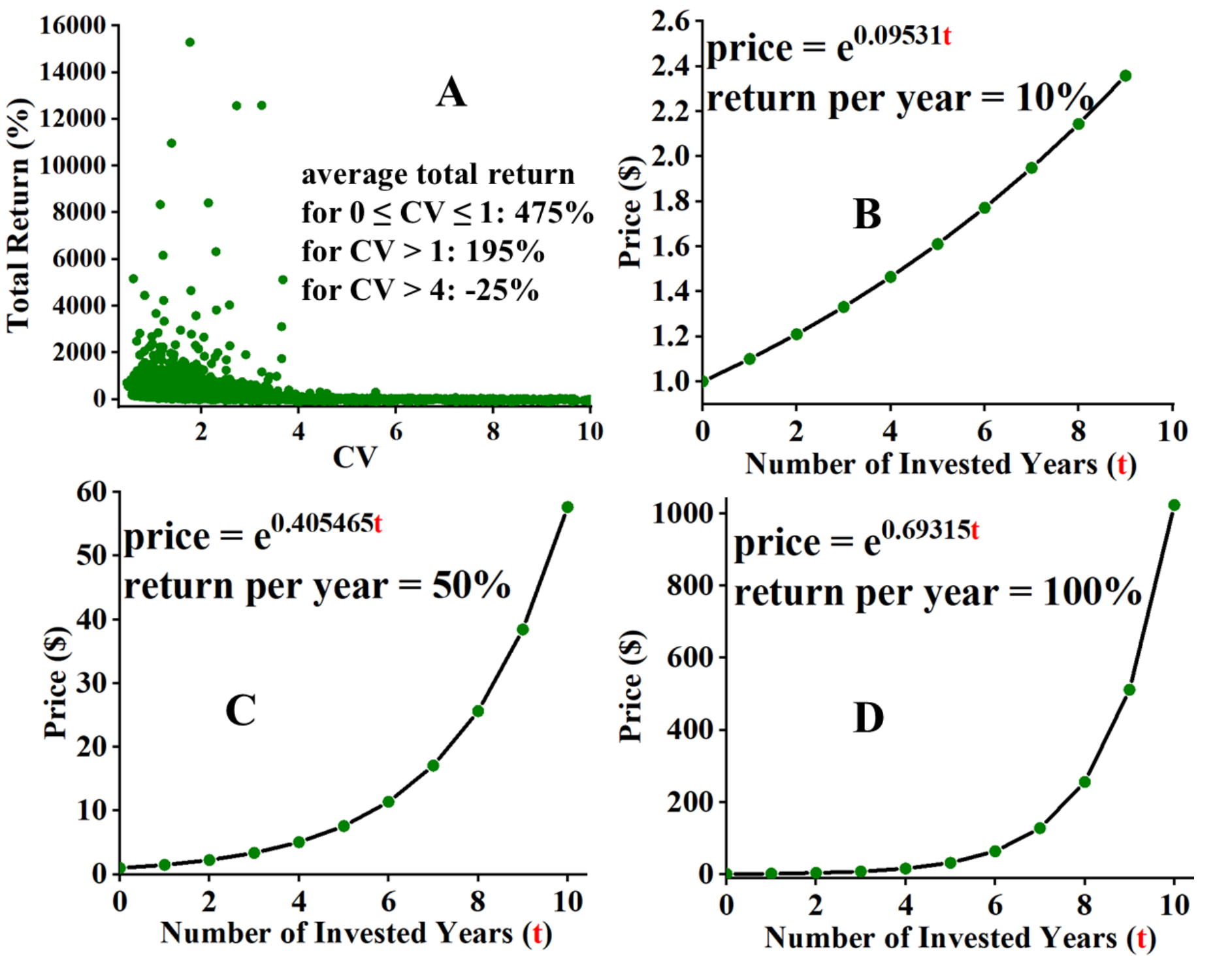

Figure 5. A. Plot of total return vs CV for stocks with CV ≤ 10 (see Appendix). B-D. Theoretical risk-free price growth growths.

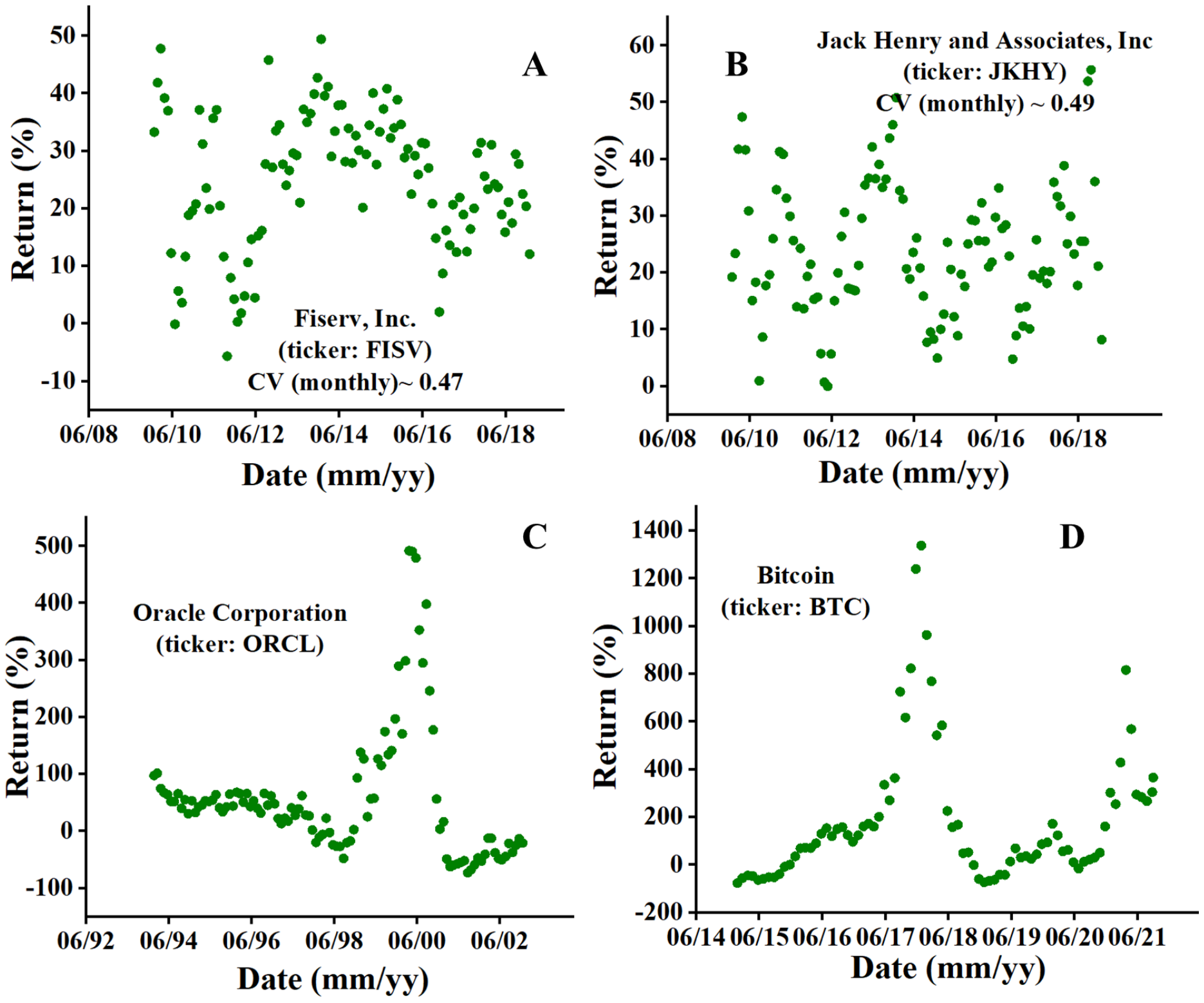

Figure 6. A-B. Stocks with normal growth. C-D. Assets in a bubble.

## VI. CV allows calculation of risk along the efficient frontier

Building a portfolio has been a practice for decades to mitigate the risk of untoward price declines, and this idea was mathematically embodied in MPT and gave rise to a very popular investment tool known as the efficient frontier (Fabozzi, Markowitz, and Gupta 2008; Markowitz 1952). MPT suggests that for any group of securities, there exists a set of portfolios offering the highest expected return for any given risk (defined in MPT as $\sigma$). These portfolios lie along the upper half of a curve (see sample curves in Figure 7) that is said to be the efficient frontier (Fabozzi, Markowitz, and Gupta 2008; Markowitz 1952). Portfolios that lie to the left of the efficient frontier are considered to have lower risks but also offer lower returns, whereas portfolios that lie to the right of the efficient frontier are considered to have higher risks but also deliver higher returns. Portfolios that lie below the curve are said to be suboptimal (Fabozzi, Markowitz, and Gupta 2008; Markowitz 1952).

We examined $\Phi_R$ along the efficient curve of a two-security portfolio (Table 1) previously described (Templeton 2013). The relationship of portfolio expected return ($\mu_p$), standard deviation ($\sigma_p$), and coefficient of variation ($CV_p$) is described as follows:

$$CV_p = \frac{\sigma_p}{\mu_p} = \frac{\sqrt{\left(\sum_{i=1}^{n} w_i^2 \sigma_i^2 + \sum_{i=1}^{n} \sum_{j=1}^{n} 2 w_i w_j \, \rho_{i,j} \sigma_i \sigma_j \quad j \neq i\right)}}{\sum_{i=1}^{n} w_n \mu_n} \qquad Eq. 11$$

For a two-security portfolio, equation 11 is reduced to

$$CV_p = \frac{\sigma_p}{\mu_p} = \frac{\sqrt{[w_1^2 \sigma_1^2 + (1 - w_1)^2 \sigma_2^2 + 2 w_1 (1 - w_1) \rho \sigma_1 \sigma_2]}}{w_1 \mu_1 + (1 - w_1) \mu_2} \qquad Eq. 12$$

where $w_1$ and $(1 - w_1)$ are the proportions of the asset allocated to securities 1 and 2, respectively. The frontiers with correlation $\rho = -1, 0,$ and 1 were constructed and described in Figure 7.

Table 1. Theoretical securities described in (Templeton 2013).

| | security A | security B |
|---|---|---|
| expected return (%) | 12 | 4 |
| standard deviation | 20 | 9 |

**Negative correlation ρ = -1**

| $w_1$ | $(1-w_1)$ | σ | μ | CV | $\Phi_R$ |
|---|---|---|---|---|---|
| 0.00 | 1.00 | 9.00 | 5.00 | 1.80 | 28.93 |
| 0.10 | 0.90 | 6.10 | 5.70 | 1.07 | 17.50 |
| 0.15 | 0.85 | 4.65 | 6.05 | 0.77 | 9.66 |
| 0.20 | 0.80 | 3.20 | 6.40 | 0.50 | 2.28 |
| 0.25 | 0.75 | 1.75 | 6.75 | 0.26 | 0.01 |
| 0.30 | 0.70 | 0.30 | 7.10 | 0.04 | 0.00 |
| 0.35 | 0.65 | 1.15 | 7.45 | 0.15 | 0.00 |
| 0.40 | 0.60 | 2.60 | 7.80 | 0.33 | 0.13 |
| 0.45 | 0.55 | 4.05 | 8.15 | 0.50 | 2.21 |
| 0.50 | 0.50 | 5.50 | 8.50 | 0.65 | 6.11 |
| 0.55 | 0.45 | 6.95 | 8.85 | 0.79 | 10.14 |
| 0.60 | 0.40 | 8.40 | 9.20 | 0.91 | 13.67 |
| 0.65 | 0.35 | 9.85 | 9.55 | 1.03 | 16.61 |
| 0.70 | 0.30 | 11.30 | 9.90 | 1.14 | 19.05 |
| 0.75 | 0.25 | 12.75 | 10.25 | 1.24 | 21.07 |
| 0.80 | 0.20 | 14.20 | 10.60 | 1.34 | 22.77 |
| 0.85 | 0.15 | 15.65 | 10.95 | 1.43 | 24.21 |
| 0.90 | 0.10 | 17.10 | 11.30 | 1.51 | 25.44 |
| 0.95 | 0.05 | 18.55 | 11.65 | 1.59 | 26.50 |
| 1.00 | 0.00 | 20.00 | 12.00 | 1.67 | 27.43 |

**No correlation ρ = 0**

| $w_1$ | $(1-w_1)$ | σ | μ | CV | $\Phi_R$ |
|---|---|---|---|---|---|
| 0.00 | 1.00 | 9.00 | 4.00 | 2.25 | 32.84 |
| 0.10 | 0.90 | 8.34 | 4.80 | 1.74 | 28.25 |
| 0.15 | 0.85 | 8.22 | 5.20 | 1.58 | 26.34 |
| 0.20 | 0.80 | 8.24 | 5.60 | 1.47 | 24.83 |
| 0.25 | 0.75 | 8.40 | 6.00 | 1.40 | 23.75 |
| 0.30 | 0.70 | 8.70 | 6.40 | 1.36 | 23.10 |
| 0.35 | 0.65 | 9.12 | 6.80 | 1.34 | 22.80 |
| 0.40 | 0.60 | 9.65 | 7.20 | 1.34 | 22.78 |
| 0.45 | 0.55 | 10.27 | 7.60 | 1.35 | 22.97 |
| 0.50 | 0.50 | 10.97 | 8.00 | 1.37 | 23.28 |
| 0.55 | 0.45 | 11.72 | 8.40 | 1.40 | 23.68 |
| 0.60 | 0.40 | 12.53 | 8.80 | 1.42 | 24.12 |
| 0.65 | 0.35 | 13.38 | 9.20 | 1.45 | 24.58 |
| 0.70 | 0.30 | 14.26 | 9.60 | 1.49 | 25.04 |
| 0.75 | 0.25 | 15.17 | 10.00 | 1.52 | 25.49 |
| 0.80 | 0.20 | 16.10 | 10.40 | 1.55 | 25.92 |
| 0.85 | 0.15 | 17.05 | 10.80 | 1.58 | 26.33 |
| 0.90 | 0.10 | 18.02 | 11.20 | 1.61 | 26.72 |
| 0.95 | 0.05 | 19.01 | 11.60 | 1.64 | 27.08 |
| 1.00 | 0.00 | 20.00 | 12.00 | 1.67 | 27.43 |

**Positive correlation ρ = 1**

| $w_1$ | $(1-w_1)$ | σ | μ | CV | $\Phi_R$ |
|---|---|---|---|---|---|
| 0.00 | 1.00 | 9.00 | 4.00 | 2.25 | 32.84 |
| 0.10 | 0.90 | 10.10 | 4.80 | 2.10 | 31.73 |
| 0.15 | 0.85 | 10.65 | 5.20 | 2.05 | 31.27 |
| 0.20 | 0.80 | 11.20 | 5.60 | 2.00 | 30.85 |
| 0.25 | 0.75 | 11.75 | 6.00 | 1.96 | 30.48 |
| 0.30 | 0.70 | 12.30 | 6.40 | 1.92 | 30.14 |
| 0.35 | 0.65 | 12.85 | 6.80 | 1.89 | 29.83 |
| 0.40 | 0.60 | 13.40 | 7.20 | 1.86 | 29.55 |
| 0.45 | 0.55 | 13.95 | 7.60 | 1.84 | 29.29 |
| 0.50 | 0.50 | 14.50 | 8.00 | 1.81 | 29.06 |
| 0.55 | 0.45 | 15.05 | 8.40 | 1.79 | 28.84 |
| 0.60 | 0.40 | 15.60 | 8.80 | 1.77 | 28.63 |
| 0.65 | 0.35 | 16.15 | 9.20 | 1.76 | 28.45 |
| 0.70 | 0.30 | 16.70 | 9.60 | 1.74 | 28.27 |
| 0.75 | 0.25 | 17.25 | 10.00 | 1.73 | 28.11 |
| 0.80 | 0.20 | 17.80 | 10.40 | 1.71 | 27.95 |
| 0.85 | 0.15 | 18.35 | 10.80 | 1.70 | 27.81 |
| 0.90 | 0.10 | 18.90 | 11.20 | 1.69 | 27.67 |
| 0.95 | 0.05 | 19.45 | 11.60 | 1.68 | 27.55 |
| 1.00 | 0.00 | 20.00 | 12.00 | 1.67 | 27.43 |

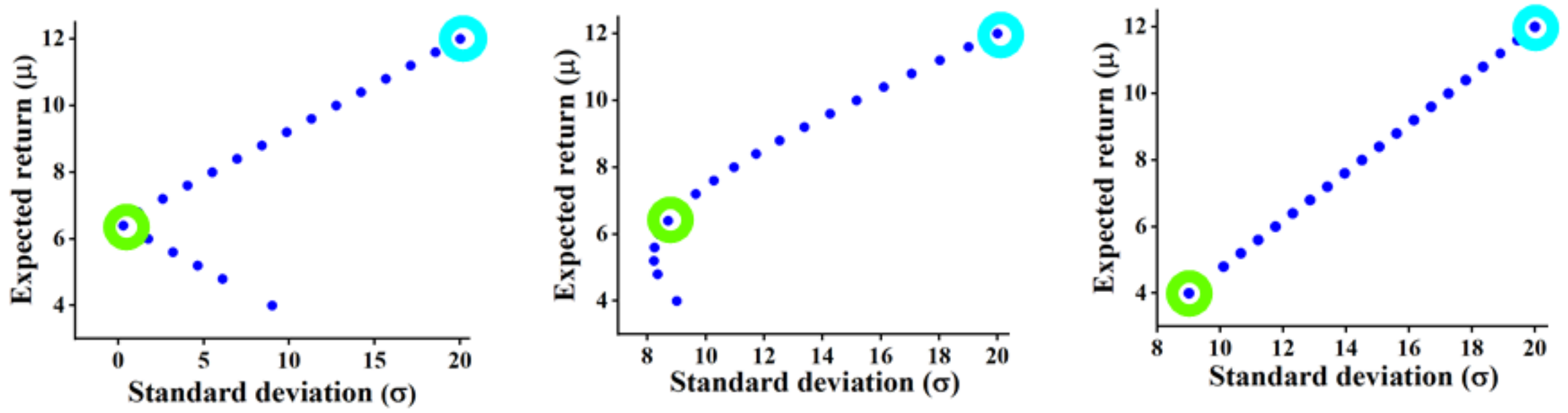

Figure 7. Calculation of CV and $\Phi_R$ along the efficient frontier. Encircled points in each figure correspond to the data of the same highlight color found in each table.

Indeed, we observe from the efficient frontier with $\rho = -1$ (Figure 7) that lower $\sigma_p$ is associated with lower $\Phi_R$ and vice versa, and this relationship is maintained even for the efficient frontier with $\rho = 0$. However at $\rho = 0$, comparison of $\mu_p$ and risk (CV or $\Phi_R$) between any two portfolios becomes an important consideration. For example, the portfolio that has an expected return $\mu_p = 6.4$ has $\Phi_R \sim 23\%$, while the portfolio with an expected return $\mu_p = 12$ has $\Phi_R \sim 27\%$. The expected return almost doubled but there is only a slight increase in $\Phi_R$, suggesting that the portfolio with $\mu_p = 12$ may be the more efficient of the two portfolios. Even more surprising is the fact that the relationship between $\sigma_p$ and $\Phi_R$ is reversed for the efficient frontier with $\rho = 1$, *i.e.*, lower $\sigma_p$ corresponds to higher $\Phi_R$, and vice versa. These analyses revealed that the direct relationship between $\sigma_p$ and $\Phi_R$ along the efficient frontier may only hold true for portfolios that have negative overall correlation. However, the reader is invited to verify that even at slightly negative correlation (*e.g.*, $\rho = -0.3$), $CV_p$ remains an important consideration. Interestingly, it had been found that inter- and intra-assets are likely to be positively correlated even for assets from diversified international origin (Hunter and Coggin 1990; Van Landschoot 2007; Chernih, Vanduffel, and Henrard 2006; Newell and Peng 2008), indicating that $CV_p$ calculation may be the more superior tool than the efficient frontier. Judging from the observations from this and the previous section, we can posit that to maximize portfolio return, one only needs to minimize portfolio CV.

## VII. Conclusion

In addition to presenting a method for generating CV that strongly correlates with stock performance, we also presented here for the first time a mathematical equation that allows us to calculate risk using CV. This equation perfectly describes known market phenomena which provides confidence for its practical application in risk analysis. The equation suggests that stocks with high return are not necessarily risky. In fact, any stock with high return can be a risk-free if its volatility is no more than 25% of its return. This observation appears to contradict the notion that volatility is a measure of risk as proposed in the mean-variance analysis. Finally, we found that stocks with low but positive CV of up to 1.00 have approximately exponential price growths, and that these stocks appear to be outperforming stocks with high CV (high risk) by hundreds and even thousands of percent as opposed to only tens of percent when minimizing risk by minimizing volatility. Along with the empirical data from previous studies, it appears that the low risk/high return may indeed be the true risk/return correlation in the equity market, which appears to contradict the notion of high risk/high return in modern finance as espoused by the Modern Portfolio Theory. Consequently, it appears that in building a portfolio, one must minimize CV to maximize returns, but it needs to be borne in mind that CV comparison must be done only within an asset class as previously illustrated (Haugen and Heins 1975; Baker and Haugen 2012).

## VIII. Methodology

The historical and raw monthly share prices of US domestic stocks between December 2018 to December 2019 were downloaded from the Center for Research in Security Prices database which is available at Wharton Research Data Services. The prices were adjusted to reflect stock splits, repurchases, etc. by dividing the raw prices with the corresponding cumulative factor (CFACPR) also downloaded from the same database. Only stocks that have the complete historical monthly closing prices within the aforementioned period were included in the calculation. Bitcoin historical monthly closing prices were downloaded from www.investing.com. All plots were generated using OriginPro 2021 (OriginLab®).

## Appendix

## Annual and total return ($\mu$), annual standard deviation ($\sigma$), and CV of stocks between Dec 2008 – Dec 2018.

| based on monthly closing prices | | | | | | | based on end-of-year closing prices | | | |
|---|---|---|---|---|---|---|---|---|---|---|
| PERMNO | Ticker | σ | μ | total μ | CV | | PERMNO | σ | μ | CV |
| 10696 | FISV | 11.67 | 25.02 | 708 | 0.47 | | 10696 | 13.57 | 23.91 | 0.57 |
| 88664 | JKHY | 11.80 | 23.84 | 552 | 0.49 | | 88664 | 13.24 | 21.23 | 0.62 |
| 23660 | CTAS | 13.65 | 25.26 | 623 | 0.54 | | 23660 | 12.55 | 22.44 | 0.56 |
| 92655 | UNH | 15.99 | 29.47 | 837 | 0.54 | | 92655 | 12.85 | 25.67 | 0.50 |
| 82515 | POOL | 15.36 | 27.17 | 727 | 0.57 | | 82515 | 11.76 | 24.04 | 0.49 |
| 91055 | SPHQ | 7.67 | 13.49 | 178 | 0.57 | | 91055 | 10.80 | 11.25 | 0.96 |
| 10933 | MKL | 9.20 | 16.11 | 247 | 0.57 | | 10933 | 13.18 | 13.95 | 0.94 |
| 44644 | ADP | 9.90 | 16.96 | 282 | 0.58 | | 44644 | 11.28 | 14.79 | 0.76 |
| 49680 | DHR | 10.65 | 18.18 | 374 | 0.59 | | 49680 | 11.93 | 17.38 | 0.69 |
| 36003 | ROL | 14.59 | 24.74 | 574 | 0.59 | | 36003 | 17.06 | 22.06 | 0.77 |
| 76261 | EXPO | 14.71 | 24.89 | 574 | 0.59 | | 76261 | 14.83 | 21.88 | 0.68 |
| 77338 | ROP | 13.65 | 22.94 | 514 | 0.60 | | 77338 | 15.48 | 20.79 | 0.74 |
| 90248 | DPZ | 30.19 | 50.36 | 5165 | 0.60 | | 90248 | 32.52 | 51.57 | 0.63 |
| 22825 | CMD | 22.41 | 36.56 | 1042 | 0.61 | | 22825 | 26.79 | 30.46 | 0.88 |
| 11533 | FICO | 19.65 | 31.99 | 1009 | 0.61 | | 11533 | 14.02 | 27.87 | 0.50 |
| 88291 | IYC | 11.13 | 17.93 | 327 | 0.62 | | 88291 | 12.76 | 16.24 | 0.79 |
| 91225 | IHI | 11.92 | 19.13 | 422 | 0.62 | | 91225 | 13.03 | 18.60 | 0.70 |
| 45751 | MMC | 10.39 | 16.64 | 229 | 0.62 | | 45751 | 14.97 | 13.53 | 1.11 |
| 78947 | ELS | 11.86 | 18.86 | 406 | 0.63 | | 78947 | 13.16 | 18.25 | 0.72 |
| 88216 | IVW | 9.38 | 14.84 | 235 | 0.63 | | 88216 | 11.42 | 13.37 | 0.85 |
| 89071 | ACN | 12.36 | 19.53 | 330 | 0.63 | | 89071 | 11.29 | 16.23 | 0.70 |
| 66181 | HD | 16.67 | 26.24 | 646 | 0.64 | | 66181 | 16.94 | 23.39 | 0.72 |
| 86755 | QQQ | 12.40 | 19.49 | 419 | 0.64 | | 86755 | 16.96 | 18.92 | 0.90 |
| 90619 | PWB | 10.54 | 16.43 | 285 | 0.64 | | 90619 | 14.01 | 15.19 | 0.92 |
| 17743 | WRB | 8.76 | 13.63 | 138 | 0.64 | | 17743 | 12.74 | 9.82 | 1.30 |
| 38762 | NI | 14.73 | 22.90 | 544 | 0.64 | | 38762 | 13.73 | 21.18 | 0.65 |
| 78975 | INTU | 16.72 | 25.78 | 727 | 0.65 | | 78975 | 16.29 | 24.44 | 0.67 |
| 87055 | COST | 11.48 | 17.67 | 288 | 0.65 | | 87055 | 6.66 | 14.70 | 0.45 |
| 23229 | CMS | 10.84 | 16.59 | 391 | 0.65 | | 23229 | 14.99 | 18.02 | 0.83 |
| 23393 | CHD | 11.14 | 16.94 | 369 | 0.66 | | 23393 | 9.71 | 17.06 | 0.57 |
| 88604 | SPYG | 10.24 | 15.46 | 258 | 0.66 | | 88604 | 12.58 | 14.20 | 0.89 |
| 84588 | MTN | 18.96 | 28.55 | 693 | 0.66 | | 84588 | 19.78 | 24.65 | 0.80 |
| 92614 | AWK | 12.20 | 18.36 | 335 | 0.66 | | 92614 | 8.97 | 16.15 | 0.56 |
| 21928 | IDA | 9.98 | 15.01 | 216 | 0.66 | | 21928 | 8.58 | 12.49 | 0.69 |
| 23182 | CPK | 10.94 | 16.45 | 287 | 0.67 | | 23182 | 11.20 | 14.99 | 0.75 |
| 91152 | TDG | 18.97 | 28.43 | 913 | 0.67 | | 91152 | 14.62 | 26.80 | 0.55 |
| 90454 | MKTX | 28.31 | 42.36 | 2490 | 0.67 | | 90454 | 27.19 | 40.81 | 0.67 |
| 11403 | CDNS | 18.27 | 27.21 | 1088 | 0.67 | | 11403 | 22.05 | 29.71 | 0.74 |
| 86453 | XLY | 12.83 | 18.98 | 359 | 0.68 | | 86453 | 14.48 | 17.24 | 0.84 |
| 92512 | MGK | 10.18 | 15.04 | 240 | 0.68 | | 92512 | 13.02 | 13.67 | 0.95 |
| 91233 | MA | 21.31 | 31.46 | 1220 | 0.68 | | 91233 | 31.54 | 32.82 | 0.96 |
| 88220 | IWF | 10.46 | 15.37 | 253 | 0.68 | | 88220 | 13.16 | 14.12 | 0.93 |
| 89807 | FVD | 8.06 | 11.84 | 151 | 0.68 | | 89807 | 8.68 | 9.94 | 0.87 |
| 89060 | IGV | 13.59 | 19.90 | 446 | 0.68 | | 89060 | 16.77 | 19.55 | 0.86 |
| 92611 | V | 19.04 | 27.84 | 906 | 0.68 | | 92611 | 26.38 | 28.62 | 0.92 |
| 86457 | XLK | 11.41 | 16.60 | 302 | 0.69 | | 86457 | 15.63 | 15.83 | 0.99 |
| 90256 | JKE | 10.99 | 15.93 | 285 | 0.69 | | 90256 | 14.48 | 15.22 | 0.95 |
| 88403 | IUSG | 10.50 | 15.20 | 251 | 0.69 | | 88403 | 12.96 | 14.02 | 0.92 |
| 81294 | WDFC | 14.88 | 21.43 | 548 | 0.69 | | 81294 | 19.90 | 21.89 | 0.91 |
| 90003 | VUG | 10.48 | 15.06 | 240 | 0.70 | | 90003 | 13.34 | 13.71 | 0.97 |
| 17478 | SPGI | 17.67 | 25.31 | 633 | 0.70 | | 17478 | 18.87 | 23.28 | 0.81 |
| 92523 | EBSB | 14.73 | 20.96 | 279 | 0.70 | | 92523 | 21.93 | 16.44 | 1.33 |
| 89003 | FIS | 14.52 | 20.63 | 530 | 0.70 | | 89003 | 18.48 | 21.46 | 0.86 |
| 88927 | IGM | 13.46 | 19.11 | 412 | 0.70 | | 88927 | 19.20 | 19.02 | 1.01 |
| 32986 | ATO | 11.00 | 15.39 | 291 | 0.71 | | 32986 | 8.47 | 14.89 | 0.57 |
| 44329 | TFX | 15.56 | 21.77 | 416 | 0.71 | | 44329 | 15.69 | 18.70 | 0.84 |
| 91556 | ROST | 20.69 | 28.94 | 1019 | 0.71 | | 91556 | 16.04 | 28.23 | 0.57 |
| 89869 | ONEQ | 12.57 | 17.56 | 319 | 0.72 | | 89869 | 16.18 | 16.38 | 0.99 |
| 86111 | AMT | 13.46 | 18.80 | 440 | 0.72 | | 86111 | 15.05 | 19.20 | 0.78 |
| 77357 | SNPS | 13.17 | 18.37 | 355 | 0.72 | | 77357 | 14.45 | 17.14 | 0.84 |
| 60687 | AFG | 14.75 | 20.56 | 296 | 0.72 | | 60687 | 16.63 | 15.86 | 1.05 |
| 10026 | JJSF | 11.89 | 16.56 | 303 | 0.72 | | 10026 | 11.36 | 15.45 | 0.74 |
| 52090 | MKC | 11.64 | 16.17 | 337 | 0.72 | | 52090 | 10.39 | 16.29 | 0.64 |
| 23536 | WEC | 9.90 | 13.67 | 230 | 0.72 | | 23536 | 8.74 | 12.99 | 0.67 |
| 92795 | OEF | 8.93 | 12.31 | 158 | 0.72 | | 92795 | 10.12 | 10.37 | 0.98 |
| 85945 | HEI | 23.19 | 31.91 | 937 | 0.73 | | 85945 | 19.19 | 27.63 | 0.69 |
| 84234 | PLUS | 25.10 | 34.53 | 1254 | 0.73 | | 84234 | 17.24 | 30.87 | 0.56 |
| 77037 | ODFL | 24.08 | 33.07 | 876 | 0.73 | | 77037 | 28.00 | 28.74 | 0.97 |
| 84788 | AMZN | 31.05 | 42.54 | 2829 | 0.73 | | 84788 | 55.47 | 48.71 | 1.14 |
| 91308 | QLD | 29.98 | 40.96 | 1899 | 0.73 | | 91308 | 40.53 | 39.82 | 1.02 |
| 91298 | FDN | 18.61 | 25.41 | 731 | 0.73 | | 91298 | 26.24 | 25.89 | 1.01 |
| 79026 | CHDN | 19.61 | 26.72 | 504 | 0.73 | | 79026 | 20.16 | 21.17 | 0.95 |
| 92514 | MGC | 9.59 | 12.96 | 176 | 0.74 | | 92514 | 10.94 | 11.16 | 0.98 |
| 91209 | VIG | 8.43 | 11.39 | 144 | 0.74 | | 91209 | 9.73 | 9.72 | 1.00 |
| 90766 | PSJ | 14.57 | 19.58 | 455 | 0.74 | | 90766 | 17.53 | 19.80 | 0.89 |
| 24766 | NOC | 19.15 | 25.72 | 504 | 0.74 | | 24766 | 23.05 | 21.70 | 1.06 |
| 24205 | NEE | 10.51 | 14.02 | 245 | 0.75 | | 24205 | 10.97 | 13.68 | 0.80 |
| 89996 | VGT | 13.88 | 18.51 | 390 | 0.75 | | 89996 | 18.91 | 18.46 | 1.02 |
| 85765 | DIA | 9.53 | 12.70 | 166 | 0.75 | | 85765 | 10.70 | 10.76 | 0.99 |
| 90892 | PFM | 7.84 | 10.44 | 114 | 0.75 | | 90892 | 8.89 | 8.22 | 1.08 |
| 10145 | HON | 14.36 | 19.13 | 322 | 0.75 | | 10145 | 16.77 | 16.56 | 1.01 |
| 84398 | SPY | 9.85 | 13.07 | 177 | 0.75 | | 84398 | 11.32 | 11.25 | 1.01 |
| 88215 | IVV | 9.90 | 13.12 | 179 | 0.75 | | 88215 | 11.40 | 11.32 | 1.01 |
| 85753 | VRSN | 18.42 | 24.39 | 677 | 0.76 | | 85753 | 24.12 | 24.95 | 0.97 |
| 23931 | XEL | 8.27 | 10.95 | 166 | 0.76 | | 23931 | 9.77 | 10.65 | 0.92 |
| 65402 | AOS | 23.93 | 31.61 | 768 | 0.76 | | 65402 | 29.10 | 27.53 | 1.06 |
| 91226 | IHF | 14.77 | 19.45 | 362 | 0.76 | | 91226 | 12.96 | 17.17 | 0.75 |
| 84769 | APH | 16.67 | 21.90 | 576 | 0.76 | | 84769 | 30.88 | 24.26 | 1.27 |
| 91326 | DLN | 8.77 | 11.48 | 132 | 0.76 | | 91326 | 9.36 | 9.17 | 1.02 |
| 90002 | VV | 10.14 | 13.26 | 182 | 0.76 | | 90002 | 11.60 | 11.47 | 1.01 |
| 89992 | VCR | 14.65 | 19.16 | 367 | 0.76 | | 89992 | 17.12 | 17.75 | 0.96 |
| 76639 | | 18.02 | 23.56 | 481 | 0.76 | | 76639 | 25.51 | 21.57 | 1.18 |
| 89988 | ITOT | 10.26 | 13.37 | 181 | 0.77 | | 89988 | 11.65 | 11.42 | 1.02 |
| 89895 | DVY | 8.91 | 11.56 | 116 | 0.77 | | 89895 | 10.06 | 8.45 | 1.19 |
| 73139 | SYK | 13.70 | 17.67 | 292 | 0.78 | | 73139 | 15.53 | 15.61 | 1.00 |
| 87445 | TDY | 18.87 | 24.34 | 365 | 0.78 | | 87445 | 20.95 | 18.37 | 1.14 |
| 89995 | VHT | 12.06 | 15.55 | 251 | 0.78 | | 89995 | 13.19 | 14.06 | 0.94 |
| 91597 | VYM | 8.96 | 11.53 | 132 | 0.78 | | 91597 | 9.55 | 9.17 | 1.04 |
| 90427 | TXRH | 18.64 | 23.99 | 670 | 0.78 | | 90427 | 24.22 | 24.76 | 0.98 |
| 88221 | IWB | 10.37 | 13.33 | 184 | 0.78 | | 88221 | 11.87 | 11.57 | 1.03 |
| 86451 | XLV | 11.34 | 14.54 | 226 | 0.78 | | 86451 | 12.57 | 13.15 | 0.96 |
| 77629 | USPH | 20.48 | 26.25 | 668 | 0.78 | | 77629 | 14.07 | 23.37 | 0.60 |
| 91309 | DDM | 22.36 | 28.63 | 611 | 0.78 | | 91309 | 24.80 | 23.91 | 1.04 |
| 58819 | LNT | 11.25 | 14.35 | 190 | 0.78 | | 58819 | 11.85 | 11.79 | 1.00 |
| 76185 | TYL | 28.62 | 36.48 | 1451 | 0.78 | | 76185 | 38.91 | 36.45 | 1.07 |
| 51633 | | 10.39 | 13.22 | 188 | 0.79 | | 51633 | 13.46 | 11.89 | 1.13 |
| 90257 | JKD | 10.22 | 13.00 | 163 | 0.79 | | 90257 | 12.20 | 10.79 | 1.13 |
| 88407 | IJT | 13.55 | 17.23 | 262 | 0.79 | | 88407 | 14.61 | 14.55 | 1.00 |
| 76709 | IDXX | 24.40 | 30.98 | 931 | 0.79 | | 76709 | 18.61 | 27.49 | 0.68 |
| 64194 | CHE | 20.50 | 26.01 | 612 | 0.79 | | 64194 | 20.59 | 23.41 | 0.88 |
| 91324 | DTD | 9.26 | 11.74 | 137 | 0.79 | | 91324 | 9.84 | 9.43 | 1.04 |
| 88302 | IYY | 10.51 | 13.32 | 182 | 0.79 | | 88302 | 11.95 | 11.52 | 1.04 |
| 88296 | IYH | 11.91 | 15.09 | 237 | 0.79 | | 88296 | 12.91 | 13.55 | 0.95 |
| 89993 | VDC | 8.11 | 10.25 | 127 | 0.79 | | 89993 | 9.13 | 8.89 | 1.03 |
| 70578 | ECL | 12.89 | 16.28 | 319 | 0.79 | | 70578 | 13.14 | 16.04 | 0.82 |
| 90878 | VTI | 10.68 | 13.46 | 185 | 0.79 | | 90878 | 12.17 | 11.65 | 1.04 |
| 88223 | IWV | 10.63 | 13.37 | 183 | 0.79 | | 88223 | 12.09 | 11.54 | 1.05 |
| 91876 | PDP | 12.68 | 15.93 | 231 | 0.80 | | 91876 | 13.37 | 13.44 | 1.00 |
| 36468 | SHW | 20.95 | 26.30 | 559 | 0.80 | | 36468 | 26.38 | 23.13 | 1.14 |
| 92008 | FTC | 12.11 | 15.20 | 218 | 0.80 | | 92008 | 14.64 | 13.10 | 1.12 |
| 86372 | GIB | 19.42 | 24.25 | 684 | 0.80 | | 86372 | 20.74 | 24.24 | 0.86 |
| 86452 | XLP | 7.77 | 9.68 | 113 | 0.80 | | 86452 | 8.68 | 8.17 | 1.06 |
| 90305 | EXR | 26.05 | 32.46 | 777 | 0.80 | | 90305 | 22.58 | 26.16 | 0.86 |
| 79103 | ORLY | 22.81 | 28.27 | 1020 | 0.81 | | 79103 | 21.70 | 29.12 | 0.75 |
| 91221 | ITA | 16.07 | 19.85 | 319 | 0.81 | | 91221 | 17.82 | 16.55 | 1.08 |
| 88225 | IYW | 14.31 | 17.66 | 353 | 0.81 | | 88225 | 19.63 | 17.63 | 1.11 |
| 91307 | SSO | 23.12 | 28.44 | 607 | 0.81 | | 91307 | 25.75 | 24.05 | 1.07 |
| 85442 | TSM | 14.90 | 18.32 | 370 | 0.81 | | 85442 | 18.28 | 18.00 | 1.02 |
| 89269 | CNC | 29.38 | 36.00 | 1070 | 0.82 | | 89269 | 31.39 | 31.23 | 1.01 |
| 90870 | ROLL | 20.48 | 25.07 | 546 | 0.82 | | 90870 | 22.46 | 22.36 | 1.00 |
| 91810 | EPS | 10.52 | 12.87 | 170 | 0.82 | | 91810 | 12.34 | 11.07 | 1.11 |
| 79057 | CB | 10.58 | 12.95 | 144 | 0.82 | | 79057 | 12.28 | 9.97 | 1.23 |
| 82276 | ACGL | 14.68 | 17.95 | 243 | 0.82 | | 82276 | 14.71 | 14.01 | 1.05 |
| 71271 | Y | 8.55 | 10.44 | 135 | 0.82 | | 71271 | 10.70 | 9.37 | 1.14 |
| 89447 | CVGW | 18.93 | 23.10 | 534 | 0.82 | | 89447 | 22.52 | 22.21 | 1.01 |
| 85040 | SLP | 31.19 | 37.85 | 2063 | 0.82 | | 85040 | 28.94 | 38.64 | 0.75 |
| 79698 | IT | 23.93 | 29.02 | 617 | 0.82 | | 79698 | 26.57 | 24.00 | 1.11 |
| 91577 | DSI | 10.70 | 12.97 | 180 | 0.83 | | 91577 | 12.60 | 11.48 | 1.10 |
| 29938 | BF | 14.56 | 17.63 | 257 | 0.83 | | 29938 | 17.66 | 14.88 | 1.19 |
| 29946 | BF | 15.12 | 18.30 | 247 | 0.83 | | 29946 | 18.91 | 14.62 | 1.29 |
| 85348 | YUM | 13.81 | 16.71 | 318 | 0.83 | | 85348 | 13.02 | 16.02 | 0.81 |
| 39642 | BDX | 12.81 | 15.49 | 229 | 0.83 | | 39642 | 15.06 | 13.55 | 1.11 |
| 66376 | WSO | 13.69 | 16.53 | 262 | 0.83 | | 66376 | 14.60 | 14.67 | 1.00 |
| 90526 | SUSA | 10.42 | 12.59 | 169 | 0.83 | | 90526 | 12.36 | 11.03 | 1.12 |
| 83382 | TCX | 44.34 | 53.42 | 4450 | 0.83 | | 83382 | 53.71 | 54.99 | 0.98 |
| 44206 | ES | 9.96 | 11.96 | 170 | 0.83 | | 44206 | 9.14 | 10.79 | 0.85 |
| 86196 | | 29.49 | 35.33 | 1577 | 0.83 | | 86196 | 33.63 | 36.21 | 0.93 |
| 85621 | MTD | 23.15 | 27.65 | 739 | 0.84 | | 85621 | 20.88 | 25.34 | 0.82 |
| 89232 | IXN | 13.76 | 16.39 | 289 | 0.84 | | 89232 | 18.33 | 15.80 | 1.16 |
| 48653 | HUM | 25.39 | 30.18 | 668 | 0.84 | | 48653 | 22.41 | 24.60 | 0.91 |
| 91643 | PKW | 12.92 | 15.32 | 226 | 0.84 | | 91643 | 15.85 | 13.53 | 1.17 |
| 91794 | RXL | 27.61 | 32.61 | 898 | 0.85 | | 91794 | 30.83 | 28.97 | 1.06 |
| 90936 | PPA | 14.74 | 17.37 | 252 | 0.85 | | 90936 | 16.30 | 14.42 | 1.13 |
| 83728 | RNR | 9.87 | 11.63 | 159 | 0.85 | | 83728 | 9.81 | 10.40 | 0.94 |
| 88299 | IYK | 9.36 | 10.97 | 130 | 0.85 | | 88299 | 11.84 | 9.30 | 1.27 |
| 57665 | NKE | 18.95 | 22.19 | 481 | 0.85 | | 57665 | 18.42 | 20.63 | 0.89 |
| 10252 | INDB | 14.52 | 16.95 | 169 | 0.86 | | 10252 | 20.85 | 12.09 | 1.73 |
| 88229 | IJR | 13.73 | 16.02 | 215 | 0.86 | | 88229 | 15.54 | 13.13 | 1.18 |
| 42877 | JBHT | 15.59 | 18.19 | 254 | 0.86 | | 42877 | 18.31 | 14.94 | 1.23 |
| 91849 | BR | 20.75 | 24.18 | 668 | 0.86 | | 91849 | 29.32 | 25.36 | 1.16 |
| 69796 | STZ | 32.98 | 38.41 | 920 | 0.86 | | 69796 | 38.62 | 31.51 | 1.23 |
| 91181 | QQEW | 15.06 | 17.47 | 338 | 0.86 | | 91181 | 19.96 | 17.35 | 1.15 |
| 46578 | CLX | 9.51 | 11.03 | 177 | 0.86 | | 46578 | 10.22 | 11.16 | 0.92 |
| 59459 | TRV | 11.59 | 13.42 | 165 | 0.86 | | 59459 | 10.19 | 10.68 | 0.95 |
| 80320 | CPRT | 21.71 | 25.07 | 603 | 0.87 | | 80320 | 19.04 | 22.84 | 0.83 |
| 11587 | ATRI | 20.24 | 23.30 | 663 | 0.87 | | 11587 | 22.06 | 24.40 | 0.90 |
| 88398 | IJK | 13.69 | 15.75 | 245 | 0.87 | | 88398 | 16.27 | 14.23 | 1.14 |
| 90993 | ICE | 15.89 | 18.26 | 357 | 0.87 | | 90993 | 25.17 | 18.41 | 1.37 |
| 57568 | BLL | 14.34 | 16.46 | 342 | 0.87 | | 57568 | 11.94 | 16.58 | 0.72 |
| 62498 | WST | 19.68 | 22.56 | 419 | 0.87 | | 62498 | 26.69 | 20.26 | 1.32 |
| 90319 | GOOGL | 18.97 | 21.74 | 578 | 0.87 | | 90319 | 34.94 | 24.92 | 1.40 |
| 92513 | MGV | 9.82 | 11.24 | 128 | 0.87 | | 92513 | 10.27 | 9.01 | 1.14 |
| 92573 | | 12.02 | 13.75 | 190 | 0.87 | | 92573 | 13.93 | 12.01 | 1.16 |
| 89998 | VPU | 7.08 | 8.10 | 93 | 0.87 | | 89998 | 9.08 | 7.14 | 1.27 |
| 60580 | TTC | 23.70 | 27.07 | 577 | 0.88 | | 60580 | 23.85 | 23.25 | 1.03 |
| 52476 | EFX | 16.49 | 18.83 | 251 | 0.88 | | 52476 | 18.07 | 14.75 | 1.22 |
| 90511 | PEY | 10.45 | 11.90 | 103 | 0.88 | | 90511 | 12.35 | 7.99 | 1.55 |
| 75591 | IEX | 19.77 | 22.50 | 423 | 0.88 | | 75591 | 21.73 | 19.71 | 1.10 |
| 88292 | IDU | 7.07 | 8.04 | 94 | 0.88 | | 88292 | 9.23 | 7.19 | 1.28 |
| 61735 | AON | 14.33 | 16.29 | 218 | 0.88 | | 61735 | 17.85 | 13.52 | 1.32 |
| 15579 | TXN | 19.22 | 21.78 | 509 | 0.88 | | 15579 | 25.32 | 22.15 | 1.14 |
| 91999 | FXH | 17.16 | 19.44 | 382 | 0.88 | | 91999 | 19.53 | 18.43 | 1.06 |
| 89058 | IWP | 13.87 | 15.70 | 264 | 0.88 | | 89058 | 16.88 | 14.86 | 1.14 |
| 11674 | DTE | 12.89 | 14.59 | 209 | 0.88 | | 11674 | 11.43 | 12.48 | 0.92 |
| 86288 | CSGP | 29.16 | 32.92 | 924 | 0.89 | | 86288 | 33.18 | 29.54 | 1.12 |
| 89525 | CMCSA | 17.88 | 20.18 | 303 | 0.89 | | 89525 | 21.49 | 16.71 | 1.29 |
| 90259 | JKG | 13.40 | 15.12 | 224 | 0.89 | | 90259 | 15.46 | 13.44 | 1.15 |
| 40539 | TJX | 21.61 | 24.35 | 770 | 0.89 | | 40539 | 24.94 | 26.20 | 0.95 |
| 65541 | COO | 27.80 | 31.26 | 1452 | 0.89 | | 65541 | 37.96 | 35.58 | 1.07 |
| 24109 | AEP | 9.33 | 10.49 | 125 | 0.89 | | 24109 | 9.67 | 8.80 | 1.10 |
| 85726 | DENN | 23.52 | 26.40 | 715 | 0.89 | | 85726 | 22.00 | 25.06 | 0.88 |
| 86572 | DSGX | 24.78 | 27.78 | 785 | 0.89 | | 86572 | 29.86 | 27.12 | 1.10 |
| 18956 | TY | 10.93 | 12.24 | 139 | 0.89 | | 18956 | 12.20 | 9.73 | 1.25 |
| 75976 | NEOG | 24.86 | 27.79 | 585 | 0.89 | | 75976 | 29.68 | 24.57 | 1.21 |
| 90004 | VTV | 10.50 | 11.73 | 138 | 0.90 | | 90004 | 11.10 | 9.57 | 1.16 |
| 91828 | QQXT | 14.28 | 15.95 | 278 | 0.90 | | 91828 | 17.88 | 15.44 | 1.16 |
| 79382 | PFC | 20.26 | 22.63 | 534 | 0.90 | | 79382 | 17.08 | 21.40 | 0.80 |
| 32678 | HEI | 29.58 | 33.00 | 851 | 0.90 | | 32678 | 25.14 | 27.61 | 0.91 |
| 91137 | FDL | 9.66 | 10.76 | 107 | 0.90 | | 91137 | 8.27 | 7.82 | 1.06 |
| 77447 | ICUI | 24.26 | 27.00 | 593 | 0.90 | | 77447 | 16.06 | 22.32 | 0.72 |
| 11955 | WM | 12.16 | 13.52 | 169 | 0.90 | | 11955 | 14.30 | 11.20 | 1.28 |
| 79133 | ATR | 11.78 | 13.09 | 167 | 0.90 | | 79133 | 15.75 | 11.26 | 1.40 |
| 62092 | TMO | 20.28 | 22.50 | 557 | 0.90 | | 62092 | 25.85 | 23.13 | 1.12 |
| 86810 | TREX | 38.56 | 42.66 | 1343 | 0.90 | | 86810 | 40.41 | 35.66 | 1.13 |
| 38093 | AJG | 13.27 | 14.67 | 184 | 0.90 | | 38093 | 17.18 | 12.26 | 1.40 |
| 89748 | PWC | 12.31 | 13.62 | 171 | 0.90 | | 89748 | 14.22 | 11.29 | 1.26 |
| 41292 | HCSG | 14.89 | 16.43 | 278 | 0.91 | | 41292 | 17.26 | 15.55 | 1.11 |
| 88295 | IYJ | 13.87 | 15.29 | 198 | 0.91 | | 88295 | 15.63 | 12.56 | 1.24 |
| 91001 | SDY | 9.53 | 10.51 | 121 | 0.91 | | 91001 | 9.55 | 8.63 | 1.11 |
| 91424 | VOT | 14.09 | 15.54 | 253 | 0.91 | | 91424 | 16.59 | 14.51 | 1.14 |
| 91182 | QTEC | 18.86 | 20.77 | 476 | 0.91 | | 91182 | 26.09 | 21.45 | 1.22 |
| 17778 | BRK | 13.36 | 14.70 | 217 | 0.91 | | 17778 | 15.02 | 13.16 | 1.14 |
| 60871 | ADI | 15.62 | 17.17 | 351 | 0.91 | | 60871 | 20.90 | 17.79 | 1.17 |
| 67360 | UTL | 9.29 | 10.21 | 145 | 0.91 | | 67360 | 12.29 | 10.01 | 1.23 |
| 75510 | ADBE | 27.32 | 30.02 | 963 | 0.91 | | 75510 | 30.52 | 30.00 | 1.02 |
| 55511 | PNM | 15.61 | 17.13 | 308 | 0.91 | | 55511 | 11.97 | 15.63 | 0.77 |
| 87034 | LII | 23.24 | 25.49 | 578 | 0.91 | | 87034 | 25.68 | 23.78 | 1.08 |
| 24985 | AEE | 9.41 | 10.32 | 96 | 0.91 | | 24985 | 14.28 | 7.85 | 1.82 |
| 80498 | ABCB | 24.80 | 27.19 | 175 | 0.91 | | 80498 | 33.79 | 15.70 | 2.15 |
| 81659 | MDY | 13.54 | 14.80 | 212 | 0.91 | | 81659 | 15.82 | 13.06 | 1.21 |
| 83443 | BRK | 13.64 | 14.90 | 218 | 0.92 | | 83443 | 14.89 | 13.18 | 1.13 |
| 91993 | FEX | 13.35 | 14.56 | 206 | 0.92 | | 91993 | 15.54 | 12.80 | 1.21 |
| 88218 | IJH | 13.59 | 14.79 | 211 | 0.92 | | 88218 | 15.92 | 13.07 | 1.22 |
| 62308 | GL | 20.05 | 21.77 | 275 | 0.92 | | 62308 | 20.08 | 15.72 | 1.28 |
| 42200 | PKI | 19.49 | 21.12 | 465 | 0.92 | | 42200 | 24.68 | 21.30 | 1.16 |
| 87000 | STMP | 44.72 | 48.45 | 1483 | 0.92 | | 87000 | 49.32 | 39.32 | 1.25 |
| 90175 | XPO | 47.53 | 51.46 | 1140 | 0.92 | | 90175 | 49.55 | 37.89 | 1.31 |
| 48506 | MCO | 24.13 | 26.09 | 597 | 0.92 | | 48506 | 24.88 | 23.69 | 1.05 |
| 90765 | PEJ | 16.93 | 18.30 | 323 | 0.93 | | 90765 | 19.83 | 16.97 | 1.17 |
| 82830 | NICE | 16.67 | 18.02 | 382 | 0.93 | | 82830 | 13.19 | 17.70 | 0.74 |
| 90264 | JKK | 15.07 | 16.25 | 242 | 0.93 | | 90264 | 16.83 | 14.17 | 1.19 |
| 21742 | CASY | 17.61 | 18.97 | 463 | 0.93 | | 21742 | 16.41 | 19.92 | 0.82 |
| 32870 | HRL | 17.98 | 19.33 | 449 | 0.93 | | 32870 | 19.20 | 19.96 | 0.96 |
| 88610 | XNTK | 16.64 | 17.88 | 251 | 0.93 | | 88610 | 26.90 | 16.22 | 1.66 |
| 88873 | GPN | 22.01 | 23.61 | 529 | 0.93 | | 88873 | 28.15 | 23.06 | 1.22 |
| 14752 | TPL | 44.57 | 47.80 | 2245 | 0.93 | | 14752 | 37.78 | 40.89 | 0.92 |
| 91328 | DON | 13.26 | 14.21 | 182 | 0.93 | | 91328 | 12.77 | 11.59 | 1.10 |
| 91790 | ROM | 34.42 | 36.85 | 1367 | 0.93 | | 91790 | 47.47 | 37.05 | 1.28 |
| 71475 | AWR | 14.44 | 15.46 | 307 | 0.93 | | 71475 | 13.09 | 15.72 | 0.83 |
| 25129 | CBSH | 11.36 | 12.15 | 109 | 0.93 | | 25129 | 16.47 | 8.66 | 1.90 |
| 89792 | CVCO | 23.92 | 25.58 | 385 | 0.94 | | 89792 | 21.83 | 19.02 | 1.15 |
| 92402 | MSCI | 24.17 | 25.84 | 730 | 0.94 | | 92402 | 30.44 | 26.82 | 1.14 |
| 91325 | DHS | 10.18 | 10.87 | 112 | 0.94 | | 91325 | 9.06 | 8.15 | 1.11 |
| 82542 | CCF | 31.09 | 33.19 | 786 | 0.94 | | 82542 | 41.39 | 29.90 | 1.38 |
| 87525 | | 20.85 | 22.25 | 294 | 0.94 | | 87525 | 32.85 | 19.16 | 1.71 |
| 90601 | NDAQ | 17.34 | 18.49 | 230 | 0.94 | | 90601 | 20.11 | 14.23 | 1.41 |
| 86996 | SBAC | 24.07 | 25.66 | 892 | 0.94 | | 86996 | 36.98 | 29.96 | 1.23 |
| 51625 | SCI | 25.00 | 26.62 | 710 | 0.94 | | 51625 | 18.07 | 24.38 | 0.74 |
| 26463 | | 12.39 | 13.19 | 183 | 0.94 | | 26463 | 14.13 | 11.69 | 1.21 |
| 92818 | SPXL | 37.70 | 40.10 | 985 | 0.94 | | 92818 | 41.91 | 32.88 | 1.27 |
| 64390 | PGR | 17.19 | 18.25 | 307 | 0.94 | | 64390 | 17.72 | 16.17 | 1.10 |
| 61399 | LOW | 19.27 | 20.42 | 329 | 0.94 | | 61399 | 18.30 | 16.95 | 1.08 |
| 40272 | IFF | 17.16 | 18.17 | 352 | 0.94 | | 40272 | 17.92 | 17.59 | 1.02 |
| 68292 | SIGI | 16.87 | 17.86 | 166 | 0.94 | | 68292 | 20.65 | 12.12 | 1.70 |
| 77917 | ASGN | 40.55 | 42.90 | 861 | 0.95 | | 77917 | 32.11 | 28.99 | 1.11 |
| 91996 | FXG | 13.14 | 13.90 | 198 | 0.95 | | 91996 | 14.73 | 12.41 | 1.19 |
| 48725 | UNP | 21.94 | 23.20 | 478 | 0.95 | | 48725 | 23.48 | 21.78 | 1.08 |
| 91451 | RXI | 13.71 | 14.49 | 212 | 0.95 | | 91451 | 16.53 | 13.12 | 1.26 |
| 90001 | VO | 14.11 | 14.90 | 220 | 0.95 | | 90001 | 16.06 | 13.36 | 1.20 |
| 92550 | BIP | 18.10 | 19.05 | 362 | 0.95 | | 92550 | 22.09 | 18.60 | 1.19 |
| 83621 | ANSS | 21.03 | 22.09 | 413 | 0.95 | | 83621 | 22.87 | 19.60 | 1.17 |
| 22779 | MSI | 19.30 | 20.24 | 541 | 0.95 | | 22779 | 21.03 | 21.84 | 0.96 |
| 89059 | IWR | 13.84 | 14.51 | 211 | 0.95 | | 89059 | 15.85 | 13.04 | 1.22 |
| 91007 | MDYG | 15.11 | 15.82 | 260 | 0.96 | | 91007 | 19.23 | 15.04 | 1.28 |
| 88401 | IWO | 15.37 | 16.08 | 230 | 0.96 | | 88401 | 17.29 | 13.86 | 1.25 |
| 10107 | MSFT | 19.05 | 19.91 | 422 | 0.96 | | 10107 | 21.03 | 19.63 | 1.07 |
| 90621 | XMMO | 14.65 | 15.30 | 231 | 0.96 | | 90621 | 14.74 | 13.59 | 1.08 |
| 77649 | STE | 16.79 | 17.54 | 347 | 0.96 | | 77649 | 18.69 | 17.66 | 1.06 |
| 61567 | HXL | 25.35 | 26.48 | 676 | 0.96 | | 61567 | 28.22 | 25.44 | 1.11 |
| 88860 | ALGN | 47.50 | 49.58 | 2293 | 0.96 | | 88860 | 50.22 | 44.46 | 1.13 |
| 61621 | PAYX | 10.73 | 11.19 | 148 | 0.96 | | 61621 | 14.84 | 10.32 | 1.44 |
| 89006 | MDLZ | 10.70 | 11.15 | 130 | 0.96 | | 89006 | 14.29 | 9.47 | 1.51 |
| 48523 | LANC | 14.26 | 14.84 | 416 | 0.96 | | 48523 | 16.34 | 18.86 | 0.87 |
| 85067 | MMS | 24.06 | 25.02 | 642 | 0.96 | | 85067 | 20.13 | 23.74 | 0.85 |
| 64186 | CI | 28.20 | 29.26 | 1027 | 0.96 | | 64186 | 36.45 | 31.55 | 1.16 |
| 86458 | XLU | 7.11 | 7.37 | 82 | 0.96 | | 86458 | 9.46 | 6.56 | 1.44 |
| 90458 | NWE | 10.62 | 11.00 | 153 | 0.97 | | 90458 | 12.34 | 10.34 | 1.19 |
| 88406 | IJS | 14.33 | 14.83 | 171 | 0.97 | | 88406 | 16.73 | 11.64 | 1.44 |
| 24969 | AVA | 12.49 | 12.92 | 119 | 0.97 | | 24969 | 14.19 | 9.04 | 1.57 |
| 60206 | SNA | 20.24 | 20.94 | 269 | 0.97 | | 60206 | 23.36 | 16.06 | 1.45 |
| 76201 | XLNX | 16.10 | 16.63 | 378 | 0.97 | | 76201 | 13.26 | 17.61 | 0.75 |
| 11397 | WASH | 14.31 | 14.77 | 141 | 0.97 | | 11397 | 22.80 | 11.25 | 2.03 |
| 90620 | PWV | 11.05 | 11.40 | 126 | 0.97 | | 90620 | 12.90 | 9.20 | 1.40 |
| 89233 | IXJ | 11.60 | 11.95 | 150 | 0.97 | | 89233 | 11.85 | 10.18 | 1.16 |
| 39571 | GGG | 22.68 | 23.37 | 429 | 0.97 | | 39571 | 24.70 | 20.34 | 1.21 |
| 52038 | SYY | 12.14 | 12.51 | 173 | 0.97 | | 52038 | 10.52 | 10.99 | 0.96 |
| 88607 | SLYG | 16.35 | 16.84 | 276 | 0.97 | | 88607 | 18.55 | 15.45 | 1.20 |
| 89061 | SOXX | 20.34 | 20.95 | 453 | 0.97 | | 89061 | 26.79 | 21.23 | 1.26 |
| 81472 | ASML | 24.73 | 25.44 | 563 | 0.97 | | 81472 | 31.62 | 24.25 | 1.30 |
| 13901 | MO | 15.42 | 15.86 | 228 | 0.97 | | 13901 | 17.88 | 14.18 | 1.26 |
| 11896 | MXIM | 16.37 | 16.83 | 345 | 0.97 | | 11896 | 24.13 | 18.00 | 1.34 |
| 91327 | DTN | 11.43 | 11.75 | 126 | 0.97 | | 91327 | 11.68 | 9.07 | 1.29 |
| 20512 | CACI | 16.39 | 16.85 | 219 | 0.97 | | 20512 | 11.88 | 12.85 | 0.92 |
| 86456 | XLI | 14.39 | 14.78 | 175 | 0.97 | | 86456 | 16.06 | 11.72 | 1.37 |
| 85257 | AME | 20.97 | 21.53 | 404 | 0.97 | | 85257 | 23.82 | 19.68 | 1.21 |
| 78077 | KAI | 28.22 | 28.92 | 504 | 0.98 | | 78077 | 29.02 | 22.85 | 1.27 |
| 90771 | PBJ | 10.92 | 11.18 | 128 | 0.98 | | 90771 | 13.16 | 9.27 | 1.42 |
| 77660 | CACC | 39.19 | 40.06 | 2687 | 0.98 | | 77660 | 59.29 | 46.91 | 1.26 |
| 12062 | LH | 11.29 | 11.54 | 96 | 0.98 | | 12062 | 13.23 | 7.77 | 1.70 |
| 56573 | ITW | 17.05 | 17.42 | 261 | 0.98 | | 56573 | 22.58 | 15.89 | 1.42 |
| 90258 | JKF | 9.27 | 9.47 | 93 | 0.98 | | 90258 | 9.96 | 7.24 | 1.38 |
| 27991 | PNW | 11.70 | 11.95 | 165 | 0.98 | | 27991 | 10.25 | 10.67 | 0.96 |
| 84181 | CNI | 16.99 | 17.35 | 303 | 0.98 | | 84181 | 18.66 | 16.43 | 1.14 |
| 79210 | RGA | 16.81 | 17.13 | 227 | 0.98 | | 79210 | 19.44 | 14.01 | 1.39 |
| 82651 | WAT | 18.41 | 18.74 | 415 | 0.98 | | 82651 | 22.61 | 19.56 | 1.16 |
| 89999 | VBK | 16.14 | 16.37 | 258 | 0.99 | | 89999 | 17.58 | 14.78 | 1.19 |
| 90348 | VIS | 14.84 | 15.03 | 181 | 0.99 | | 90348 | 16.69 | 12.04 | 1.39 |
| 92855 | TECL | 49.20 | 49.79 | 2392 | 0.99 | | 92855 | 67.08 | 50.04 | 1.34 |
| 91297 | FBT | 24.75 | 25.00 | 531 | 0.99 | | 91297 | 27.11 | 23.22 | 1.17 |
| 86228 | RSG | 13.23 | 13.37 | 191 | 0.99 | | 86228 | 10.25 | 11.69 | 0.88 |
| 88402 | IUSV | 11.47 | 11.58 | 128 | 0.99 | | 88402 | 12.37 | 9.24 | 1.34 |
| 23473 | CINF | 14.01 | 14.13 | 166 | 0.99 | | 23473 | 15.71 | 11.28 | 1.39 |
| 90963 | AMSF | 16.04 | 16.17 | 176 | 0.99 | | 90963 | 21.11 | 12.38 | 1.71 |
| 50789 | MGEE | 12.34 | 12.44 | 173 | 0.99 | | 50789 | 13.26 | 11.22 | 1.18 |
| 88219 | IWD | 11.23 | 11.32 | 124 | 0.99 | | 88219 | 11.98 | 9.02 | 1.33 |
| 89110 | OMCL | 23.39 | 23.51 | 402 | 1.00 | | 89110 | 25.11 | 19.74 | 1.27 |
| 89286 | VXF | 15.14 | 15.18 | 216 | 1.00 | | 89286 | 16.80 | 13.32 | 1.26 |
| 85198 | OZK | 25.72 | 25.78 | 208 | 1.00 | | 85198 | 34.16 | 17.56 | 1.94 |
| 24248 | ALE | 11.07 | 11.09 | 136 | 1.00 | | 24248 | 10.85 | 9.46 | 1.15 |
| 88217 | IVE | 11.22 | 11.23 | 124 | 1.00 | | 88217 | 12.12 | 9.02 | 1.34 |
| 91204 | POR | 10.29 | 10.28 | 135 | 1.00 | | 91204 | 8.83 | 9.26 | 0.95 |
| 92002 | FXL | 19.24 | 19.21 | 389 | 1.00 | | 92002 | 23.20 | 19.13 | 1.21 |
| 84761 | TTWO | 37.58 | 37.53 | 1262 | 1.00 | | 84761 | 40.04 | 34.81 | 1.15 |
| 90884 | FDM | 15.19 | 15.17 | 173 | 1.00 | | 90884 | 18.58 | 11.96 | 1.55 |
| 92058 | FIW | 13.87 | 13.80 | 179 | 1.00 | | 92058 | 16.91 | 12.01 | 1.41 |
| 75470 | MIDD | 31.33 | 31.10 | 1030 | 1.01 | | 75470 | 36.49 | 31.99 | 1.14 |
| 75107 | ABMD | 61.10 | 60.66 | 1880 | 1.01 | | 75107 | 57.87 | 47.13 | 1.23 |
| 82581 | HSIC | 15.56 | 15.45 | 328 | 1.01 | | 82581 | 17.06 | 16.77 | 1.02 |
| 91006 | SLY | 15.55 | 15.42 | 218 | 1.01 | | 91006 | 17.13 | 13.41 | 1.28 |
| 91450 | KXI | 8.80 | 8.72 | 93 | 1.01 | | 91450 | 9.55 | 7.20 | 1.33 |
| 87184 | | 28.59 | 28.33 | 1229 | 1.01 | | 87184 | 43.79 | 35.19 | 1.24 |
| 88281 | CRL | 17.35 | 17.19 | 332 | 1.01 | | 88281 | 22.01 | 17.78 | 1.24 |
| 81655 | DRI | 16.26 | 16.10 | 298 | 1.01 | | 81655 | 12.82 | 15.47 | 0.83 |
| 21178 | LMT | 18.61 | 18.41 | 211 | 1.01 | | 21178 | 23.02 | 14.07 | 1.64 |
| 79166 | GABC | 16.81 | 16.63 | 266 | 1.01 | | 79166 | 22.99 | 15.88 | 1.45 |
| 90005 | VB | 15.34 | 15.15 | 211 | 1.01 | | 90005 | 16.51 | 13.10 | 1.26 |
| 81481 | DLTR | 24.79 | 24.46 | 550 | 1.01 | | 81481 | 27.25 | 23.25 | 1.17 |
| 90768 | PJP | 19.30 | 19.03 | 287 | 1.01 | | 90768 | 19.76 | 16.04 | 1.23 |
| 11348 | CNBKA | 18.59 | 18.31 | 330 | 1.02 | | 11348 | 16.91 | 16.85 | 1.00 |
| 65307 | RPM | 17.54 | 17.24 | 342 | 1.02 | | 65307 | 19.34 | 17.45 | 1.11 |
| 75912 | PTC | 27.94 | 27.44 | 555 | 1.02 | | 75912 | 22.96 | 22.80 | 1.01 |
| 61188 | SWX | 14.98 | 14.72 | 203 | 1.02 | | 61188 | 16.35 | 12.80 | 1.28 |
| 13856 | PEP | 8.63 | 8.47 | 102 | 1.02 | | 13856 | 8.13 | 7.55 | 1.08 |
| 90215 | CRM | 37.88 | 37.12 | 1612 | 1.02 | | 90215 | 45.26 | 39.35 | 1.15 |
| 91423 | VOE | 14.67 | 14.33 | 191 | 1.02 | | 91423 | 16.17 | 12.34 | 1.31 |
| 91310 | MVV | 31.47 | 30.71 | 631 | 1.02 | | 91310 | 33.62 | 26.37 | 1.27 |
| 88404 | IJJ | 14.02 | 13.68 | 175 | 1.03 | | 88404 | 16.10 | 11.74 | 1.37 |
| 91124 | XBI | 23.21 | 22.61 | 301 | 1.03 | | 91124 | 23.92 | 17.18 | 1.39 |
| 79678 | ATVI | 25.58 | 24.90 | 439 | 1.03 | | 79678 | 40.87 | 24.10 | 1.70 |
| 43449 | MCD | 13.67 | 13.28 | 186 | 1.03 | | 43449 | 17.15 | 12.22 | 1.40 |
| 78950 | ALG | 28.73 | 27.91 | 417 | 1.03 | | 78950 | 37.27 | 23.11 | 1.61 |
| 85299 | LKFN | 16.67 | 16.18 | 153 | 1.03 | | 85299 | 25.93 | 12.45 | 2.08 |
| 78981 | LSTR | 14.33 | 13.89 | 149 | 1.03 | | 78981 | 18.21 | 10.90 | 1.67 |
| 76605 | AZO | 21.72 | 21.04 | 501 | 1.03 | | 76605 | 21.84 | 21.27 | 1.03 |
| 24942 | | 19.40 | 18.79 | 200 | 1.03 | | 24942 | 21.53 | 13.42 | 1.60 |
| 91227 | IHE | 15.78 | 15.27 | 212 | 1.03 | | 91227 | 16.18 | 13.13 | 1.23 |
| 91538 | PTH | 19.92 | 19.24 | 263 | 1.04 | | 91538 | 19.46 | 15.19 | 1.28 |
| 87476 | BMRC | 14.02 | 13.53 | 244 | 1.04 | | 87476 | 13.23 | 13.82 | 0.96 |
| 75041 | FUN | 24.25 | 23.39 | 277 | 1.04 | | 75041 | 27.15 | 17.18 | 1.58 |
| 90776 | WAL | 32.38 | 31.20 | 291 | 1.04 | | 90776 | 57.40 | 27.98 | 2.05 |
| 80102 | SUI | 26.37 | 25.41 | 626 | 1.04 | | 80102 | 20.53 | 23.31 | 0.88 |
| 88222 | IWM | 14.70 | 14.16 | 172 | 1.04 | | 88222 | 16.21 | 11.60 | 1.40 |
| 82303 | RE | 14.31 | 13.79 | 186 | 1.04 | | 82303 | 14.60 | 11.88 | 1.23 |
| 43553 | VFC | 23.49 | 22.62 | 421 | 1.04 | | 43553 | 26.71 | 20.69 | 1.29 |
| 27334 | CSL | 19.81 | 19.06 | 386 | 1.04 | | 27334 | 21.98 | 18.85 | 1.17 |
| 89781 | MOH | 29.85 | 28.70 | 890 | 1.04 | | 89781 | 18.92 | 27.10 | 0.70 |
| 17929 | UGI | 14.87 | 14.27 | 228 | 1.04 | | 17929 | 18.06 | 13.90 | 1.30 |
| 78840 | IAC | 34.88 | 33.47 | 1064 | 1.04 | | 78840 | 29.91 | 30.86 | 0.97 |
| 85992 | MANH | 36.20 | 34.64 | 972 | 1.04 | | 85992 | 36.45 | 31.56 | 1.15 |
| 90272 | | 39.83 | 38.01 | 1736 | 1.05 | | 90272 | 59.87 | 43.04 | 1.39 |
| 92005 | FNX | 15.97 | 15.23 | 226 | 1.05 | | 92005 | 18.73 | 13.93 | 1.34 |
| 81736 | RMD | 22.82 | 21.73 | 508 | 1.05 | | 81736 | 25.16 | 22.36 | 1.13 |
| 86144 | DOX | 14.59 | 13.90 | 220 | 1.05 | | 86144 | 18.03 | 13.53 | 1.33 |
| 65306 | UNF | 21.42 | 20.39 | 382 | 1.05 | | 65306 | 24.72 | 19.34 | 1.28 |
| 81677 | WAB | 23.88 | 22.67 | 253 | 1.05 | | 81677 | 25.73 | 15.93 | 1.62 |
| 82694 | EME | 16.81 | 15.95 | 166 | 1.05 | | 82694 | 20.31 | 12.05 | 1.68 |
| 91053 | PRF | 14.55 | 13.80 | 198 | 1.05 | | 91053 | 15.77 | 12.53 | 1.26 |
| 11775 | JOUT | 36.65 | 34.73 | 954 | 1.06 | | 11775 | 34.16 | 30.94 | 1.10 |
| 26403 | DIS | 21.35 | 20.21 | 383 | 1.06 | | 26403 | 19.10 | 18.38 | 1.04 |
| 89057 | IWS | 14.37 | 13.58 | 168 | 1.06 | | 89057 | 15.29 | 11.37 | 1.34 |
| 24053 | EVRG | 13.98 | 13.20 | 177 | 1.06 | | 24053 | 12.24 | 11.31 | 1.08 |
| 85082 | EGP | 13.35 | 12.57 | 158 | 1.06 | | 85082 | 12.45 | 10.57 | 1.18 |
| 68419 | WTM | 17.10 | 16.08 | 221 | 1.06 | | 68419 | 11.27 | 12.87 | 0.88 |
| 46674 | GPC | 13.72 | 12.90 | 154 | 1.06 | | 46674 | 17.21 | 10.99 | 1.57 |
| 92432 | ENSG | 30.57 | 28.70 | 688 | 1.06 | | 92432 | 35.24 | 27.12 | 1.30 |
| 85177 | EGHT | 56.06 | 52.61 | 3674 | 1.07 | | 85177 | 68.29 | 54.27 | 1.26 |
| 86896 | CBU | 16.26 | 15.23 | 139 | 1.07 | | 86896 | 26.40 | 11.75 | 2.25 |
| 89297 | CCOI | 26.12 | 24.44 | 592 | 1.07 | | 89297 | 27.92 | 24.05 | 1.16 |
| 90455 | MPWR | 29.48 | 27.59 | 822 | 1.07 | | 90455 | 34.57 | 29.44 | 1.17 |
| 92004 | FYX | 16.67 | 15.57 | 211 | 1.07 | | 92004 | 18.71 | 13.42 | 1.39 |
| 81208 | CIGI | 33.47 | 31.25 | 600 | 1.07 | | 81208 | 31.58 | 25.26 | 1.25 |
| 85886 | CNXN | 26.02 | 24.27 | 481 | 1.07 | | 85886 | 36.05 | 22.99 | 1.57 |
| 85913 | MAR | 27.89 | 26.01 | 500 | 1.07 | | 85913 | 33.01 | 24.04 | 1.37 |
| 11707 | | 42.34 | 39.47 | 717 | 1.07 | | 11707 | 41.38 | 28.88 | 1.43 |
| 87657 | EW | 34.29 | 31.96 | 1015 | 1.07 | | 87657 | 38.45 | 32.47 | 1.18 |
| 75654 | CGNX | 42.62 | 39.66 | 945 | 1.07 | | 75654 | 50.02 | 35.16 | 1.42 |
| 21936 | PFE | 12.48 | 11.60 | 146 | 1.08 | | 21936 | 10.10 | 9.86 | 1.03 |
| 88893 | IBB | 21.84 | 20.30 | 307 | 1.08 | | 88893 | 23.84 | 17.26 | 1.38 |
| 90591 | EFSC | 23.67 | 21.98 | 147 | 1.08 | | 90591 | 35.43 | 15.26 | 2.32 |
| 10892 | WSFS | 24.23 | 22.49 | 137 | 1.08 | | 10892 | 44.11 | 16.64 | 2.65 |
| 92835 | AOR | 6.94 | 6.44 | 66 | 1.08 | | 92835 | 7.73 | 5.47 | 1.41 |
| 82179 | IART | 21.03 | 19.51 | 179 | 1.08 | | 82179 | 21.15 | 12.97 | 1.63 |
| 81282 | ACIW | 22.51 | 20.88 | 422 | 1.08 | | 81282 | 25.26 | 20.34 | 1.24 |
| 92729 | JBT | 31.85 | 29.53 | 779 | 1.08 | | 92729 | 43.96 | 31.35 | 1.40 |
| 79841 | FWRD | 14.35 | 13.30 | 126 | 1.08 | | 79841 | 11.84 | 9.11 | 1.30 |
| 83597 | FDS | 19.26 | 17.84 | 352 | 1.08 | | 83597 | 18.66 | 17.59 | 1.06 |
| 90000 | VBR | 15.05 | 13.94 | 169 | 1.08 | | 90000 | 16.23 | 11.49 | 1.41 |
| 14593 | AAPL | 36.15 | 33.47 | 1194 | 1.08 | | 14593 | 44.59 | 34.48 | 1.29 |
| 69649 | RJF | 21.87 | 20.22 | 334 | 1.08 | | 69649 | 19.36 | 17.36 | 1.12 |
| 82855 | | 17.23 | 15.90 | 177 | 1.08 | | 82855 | 14.89 | 11.64 | 1.28 |
| 90925 | FNF | 19.58 | 18.06 | 190 | 1.08 | | 90925 | 28.04 | 14.37 | 1.95 |
| 85259 | BAM | 18.04 | 16.64 | 295 | 1.08 | | 85259 | 24.13 | 17.04 | 1.42 |
| 80681 | ESS | 17.62 | 16.24 | 219 | 1.08 | | 80681 | 16.36 | 13.31 | 1.23 |
| 92121 | DFS | 29.74 | 27.38 | 519 | 1.09 | | 92121 | 28.30 | 23.24 | 1.22 |
| 82642 | EL | 32.91 | 30.30 | 740 | 1.09 | | 82642 | 29.02 | 26.69 | 1.09 |
| 85631 | ADSK | 28.47 | 26.19 | 555 | 1.09 | | 85631 | 20.89 | 22.45 | 0.93 |
| 14008 | AMGN | 17.45 | 16.02 | 237 | 1.09 | | 14008 | 17.33 | 14.10 | 1.23 |
| 25582 | LHX | 21.69 | 19.87 | 273 | 1.09 | | 25582 | 21.70 | 16.00 | 1.36 |
| 91403 | KALU | 17.07 | 15.63 | 296 | 1.09 | | 91403 | 29.67 | 17.79 | 1.67 |
| 88031 | MNST | 36.66 | 33.57 | 781 | 1.09 | | 88031 | 29.99 | 27.70 | 1.08 |
| 77702 | SBUX | 31.72 | 29.03 | 1262 | 1.09 | | 77702 | 43.15 | 34.81 | 1.24 |
| 64450 | NJR | 11.83 | 10.83 | 132 | 1.09 | | 64450 | 13.80 | 9.63 | 1.43 |
| 90261 | JKI | 15.64 | 14.30 | 183 | 1.09 | | 90261 | 16.43 | 12.05 | 1.36 |
| 89179 | ANTM | 25.69 | 23.44 | 523 | 1.10 | | 89179 | 22.54 | 21.93 | 1.03 |
| 89880 | IYT | 16.41 | 14.94 | 161 | 1.10 | | 89880 | 18.44 | 11.50 | 1.60 |
| 91818 | ESGR | 16.99 | 15.43 | 183 | 1.10 | | 91818 | 14.23 | 11.85 | 1.20 |
| 41452 | HELE | 27.86 | 25.29 | 656 | 1.10 | | 41452 | 19.55 | 23.88 | 0.82 |
| 90263 | JKJ | 15.87 | 14.39 | 190 | 1.10 | | 90263 | 18.27 | 12.60 | 1.45 |
| 90569 | WEX | 31.86 | 28.87 | 1012 | 1.10 | | 90569 | 45.93 | 32.67 | 1.41 |
| 92003 | FXU | 9.43 | 8.52 | 107 | 1.11 | | 92003 | 10.59 | 8.00 | 1.32 |
| 63467 | BRO | 14.95 | 13.51 | 164 | 1.11 | | 63467 | 17.06 | 11.35 | 1.50 |
| 81073 | UHAL | 35.72 | 32.26 | 850 | 1.11 | | 81073 | 38.22 | 30.07 | 1.27 |
| 22509 | PPG | 23.18 | 20.93 | 391 | 1.11 | | 22509 | 26.76 | 19.99 | 1.34 |
| 76602 | PSB | 13.98 | 12.63 | 193 | 1.11 | | 76602 | 9.48 | 11.71 | 0.81 |
| 92322 | ULTA | 57.70 | 52.08 | 2857 | 1.11 | | 92322 | 42.67 | 45.92 | 0.93 |
| 91995 | FXD | 18.66 | 16.80 | 282 | 1.11 | | 91995 | 21.68 | 16.10 | 1.35 |
| 89425 | WTBA | 19.19 | 17.27 | 56 | 1.11 | | 89425 | 33.65 | 10.69 | 3.15 |
| 50876 | LLY | 13.95 | 12.55 | 187 | 1.11 | | 50876 | 17.59 | 12.40 | 1.42 |
| 92574 | | 16.56 | 14.89 | 220 | 1.11 | | 92574 | 20.43 | 13.98 | 1.46 |
| 76582 | DORM | 41.14 | 37.00 | 1264 | 1.11 | | 76582 | 48.52 | 37.12 | 1.31 |
| 19561 | BA | 32.83 | 29.49 | 656 | 1.11 | | 19561 | 32.65 | 25.66 | 1.27 |
| 92837 | AOK | 3.42 | 3.07 | 26 | 1.11 | | 92837 | 3.89 | 2.43 | 1.60 |
| 76795 | ZBRA | 31.19 | 28.00 | 686 | 1.11 | | 76795 | 21.14 | 24.62 | 0.86 |
| 22592 | MMM | 17.46 | 15.67 | 231 | 1.11 | | 22592 | 22.73 | 14.76 | 1.54 |
| 22111 | JNJ | 11.37 | 10.20 | 116 | 1.11 | | 22111 | 11.69 | 8.55 | 1.37 |
| 91787 | USD | 44.13 | 39.57 | 1151 | 1.12 | | 91787 | 54.28 | 38.33 | 1.42 |
| 88362 | NTES | 32.81 | 29.37 | 965 | 1.12 | | 88362 | 40.33 | 32.62 | 1.24 |
| 76868 | AAON | 28.38 | 25.37 | 467 | 1.12 | | 76868 | 41.27 | 23.63 | 1.75 |
| 46886 | KLAC | 20.60 | 18.40 | 311 | 1.12 | | 46886 | 23.35 | 17.15 | 1.36 |
| 91998 | FXO | 16.62 | 14.83 | 187 | 1.12 | | 91998 | 16.79 | 12.27 | 1.37 |
| 87541 | PKG | 30.10 | 26.84 | 520 | 1.12 | | 87541 | 34.61 | 24.78 | 1.40 |
| 77918 | LFUS | 38.01 | 33.85 | 933 | 1.12 | | 77918 | 32.47 | 29.91 | 1.09 |
| 54253 | MSA | 20.39 | 18.16 | 294 | 1.12 | | 54253 | 19.88 | 16.20 | 1.23 |
| 11547 | CNMD | 20.54 | 18.26 | 168 | 1.13 | | 11547 | 17.35 | 11.46 | 1.51 |
| 92836 | AOM | 4.83 | 4.29 | 40 | 1.13 | | 92836 | 5.38 | 3.56 | 1.51 |
| 89575 | SAFT | 13.06 | 11.58 | 115 | 1.13 | | 89575 | 16.56 | 9.11 | 1.82 |
| 92834 | AOA | 10.20 | 9.02 | 99 | 1.13 | | 92834 | 10.94 | 7.64 | 1.43 |
| 91503 | LMAT | 40.23 | 35.53 | 924 | 1.13 | | 91503 | 51.70 | 34.38 | 1.50 |
| 90090 | SBNY | 22.40 | 19.78 | 258 | 1.13 | | 90090 | 24.88 | 16.07 | 1.55 |
| 10318 | BCPC | 26.37 | 23.25 | 372 | 1.13 | | 10318 | 25.96 | 19.28 | 1.35 |
| 90276 | BLKB | 27.60 | 24.29 | 366 | 1.14 | | 90276 | 36.38 | 21.75 | 1.67 |
| 85863 | COLM | 25.26 | 22.21 | 375 | 1.14 | | 85863 | 21.24 | 18.67 | 1.14 |
| 28804 | ALK | 41.98 | 36.88 | 732 | 1.14 | | 28804 | 31.69 | 27.31 | 1.16 |
| 11786 | SIVB | 37.63 | 33.05 | 624 | 1.14 | | 11786 | 32.41 | 25.57 | 1.27 |
| 89866 | LKQ | 21.75 | 19.09 | 307 | 1.14 | | 89866 | 33.03 | 19.79 | 1.67 |
| 85645 | INGR | 22.21 | 19.50 | 217 | 1.14 | | 85645 | 23.35 | 14.63 | 1.60 |
| 92000 | FXR | 18.65 | 16.37 | 225 | 1.14 | | 92000 | 21.82 | 14.45 | 1.51 |
| 91808 | EES | 18.78 | 16.48 | 243 | 1.14 | | 91808 | 21.06 | 14.82 | 1.42 |
| 91500 | EXLS | 26.23 | 22.98 | 514 | 1.14 | | 91500 | 35.78 | 23.65 | 1.51 |
| 90634 | AX | 57.64 | 50.48 | 2020 | 1.14 | | 90634 | 61.75 | 45.55 | 1.36 |
| 92007 | FTA | 15.89 | 13.91 | 182 | 1.14 | | 92007 | 17.35 | 12.19 | 1.42 |
| 90834 | IWC | 16.34 | 14.24 | 159 | 1.15 | | 90834 | 18.47 | 11.36 | 1.63 |
| 86783 | BKNG | 50.49 | 43.96 | 2239 | 1.15 | | 86783 | 61.59 | 45.91 | 1.34 |
| 27562 | CBRL | 27.53 | 23.92 | 676 | 1.15 | | 27562 | 33.13 | 26.49 | 1.25 |
| 92775 | TREE | 78.77 | 68.37 | 8345 | 1.15 | | 92775 | 112.11 | 86.49 | 1.30 |
| 80286 | TSCO | 34.74 | 30.13 | 824 | 1.15 | | 80286 | 32.80 | 28.48 | 1.15 |
| 36346 | SRCE | 17.33 | 15.02 | 88 | 1.15 | | 36346 | 26.01 | 9.42 | 2.76 |
| 90952 | ISBC | 17.72 | 15.34 | 97 | 1.16 | | 90952 | 20.90 | 8.92 | 2.34 |
| 81678 | LECO | 22.28 | 19.28 | 210 | 1.16 | | 81678 | 23.81 | 14.32 | 1.66 |
| 61516 | BIO | 19.48 | 16.85 | 208 | 1.16 | | 61516 | 14.39 | 12.75 | 1.13 |
| 91722 | UWM | 33.75 | 29.17 | 424 | 1.16 | | 91722 | 35.41 | 22.73 | 1.56 |
| 90622 | XSVM | 15.51 | 13.39 | 145 | 1.16 | | 90622 | 18.39 | 10.72 | 1.72 |
| 92576 | | 19.13 | 16.50 | 223 | 1.16 | | 92576 | 21.81 | 14.34 | 1.52 |
| 11308 | KO | 8.99 | 7.73 | 109 | 1.16 | | 11308 | 8.61 | 7.96 | 1.08 |
| 64653 | PSA | 16.54 | 14.21 | 155 | 1.16 | | 64653 | 16.29 | 10.86 | 1.50 |
| 27043 | VAR | 19.68 | 16.87 | 269 | 1.17 | | 27043 | 18.96 | 15.27 | 1.24 |
| 29612 | AIT | 19.45 | 16.63 | 185 | 1.17 | | 29612 | 22.44 | 13.06 | 1.72 |
| 76614 | REGN | 59.26 | 50.64 | 1934 | 1.17 | | 76614 | 64.90 | 45.66 | 1.42 |
| 89876 | CRI | 24.55 | 20.96 | 324 | 1.17 | | 89876 | 22.70 | 17.85 | 1.27 |
| 91340 | DFJ | 10.43 | 8.90 | 66 | 1.17 | | 91340 | 14.43 | 6.14 | 2.35 |
| 85663 | URI | 60.62 | 51.58 | 1024 | 1.18 | | 85663 | 49.99 | 36.46 | 1.37 |
| 78527 | | 38.16 | 32.44 | 1306 | 1.18 | | 78527 | 44.59 | 36.60 | 1.22 |
| 92221 | MELI | 47.50 | 40.34 | 1685 | 1.18 | | 92221 | 68.42 | 43.91 | 1.56 |
| 91519 | DEI | 19.80 | 16.79 | 161 | 1.18 | | 91519 | 12.36 | 10.76 | 1.15 |
| 87447 | UPS | 11.35 | 9.61 | 77 | 1.18 | | 87447 | 18.06 | 7.19 | 2.51 |
| 20482 | ABT | 15.60 | 13.20 | 184 | 1.18 | | 20482 | 19.02 | 12.45 | 1.53 |
| 64929 | KWR | 42.95 | 36.35 | 980 | 1.18 | | 64929 | 34.27 | 30.70 | 1.12 |
| 83639 | CHKP | 23.02 | 19.44 | 441 | 1.18 | | 83639 | 25.39 | 20.55 | 1.24 |
| 86339 | CCI | 23.43 | 19.77 | 518 | 1.19 | | 86339 | 39.21 | 24.26 | 1.62 |
| 91330 | DES | 14.95 | 12.61 | 122 | 1.19 | | 91330 | 16.05 | 9.37 | 1.71 |
| 81046 | RELX | 15.80 | 13.31 | 175 | 1.19 | | 81046 | 17.82 | 11.89 | 1.50 |
| 86356 | EBAY | 25.63 | 21.54 | 386 | 1.19 | | 86356 | 28.95 | 20.17 | 1.44 |
| 79323 | ALL | 20.64 | 17.32 | 152 | 1.19 | | 79323 | 25.23 | 12.29 | 2.05 |
| 91223 | IAK | 16.22 | 13.61 | 142 | 1.19 | | 91223 | 15.77 | 10.24 | 1.54 |
| 44506 | | 31.78 | 26.65 | 763 | 1.19 | | 44506 | 33.31 | 27.89 | 1.19 |
| 77637 | FC | 25.69 | 21.55 | 269 | 1.19 | | 77637 | 23.82 | 16.03 | 1.49 |
| 59774 | LDL | 40.57 | 34.01 | 253 | 1.19 | | 59774 | 45.07 | 22.78 | 1.98 |
| 59619 | SXT | 16.01 | 13.40 | 134 | 1.19 | | 59619 | 20.47 | 10.62 | 1.93 |
| 83534 | NBIX | 71.31 | 59.69 | 2132 | 1.19 | | 83534 | 80.50 | 54.34 | 1.48 |
| 76841 | BIIB | 33.94 | 28.36 | 586 | 1.20 | | 76841 | 32.19 | 24.64 | 1.31 |
| 12781 | SR | 10.34 | 8.64 | 58 | 1.20 | | 12781 | 13.95 | 5.65 | 2.47 |
| 79785 | NVR | 27.16 | 22.68 | 434 | 1.20 | | 79785 | 38.76 | 23.23 | 1.67 |
| 91315 | KRE | 15.88 | 13.26 | 60 | 1.20 | | 91315 | 21.83 | 6.84 | 3.19 |
| 91618 | ORBC | 26.98 | 22.50 | 282 | 1.20 | | 91618 | 21.77 | 16.15 | 1.35 |
| 76529 | NRIM | 16.40 | 13.66 | 219 | 1.20 | | 76529 | 21.08 | 13.87 | 1.52 |
| 39693 | B | 24.60 | 20.49 | 270 | 1.20 | | 39693 | 25.68 | 16.36 | 1.57 |
| 48486 | LRCX | 32.52 | 27.05 | 540 | 1.20 | | 48486 | 39.37 | 26.31 | 1.50 |
| 80784 | BTO | 16.87 | 14.03 | 103 | 1.20 | | 80784 | 21.37 | 9.45 | 2.26 |
| 88228 | IYF | 15.54 | 12.92 | 135 | 1.20 | | 88228 | 14.82 | 9.87 | 1.50 |
| 16600 | HSY | 15.46 | 12.84 | 209 | 1.20 | | 16600 | 16.31 | 13.01 | 1.25 |
| 78987 | MCHP | 20.91 | 17.36 | 268 | 1.20 | | 78987 | 23.14 | 16.05 | 1.44 |
| 88949 | SGEN | 30.96 | 25.69 | 534 | 1.21 | | 88949 | 26.51 | 22.85 | 1.16 |
| 84319 | LAD | 55.64 | 46.09 | 2241 | 1.21 | | 84319 | 53.06 | 45.89 | 1.16 |
| 85269 | VLO | 29.66 | 24.56 | 278 | 1.21 | | 85269 | 32.74 | 18.35 | 1.78 |
| 81675 | LSI | 19.74 | 16.33 | 158 | 1.21 | | 81675 | 18.86 | 11.39 | 1.66 |
| 43350 | SWK | 21.43 | 17.72 | 251 | 1.21 | | 43350 | 23.39 | 15.66 | 1.49 |
| 89393 | NFLX | 102.91 | 85.07 | 6168 | 1.21 | | 89393 | 108.82 | 80.38 | 1.35 |
| 85592 | PLD | 19.14 | 15.81 | 151 | 1.21 | | 85592 | 14.15 | 10.47 | 1.35 |
| 91483 | PRFZ | 20.18 | 16.67 | 253 | 1.21 | | 91483 | 22.40 | 15.34 | 1.46 |
| 57809 | NDSN | 32.34 | 26.70 | 639 | 1.21 | | 57809 | 38.66 | 27.43 | 1.41 |
| 47002 | KAMN | 20.15 | 16.60 | 209 | 1.21 | | 47002 | 14.59 | 12.81 | 1.14 |
| 88399 | IWN | 14.88 | 12.25 | 119 | 1.21 | | 88399 | 16.55 | 9.29 | 1.78 |
| 86136 | SRE | 12.86 | 10.59 | 154 | 1.21 | | 86136 | 16.07 | 10.83 | 1.48 |
| 52898 | WTRG | 13.51 | 11.11 | 108 | 1.22 | | 52898 | 15.60 | 8.61 | 1.81 |
| 87114 | MBWM | 37.77 | 31.05 | 557 | 1.22 | | 87114 | 57.00 | 29.83 | 1.91 |
| 10693 | EHC | 28.52 | 23.42 | 463 | 1.22 | | 10693 | 26.36 | 21.33 | 1.24 |
| 86778 | MKSI | 30.80 | 25.28 | 337 | 1.22 | | 86778 | 29.69 | 19.39 | 1.53 |
| 77730 | TSN | 30.81 | 25.29 | 510 | 1.22 | | 77730 | 28.66 | 23.25 | 1.23 |
| 90770 | PBS | 20.45 | 16.78 | 294 | 1.22 | | 90770 | 25.01 | 16.85 | 1.48 |
| 65294 | RLI | 15.49 | 12.71 | 126 | 1.22 | | 65294 | 21.44 | 10.22 | 2.10 |
| 90664 | DXCM | 60.87 | 49.85 | 4241 | 1.22 | | 90664 | 74.91 | 61.83 | 1.21 |
| 91989 | FAB | 17.44 | 14.26 | 194 | 1.22 | | 91989 | 19.98 | 13.01 | 1.54 |
| 91316 | XRT | 19.92 | 16.27 | 303 | 1.22 | | 91316 | 26.40 | 17.35 | 1.52 |
| 92050 | PODD | 45.72 | 37.21 | 927 | 1.23 | | 92050 | 36.76 | 30.65 | 1.20 |
| 82486 | ERIE | 19.80 | 16.10 | 254 | 1.23 | | 82486 | 21.03 | 14.99 | 1.40 |
| 43481 | SXI | 32.53 | 26.43 | 239 | 1.23 | | 43481 | 24.12 | 15.51 | 1.55 |
| 85187 | HSKA | 71.92 | 58.43 | 3343 | 1.23 | | 85187 | 49.47 | 49.65 | 1.00 |
| 89125 | CP | 28.06 | 22.78 | 428 | 1.23 | | 89125 | 28.31 | 21.48 | 1.32 |
| 89626 | CME | 19.05 | 15.47 | 352 | 1.23 | | 89626 | 26.38 | 18.97 | 1.39 |
| 85390 | PLCE | 26.70 | 21.67 | 316 | 1.23 | | 85390 | 37.08 | 20.74 | 1.79 |
| 89994 | VFH | 16.39 | 13.29 | 129 | 1.23 | | 89994 | 15.99 | 9.78 | 1.64 |
| 89056 | HDB | 30.57 | 24.76 | 626 | 1.23 | | 89056 | 35.75 | 26.55 | 1.35 |
| 84392 | KRC | 17.35 | 14.01 | 88 | 1.24 | | 84392 | 16.60 | 7.65 | 2.17 |
| 84828 | RYAAY | 21.78 | 17.57 | 139 | 1.24 | | 84828 | 24.82 | 11.75 | 2.11 |
| 60628 | FDX | 22.39 | 18.04 | 151 | 1.24 | | 60628 | 26.83 | 12.81 | 2.09 |
| 91004 | KIE | 19.59 | 15.79 | 208 | 1.24 | | 91004 | 16.78 | 13.05 | 1.29 |
| 28222 | APD | 16.07 | 12.94 | 250 | 1.24 | | 28222 | 22.00 | 15.12 | 1.45 |
| 77120 | HMN | 24.08 | 19.37 | 308 | 1.24 | | 77120 | 28.03 | 18.21 | 1.54 |
| 85464 | CWST | 46.46 | 37.36 | 598 | 1.24 | | 85464 | 49.00 | 30.05 | 1.63 |
| 88608 | SLYV | 16.76 | 13.47 | 154 | 1.24 | | 88608 | 18.91 | 11.23 | 1.68 |
| 84010 | USNA | 38.21 | 30.70 | 588 | 1.24 | | 84010 | 44.04 | 27.33 | 1.61 |
| 76839 | FIZZ | 42.09 | 33.82 | 697 | 1.24 | | 76839 | 42.14 | 29.00 | 1.45 |
| 87532 | RDI | 23.57 | 18.93 | 269 | 1.24 | | 87532 | 28.31 | 16.80 | 1.69 |
| 82292 | THG | 18.45 | 14.82 | 172 | 1.24 | | 82292 | 19.44 | 12.05 | 1.61 |
| 12008 | DY | 44.44 | 35.58 | 557 | 1.25 | | 12008 | 43.69 | 28.59 | 1.53 |
| 64311 | NSC | 22.97 | 18.39 | 218 | 1.25 | | 64311 | 21.80 | 14.25 | 1.53 |
| 91545 | PYZ | 18.32 | 14.66 | 172 | 1.25 | | 91545 | 21.41 | 12.50 | 1.71 |
| 89382 | SAP | 15.77 | 12.61 | 175 | 1.25 | | 89382 | 20.74 | 12.36 | 1.68 |
| 87075 | JCOM | 21.70 | 17.32 | 246 | 1.25 | | 87075 | 24.30 | 15.35 | 1.58 |
| 10777 | FCNCA | 18.20 | 14.53 | 147 | 1.25 | | 10777 | 16.36 | 10.50 | 1.56 |
| 10606 | WTS | 18.93 | 15.11 | 158 | 1.25 | | 10606 | 21.28 | 11.88 | 1.79 |
| 64936 | D | 11.60 | 9.26 | 99 | 1.25 | | 64936 | 13.36 | 7.91 | 1.69 |
| 91009 | SPMD | 16.07 | 12.82 | 161 | 1.25 | | 91009 | 16.86 | 11.20 | 1.51 |
| 90038 | AIZ | 21.58 | 17.21 | 198 | 1.25 | | 90038 | 30.22 | 14.47 | 2.09 |
| 89617 | EQIX | 31.45 | 25.02 | 563 | 1.26 | | 89617 | 44.55 | 27.42 | 1.62 |
| 79637 | UHS | 28.30 | 22.52 | 520 | 1.26 | | 79637 | 28.57 | 22.94 | 1.25 |
| 88593 | ASR | 26.48 | 21.02 | 303 | 1.26 | | 88593 | 33.60 | 18.39 | 1.83 |
| 90935 | PKB | 16.14 | 12.80 | 102 | 1.26 | | 90935 | 21.48 | 9.29 | 2.31 |
| 89765 | FVL | 14.57 | 11.56 | 104 | 1.26 | | 89765 | 16.93 | 8.56 | 1.98 |
| 90312 | WLK | 45.74 | 36.25 | 712 | 1.26 | | 90312 | 47.72 | 31.54 | 1.51 |
| 88901 | MMP | 22.61 | 17.85 | 278 | 1.27 | | 88901 | 23.92 | 16.60 | 1.44 |
| 86048 | HCKT | 34.89 | 27.53 | 448 | 1.27 | | 86048 | 28.45 | 21.27 | 1.34 |
| 92157 | TEL | 28.74 | 22.67 | 367 | 1.27 | | 92157 | 25.61 | 19.25 | 1.33 |
| 90441 | | 29.95 | 23.61 | 496 | 1.27 | | 90441 | 25.93 | 22.33 | 1.16 |
| 44601 | AVY | 22.62 | 17.81 | 174 | 1.27 | | 44601 | 27.88 | 13.87 | 2.01 |
| 89653 | PFS | 13.80 | 10.86 | 58 | 1.27 | | 89653 | 24.18 | 7.14 | 3.39 |
| 60097 | MDT | 15.24 | 11.98 | 189 | 1.27 | | 60097 | 18.33 | 12.55 | 1.46 |
| 81540 | ABC | 27.13 | 21.31 | 317 | 1.27 | | 81540 | 26.57 | 18.21 | 1.46 |
| 10395 | | 14.66 | 11.52 | 153 | 1.27 | | 10395 | 20.13 | 11.39 | 1.77 |
| 12052 | GD | 21.37 | 16.77 | 173 | 1.27 | | 12052 | 20.48 | 12.29 | 1.67 |
| 83738 | CASS | 17.02 | 13.35 | 177 | 1.28 | | 83738 | 25.30 | 13.21 | 1.92 |
| 76504 | CATY | 23.62 | 18.51 | 41 | 1.28 | | 76504 | 48.67 | 14.00 | 3.48 |
| 27959 | DUK | 9.00 | 7.05 | 92 | 1.28 | | 27959 | 11.24 | 7.27 | 1.55 |
| 91594 | SPR | 35.03 | 27.32 | 609 | 1.28 | | 91594 | 42.29 | 27.37 | 1.55 |
| 80034 | SHOO | 33.41 | 26.02 | 619 | 1.28 | | 80034 | 30.96 | 24.96 | 1.24 |
| 68196 | SEIC | 24.23 | 18.85 | 194 | 1.29 | | 68196 | 29.68 | 15.38 | 1.93 |
| 32942 | HUBB | 20.61 | 16.02 | 204 | 1.29 | | 32942 | 20.50 | 13.59 | 1.51 |
| 66157 | USB | 16.14 | 12.53 | 83 | 1.29 | | 66157 | 14.23 | 7.09 | 2.01 |
| 11442 | CHCO | 14.41 | 11.18 | 94 | 1.29 | | 11442 | 18.25 | 8.10 | 2.25 |
| 82833 | OTEX | 24.72 | 19.17 | 333 | 1.29 | | 82833 | 23.06 | 17.78 | 1.30 |
| 76592 | DEO | 15.10 | 11.70 | 150 | 1.29 | | 76592 | 17.82 | 10.86 | 1.64 |
| 11203 | ARGO | 18.38 | 14.22 | 203 | 1.29 | | 11203 | 22.26 | 13.81 | 1.61 |
| 15456 | FL | 30.83 | 23.83 | 625 | 1.29 | | 15456 | 28.88 | 25.31 | 1.14 |
| 11404 | ED | 10.92 | 8.44 | 96 | 1.29 | | 11404 | 12.68 | 7.67 | 1.65 |
| 12431 | TT | 30.35 | 23.44 | 562 | 1.29 | | 12431 | 41.08 | 26.85 | 1.53 |
| 91449 | EXI | 14.87 | 11.48 | 116 | 1.30 | | 91449 | 16.05 | 9.13 | 1.76 |
| 89704 | IHG | 25.23 | 19.45 | 348 | 1.30 | | 89704 | 28.04 | 19.05 | 1.47 |
| 91112 | PAC | 24.69 | 19.03 | 254 | 1.30 | | 91112 | 28.91 | 16.71 | 1.73 |
| 75341 | DRE | 17.69 | 13.57 | 136 | 1.30 | | 75341 | 12.62 | 9.60 | 1.31 |
| 90773 | PBE | 23.15 | 17.74 | 243 | 1.30 | | 90773 | 24.60 | 15.48 | 1.59 |
| 90265 | JKL | 17.84 | 13.66 | 151 | 1.31 | | 90265 | 18.68 | 11.12 | 1.68 |
| 91678 | TAST | 43.58 | 33.34 | 1208 | 1.31 | | 91678 | 55.02 | 37.87 | 1.45 |
| 60442 | PNC | 21.56 | 16.49 | 139 | 1.31 | | 60442 | 15.40 | 10.11 | 1.52 |
| 75605 | FRME | 28.92 | 22.11 | 54 | 1.31 | | 75605 | 43.30 | 15.30 | 2.83 |
| 77496 | MHK | 29.54 | 22.54 | 172 | 1.31 | | 77496 | 33.23 | 16.34 | 2.03 |
| 10200 | RGEN | 48.94 | 37.35 | 1295 | 1.31 | | 10200 | 40.78 | 35.51 | 1.15 |
| 76856 | FCFS | 29.38 | 22.42 | 280 | 1.31 | | 76856 | 24.17 | 16.89 | 1.43 |
| 63263 | NVO | 27.53 | 21.00 | 348 | 1.31 | | 63263 | 32.96 | 20.68 | 1.59 |
| 29209 | ARW | 20.94 | 15.96 | 266 | 1.31 | | 29209 | 22.11 | 15.67 | 1.41 |
| 53859 | MSEX | 17.75 | 13.53 | 210 | 1.31 | | 53859 | 20.13 | 13.36 | 1.51 |
| 62148 | CSX | 32.91 | 25.06 | 474 | 1.31 | | 62148 | 27.29 | 22.19 | 1.23 |
| 87432 | A | 27.53 | 20.95 | 496 | 1.31 | | 87432 | 33.24 | 23.10 | 1.44 |
| 41355 | PH | 24.44 | 18.58 | 251 | 1.31 | | 41355 | 31.96 | 17.51 | 1.83 |
| 53110 | BMI | 18.26 | 13.88 | 239 | 1.32 | | 53110 | 25.28 | 15.69 | 1.61 |
| 20765 | AJRD | 45.90 | 34.84 | 857 | 1.32 | | 20765 | 45.86 | 32.44 | 1.41 |
| 84403 | WWD | 23.38 | 17.73 | 223 | 1.32 | | 84403 | 17.11 | 13.52 | 1.27 |
| 17750 | KMB | 12.30 | 9.31 | 125 | 1.32 | | 17750 | 11.49 | 9.03 | 1.27 |
| 84606 | EEFT | 32.80 | 24.81 | 782 | 1.32 | | 84606 | 38.45 | 28.94 | 1.33 |
| 91222 | IAT | 16.41 | 12.41 | 66 | 1.32 | | 91222 | 18.65 | 6.71 | 2.78 |
| 61743 | DLX | 28.24 | 21.32 | 157 | 1.32 | | 61743 | 33.98 | 15.24 | 2.23 |
| 28302 | UL | 13.66 | 10.31 | 127 | 1.32 | | 28302 | 16.14 | 9.54 | 1.69 |
| 52978 | HAS | 22.55 | 17.02 | 179 | 1.32 | | 52978 | 25.15 | 13.46 | 1.87 |
| 75694 | TECH | 16.33 | 12.31 | 124 | 1.33 | | 75694 | 13.93 | 9.16 | 1.52 |
| 91795 | UYG | 34.43 | 25.95 | 225 | 1.33 | | 91795 | 34.01 | 17.15 | 1.98 |
| 78693 | UFCS | 17.06 | 12.82 | 78 | 1.33 | | 78693 | 23.03 | 8.63 | 2.67 |
| 89017 | WLTW | 13.51 | 10.16 | 130 | 1.33 | | 89017 | 15.54 | 9.68 | 1.60 |
| 87056 | BMRN | 35.05 | 26.28 | 378 | 1.33 | | 87056 | 22.03 | 18.92 | 1.16 |
| 37568 | FELE | 21.91 | 16.39 | 205 | 1.34 | | 37568 | 26.26 | 14.70 | 1.79 |
| 86021 | | 18.09 | 13.52 | 145 | 1.34 | | 86021 | 19.95 | 11.06 | 1.80 |
| 86580 | NVDA | 71.67 | 53.56 | 1554 | 1.34 | | 86580 | 79.95 | 48.67 | 1.64 |
| 80683 | FR | 38.53 | 28.78 | 282 | 1.34 | | 80683 | 26.14 | 17.12 | 1.53 |
| 53065 | IPG | 27.85 | 20.79 | 421 | 1.34 | | 53065 | 32.26 | 21.42 | 1.51 |
| 51369 | TER | 34.58 | 25.82 | 644 | 1.34 | | 51369 | 49.97 | 28.99 | 1.72 |
| 77526 | RNST | 18.99 | 14.17 | 77 | 1.34 | | 77526 | 27.63 | 8.91 | 3.10 |
| 89002 | ADS | 30.13 | 22.47 | 223 | 1.34 | | 89002 | 35.11 | 17.43 | 2.01 |
| 25953 | DOV | 24.19 | 18.03 | 225 | 1.34 | | 25953 | 22.95 | 14.62 | 1.57 |
| 91514 | | 25.89 | 19.29 | 180 | 1.34 | | 91514 | 31.59 | 14.46 | 2.18 |
| 86455 | XLF | 18.26 | 13.60 | 134 | 1.34 | | 86455 | 17.16 | 10.18 | 1.69 |
| 47896 | JPM | 20.87 | 15.50 | 210 | 1.35 | | 47896 | 19.59 | 13.60 | 1.44 |
| 18091 | CW | 25.82 | 19.15 | 206 | 1.35 | | 18091 | 31.35 | 15.04 | 2.08 |
| 85517 | CHH | 17.89 | 13.26 | 138 | 1.35 | | 85517 | 19.89 | 10.60 | 1.88 |
| 11884 | IPAR | 38.15 | 28.26 | 754 | 1.35 | | 11884 | 36.25 | 28.95 | 1.25 |
| 92123 | GRX | 13.07 | 9.68 | 106 | 1.35 | | 92123 | 14.12 | 8.33 | 1.70 |
| 56653 | BIF | 14.13 | 10.46 | 124 | 1.35 | | 56653 | 14.43 | 9.29 | 1.55 |
| 92852 | LOPE | 38.71 | 28.62 | 412 | 1.35 | | 92852 | 33.99 | 21.78 | 1.56 |
| 19393 | BMY | 17.31 | 12.79 | 124 | 1.35 | | 19393 | 23.61 | 10.44 | 2.26 |
| 79072 | BANF | 19.54 | 14.44 | 89 | 1.35 | | 79072 | 24.27 | 8.93 | 2.72 |
| 89826 | TCBI | 35.36 | 26.12 | 282 | 1.35 | | 89826 | 33.79 | 19.38 | 1.74 |
| 85731 | HFWA | 17.29 | 12.77 | 143 | 1.35 | | 85731 | 13.40 | 9.99 | 1.34 |
| 84381 | ROK | 32.41 | 23.88 | 367 | 1.36 | | 84381 | 27.11 | 19.59 | 1.38 |
| 90396 | CUBE | 37.73 | 27.79 | 545 | 1.36 | | 90396 | 23.14 | 22.45 | 1.03 |
| 20395 | CVBF | 19.94 | 14.69 | 70 | 1.36 | | 20395 | 25.86 | 8.01 | 3.23 |
| 28310 | | 13.88 | 10.22 | 119 | 1.36 | | 28310 | 14.91 | 9.02 | 1.65 |
| 91123 | XHB | 20.04 | 14.74 | 171 | 1.36 | | 91123 | 22.85 | 12.62 | 1.81 |
| 75182 | PII | 41.14 | 30.24 | 435 | 1.36 | | 75182 | 44.08 | 26.69 | 1.65 |
| 86818 | DHIL | 24.95 | 18.33 | 130 | 1.36 | | 86818 | 27.84 | 11.52 | 2.42 |
| 92816 | TNA | 52.83 | 38.77 | 394 | 1.36 | | 92816 | 55.85 | 28.26 | 1.98 |
| 12758 | ECOL | 23.74 | 17.42 | 211 | 1.36 | | 12758 | 21.85 | 13.83 | 1.58 |
| 80050 | UFPT | 36.66 | 26.85 | 468 | 1.37 | | 80050 | 26.79 | 21.21 | 1.26 |
| 84385 | NSP | 42.23 | 30.92 | 761 | 1.37 | | 84385 | 26.79 | 26.66 | 1.00 |
| 82634 | SAM | 46.02 | 33.64 | 748 | 1.37 | | 82634 | 41.61 | 29.81 | 1.40 |
| 84723 | LNG | 78.17 | 57.13 | 1977 | 1.37 | | 84723 | 61.63 | 48.34 | 1.27 |
| 10909 | CERN | 28.56 | 20.85 | 446 | 1.37 | | 10909 | 39.75 | 23.69 | 1.68 |
| 91617 | OMAB | 27.73 | 20.24 | 243 | 1.37 | | 91617 | 26.66 | 15.69 | 1.70 |
| 89070 | ZBH | 17.17 | 12.53 | 157 | 1.37 | | 89070 | 21.35 | 11.68 | 1.83 |
| 21259 | CSWC | 16.88 | 12.30 | 99 | 1.37 | | 21259 | 21.27 | 9.16 | 2.32 |
| 91915 | CNK | 25.29 | 18.43 | 382 | 1.37 | | 91915 | 29.95 | 19.85 | 1.51 |
| 80575 | SSD | 17.91 | 13.02 | 95 | 1.38 | | 80575 | 13.78 | 7.67 | 1.80 |
| 91356 | HOMB | 29.07 | 21.13 | 167 | 1.38 | | 91356 | 44.47 | 16.44 | 2.70 |
| 92597 | HI | 16.85 | 12.25 | 127 | 1.38 | | 92597 | 15.57 | 9.61 | 1.62 |
| 88297 | IYG | 19.51 | 14.16 | 139 | 1.38 | | 88297 | 19.34 | 10.71 | 1.81 |
| 83642 | CNOB | 20.70 | 15.01 | 126 | 1.38 | | 83642 | 25.47 | 11.03 | 2.31 |
| 90808 | EXPE | 43.51 | 31.51 | 1262 | 1.38 | | 90808 | 69.88 | 40.91 | 1.71 |
| 11618 | FAST | 20.94 | 15.17 | 200 | 1.38 | | 11618 | 19.55 | 13.08 | 1.49 |
| 92712 | VT | 12.66 | 9.17 | 98 | 1.38 | | 92712 | 14.16 | 7.94 | 1.78 |
| 90440 | NP | 46.76 | 33.84 | 567 | 1.38 | | 90440 | 27.68 | 24.24 | 1.14 |
| 59328 | INTC | 20.12 | 14.55 | 220 | 1.38 | | 59328 | 18.76 | 13.74 | 1.37 |
| 92585 | ACWI | 12.66 | 9.15 | 98 | 1.38 | | 92585 | 14.00 | 7.89 | 1.78 |
| 88373 | NNN | 15.73 | 11.36 | 182 | 1.38 | | 88373 | 12.21 | 11.53 | 1.06 |
| 85763 | WPC | 17.56 | 12.68 | 179 | 1.39 | | 85763 | 14.52 | 11.71 | 1.24 |
| 88863 | EXAS | 103.52 | 74.73 | 10970 | 1.39 | | 88863 | 166.10 | 107.36 | 1.55 |
| 11731 | AIN | 32.37 | 23.36 | 386 | 1.39 | | 11731 | 27.89 | 19.74 | 1.41 |
| 77661 | DHI | 28.02 | 20.17 | 390 | 1.39 | | 77661 | 35.44 | 21.89 | 1.62 |
| 88442 | GGAL | 72.81 | 52.39 | 1120 | 1.39 | | 88442 | 82.87 | 53.73 | 1.54 |
| 77057 | NHI | 16.86 | 12.11 | 175 | 1.39 | | 77057 | 16.28 | 11.74 | 1.39 |
| 76095 | HOLX | 19.47 | 13.98 | 214 | 1.39 | | 76095 | 15.45 | 13.05 | 1.18 |
| 92229 | VRTU | 42.35 | 30.40 | 655 | 1.39 | | 92229 | 52.90 | 31.65 | 1.67 |
| 90880 | AMP | 34.64 | 24.85 | 347 | 1.39 | | 90880 | 40.04 | 22.46 | 1.78 |
| 86302 | HTBK | 21.05 | 15.10 | 1 | 1.39 | | 86302 | 31.48 | 6.18 | 5.09 |
| 80412 | ADC | 19.54 | 14.01 | 226 | 1.39 | | 80412 | 11.67 | 13.08 | 0.89 |
| 20598 | CALM | 24.22 | 17.36 | 195 | 1.39 | | 20598 | 17.97 | 12.64 | 1.42 |
| 23712 | PEG | 8.96 | 6.41 | 78 | 1.40 | | 23712 | 11.82 | 6.54 | 1.81 |
| 89506 | SAIA | 49.27 | 35.28 | 671 | 1.40 | | 89506 | 57.87 | 36.73 | 1.58 |
| 89736 | HTLF | 25.44 | 18.18 | 113 | 1.40 | | 89736 | 31.29 | 11.64 | 2.69 |
| 58683 | LUV | 44.20 | 31.57 | 439 | 1.40 | | 58683 | 47.93 | 26.02 | 1.84 |
| 78081 | LGND | 59.94 | 42.80 | 725 | 1.40 | | 78081 | 59.67 | 34.16 | 1.75 |
| 83469 | ARTNA | 14.78 | 10.55 | 119 | 1.40 | | 83469 | 11.19 | 8.70 | 1.29 |
| 90564 | PBH | 35.61 | 25.41 | 193 | 1.40 | | 90564 | 42.07 | 17.85 | 2.36 |
| 79299 | PZZA | 31.11 | 22.16 | 332 | 1.40 | | 79299 | 33.15 | 20.52 | 1.62 |
| 85059 | FIX | 29.86 | 21.27 | 310 | 1.40 | | 85059 | 27.67 | 17.94 | 1.54 |
| 92188 | | 43.04 | 30.65 | 580 | 1.40 | | 92188 | 48.94 | 28.79 | 1.70 |
| 12036 | GATX | 20.32 | 14.47 | 129 | 1.40 | | 12036 | 19.65 | 10.28 | 1.91 |
| 90708 | MORN | 19.51 | 13.88 | 209 | 1.41 | | 90708 | 16.93 | 13.17 | 1.29 |
| 86124 | BMTC | 16.52 | 11.75 | 71 | 1.41 | | 86124 | 22.75 | 7.71 | 2.95 |
| 76744 | VRTX | 34.59 | 24.60 | 445 | 1.41 | | 76744 | 44.88 | 25.94 | 1.73 |
| 42550 | NEU | 38.33 | 27.26 | 1080 | 1.41 | | 42550 | 69.94 | 38.05 | 1.84 |
| 35554 | MTB | 22.33 | 15.86 | 149 | 1.41 | | 35554 | 17.03 | 10.80 | 1.58 |
| 79824 | FFIN | 22.44 | 15.93 | 213 | 1.41 | | 79824 | 26.63 | 14.56 | 1.83 |
| 16548 | BOH | 13.04 | 9.25 | 49 | 1.41 | | 16548 | 18.61 | 5.45 | 3.41 |
| 84563 | SLGN | 13.88 | 9.82 | 98 | 1.41 | | 84563 | 13.00 | 7.81 | 1.66 |
| 19828 | WEN | 25.12 | 17.77 | 216 | 1.41 | | 19828 | 28.08 | 14.76 | 1.90 |
| 82686 | CTXS | 29.57 | 20.88 | 446 | 1.42 | | 82686 | 28.34 | 21.20 | 1.34 |
| 89031 | AAXN | 56.88 | 40.12 | 729 | 1.42 | | 89031 | 40.46 | 29.81 | 1.36 |
| 82311 | BAP | 27.68 | 19.50 | 344 | 1.42 | | 82311 | 32.96 | 20.70 | 1.59 |
| 87426 | LCII | 39.28 | 27.66 | 457 | 1.42 | | 87426 | 37.60 | 24.83 | 1.51 |
| 90442 | | 30.83 | 21.70 | 462 | 1.42 | | 90442 | 27.91 | 21.86 | 1.28 |
| 89327 | ABG | 48.57 | 34.19 | 1359 | 1.42 | | 89327 | 49.35 | 37.51 | 1.32 |
| 11293 | WSBC | 17.12 | 12.05 | 35 | 1.42 | | 11293 | 32.66 | 8.29 | 3.94 |
| 52716 | HRC | 36.16 | 25.46 | 438 | 1.42 | | 52716 | 28.10 | 21.28 | 1.32 |
| 88857 | YORW | 14.83 | 10.44 | 165 | 1.42 | | 88857 | 18.19 | 11.47 | 1.59 |
| 80990 | | 50.46 | 35.49 | 731 | 1.42 | | 80990 | 39.65 | 28.66 | 1.38 |
| 90871 | RUTH | 53.97 | 37.95 | 1737 | 1.42 | | 90871 | 47.66 | 39.76 | 1.20 |
| 78876 | CRY | 29.89 | 21.01 | 192 | 1.42 | | 78876 | 39.42 | 17.01 | 2.32 |
| 92064 | REZ | 17.26 | 12.13 | 131 | 1.42 | | 92064 | 12.39 | 9.37 | 1.32 |
| 20204 | CR | 28.76 | 20.18 | 319 | 1.43 | | 20204 | 32.78 | 19.40 | 1.69 |
| 86158 | CTSH | 36.26 | 25.39 | 603 | 1.43 | | 86158 | 48.97 | 27.97 | 1.75 |
| 90948 | | 24.90 | 17.43 | 108 | 1.43 | | 90948 | 26.43 | 10.68 | 2.48 |
| 90971 | IRBT | 47.52 | 33.27 | 827 | 1.43 | | 90971 | 41.36 | 31.19 | 1.33 |
| 91018 | CORE | 30.76 | 21.53 | 332 | 1.43 | | 91018 | 32.51 | 19.98 | 1.63 |
| 89395 | PNFP | 26.13 | 18.29 | 55 | 1.43 | | 89395 | 35.17 | 10.30 | 3.41 |
| 80563 | PENN | 32.62 | 22.82 | 288 | 1.43 | | 80563 | 43.77 | 20.81 | 2.10 |
| 90361 | JOBS | 70.09 | 49.03 | 1927 | 1.43 | | 90361 | 77.22 | 50.05 | 1.54 |
| 76750 | MNRO | 25.29 | 17.66 | 304 | 1.43 | | 76750 | 25.51 | 17.37 | 1.47 |
| 24476 | COHR | 50.61 | 35.25 | 393 | 1.44 | | 24476 | 53.54 | 29.27 | 1.83 |
| 82307 | DVA | 18.88 | 13.14 | 108 | 1.44 | | 82307 | 20.24 | 9.38 | 2.16 |
| 79452 | CPT | 20.73 | 14.43 | 181 | 1.44 | | 79452 | 15.89 | 11.93 | 1.33 |
| 89997 | VAW | 16.73 | 11.64 | 144 | 1.44 | | 89997 | 20.86 | 11.12 | 1.88 |
| 10443 | FRPH | 19.61 | 13.63 | 146 | 1.44 | | 10443 | 22.11 | 11.52 | 1.92 |
| 90516 | HLF | 51.61 | 35.82 | 988 | 1.44 | | 90516 | 58.62 | 40.42 | 1.45 |
| 90979 | UAA | 50.12 | 34.77 | 496 | 1.44 | | 90979 | 45.19 | 28.04 | 1.61 |
| 83111 | ALXN | 40.49 | 28.05 | 438 | 1.44 | | 83111 | 36.19 | 23.59 | 1.53 |
| 76136 | KMPR | 31.10 | 21.54 | 316 | 1.44 | | 76136 | 21.45 | 17.05 | 1.26 |
| 75828 | EA | 42.19 | 29.22 | 392 | 1.44 | | 75828 | 40.56 | 23.14 | 1.75 |
| 76892 | BOKF | 15.73 | 10.88 | 82 | 1.45 | | 76892 | 16.86 | 7.34 | 2.30 |
| 61313 | DCI | 20.88 | 14.43 | 158 | 1.45 | | 61313 | 24.15 | 12.39 | 1.95 |
| 19350 | DE | 26.38 | 18.20 | 289 | 1.45 | | 19350 | 25.85 | 17.06 | 1.51 |
| 82547 | AEIS | 40.35 | 27.82 | 331 | 1.45 | | 82547 | 40.30 | 21.85 | 1.84 |
| 78045 | MGA | 41.14 | 28.35 | 507 | 1.45 | | 78045 | 46.17 | 27.78 | 1.66 |
| 28118 | NC | 47.54 | 32.75 | 679 | 1.45 | | 28118 | 56.07 | 32.63 | 1.72 |
| 10104 | ORCL | 16.82 | 11.58 | 155 | 1.45 | | 10104 | 19.95 | 11.51 | 1.73 |
| 92001 | FXZ | 21.25 | 14.60 | 180 | 1.46 | | 92001 | 24.41 | 13.25 | 1.84 |
| 45911 | SWKS | 59.24 | 40.68 | 1110 | 1.46 | | 45911 | 71.08 | 43.56 | 1.63 |
| 17137 | BSET | 52.85 | 36.30 | 498 | 1.46 | | 17137 | 33.69 | 24.64 | 1.37 |
| 91052 | PHO | 13.90 | 9.54 | 96 | 1.46 | | 91052 | 15.58 | 8.02 | 1.94 |
| 86719 | EWBC | 42.26 | 29.01 | 173 | 1.46 | | 86719 | 23.17 | 12.68 | 1.83 |
| 89900 | SNX | 31.16 | 21.37 | 614 | 1.46 | | 89900 | 59.60 | 31.63 | 1.88 |
| 90249 | ENS | 33.80 | 23.17 | 606 | 1.46 | | 90249 | 42.40 | 27.66 | 1.53 |
| 36898 | SCL | 26.83 | 18.37 | 215 | 1.46 | | 36898 | 28.60 | 15.71 | 1.82 |
| 81501 | NATI | 21.48 | 14.71 | 179 | 1.46 | | 81501 | 14.44 | 11.65 | 1.24 |
| 37154 | FONR | 77.61 | 53.06 | 2339 | 1.46 | | 37154 | 127.50 | 68.32 | 1.87 |
| 82743 | SGU | 22.07 | 15.08 | 291 | 1.46 | | 82743 | 28.18 | 17.52 | 1.61 |
| 55029 | MNR | 16.87 | 11.50 | 77 | 1.47 | | 55029 | 20.61 | 7.76 | 2.66 |
| 56274 | CAG | 16.55 | 11.29 | 68 | 1.47 | | 56274 | 21.81 | 7.78 | 2.80 |
| 89946 | AEL | 38.78 | 26.44 | 299 | 1.47 | | 89946 | 42.78 | 20.53 | 2.08 |
| 12650 | KSU | 37.59 | 25.61 | 401 | 1.47 | | 12650 | 32.75 | 21.98 | 1.49 |
| 91688 | EIG | 24.56 | 16.71 | 154 | 1.47 | | 91688 | 23.63 | 12.01 | 1.97 |
| 91788 | URE | 37.25 | 25.32 | 250 | 1.47 | | 91788 | 23.15 | 15.33 | 1.51 |
| 92443 | NFBK | 11.73 | 7.96 | 69 | 1.47 | | 92443 | 15.07 | 6.41 | 2.35 |
| 88446 | ILMN | 54.67 | 37.06 | 1051 | 1.48 | | 88446 | 54.73 | 39.92 | 1.37 |
| 80080 | EMN | 33.58 | 22.75 | 361 | 1.48 | | 80080 | 37.04 | 21.18 | 1.75 |
| 89216 | AYI | 35.06 | 23.73 | 229 | 1.48 | | 89216 | 36.85 | 18.03 | 2.04 |
| 91692 | | 59.09 | 39.99 | 1253 | 1.48 | | 91692 | 54.25 | 38.76 | 1.40 |
| 83856 | CSV | 46.79 | 31.61 | 671 | 1.48 | | 83856 | 46.91 | 30.21 | 1.55 |
| 77165 | JBSS | 52.49 | 35.44 | 933 | 1.48 | | 77165 | 74.69 | 41.79 | 1.79 |
| 92812 | FAS | 55.66 | 37.57 | 153 | 1.48 | | 92812 | 59.11 | 23.39 | 2.53 |
| 88352 | ISRG | 50.60 | 34.15 | 1031 | 1.48 | | 88352 | 49.62 | 34.87 | 1.42 |
| 80183 | MAA | 18.38 | 12.40 | 158 | 1.48 | | 80183 | 14.41 | 10.75 | 1.34 |
| 77453 | JACK | 29.54 | 19.92 | 251 | 1.48 | | 77453 | 33.86 | 17.54 | 1.93 |
| 30681 | OMC | 16.73 | 11.26 | 172 | 1.49 | | 30681 | 20.58 | 12.11 | 1.70 |
| 71176 | SON | 14.74 | 9.91 | 129 | 1.49 | | 71176 | 16.97 | 9.79 | 1.73 |
| 89462 | CAE | 21.48 | 14.44 | 176 | 1.49 | | 89462 | 19.89 | 12.34 | 1.61 |
| 36281 | SEB | 26.34 | 17.71 | 196 | 1.49 | | 36281 | 26.61 | 14.51 | 1.83 |
| 54199 | SJW | 17.78 | 11.94 | 86 | 1.49 | | 54199 | 31.19 | 9.66 | 3.23 |
| 52695 | GWW | 27.38 | 18.37 | 258 | 1.49 | | 52695 | 18.59 | 15.05 | 1.24 |
| 87415 | | 36.91 | 24.74 | 480 | 1.49 | | 87415 | 35.70 | 23.62 | 1.51 |
| 76392 | MMSI | 31.47 | 21.09 | 289 | 1.49 | | 76392 | 22.79 | 16.45 | 1.39 |
| 92807 | ASRV | 16.44 | 11.01 | 103 | 1.49 | | 92807 | 19.56 | 8.74 | 2.24 |
| 11481 | EBIX | 45.60 | 30.53 | 434 | 1.49 | | 11481 | 51.47 | 28.24 | 1.82 |
| 84389 | SSB | 24.05 | 16.10 | 74 | 1.49 | | 84389 | 28.52 | 8.92 | 3.20 |
| 83690 | RUSHB | 32.44 | 21.72 | 335 | 1.49 | | 83690 | 33.81 | 20.27 | 1.67 |
| 87162 | FLWS | 35.53 | 23.78 | 220 | 1.49 | | 87162 | 34.02 | 16.88 | 2.02 |
| 15457 | ATRO | 60.51 | 40.48 | 856 | 1.49 | | 15457 | 70.48 | 38.92 | 1.81 |
| 89893 | GDV | 15.78 | 10.55 | 87 | 1.50 | | 89893 | 17.88 | 7.86 | 2.27 |
| 18729 | CL | 12.01 | 8.03 | 74 | 1.50 | | 18729 | 13.81 | 6.54 | 2.11 |
| 91220 | ITB | 26.10 | 17.44 | 206 | 1.50 | | 91220 | 31.94 | 15.59 | 2.05 |
| 35991 | ROG | 38.07 | 25.41 | 257 | 1.50 | | 35991 | 43.27 | 20.69 | 2.09 |
| 89062 | IGN | 17.20 | 11.46 | 180 | 1.50 | | 89062 | 20.73 | 12.46 | 1.66 |
| 86449 | XLB | 15.14 | 10.09 | 122 | 1.50 | | 86449 | 19.09 | 9.81 | 1.95 |
| 42286 | WWW | 24.84 | 16.55 | 203 | 1.50 | | 42286 | 30.53 | 15.91 | 1.92 |
| 81045 | O | 18.14 | 12.07 | 172 | 1.50 | | 81045 | 11.99 | 11.11 | 1.08 |
| 76721 | CRVL | 35.16 | 23.40 | 462 | 1.50 | | 76721 | 39.65 | 24.10 | 1.65 |
| 78916 | | 38.65 | 25.71 | 403 | 1.50 | | 78916 | 39.66 | 23.56 | 1.68 |
| 22752 | MRK | 15.45 | 10.27 | 151 | 1.50 | | 22752 | 13.29 | 10.36 | 1.28 |
| 88152 | CCMP | 28.77 | 19.12 | 266 | 1.50 | | 88152 | 23.20 | 16.07 | 1.44 |
| 89215 | AMN | 52.05 | 34.57 | 570 | 1.51 | | 89215 | 53.67 | 29.37 | 1.83 |
| 19721 | VVI | 22.83 | 15.11 | 102 | 1.51 | | 19721 | 29.23 | 10.73 | 2.73 |
| 36768 | FLXS | 35.27 | 23.33 | 230 | 1.51 | | 36768 | 41.66 | 20.72 | 2.01 |
| 25232 | OKE | 37.73 | 24.95 | 324 | 1.51 | | 25232 | 50.44 | 24.59 | 2.05 |
| 76226 | VIAC | 45.35 | 29.98 | 434 | 1.51 | | 76226 | 35.53 | 23.11 | 1.54 |
| 68144 | | 28.95 | 19.12 | 71 | 1.51 | | 68144 | 33.09 | 10.41 | 3.18 |
| 13936 | PVH | 35.61 | 23.51 | 362 | 1.51 | | 13936 | 44.21 | 24.29 | 1.82 |
| 90744 | SIMO | 71.44 | 47.15 | 1407 | 1.52 | | 90744 | 119.37 | 54.94 | 2.17 |
| 90125 | CORT | 79.90 | 52.71 | 1210 | 1.52 | | 90125 | 78.29 | 49.43 | 1.58 |
| 85738 | MRCY | 43.05 | 28.38 | 649 | 1.52 | | 85738 | 40.43 | 28.76 | 1.41 |
| 81584 | CBZ | 21.25 | 14.01 | 128 | 1.52 | | 81584 | 23.57 | 10.73 | 2.20 |
| 91926 | DAL | 51.96 | 34.25 | 335 | 1.52 | | 91926 | 49.37 | 23.45 | 2.11 |
| 91992 | FRI | 18.85 | 12.41 | 130 | 1.52 | | 91992 | 12.53 | 9.32 | 1.34 |
| 83987 | FMS | 14.58 | 9.60 | 37 | 1.52 | | 83987 | 16.96 | 4.77 | 3.55 |
| 91317 | XPH | 22.13 | 14.55 | 143 | 1.52 | | 91317 | 23.09 | 11.41 | 2.02 |
| 11992 | TCF | 17.66 | 11.61 | 31 | 1.52 | | 11992 | 25.30 | 5.33 | 4.74 |
| 90568 | MNTX | 59.23 | 38.88 | 457 | 1.52 | | 90568 | 60.94 | 33.09 | 1.84 |
| 78015 | MTX | 25.70 | 16.87 | 151 | 1.52 | | 78015 | 34.44 | 14.52 | 2.37 |
| 84438 | HLIO | 33.06 | 21.67 | 164 | 1.53 | | 84438 | 35.82 | 16.03 | 2.23 |
| 84597 | | 26.51 | 17.37 | 244 | 1.53 | | 84597 | 28.25 | 15.85 | 1.78 |
| 42585 | SJM | 17.52 | 11.47 | 116 | 1.53 | | 42585 | 18.15 | 9.40 | 1.93 |
| 86026 | IOSP | 52.96 | 34.66 | 949 | 1.53 | | 86026 | 34.98 | 30.42 | 1.15 |
| 17830 | RTX | 17.46 | 11.42 | 99 | 1.53 | | 17830 | 18.37 | 8.53 | 2.15 |
| 89508 | KMX | 36.95 | 24.16 | 696 | 1.53 | | 89508 | 64.46 | 32.26 | 2.00 |
| 34032 | MAS | 31.17 | 20.38 | 196 | 1.53 | | 34032 | 28.19 | 14.86 | 1.90 |
| 37875 | FUL | 22.68 | 14.82 | 165 | 1.53 | | 37875 | 28.49 | 13.48 | 2.11 |
| 15720 | EIX | 13.78 | 9.00 | 77 | 1.53 | | 15720 | 16.26 | 6.92 | 2.35 |
| 88854 | IOO | 11.75 | 7.66 | 69 | 1.53 | | 88854 | 11.55 | 5.96 | 1.94 |
| 76230 | TRMB | 30.18 | 19.60 | 205 | 1.54 | | 76230 | 28.49 | 15.12 | 1.88 |
| 27887 | BAX | 17.29 | 11.22 | 122 | 1.54 | | 27887 | 18.29 | 9.60 | 1.90 |
| 51423 | MCS | 23.27 | 15.09 | 143 | 1.54 | | 51423 | 27.89 | 12.17 | 2.29 |
| 91457 | EDU | 43.14 | 27.92 | 299 | 1.55 | | 91457 | 51.06 | 24.55 | 2.08 |
| 90711 | | 16.67 | 10.78 | 60 | 1.55 | | 90711 | 21.16 | 6.70 | 3.16 |
| 91668 | IPGP | 76.52 | 49.49 | 760 | 1.55 | | 91668 | 51.28 | 33.24 | 1.54 |
| 78211 | UVE | 58.24 | 37.66 | 1460 | 1.55 | | 78211 | 79.69 | 46.30 | 1.72 |
| 79036 | DDF | 14.64 | 9.47 | 122 | 1.55 | | 79036 | 15.38 | 9.23 | 1.67 |
| 90866 | LBTYK | 31.01 | 20.05 | 226 | 1.55 | | 90866 | 31.59 | 16.92 | 1.87 |
| 52708 | LEN | 37.30 | 24.11 | 359 | 1.55 | | 52708 | 38.49 | 21.99 | 1.75 |
| 79249 | BBSI | 56.92 | 36.77 | 425 | 1.55 | | 79249 | 58.31 | 32.70 | 1.78 |
| 92261 | G | 22.12 | 14.28 | 228 | 1.55 | | 92261 | 27.59 | 15.17 | 1.82 |
| 75825 | EOG | 25.63 | 16.55 | 162 | 1.55 | | 75825 | 24.79 | 12.64 | 1.96 |
| 81541 | | 22.52 | 14.52 | 42 | 1.55 | | 81541 | 27.86 | 6.51 | 4.28 |
| 11638 | | 36.50 | 23.50 | 122 | 1.55 | | 11638 | 36.81 | 14.70 | 2.50 |
| 59300 | UDR | 25.08 | 16.14 | 187 | 1.55 | | 59300 | 16.25 | 12.14 | 1.34 |
| 60943 | LEG | 20.67 | 13.29 | 136 | 1.56 | | 60943 | 18.43 | 10.43 | 1.77 |
| 79248 | THRM | 54.04 | 34.70 | 1126 | 1.56 | | 79248 | 51.44 | 36.37 | 1.41 |
| 10180 | | 81.34 | 52.23 | 47 | 1.56 | | 10180 | 89.77 | 37.13 | 2.42 |
| 10065 | ADX | 11.15 | 7.16 | 57 | 1.56 | | 10065 | 14.16 | 5.49 | 2.58 |
| 28564 | CTS | 36.00 | 23.09 | 370 | 1.56 | | 28564 | 34.53 | 20.66 | 1.67 |
| 89731 | LEN | 39.55 | 25.36 | 393 | 1.56 | | 89731 | 40.33 | 23.42 | 1.72 |
| 76736 | ALKS | 38.42 | 24.63 | 177 | 1.56 | | 76736 | 46.89 | 18.84 | 2.49 |
| 86091 | SBSI | 17.47 | 11.19 | 105 | 1.56 | | 86091 | 24.39 | 9.56 | 2.55 |
| 80100 | SPG | 26.67 | 17.08 | 236 | 1.56 | | 80100 | 19.31 | 14.31 | 1.35 |
| 52230 | RHI | 23.70 | 15.17 | 175 | 1.56 | | 52230 | 17.95 | 11.97 | 1.50 |
| 41187 | HE | 13.53 | 8.66 | 65 | 1.56 | | 41187 | 12.30 | 5.80 | 2.12 |
| 79507 | MCRI | 39.62 | 25.33 | 227 | 1.56 | | 79507 | 41.24 | 18.85 | 2.19 |
| 88961 | RWR | 18.36 | 11.72 | 112 | 1.57 | | 88961 | 12.33 | 8.43 | 1.46 |
| 80828 | MLR | 25.50 | 16.28 | 409 | 1.57 | | 80828 | 34.25 | 20.92 | 1.64 |
| 90868 | OFLX | 33.73 | 21.53 | 158 | 1.57 | | 90868 | 43.13 | 16.82 | 2.56 |
| 92203 | LULU | 86.92 | 55.39 | 2967 | 1.57 | | 92203 | 89.16 | 57.22 | 1.56 |
| 86594 | KFY | 35.91 | 22.87 | 246 | 1.57 | | 86594 | 29.57 | 16.64 | 1.78 |
| 21792 | CNP | 17.69 | 11.27 | 124 | 1.57 | | 21792 | 16.51 | 9.56 | 1.73 |
| 90520 | CE | 39.89 | 25.37 | 624 | 1.57 | | 90520 | 48.19 | 27.65 | 1.74 |
| 80381 | AVB | 22.99 | 14.61 | 187 | 1.57 | | 80381 | 18.71 | 12.50 | 1.50 |
| 24440 | OGE | 19.65 | 12.48 | 204 | 1.58 | | 24440 | 19.51 | 13.42 | 1.45 |
| 34948 | OXM | 50.90 | 32.31 | 710 | 1.58 | | 34948 | 50.11 | 31.04 | 1.61 |
| 89406 | NPO | 31.43 | 19.96 | 179 | 1.58 | | 89406 | 34.28 | 15.98 | 2.15 |
| 81067 | TGS | 60.01 | 38.05 | 639 | 1.58 | | 81067 | 62.61 | 36.87 | 1.70 |
| 64995 | KEY | 24.75 | 15.68 | 73 | 1.58 | | 64995 | 33.10 | 10.10 | 3.28 |
| 87444 | SNN | 18.05 | 11.42 | 189 | 1.58 | | 87444 | 21.80 | 13.00 | 1.68 |
| 91571 | EBS | 32.01 | 20.21 | 138 | 1.58 | | 91571 | 38.12 | 15.54 | 2.45 |
| 88310 | PUK | 27.72 | 17.46 | 180 | 1.59 | | 88310 | 31.25 | 14.61 | 2.14 |
| 75906 | GSBC | 24.48 | 15.42 | 302 | 1.59 | | 75906 | 27.42 | 17.34 | 1.58 |
| 91413 | EVR | 38.49 | 24.25 | 473 | 1.59 | | 91413 | 53.50 | 27.34 | 1.96 |
| 88894 | ICF | 19.01 | 11.97 | 117 | 1.59 | | 88894 | 12.33 | 8.65 | 1.43 |
| 58413 | FRT | 16.63 | 10.47 | 90 | 1.59 | | 58413 | 13.01 | 7.34 | 1.77 |
| 90227 | LBTYA | 32.28 | 20.32 | 214 | 1.59 | | 90227 | 32.84 | 16.81 | 1.95 |
| 89789 | AXS | 15.10 | 9.50 | 77 | 1.59 | | 89789 | 17.37 | 7.19 | 2.41 |
| 81241 | BRKS | 41.47 | 26.09 | 351 | 1.59 | | 81241 | 26.33 | 18.99 | 1.39 |
| 61807 | MOG | 24.92 | 15.67 | 112 | 1.59 | | 61807 | 27.23 | 10.61 | 2.57 |
| 89217 | AAP | 29.90 | 18.80 | 368 | 1.59 | | 89217 | 33.13 | 21.35 | 1.55 |
| 89942 | GBLI | 22.95 | 14.42 | 41 | 1.59 | | 89942 | 20.50 | 5.61 | 3.65 |
| 55001 | TRN | 41.44 | 26.03 | 267 | 1.59 | | 55001 | 25.07 | 16.37 | 1.53 |
| 39220 | SMP | 54.88 | 34.41 | 1300 | 1.59 | | 39220 | 47.39 | 36.53 | 1.30 |
| 90373 | DLR | 23.88 | 14.97 | 224 | 1.60 | | 90373 | 23.30 | 14.75 | 1.58 |
| 81666 | HA | 51.99 | 32.57 | 314 | 1.60 | | 81666 | 60.27 | 25.92 | 2.33 |
| 88895 | MPX | 32.34 | 20.26 | 201 | 1.60 | | 88895 | 50.64 | 19.81 | 2.56 |
| 45356 | JCI | 22.75 | 14.24 | 193 | 1.60 | | 45356 | 28.92 | 14.75 | 1.96 |
| 76266 | PRK | 10.88 | 6.81 | 18 | 1.60 | | 76266 | 19.55 | 3.31 | 5.91 |
| 23297 | EAT | 25.41 | 15.89 | 317 | 1.60 | | 23297 | 24.35 | 17.83 | 1.37 |
| 83774 | OCFC | 17.42 | 10.89 | 36 | 1.60 | | 83774 | 22.87 | 5.34 | 4.29 |
| 88197 | FULT | 18.80 | 11.74 | 61 | 1.60 | | 88197 | 20.01 | 6.41 | 3.12 |
| 91659 | ALGT | 28.48 | 17.77 | 106 | 1.60 | | 91659 | 24.97 | 10.22 | 2.44 |
| 83414 | SASR | 22.06 | 13.76 | 44 | 1.60 | | 83414 | 45.38 | 12.12 | 3.74 |
| 86299 | | 55.92 | 34.85 | 883 | 1.60 | | 86299 | 42.59 | 31.03 | 1.37 |
| 85488 | OSIS | 33.48 | 20.85 | 429 | 1.61 | | 85488 | 34.02 | 22.14 | 1.54 |
| 11762 | ETN | 26.98 | 16.80 | 176 | 1.61 | | 11762 | 27.61 | 13.77 | 2.01 |
| 91977 | BGS | 36.34 | 22.60 | 435 | 1.61 | | 91977 | 34.04 | 22.51 | 1.51 |
| 77366 | ELY | 20.77 | 12.91 | 65 | 1.61 | | 77366 | 20.42 | 7.09 | 2.88 |
| 91224 | IAI | 22.01 | 13.68 | 181 | 1.61 | | 91224 | 26.16 | 13.59 | 1.92 |
| 92060 | USRT | 16.95 | 10.53 | 96 | 1.61 | | 92060 | 11.80 | 7.56 | 1.56 |
| 73219 | RGR | 43.03 | 26.74 | 791 | 1.61 | | 73219 | 50.13 | 34.48 | 1.45 |
| 56856 | BRT | 26.06 | 16.19 | 270 | 1.61 | | 56856 | 26.31 | 16.50 | 1.59 |
| 92239 | CXO | 38.87 | 24.10 | 350 | 1.61 | | 92239 | 44.36 | 22.89 | 1.94 |
| 85426 | RHP | 44.78 | 27.76 | 515 | 1.61 | | 85426 | 38.15 | 25.16 | 1.52 |
| 89964 | NNI | 38.04 | 23.56 | 265 | 1.61 | | 89964 | 23.51 | 16.13 | 1.46 |
| 90506 | SOHO | 37.62 | 23.28 | 343 | 1.62 | | 90506 | 31.73 | 19.60 | 1.62 |
| 90041 | HTD | 17.04 | 10.54 | 98 | 1.62 | | 90041 | 13.30 | 7.86 | 1.69 |
| 90902 | ITRN | 30.77 | 19.03 | 339 | 1.62 | | 90902 | 32.91 | 19.93 | 1.65 |
| 16126 | AZZ | 24.75 | 15.30 | 222 | 1.62 | | 16126 | 26.38 | 15.09 | 1.75 |
| 92296 | | 34.34 | 21.20 | 251 | 1.62 | | 92296 | 32.27 | 17.32 | 1.86 |
| 79878 | UFPI | 26.59 | 16.42 | 189 | 1.62 | | 79878 | 25.93 | 14.16 | 1.83 |
| 15070 | AROW | 11.02 | 6.80 | 65 | 1.62 | | 15070 | 18.74 | 6.41 | 2.92 |
| 91602 | AER | 60.75 | 37.47 | 1216 | 1.62 | | 91602 | 80.25 | 44.84 | 1.79 |
| 91463 | CVLT | 36.52 | 22.51 | 341 | 1.62 | | 91463 | 35.10 | 20.78 | 1.69 |
| 84656 | ALV | 40.17 | 24.76 | 364 | 1.62 | | 84656 | 42.43 | 22.87 | 1.86 |
| 80204 | MLM | 18.45 | 11.37 | 77 | 1.62 | | 80204 | 24.57 | 8.20 | 3.00 |
| 84042 | PAG | 32.09 | 19.78 | 425 | 1.62 | | 84042 | 36.81 | 22.53 | 1.63 |
| 88294 | IYR | 17.09 | 10.52 | 101 | 1.62 | | 88294 | 11.49 | 7.79 | 1.47 |
| 89915 | MGRC | 24.00 | 14.77 | 141 | 1.63 | | 89915 | 23.62 | 11.50 | 2.05 |
| 80169 | AGM | 74.87 | 46.06 | 1627 | 1.63 | | 80169 | 53.25 | 41.72 | 1.28 |
| 84767 | ARE | 21.89 | 13.46 | 91 | 1.63 | | 84767 | 15.93 | 7.70 | 2.07 |
| 80223 | BPFH | 24.44 | 15.03 | 55 | 1.63 | | 80223 | 24.96 | 7.13 | 3.50 |
| 13777 | AMSWA | 18.45 | 11.34 | 122 | 1.63 | | 13777 | 18.47 | 9.73 | 1.90 |
| 89800 | NTGR | 33.83 | 20.78 | 356 | 1.63 | | 89800 | 32.11 | 19.82 | 1.62 |
| 19166 | FMC | 30.52 | 18.75 | 231 | 1.63 | | 19166 | 33.67 | 17.48 | 1.93 |
| 92043 | IBKR | 31.92 | 19.58 | 205 | 1.63 | | 92043 | 34.74 | 16.05 | 2.17 |
| 75100 | TIF | 33.42 | 20.49 | 241 | 1.63 | | 75100 | 36.84 | 18.13 | 2.03 |
| 77114 | WRLD | 46.13 | 28.24 | 418 | 1.63 | | 77114 | 39.61 | 24.97 | 1.59 |
| 25081 | CMA | 35.03 | 21.44 | 246 | 1.63 | | 25081 | 35.16 | 18.44 | 1.91 |
| 90011 | MGLN | 18.99 | 11.61 | 45 | 1.64 | | 90011 | 19.54 | 5.82 | 3.36 |
| 86717 | EPAY | 43.33 | 26.44 | 576 | 1.64 | | 86717 | 47.90 | 27.74 | 1.73 |
| 85418 | CAC | 17.20 | 10.49 | 100 | 1.64 | | 85418 | 19.64 | 8.66 | 2.27 |
| 90350 | VNQ | 18.38 | 11.19 | 105 | 1.64 | | 90350 | 12.49 | 8.06 | 1.55 |
| 88343 | PACW | 23.81 | 14.49 | 24 | 1.64 | | 88343 | 30.46 | 5.82 | 5.23 |
| 79758 | BYD | 39.50 | 24.02 | 339 | 1.64 | | 79758 | 43.79 | 23.61 | 1.85 |
| 78908 | RCKY | 47.13 | 28.65 | 544 | 1.64 | | 78908 | 34.56 | 24.58 | 1.41 |
| 92723 | WOOD | 20.09 | 12.22 | 94 | 1.64 | | 92723 | 19.53 | 8.51 | 2.29 |
| 76946 | TRNS | 30.47 | 18.52 | 144 | 1.65 | | 76946 | 25.90 | 12.92 | 2.00 |
| 60098 | OTTR | 15.82 | 9.61 | 113 | 1.65 | | 60098 | 18.41 | 9.11 | 2.02 |
| 75517 | MRTN | 24.30 | 14.75 | 113 | 1.65 | | 75517 | 28.86 | 11.07 | 2.61 |
| 20750 | CWT | 15.86 | 9.62 | 105 | 1.65 | | 20750 | 20.01 | 9.05 | 2.21 |
| 89455 | RUSHA | 34.34 | 20.83 | 302 | 1.65 | | 89455 | 36.65 | 20.45 | 1.79 |
| 87267 | BLK | 25.65 | 15.56 | 193 | 1.65 | | 87267 | 31.02 | 14.92 | 2.08 |
| 92490 | SCZ | 16.40 | 9.94 | 102 | 1.65 | | 92490 | 20.18 | 8.99 | 2.24 |
| 86287 | RBCAA | 15.15 | 9.17 | 42 | 1.65 | | 86287 | 19.53 | 5.11 | 3.82 |
| 40125 | DXC | 39.98 | 24.17 | 248 | 1.65 | | 40125 | 46.88 | 22.88 | 2.05 |
| 79906 | INCY | 97.52 | 58.93 | 1578 | 1.65 | | 79906 | 75.61 | 47.51 | 1.59 |
| 82573 | FFIC | 21.69 | 13.10 | 80 | 1.66 | | 82573 | 20.14 | 7.74 | 2.60 |
| 80969 | CVLG | 77.77 | 46.96 | 860 | 1.66 | | 80969 | 90.47 | 52.34 | 1.73 |
| 53613 | MU | 81.88 | 49.42 | 1102 | 1.66 | | 53613 | 120.30 | 61.92 | 1.94 |
| 91658 | AIMC | 44.93 | 27.12 | 218 | 1.66 | | 91658 | 38.21 | 18.90 | 2.02 |
| 68830 | REX | 59.34 | 35.81 | 744 | 1.66 | | 68830 | 51.43 | 32.10 | 1.60 |
| 43772 | | 13.07 | 7.88 | 94 | 1.66 | | 43772 | 10.88 | 7.33 | 1.48 |
| 65875 | VZ | 12.54 | 7.55 | 77 | 1.66 | | 65875 | 7.85 | 6.12 | 1.28 |
| 90857 | BIDU | 66.65 | 40.10 | 1115 | 1.66 | | 90857 | 78.54 | 44.20 | 1.78 |
| 11006 | FLIC | 18.18 | 10.93 | 89 | 1.66 | | 11006 | 23.53 | 8.83 | 2.67 |
| 75241 | PXD | 55.10 | 33.12 | 713 | 1.66 | | 75241 | 68.14 | 35.36 | 1.93 |
| 88944 | FCBC | 18.12 | 10.89 | -10 | 1.66 | | 88944 | 32.92 | 5.30 | 6.21 |
| 89810 | GPK | 56.59 | 34.01 | 833 | 1.66 | | 89810 | 64.95 | 35.23 | 1.84 |
| 14702 | AMAT | 35.69 | 21.35 | 223 | 1.67 | | 14702 | 39.18 | 18.72 | 2.09 |
| 81254 | SNEX | 32.04 | 19.16 | 326 | 1.67 | | 81254 | 33.62 | 19.74 | 1.70 |
| 84262 | STLD | 29.45 | 17.57 | 169 | 1.68 | | 84262 | 40.41 | 16.19 | 2.50 |
| 35044 | RF | 34.24 | 20.41 | 68 | 1.68 | | 35044 | 36.20 | 10.98 | 3.30 |
| 12570 | ITT | 26.29 | 15.67 | 170 | 1.68 | | 12570 | 29.43 | 13.28 | 2.22 |
| 14761 | TREC | 48.81 | 29.07 | 388 | 1.68 | | 14761 | 43.66 | 24.35 | 1.79 |
| 90729 | GAIN | 18.73 | 11.15 | 90 | 1.68 | | 90729 | 25.38 | 8.94 | 2.84 |
| 38659 | GNTX | 34.29 | 20.41 | 358 | 1.68 | | 38659 | 43.67 | 23.05 | 1.89 |
| 23579 | TXT | 40.64 | 24.18 | 232 | 1.68 | | 23579 | 22.94 | 14.96 | 1.53 |
| 83779 | PEGA | 44.30 | 26.36 | 674 | 1.68 | | 83779 | 64.21 | 33.73 | 1.90 |
| 82176 | | 70.48 | 41.94 | 1092 | 1.68 | | 82176 | 86.06 | 48.45 | 1.78 |
| 77610 | OLED | 76.40 | 45.45 | 890 | 1.68 | | 77610 | 81.69 | 44.36 | 1.84 |
| 10138 | TROW | 24.39 | 14.49 | 160 | 1.68 | | 10138 | 22.82 | 12.11 | 1.88 |
| 90047 | UTG | 19.41 | 11.52 | 167 | 1.69 | | 90047 | 22.05 | 12.20 | 1.81 |
| 90349 | VOX | 11.87 | 7.04 | 65 | 1.69 | | 90349 | 14.70 | 6.10 | 2.41 |
| 75603 | CRUS | 76.38 | 45.29 | 1138 | 1.69 | | 75603 | 68.17 | 42.91 | 1.59 |
| 92598 | THD | 26.66 | 15.80 | 248 | 1.69 | | 92598 | 32.98 | 17.35 | 1.90 |
| 79862 | ROCK | 38.03 | 22.54 | 198 | 1.69 | | 79862 | 28.74 | 14.64 | 1.96 |
| 91237 | VG | 94.86 | 56.21 | 1223 | 1.69 | | 91237 | 36.85 | 33.75 | 1.09 |
| 79668 | BDC | 33.85 | 20.03 | 100 | 1.69 | | 79668 | 39.74 | 14.16 | 2.81 |
| 87501 | MCBC | 39.26 | 23.22 | 177 | 1.69 | | 87501 | 47.85 | 19.68 | 2.43 |
| 84636 | WTFC | 35.49 | 20.97 | 223 | 1.69 | | 84636 | 24.07 | 14.73 | 1.63 |
| 91486 | | 53.95 | 31.88 | 476 | 1.69 | | 91486 | 59.88 | 29.91 | 2.00 |
| 83601 | WOR | 34.48 | 20.36 | 216 | 1.69 | | 83601 | 35.13 | 17.00 | 2.07 |
| 91005 | KBE | 21.33 | 12.60 | 70 | 1.69 | | 91005 | 20.60 | 7.28 | 2.83 |
| 85337 | IBA | 28.19 | 16.63 | 173 | 1.70 | | 85337 | 29.51 | 14.13 | 2.09 |
| 92220 | MASI | 35.34 | 20.85 | 260 | 1.70 | | 92220 | 30.86 | 17.59 | 1.75 |
| 61946 | BKH | 21.85 | 12.88 | 133 | 1.70 | | 61946 | 16.93 | 9.92 | 1.71 |
| 80719 | GBX | 62.38 | 36.77 | 476 | 1.70 | | 80719 | 51.94 | 29.28 | 1.77 |
| 71563 | TFC | 18.12 | 10.68 | 58 | 1.70 | | 71563 | 13.55 | 5.43 | 2.50 |
| 69331 | ECF | 11.48 | 6.76 | 76 | 1.70 | | 69331 | 13.91 | 6.62 | 2.10 |
| 76081 | THO | 42.56 | 25.07 | 295 | 1.70 | | 76081 | 56.25 | 27.62 | 2.04 |
| 11752 | IIVI | 33.93 | 19.98 | 240 | 1.70 | | 11752 | 38.23 | 18.82 | 2.03 |
| 10421 | SKYW | 41.16 | 24.22 | 139 | 1.70 | | 10421 | 36.15 | 13.57 | 2.66 |
| 10501 | AMWD | 42.47 | 24.96 | 205 | 1.70 | | 10501 | 55.20 | 24.41 | 2.26 |
| 91821 | | 59.06 | 34.70 | 1075 | 1.70 | | 91821 | 49.70 | 35.67 | 1.39 |
| 72304 | STAA | 96.50 | 56.68 | 1241 | 1.70 | | 72304 | 68.09 | 45.79 | 1.49 |
| 10874 | BC | 66.39 | 38.98 | 1003 | 1.70 | | 10874 | 64.16 | 36.83 | 1.74 |
| 87505 | NKSH | 14.63 | 8.59 | 87 | 1.70 | | 87505 | 20.03 | 8.17 | 2.45 |
| 85394 | PSMT | 39.41 | 23.09 | 186 | 1.71 | | 85394 | 41.53 | 17.10 | 2.43 |
| 80864 | RS | 22.99 | 13.47 | 257 | 1.71 | | 80864 | 39.46 | 18.34 | 2.15 |
| 88597 | AVNT | 73.23 | 42.90 | 808 | 1.71 | | 88597 | 53.57 | 34.01 | 1.57 |
| 76752 | PRGS | 24.93 | 14.61 | 176 | 1.71 | | 76752 | 27.84 | 14.00 | 1.99 |
| 79747 | AUB | 21.83 | 12.78 | 14 | 1.71 | | 79747 | 30.77 | 5.79 | 5.32 |
| 84373 | DGX | 14.44 | 8.45 | 60 | 1.71 | | 84373 | 14.85 | 5.78 | 2.57 |
| 79145 | RCL | 61.24 | 35.81 | 611 | 1.71 | | 79145 | 46.01 | 30.43 | 1.51 |
| 80191 | ALB | 33.29 | 19.46 | 246 | 1.71 | | 80191 | 34.80 | 18.20 | 1.91 |
| 76573 | MLI | 21.89 | 12.79 | 86 | 1.71 | | 76573 | 25.92 | 9.41 | 2.75 |
| 77605 | BSX | 34.15 | 19.93 | 357 | 1.71 | | 77605 | 37.95 | 21.19 | 1.79 |
| 88908 | | 26.17 | 15.27 | 24 | 1.71 | | 88908 | 30.74 | 6.29 | 4.89 |
| 91215 | DK | 56.48 | 32.95 | 515 | 1.71 | | 91215 | 42.59 | 25.57 | 1.67 |
| 77528 | STBA | 20.71 | 12.08 | 7 | 1.71 | | 77528 | 26.88 | 4.46 | 6.02 |
| 91068 | CMG | 52.87 | 30.77 | 597 | 1.72 | | 91068 | 54.73 | 31.27 | 1.75 |
| 35917 | FMBI | 21.98 | 12.79 | -1 | 1.72 | | 35917 | 25.95 | 3.17 | 8.17 |
| 87717 | EXPD | 18.02 | 10.48 | 105 | 1.72 | | 87717 | 21.17 | 9.17 | 2.31 |
| 76128 | VICR | 64.13 | 37.28 | 472 | 1.72 | | 76128 | 62.32 | 33.16 | 1.88 |
| 77785 | WIRE | 21.23 | 12.34 | 165 | 1.72 | | 77785 | 27.34 | 12.98 | 2.11 |
| 77735 | MEOH | 48.34 | 28.08 | 329 | 1.72 | | 77735 | 43.53 | 22.99 | 1.89 |
| 77878 | LTC | 14.83 | 8.60 | 106 | 1.72 | | 77878 | 12.12 | 8.06 | 1.50 |
| 80276 | SILC | 58.29 | 33.79 | 732 | 1.72 | | 80276 | 69.25 | 39.26 | 1.76 |
| 92257 | VMW | 46.09 | 26.72 | 479 | 1.72 | | 92257 | 44.70 | 25.90 | 1.73 |
| 66325 | SLM | 37.04 | 21.45 | 169 | 1.73 | | 66325 | 32.08 | 14.70 | 2.18 |
| 75460 | TNC | 34.55 | 20.00 | 238 | 1.73 | | 75460 | 32.24 | 17.00 | 1.90 |
| 83862 | | 40.56 | 23.48 | 282 | 1.73 | | 83862 | 45.13 | 21.98 | 2.05 |
| 91363 | PGTI | 83.54 | 48.34 | 1303 | 1.73 | | 91363 | 111.15 | 56.15 | 1.98 |
| 59176 | AXP | 36.80 | 21.29 | 414 | 1.73 | | 59176 | 40.35 | 22.78 | 1.77 |
| 75186 | SCHW | 28.10 | 16.23 | 157 | 1.73 | | 75186 | 31.56 | 13.77 | 2.29 |
| 50948 | SMTC | 29.33 | 16.93 | 307 | 1.73 | | 50948 | 29.00 | 18.46 | 1.57 |
| 77928 | CCU | 21.61 | 12.46 | 178 | 1.73 | | 77928 | 27.82 | 13.81 | 2.02 |
| 88182 | ON | 34.89 | 20.12 | 386 | 1.73 | | 88182 | 53.87 | 25.06 | 2.15 |
| 11844 | CTBI | 13.99 | 8.06 | 19 | 1.74 | | 11844 | 23.61 | 4.14 | 5.71 |
| 79150 | ANIK | 64.75 | 37.26 | 1006 | 1.74 | | 79150 | 97.70 | 47.18 | 2.07 |
| 10860 | OSUR | 57.74 | 33.21 | 217 | 1.74 | | 10860 | 49.54 | 21.39 | 2.32 |
| 90856 | ATRC | 77.52 | 44.56 | 1278 | 1.74 | | 90856 | 74.43 | 45.05 | 1.65 |
| 84129 | ZION | 34.96 | 20.08 | 66 | 1.74 | | 84129 | 42.58 | 12.67 | 3.36 |
| 87725 | IIIN | 32.02 | 18.37 | 115 | 1.74 | | 87725 | 35.78 | 12.29 | 2.91 |
| 82777 | MSM | 19.78 | 11.33 | 109 | 1.75 | | 82777 | 27.37 | 10.66 | 2.57 |
| 83011 | WSM | 40.60 | 23.26 | 542 | 1.75 | | 83011 | 54.92 | 28.51 | 1.93 |
| 91855 | SFNC | 17.50 | 10.02 | 64 | 1.75 | | 91855 | 19.35 | 6.52 | 2.97 |
| 90353 | BECN | 27.76 | 15.90 | 129 | 1.75 | | 90353 | 34.19 | 14.29 | 2.39 |
| 91534 | PZD | 17.74 | 10.15 | 103 | 1.75 | | 91534 | 19.81 | 8.95 | 2.21 |
| 92143 | PRO | 48.27 | 27.59 | 446 | 1.75 | | 92143 | 44.53 | 24.95 | 1.78 |
| 86582 | STL | 21.54 | 12.31 | 33 | 1.75 | | 86582 | 31.94 | 7.63 | 4.18 |
| 87837 | | 49.04 | 27.99 | 224 | 1.75 | | 87837 | 61.15 | 25.14 | 2.43 |
| 42906 | HBAN | 37.71 | 21.52 | 56 | 1.75 | | 42906 | 38.78 | 10.89 | 3.56 |
| 91983 | CLR | 45.13 | 25.71 | 288 | 1.76 | | 91983 | 56.04 | 25.20 | 2.22 |
| 91041 | LYV | 64.27 | 36.61 | 758 | 1.76 | | 91041 | 38.84 | 28.97 | 1.34 |
| 18542 | CAT | 35.33 | 20.12 | 184 | 1.76 | | 18542 | 33.15 | 15.10 | 2.20 |
| 76076 | CSCO | 20.53 | 11.69 | 166 | 1.76 | | 76076 | 18.46 | 11.64 | 1.59 |
| 48144 | KLIC | 54.90 | 31.24 | 1092 | 1.76 | | 48144 | 66.93 | 38.16 | 1.75 |
| 89916 | NXST | 173.76 | 98.68 | 15289 | 1.76 | | 89916 | 236.91 | 127.09 | 1.86 |
| 54114 | MLHR | 22.33 | 12.67 | 132 | 1.76 | | 54114 | 26.28 | 11.68 | 2.25 |
| 86814 | AMNB | 16.12 | 9.15 | 72 | 1.76 | | 86814 | 19.85 | 7.29 | 2.72 |
| 76591 | QDEL | 43.24 | 24.52 | 274 | 1.76 | | 76591 | 37.73 | 18.68 | 2.02 |
| 91815 | CENTA | 40.61 | 23.03 | 430 | 1.76 | | 91815 | 47.93 | 25.84 | 1.85 |
| 41080 | CMI | 42.23 | 23.94 | 400 | 1.76 | | 41080 | 53.09 | 26.82 | 1.98 |
| 83124 | CSGS | 24.23 | 13.73 | 82 | 1.76 | | 83124 | 30.08 | 9.78 | 3.08 |
| 92441 | IEUS | 17.20 | 9.75 | 90 | 1.77 | | 92441 | 21.37 | 8.63 | 2.48 |
| 72864 | UHT | 16.70 | 9.45 | 86 | 1.77 | | 72864 | 17.92 | 7.85 | 2.28 |
| 85058 | BXP | 21.03 | 11.88 | 105 | 1.77 | | 85058 | 14.53 | 8.30 | 1.75 |
| 17726 | CCK | 16.54 | 9.34 | 117 | 1.77 | | 17726 | 17.14 | 9.34 | 1.83 |
| 79686 | DECK | 59.42 | 33.57 | 381 | 1.77 | | 79686 | 60.00 | 30.16 | 1.99 |
| 10649 | PTSI | 80.36 | 45.37 | 463 | 1.77 | | 10649 | 58.17 | 29.57 | 1.97 |
| 82212 | KFRC | 33.44 | 18.85 | 303 | 1.77 | | 82212 | 24.65 | 17.32 | 1.42 |
| 54973 | BCV | 11.58 | 6.53 | 68 | 1.77 | | 54973 | 14.66 | 6.23 | 2.35 |
| 92672 | | 29.16 | 16.43 | 229 | 1.77 | | 92672 | 29.92 | 16.53 | 1.81 |
| 88742 | SHEN | 37.32 | 21.02 | 216 | 1.78 | | 88742 | 34.20 | 17.54 | 1.95 |
| 87812 | SLAB | 25.04 | 14.10 | 218 | 1.78 | | 87812 | 32.30 | 15.55 | 2.08 |
| 89913 | MRLN | 49.93 | 28.04 | 756 | 1.78 | | 89913 | 65.87 | 34.40 | 1.91 |
| 76282 | AN | 28.88 | 16.20 | 261 | 1.78 | | 76282 | 35.18 | 18.16 | 1.94 |
| 77300 | SMG | 19.64 | 11.02 | 107 | 1.78 | | 77300 | 27.40 | 11.02 | 2.49 |
| 10397 | WERN | 17.58 | 9.85 | 70 | 1.78 | | 10397 | 21.47 | 7.51 | 2.86 |
| 11174 | CRMT | 39.79 | 22.31 | 425 | 1.78 | | 11174 | 41.02 | 25.05 | 1.64 |
| 80924 | SIRI | 90.96 | 50.96 | 4658 | 1.78 | | 80924 | 126.09 | 71.58 | 1.76 |
| 83661 | | 28.44 | 15.93 | 143 | 1.79 | | 83661 | 27.15 | 12.55 | 2.16 |
| 87086 | PRFT | 36.33 | 20.33 | 366 | 1.79 | | 87086 | 40.30 | 22.06 | 1.83 |
| 11343 | SAFM | 26.93 | 15.05 | 187 | 1.79 | | 11343 | 25.76 | 13.87 | 1.86 |
| 21186 | WRK | 32.13 | 17.95 | 230 | 1.79 | | 21186 | 52.25 | 20.58 | 2.54 |
| 81740 | SBGI | 93.57 | 52.24 | 750 | 1.79 | | 81740 | 63.79 | 34.71 | 1.84 |
| 83969 | WLFC | 26.52 | 14.80 | 273 | 1.79 | | 83969 | 24.14 | 16.28 | 1.48 |
| 20670 | CAMP | 104.03 | 58.05 | 2791 | 1.79 | | 20670 | 215.48 | 97.88 | 2.20 |
| 81061 | MCK | 28.54 | 15.92 | 185 | 1.79 | | 81061 | 32.28 | 15.23 | 2.12 |
| 86889 | NLS | 77.28 | 43.08 | 393 | 1.79 | | 86889 | 57.98 | 27.33 | 2.12 |
| 77573 | HQL | 20.60 | 11.48 | 78 | 1.80 | | 77573 | 21.54 | 8.03 | 2.68 |
| 77902 | CAKE | 32.07 | 17.85 | 331 | 1.80 | | 77902 | 39.84 | 20.70 | 1.92 |
| 91332 | DLS | 15.97 | 8.88 | 79 | 1.80 | | 91332 | 18.91 | 7.57 | 2.50 |
| 91080 | HEES | 45.49 | 25.30 | 165 | 1.80 | | 91080 | 45.09 | 18.64 | 2.42 |
| 91531 | OC | 37.84 | 21.03 | 154 | 1.80 | | 91531 | 35.63 | 15.58 | 2.29 |
| 85888 | | 37.02 | 20.57 | 289 | 1.80 | | 85888 | 26.76 | 17.32 | 1.55 |
| 11552 | | 33.68 | 18.70 | 132 | 1.80 | | 11552 | 40.18 | 14.03 | 2.86 |
| 81095 | NWBI | 15.93 | 8.84 | 78 | 1.80 | | 81095 | 14.69 | 6.84 | 2.15 |
| 85246 | JLL | 45.74 | 25.38 | 357 | 1.80 | | 85246 | 45.23 | 23.77 | 1.90 |
| 80167 | HAIN | 36.31 | 20.09 | 66 | 1.81 | | 80167 | 41.53 | 13.94 | 2.98 |
| 86384 | CCNE | 18.21 | 10.08 | 105 | 1.81 | | 86384 | 20.81 | 9.07 | 2.30 |
| 79007 | SCVL | 39.66 | 21.93 | 426 | 1.81 | | 79007 | 37.15 | 22.22 | 1.67 |
| 84348 | VSAT | 25.52 | 14.11 | 145 | 1.81 | | 84348 | 25.57 | 11.87 | 2.15 |
| 81577 | WAFD | 19.90 | 11.00 | 79 | 1.81 | | 81577 | 23.61 | 8.26 | 2.86 |
| 87614 | LBAI | 24.64 | 13.62 | 52 | 1.81 | | 87614 | 36.84 | 9.76 | 3.78 |
| 58501 | GPX | 36.63 | 20.24 | 180 | 1.81 | | 58501 | 36.82 | 16.93 | 2.17 |
| 90194 | SP | 17.90 | 9.89 | 53 | 1.81 | | 90194 | 18.40 | 5.81 | 3.17 |
| 32803 | HFC | 54.94 | 30.31 | 461 | 1.81 | | 32803 | 38.81 | 24.03 | 1.62 |
| 79611 | MOV | 42.32 | 23.34 | 237 | 1.81 | | 79611 | 33.09 | 17.15 | 1.93 |
| 85875 | IBOC | 27.93 | 15.40 | 58 | 1.81 | | 85875 | 24.84 | 6.87 | 3.62 |
| 84226 | | 12.58 | 6.94 | 53 | 1.81 | | 84226 | 11.20 | 4.85 | 2.31 |
| 38703 | WFC | 21.78 | 12.00 | 56 | 1.82 | | 38703 | 17.72 | 5.95 | 2.98 |
| 86964 | FFIV | 53.43 | 29.31 | 609 | 1.82 | | 86964 | 61.53 | 32.54 | 1.89 |
| 90607 | BME | 13.68 | 7.50 | 83 | 1.82 | | 90607 | 16.33 | 7.38 | 2.21 |
| 78664 | DDD | 87.96 | 48.24 | 284 | 1.82 | | 78664 | 114.28 | 54.17 | 2.11 |
| 49429 | DDS | 78.88 | 43.25 | 1419 | 1.82 | | 49429 | 117.18 | 56.43 | 2.08 |
| 89416 | HOFT | 35.08 | 19.24 | 244 | 1.82 | | 89416 | 30.96 | 17.26 | 1.79 |
| 91743 | | 12.50 | 6.85 | 31 | 1.82 | | 91743 | 15.93 | 3.84 | 4.15 |
| 87952 | GBCI | 25.55 | 14.00 | 108 | 1.82 | | 87952 | 36.98 | 12.13 | 3.05 |
| 14795 | TKR | 35.89 | 19.67 | 170 | 1.83 | | 14795 | 38.30 | 15.75 | 2.43 |
| 29647 | BCE | 14.97 | 8.19 | 93 | 1.83 | | 29647 | 16.84 | 8.00 | 2.10 |
| 15202 | VMC | 22.06 | 12.07 | 42 | 1.83 | | 15202 | 24.65 | 6.15 | 4.01 |
| 75154 | CCL | 22.07 | 12.07 | 103 | 1.83 | | 75154 | 23.85 | 9.89 | 2.41 |
| 88615 | ITGR | 37.51 | 20.52 | 209 | 1.83 | | 88615 | 41.57 | 18.55 | 2.24 |
| 20117 | RAMP | 40.86 | 22.35 | 376 | 1.83 | | 20117 | 45.39 | 24.84 | 1.83 |
| 90734 | LHCG | 34.52 | 18.88 | 161 | 1.83 | | 90734 | 36.64 | 16.74 | 2.19 |
| 42922 | HURC | 29.82 | 16.30 | 198 | 1.83 | | 42922 | 25.44 | 14.11 | 1.80 |
| 68929 | ASG | 18.91 | 10.33 | 69 | 1.83 | | 68929 | 22.10 | 7.40 | 2.99 |
| 90603 | PFBC | 38.37 | 20.97 | 44 | 1.83 | | 90603 | 46.02 | 14.64 | 3.14 |
| 92469 | SATS | 29.44 | 16.09 | 147 | 1.83 | | 92469 | 32.67 | 14.13 | 2.31 |
| 92666 | RWO | 14.77 | 8.07 | 72 | 1.83 | | 92666 | 11.95 | 6.19 | 1.93 |
| 59185 | ODC | 21.93 | 11.97 | 41 | 1.83 | | 59185 | 25.52 | 6.43 | 3.97 |
| 87014 | BSRR | 23.35 | 12.74 | 14 | 1.83 | | 87014 | 34.45 | 8.03 | 4.29 |
| 31238 | | 15.07 | 8.21 | 86 | 1.84 | | 31238 | 13.70 | 7.23 | 1.89 |
| 35685 | FFBC | 28.18 | 15.35 | 91 | 1.84 | | 35685 | 22.17 | 8.55 | 2.59 |
| 47466 | NYT | 38.89 | 21.18 | 204 | 1.84 | | 47466 | 37.34 | 16.67 | 2.24 |
| 85372 | FARO | 39.44 | 21.46 | 141 | 1.84 | | 85372 | 36.13 | 15.50 | 2.33 |
| 83218 | EWL | 14.82 | 8.06 | 70 | 1.84 | | 83218 | 13.93 | 6.28 | 2.22 |
| 80688 | HIW | 14.17 | 7.71 | 41 | 1.84 | | 80688 | 14.69 | 4.51 | 3.26 |
| 10308 | TMP | 16.20 | 8.81 | 42 | 1.84 | | 10308 | 26.72 | 6.29 | 4.25 |
| 87510 | PFSW | 72.00 | 39.15 | 603 | 1.84 | | 87510 | 89.74 | 42.16 | 2.13 |
| 27888 | CFR | 19.83 | 10.78 | 74 | 1.84 | | 27888 | 21.30 | 7.43 | 2.87 |
| 91133 | BMA | 61.92 | 33.66 | 309 | 1.84 | | 91133 | 71.34 | 35.14 | 2.03 |
| 88504 | BRKR | 42.46 | 23.08 | 637 | 1.84 | | 88504 | 64.30 | 32.24 | 1.99 |
| 77235 | WTT | 43.86 | 23.84 | 420 | 1.84 | | 77235 | 43.03 | 24.51 | 1.76 |
| 19880 | MTZ | 47.19 | 25.62 | 250 | 1.84 | | 19880 | 43.13 | 19.54 | 2.21 |

| | | | | | | | | | |
|---|---|---|---|---|---|---|---|---|---|
| 86578 | GDEN | 45.51 | 24.71 | 99 | 1.84 | 86578 | 64.95 | 20.91 | 3.11 |
| 77679 | COLB | 34.24 | 18.59 | 204 | 1.84 | 77679 | 23.96 | 14.00 | 1.71 |
| 75885 | PDT | 17.00 | 9.22 | 95 | 1.84 | 75885 | 17.23 | 8.16 | 2.11 |
| 10932 | WBS | 57.68 | 31.30 | 258 | 1.84 | 10932 | 27.97 | 16.40 | 1.71 |
| 91850 | CQP | 45.14 | 24.47 | 873 | 1.84 | 91850 | 77.62 | 37.87 | 2.05 |
| 12211 | LNN | 28.32 | 15.34 | 203 | 1.85 | 12211 | 21.36 | 13.45 | 1.59 |
| 67029 | ITI | 38.33 | 20.75 | 141 | 1.85 | 67029 | 42.54 | 16.47 | 2.58 |
| 89842 | EVT | 15.12 | 8.18 | 63 | 1.85 | 89842 | 15.45 | 6.03 | 2.56 |
| 86121 | GIL | 42.16 | 22.82 | 416 | 1.85 | 86121 | 45.72 | 24.77 | 1.85 |
| 14526 | APOG | 35.89 | 19.42 | 188 | 1.85 | 14526 | 37.40 | 16.21 | 2.31 |
| 91468 | ICFI | 24.91 | 13.48 | 164 | 1.85 | 91468 | 23.74 | 12.26 | 1.94 |
| 88174 | LPSN | 65.60 | 35.45 | 936 | 1.85 | 88174 | 90.78 | 44.48 | 2.04 |
| 82287 | | 54.51 | 29.43 | 938 | 1.85 | 82287 | 78.39 | 41.88 | 1.87 |
| 55597 | MTRN | 35.92 | 19.38 | 254 | 1.85 | 55597 | 40.59 | 19.33 | 2.10 |
| 77191 | TTEK | 20.57 | 11.09 | 114 | 1.86 | 77191 | 22.52 | 9.70 | 2.32 |
| 91416 | HBI | 49.79 | 26.83 | 293 | 1.86 | 91416 | 49.20 | 23.50 | 2.09 |
| 17005 | CVS | 21.19 | 11.42 | 128 | 1.86 | 17005 | 20.62 | 10.29 | 2.00 |
| 80343 | NNBR | 94.02 | 50.57 | 193 | 1.86 | 80343 | 84.87 | 37.47 | 2.26 |
| 91673 | MLCO | 65.52 | 35.22 | 456 | 1.86 | 91673 | 61.70 | 32.34 | 1.91 |
| 84172 | NUS | 58.76 | 31.58 | 488 | 1.86 | 84172 | 100.20 | 45.72 | 2.19 |
| 86408 | LGF | 48.61 | 26.11 | 193 | 1.86 | 86408 | 45.90 | 19.43 | 2.36 |
| 83835 | TD | 23.02 | 12.36 | 177 | 1.86 | 83835 | 26.03 | 13.14 | 1.98 |
| 89655 | NVEC | 24.08 | 12.93 | 235 | 1.86 | 89655 | 23.16 | 14.92 | 1.55 |
| 55976 | WMT | 14.67 | 7.87 | 66 | 1.86 | 55976 | 18.44 | 6.70 | 2.75 |
| 92052 | TTGT | 50.08 | 26.87 | 183 | 1.86 | 92052 | 34.59 | 15.70 | 2.20 |
| 79116 | TCBK | 20.85 | 11.17 | 35 | 1.87 | 79116 | 28.16 | 6.16 | 4.57 |
| 62033 | RAVN | 36.27 | 19.40 | 200 | 1.87 | 62033 | 37.51 | 17.94 | 2.09 |
| 53604 | GRC | 20.91 | 11.18 | 63 | 1.87 | 53604 | 16.11 | 6.05 | 2.66 |
| 90829 | CF | 36.55 | 19.54 | 343 | 1.87 | 90829 | 33.39 | 20.20 | 1.65 |
| 86544 | BCOR | 51.36 | 27.45 | 253 | 1.87 | 86544 | 40.90 | 21.05 | 1.94 |
| 77780 | CENT | 45.59 | 24.36 | 488 | 1.87 | 77780 | 53.59 | 28.15 | 1.90 |
| 91402 | GTLS | 67.91 | 36.25 | 512 | 1.87 | 91402 | 54.84 | 34.44 | 1.59 |
| 90199 | CBRE | 58.49 | 31.20 | 827 | 1.87 | 90199 | 67.47 | 35.45 | 1.90 |
| 79013 | SYBT | 20.94 | 11.17 | 79 | 1.88 | 79013 | 33.50 | 9.93 | 3.37 |
| 92029 | TRS | 97.95 | 52.21 | 2313 | 1.88 | 92029 | 131.42 | 67.20 | 1.96 |
| 90720 | BLDR | 67.08 | 35.75 | 723 | 1.88 | 90720 | 84.93 | 45.13 | 1.88 |
| 89353 | JBLU | 40.93 | 21.81 | 126 | 1.88 | 89353 | 36.71 | 13.51 | 2.72 |
| 86223 | EPD | 20.34 | 10.83 | 137 | 1.88 | 86223 | 22.89 | 11.21 | 2.04 |
| 83272 | HUBG | 18.59 | 9.89 | 40 | 1.88 | 83272 | 18.37 | 4.84 | 3.80 |
| 11691 | MATX | 25.90 | 13.78 | 144 | 1.88 | 11691 | 18.06 | 10.71 | 1.69 |
| 92108 | BX | 38.75 | 20.62 | 364 | 1.88 | 92108 | 42.92 | 22.04 | 1.95 |
| 85593 | AMG | 29.91 | 15.91 | 132 | 1.88 | 85593 | 39.79 | 15.98 | 2.49 |
| 61269 | PKOH | 76.83 | 40.87 | 397 | 1.88 | 61269 | 96.56 | 38.14 | 2.53 |

| | | | | | | | | | | |
|---|---|---|---|---|---|---|---|---|---|---|
| 75905 | GIII | 86.93 | 46.22 | 773 | 1.88 | | 75905 | 83.78 | 41.71 | 2.01 |
| 46690 | CBT | 38.13 | 20.26 | 181 | 1.88 | | 46690 | 31.06 | 14.71 | 2.11 |
| 80711 | AIV | 42.76 | 22.72 | 280 | 1.88 | | 80711 | 24.14 | 16.39 | 1.47 |
| 83215 | EWH | 15.93 | 8.46 | 118 | 1.88 | | 83215 | 21.47 | 9.90 | 2.17 |
| 88439 | ENTG | 84.09 | 44.63 | 1174 | 1.88 | | 88439 | 43.89 | 34.19 | 1.28 |
| 90306 | FFA | 12.31 | 6.53 | 46 | 1.88 | | 90306 | 15.79 | 4.96 | 3.18 |
| 83989 | GTN | 147.54 | 78.27 | 3585 | 1.89 | | 83989 | 191.67 | 92.82 | 2.06 |
| 92247 | FFR | 14.34 | 7.61 | 71 | 1.89 | | 92247 | 12.09 | 6.12 | 1.97 |
| 87299 | AKAM | 43.17 | 22.90 | 305 | 1.89 | | 87299 | 36.60 | 19.95 | 1.83 |
| 17523 | | 67.21 | 35.59 | 996 | 1.89 | | 17523 | 86.74 | 43.36 | 2.00 |
| 65700 | USLM | 23.11 | 12.23 | 196 | 1.89 | | 65700 | 25.87 | 14.33 | 1.81 |
| 11995 | COKE | 36.43 | 19.26 | 286 | 1.89 | | 11995 | 33.55 | 17.78 | 1.89 |
| 91900 | GLUU | 87.11 | 46.04 | 1514 | 1.89 | | 91900 | 62.48 | 45.65 | 1.37 |
| 86004 | UMPQ | 19.30 | 10.20 | 10 | 1.89 | | 86004 | 23.86 | 3.04 | 7.85 |
| 92398 | AIA | 18.31 | 9.68 | 119 | 1.89 | | 92398 | 25.18 | 10.59 | 2.38 |
| 10182 | | 28.47 | 15.04 | 359 | 1.89 | | 10182 | 50.70 | 22.79 | 2.22 |
| 83210 | EWK | 16.68 | 8.81 | 87 | 1.89 | | 83210 | 21.43 | 8.39 | 2.55 |
| 88301 | IYM | 20.80 | 10.98 | 126 | 1.89 | | 88301 | 24.69 | 10.89 | 2.27 |
| 89284 | BCH | 27.83 | 14.69 | 222 | 1.89 | | 89284 | 32.59 | 16.25 | 2.01 |
| 18622 | GAM | 14.56 | 7.67 | 63 | 1.90 | | 18622 | 16.39 | 6.15 | 2.66 |
| 90066 | UTF | 16.12 | 8.49 | 92 | 1.90 | | 90066 | 21.12 | 8.46 | 2.50 |
| 11285 | FBNC | 27.67 | 14.57 | 78 | 1.90 | | 11285 | 23.35 | 8.32 | 2.81 |
| 92648 | CFX | 43.54 | 22.89 | 101 | 1.90 | | 92648 | 43.41 | 16.62 | 2.61 |
| 91426 | | 32.27 | 16.95 | 261 | 1.90 | | 91426 | 36.93 | 18.02 | 2.05 |
| 86432 | PB | 21.93 | 11.51 | 111 | 1.90 | | 86432 | 25.63 | 10.19 | 2.52 |
| 47715 | KBAL | 36.93 | 19.39 | 134 | 1.90 | | 47715 | 50.54 | 16.72 | 3.02 |
| 88837 | GRMN | 26.93 | 14.13 | 230 | 1.91 | | 88837 | 23.45 | 14.96 | 1.57 |
| 87251 | RDWR | 64.14 | 33.64 | 743 | 1.91 | | 87251 | 70.48 | 36.57 | 1.93 |
| 81055 | COF | 37.84 | 19.82 | 137 | 1.91 | | 81055 | 19.00 | 10.59 | 1.79 |
| 92128 | JTD | 15.24 | 7.97 | 54 | 1.91 | | 92128 | 17.62 | 5.80 | 3.04 |
| 11043 | PMD | 33.10 | 17.25 | 146 | 1.92 | | 11043 | 49.28 | 16.50 | 2.99 |
| 92602 | PM | 18.98 | 9.89 | 53 | 1.92 | | 92602 | 18.68 | 6.12 | 3.05 |
| 75846 | STAR | 51.45 | 26.75 | 311 | 1.92 | | 75846 | 70.77 | 27.66 | 2.56 |
| 35124 | CLGX | 30.72 | 15.97 | 101 | 1.92 | | 35124 | 39.41 | 12.46 | 3.16 |
| 51086 | WGO | 53.22 | 27.65 | 301 | 1.92 | | 51086 | 64.70 | 31.68 | 2.04 |
| 90913 | CMPR | 42.98 | 22.33 | 456 | 1.92 | | 90913 | 68.22 | 30.27 | 2.25 |
| 85796 | SCS | 30.04 | 15.61 | 164 | 1.92 | | 85796 | 33.43 | 14.40 | 2.32 |
| 90837 | EFG | 13.09 | 6.80 | 52 | 1.93 | | 90837 | 15.38 | 5.30 | 2.90 |
| 59045 | KEX | 28.67 | 14.89 | 146 | 1.93 | | 59045 | 29.99 | 13.18 | 2.28 |
| 75853 | ZIXI | 42.18 | 21.88 | 382 | 1.93 | | 75853 | 53.78 | 25.86 | 2.08 |
| 92280 | PDN | 17.38 | 9.01 | 96 | 1.93 | | 92280 | 21.04 | 8.75 | 2.40 |
| 77860 | JEQ | 13.90 | 7.20 | 31 | 1.93 | | 77860 | 17.52 | 4.17 | 4.20 |
| 84062 | BJRI | 44.96 | 23.28 | 370 | 1.93 | | 84062 | 41.33 | 22.81 | 1.81 |
| 86949 | SKX | 85.62 | 44.27 | 436 | 1.93 | | 86949 | 59.36 | 31.63 | 1.88 |
| 53197 | MEI | 64.47 | 33.29 | 246 | 1.94 | | 53197 | 80.21 | 28.29 | 2.84 |
| 79545 | BWA | 40.95 | 21.15 | 219 | 1.94 | | 79545 | 45.72 | 19.27 | 2.37 |
| 80054 | MED | 102.40 | 52.84 | 2165 | 1.94 | | 80054 | 142.56 | 69.01 | 2.07 |
| 80415 | EXP | 42.70 | 22.02 | 232 | 1.94 | | 80415 | 49.01 | 21.07 | 2.33 |
| 86929 | AUDC | 73.78 | 38.01 | 468 | 1.94 | | 86929 | 60.65 | 31.10 | 1.95 |
| 88417 | ACLS | 96.19 | 49.53 | 773 | 1.94 | | 88417 | 77.88 | 44.67 | 1.74 |
| 79791 | CNTY | 38.02 | 19.57 | 625 | 1.94 | | 79791 | 56.09 | 30.25 | 1.85 |
| 30437 | CET | 15.24 | 7.84 | 72 | 1.94 | | 30437 | 14.75 | 6.52 | 2.26 |
| 51263 | MTW | 52.77 | 27.16 | 88 | 1.94 | | 51263 | 47.89 | 17.43 | 2.75 |
| 78405 | RCI | 15.08 | 7.76 | 70 | 1.94 | | 78405 | 13.77 | 6.28 | 2.19 |
| 88233 | NVS | 16.58 | 8.53 | 72 | 1.94 | | 88233 | 12.36 | 6.26 | 1.97 |
| 10629 | FNB | 13.89 | 7.14 | -25 | 1.94 | | 10629 | 26.71 | 0.72 | 36.87 |
| 86435 | FNHC | 65.63 | 33.72 | 330 | 1.95 | | 86435 | 63.80 | 27.31 | 2.34 |
| 69550 | | 30.17 | 15.50 | 177 | 1.95 | | 69550 | 37.10 | 16.06 | 2.31 |
| 80928 | DAR | 37.19 | 19.09 | 250 | 1.95 | | 80928 | 30.82 | 17.62 | 1.75 |
| 79235 | AKR | 17.26 | 8.86 | 67 | 1.95 | | 79235 | 14.98 | 6.19 | 2.42 |
| 52250 | GENC | 30.88 | 15.85 | 135 | 1.95 | | 52250 | 37.10 | 13.36 | 2.78 |
| 85427 | GPI | 37.48 | 19.23 | 390 | 1.95 | | 85427 | 53.33 | 24.78 | 2.15 |
| 87487 | BGCP | 42.75 | 21.91 | 204 | 1.95 | | 87487 | 48.64 | 21.77 | 2.23 |
| 89572 | CEVA | 49.03 | 25.13 | 216 | 1.95 | | 89572 | 44.30 | 21.64 | 2.05 |
| 85440 | SRI | 78.17 | 40.02 | 441 | 1.95 | | 85440 | 60.66 | 30.85 | 1.97 |
| 88790 | | 69.27 | 35.46 | 252 | 1.95 | | 88790 | 44.84 | 20.38 | 2.20 |
| 43757 | | 100.69 | 51.53 | 739 | 1.95 | | 43757 | 109.79 | 45.51 | 2.41 |
| 92326 | CVI | 70.06 | 35.84 | 762 | 1.95 | | 92326 | 64.02 | 36.02 | 1.78 |
| 89044 | PDFS | 66.33 | 33.92 | 485 | 1.96 | | 89044 | 74.63 | 38.31 | 1.95 |
| 89728 | CUK | 23.52 | 12.03 | 115 | 1.96 | | 89728 | 26.30 | 11.00 | 2.39 |
| 92644 | RDOG | 13.91 | 7.11 | 65 | 1.96 | | 92644 | 12.16 | 5.78 | 2.10 |
| 83225 | EWJ | 11.66 | 5.95 | 32 | 1.96 | | 83225 | 14.14 | 3.72 | 3.80 |
| 82598 | NTAP | 45.21 | 23.04 | 327 | 1.96 | | 82598 | 53.83 | 24.94 | 2.16 |
| 81220 | NSIT | 55.21 | 28.13 | 491 | 1.96 | | 81220 | 24.30 | 21.44 | 1.13 |
| 91392 | WYND | 76.39 | 38.90 | 1137 | 1.96 | | 91392 | 65.27 | 38.97 | 1.67 |
| 56223 | LPX | 70.72 | 35.90 | 1324 | 1.97 | | 56223 | 113.28 | 53.02 | 2.14 |
| 92056 | FNI | 26.02 | 13.20 | 163 | 1.97 | | 92056 | 32.80 | 14.13 | 2.32 |
| 85710 | POWI | 30.11 | 15.27 | 207 | 1.97 | | 85710 | 35.09 | 16.09 | 2.18 |
| 91103 | UAL | 99.38 | 50.38 | 660 | 1.97 | | 91103 | 36.96 | 27.30 | 1.35 |
| 83671 | TBI | 27.98 | 14.17 | 132 | 1.97 | | 83671 | 29.27 | 12.08 | 2.42 |
| 72338 | NHC | 14.79 | 7.47 | 55 | 1.98 | | 72338 | 20.20 | 6.36 | 3.18 |
| 83222 | EWN | 17.50 | 8.84 | 76 | 1.98 | | 83222 | 19.90 | 7.49 | 2.66 |
| 87006 | UTHR | 35.92 | 18.11 | 248 | 1.98 | | 87006 | 42.30 | 19.17 | 2.21 |
| 89189 | JPXN | 11.85 | 5.97 | 31 | 1.98 | | 89189 | 14.67 | 3.72 | 3.94 |
| 15318 | ASB | 16.55 | 8.33 | -5 | 1.99 | | 15318 | 28.20 | 3.42 | 8.24 |
| 89548 | DS | 115.42 | 58.11 | 1037 | 1.99 | | 89548 | 84.18 | 45.92 | 1.83 |
| 83120 | CMCO | 36.01 | 18.12 | 121 | 1.99 | | 83120 | 37.43 | 14.27 | 2.62 |
| 75065 | HQH | 20.82 | 10.47 | 69 | 1.99 | | 75065 | 23.26 | 7.59 | 3.06 |
| 49656 | BK | 18.21 | 9.15 | 66 | 1.99 | | 49656 | 20.24 | 7.14 | 2.83 |
| 89778 | CHT | 10.48 | 5.25 | 53 | 2.00 | | 89778 | 8.77 | 4.68 | 1.87 |
| 18163 | PG | 10.44 | 5.23 | 49 | 2.00 | | 18163 | 8.75 | 4.38 | 2.00 |
| 27684 | AVT | 17.03 | 8.52 | 98 | 2.00 | | 27684 | 25.77 | 9.44 | 2.73 |
| 88545 | | 39.08 | 19.53 | 30 | 2.00 | | 88545 | 40.47 | 9.57 | 4.23 |
| 57808 | VSH | 51.71 | 25.82 | 489 | 2.00 | | 57808 | 54.56 | 28.62 | 1.91 |
| 84372 | NCR | 30.95 | 15.44 | 63 | 2.00 | | 84372 | 35.27 | 9.94 | 3.55 |
| 47706 | FSS | 39.16 | 19.54 | 142 | 2.00 | | 47706 | 42.62 | 15.79 | 2.70 |
| 80185 | PLT | 38.44 | 19.17 | 151 | 2.01 | | 80185 | 36.09 | 14.20 | 2.54 |
| 89533 | WYNN | 56.12 | 27.97 | 134 | 2.01 | | 89533 | 51.30 | 19.87 | 2.58 |
| 92274 | MOO | 16.61 | 8.28 | 104 | 2.01 | | 92274 | 20.97 | 9.05 | 2.32 |
| 90751 | DBI | 49.82 | 24.80 | 296 | 2.01 | | 90751 | 41.00 | 20.66 | 1.98 |
| 76837 | HAE | 37.57 | 18.68 | 254 | 2.01 | | 76837 | 27.60 | 16.21 | 1.70 |
| 70797 | USA | 15.16 | 7.53 | 54 | 2.01 | | 70797 | 15.94 | 5.49 | 2.90 |
| 79895 | BBAR | 67.16 | 33.34 | 330 | 2.01 | | 79895 | 64.01 | 32.40 | 1.98 |
| 79782 | REG | 15.90 | 7.89 | 26 | 2.02 | | 79782 | 19.30 | 3.91 | 4.94 |
| 77971 | | 34.73 | 17.22 | 159 | 2.02 | | 77971 | 37.52 | 14.77 | 2.54 |
| 90063 | CSQ | 14.36 | 7.12 | 69 | 2.02 | | 90063 | 16.42 | 6.51 | 2.52 |
| 76838 | PRA | 16.48 | 8.16 | 54 | 2.02 | | 76838 | 16.03 | 5.60 | 2.86 |
| 92195 | | 16.36 | 8.10 | 40 | 2.02 | | 92195 | 17.48 | 4.66 | 3.75 |
| 86822 | EXTR | 60.01 | 29.67 | 161 | 2.02 | | 86822 | 60.04 | 22.83 | 2.63 |
| 85176 | DXPE | 56.94 | 28.15 | 91 | 2.02 | | 85176 | 61.21 | 21.50 | 2.85 |
| 89641 | STX | 71.25 | 35.18 | 771 | 2.02 | | 89641 | 101.89 | 45.20 | 2.25 |
| 58246 | NTRS | 16.63 | 8.20 | 60 | 2.03 | | 58246 | 17.60 | 6.29 | 2.80 |
| 80128 | VMI | 22.64 | 11.17 | 81 | 2.03 | | 80128 | 25.32 | 8.90 | 2.84 |
| 77520 | AGCO | 26.42 | 13.04 | 136 | 2.03 | | 77520 | 26.61 | 11.90 | 2.24 |
| 88224 | IYZ | 12.87 | 6.35 | 60 | 2.03 | | 88224 | 14.85 | 5.76 | 2.58 |
| 16678 | KR | 28.78 | 14.18 | 108 | 2.03 | | 16678 | 29.33 | 10.93 | 2.68 |
| 10866 | CAL | 56.65 | 27.89 | 229 | 2.03 | | 10866 | 41.12 | 18.63 | 2.21 |
| 83683 | PROV | 35.96 | 17.70 | 243 | 2.03 | | 83683 | 59.76 | 23.36 | 2.56 |
| 91798 | UYM | 45.87 | 22.58 | 229 | 2.03 | | 91798 | 50.13 | 21.75 | 2.31 |
| 79734 | RVSB | 37.10 | 18.26 | 224 | 2.03 | | 79734 | 33.36 | 16.73 | 1.99 |
| 80307 | BHC | 70.15 | 34.45 | 95 | 2.04 | | 80307 | 57.72 | 27.91 | 2.07 |
| 79547 | EQR | 28.57 | 14.00 | 121 | 2.04 | | 79547 | 21.95 | 10.13 | 2.17 |
| 79464 | SHI | 30.37 | 14.86 | 147 | 2.04 | | 79464 | 27.30 | 12.87 | 2.12 |
| 85908 | GRA | 70.79 | 34.63 | 1258 | 2.04 | | 85908 | 100.27 | 46.93 | 2.14 |
| 85002 | SRPT | 164.13 | 80.12 | 2664 | 2.05 | | 85002 | 157.58 | 86.61 | 1.82 |
| 19502 | WBA | 25.08 | 12.20 | 177 | 2.06 | | 19502 | 24.81 | 13.04 | 1.90 |
| 90303 | CNS | 30.08 | 14.62 | 212 | 2.06 | | 90303 | 38.52 | 17.08 | 2.26 |
| 89574 | PRAA | 40.48 | 19.65 | 116 | 2.06 | | 89574 | 37.11 | 13.71 | 2.71 |
| 10547 | CLFD | 88.48 | 42.94 | 854 | 2.06 | | 10547 | 121.30 | 56.77 | 2.14 |
| 85035 | QRVO | 85.05 | 41.27 | 1846 | 2.06 | | 85035 | 169.63 | 75.65 | 2.24 |
| 91757 | NIE | 12.19 | 5.90 | 53 | 2.06 | | 91757 | 15.89 | 5.39 | 2.95 |
| 84780 | KNL | 29.38 | 14.23 | 83 | 2.06 | | 84780 | 28.87 | 9.49 | 3.04 |
| 22103 | EMR | 20.66 | 10.00 | 63 | 2.06 | | 22103 | 21.79 | 7.10 | 3.07 |
| 78756 | OFIX | 31.90 | 15.43 | 242 | 2.07 | | 78756 | 39.21 | 18.83 | 2.08 |
| 83815 | DCOM | 15.54 | 7.52 | 28 | 2.07 | | 83815 | 15.09 | 3.48 | 4.33 |
| 14816 | TR | 14.97 | 7.24 | 75 | 2.07 | | 14816 | 14.88 | 6.70 | 2.22 |
| 64785 | AGX | 44.20 | 21.38 | 247 | 2.07 | | 64785 | 48.10 | 21.60 | 2.23 |
| 78903 | PEBO | 30.32 | 14.65 | 57 | 2.07 | | 78903 | 37.88 | 10.79 | 3.51 |
| 10032 | PLXS | 34.02 | 16.43 | 201 | 2.07 | | 10032 | 34.34 | 15.83 | 2.17 |
| 57904 | AFL | 30.30 | 14.64 | 99 | 2.07 | | 57904 | 16.82 | 8.37 | 2.01 |
| 71175 | UNM | 27.62 | 13.33 | 58 | 2.07 | | 71175 | 30.74 | 8.86 | 3.47 |
| 90178 | ALNY | 91.49 | 44.17 | 195 | 2.07 | | 90178 | 117.98 | 47.05 | 2.51 |
| 21371 | CAH | 24.13 | 11.64 | 81 | 2.07 | | 21371 | 26.44 | 8.90 | 2.97 |
| 89977 | | 15.01 | 7.24 | 73 | 2.07 | | 89977 | 15.22 | 6.72 | 2.26 |
| 90102 | ABR | 67.47 | 32.52 | 241 | 2.07 | | 90102 | 68.19 | 25.13 | 2.71 |
| 54594 | AIR | 34.00 | 16.37 | 103 | 2.08 | | 54594 | 22.56 | 9.50 | 2.38 |
| 84413 | CERS | 71.42 | 34.36 | 624 | 2.08 | | 84413 | 65.48 | 33.33 | 1.96 |
| 90081 | CUTR | 57.87 | 27.84 | 92 | 2.08 | | 90081 | 57.01 | 17.64 | 3.23 |
| 83411 | EPM | 54.18 | 26.04 | 468 | 2.08 | | 83411 | 91.77 | 39.07 | 2.35 |
| 92537 | BJK | 23.36 | 11.22 | 88 | 2.08 | | 92537 | 29.01 | 10.14 | 2.86 |
| 31077 | CUB | 20.08 | 9.65 | 98 | 2.08 | | 31077 | 16.27 | 8.11 | 2.01 |
| 92303 | FFNW | 25.68 | 12.32 | 66 | 2.08 | | 92303 | 30.93 | 9.58 | 3.23 |
| 79564 | | 31.56 | 15.14 | 285 | 2.08 | | 79564 | 25.16 | 16.75 | 1.50 |
| 84207 | ESLT | 25.90 | 12.41 | 145 | 2.09 | | 84207 | 27.92 | 12.53 | 2.23 |
| 90998 | SPXX | 10.50 | 5.03 | 30 | 2.09 | | 90998 | 13.03 | 3.45 | 3.77 |
| 91365 | CSII | 74.95 | 35.90 | 206 | 2.09 | | 91365 | 77.93 | 30.47 | 2.56 |
| 89908 | CONN | 105.20 | 50.37 | 155 | 2.09 | | 89908 | 103.87 | 46.81 | 2.22 |
| 77274 | GILD | 41.45 | 19.84 | 145 | 2.09 | | 77274 | 43.69 | 15.62 | 2.80 |
| 91516 | AWI | 37.73 | 18.05 | 208 | 2.09 | | 91516 | 28.13 | 14.55 | 1.93 |
| 87321 | GAIA | 42.77 | 20.42 | 124 | 2.09 | | 87321 | 47.74 | 17.64 | 2.71 |
| 79244 | | 48.12 | 22.96 | 431 | 2.10 | | 79244 | 60.11 | 28.07 | 2.14 |
| 76732 | DIN | 54.26 | 25.89 | 483 | 2.10 | | 76732 | 50.10 | 27.76 | 1.80 |
| 42796 | MDP | 26.78 | 12.78 | 203 | 2.10 | | 42796 | 31.95 | 15.42 | 2.07 |
| 80254 | | 37.23 | 17.75 | 120 | 2.10 | | 80254 | 34.85 | 12.15 | 2.87 |
| 90850 | STN | 24.08 | 11.47 | 77 | 2.10 | | 90850 | 25.03 | 8.26 | 3.03 |
| 81127 | IRS | 56.26 | 26.78 | 217 | 2.10 | | 81127 | 58.00 | 25.78 | 2.25 |
| 83686 | QGEN | 17.63 | 8.39 | 89 | 2.10 | | 83686 | 19.75 | 8.34 | 2.37 |
| 79265 | FLIR | 24.10 | 11.46 | 42 | 2.10 | | 79265 | 19.40 | 5.11 | 3.80 |
| 50788 | | 37.15 | 17.67 | 221 | 2.10 | | 50788 | 35.01 | 16.93 | 2.07 |
| 90558 | HUN | 73.08 | 34.70 | 461 | 2.11 | | 90558 | 82.20 | 38.70 | 2.12 |
| 76948 | MGIC | 66.42 | 31.54 | 495 | 2.11 | | 76948 | 58.98 | 28.88 | 2.04 |
| 89036 | NTUS | 40.20 | 19.08 | 163 | 2.11 | | 89036 | 40.95 | 16.13 | 2.54 |
| 60506 | PCAR | 21.62 | 10.26 | 100 | 2.11 | | 60506 | 30.30 | 11.29 | 2.69 |
| 85860 | BRKL | 14.27 | 6.77 | 30 | 2.11 | | 85860 | 17.78 | 3.94 | 4.51 |
| 89841 | CNO | 70.91 | 33.54 | 187 | 2.11 | | 89841 | 35.93 | 16.04 | 2.24 |
| 77415 | NBTB | 14.07 | 6.65 | 24 | 2.12 | | 77415 | 22.51 | 4.21 | 5.35 |
| 66683 | RES | 61.58 | 29.08 | 128 | 2.12 | | 66683 | 60.88 | 21.31 | 2.86 |
| 50623 | MTSC | 22.50 | 10.62 | 51 | 2.12 | | 50623 | 20.90 | 6.05 | 3.46 |
| 84519 | CIEN | 36.83 | 17.38 | 406 | 2.12 | | 84519 | 43.08 | 24.91 | 1.73 |
| 80926 | CPE | 87.74 | 41.38 | 150 | 2.12 | | 80926 | 101.54 | 32.31 | 3.14 |
| 72726 | STT | 25.85 | 12.19 | 60 | 2.12 | | 72726 | 25.09 | 7.57 | 3.32 |
| 82518 | RICK | 59.23 | 27.92 | 460 | 2.12 | | 82518 | 45.31 | 25.40 | 1.78 |
| 62228 | FEIM | 30.55 | 14.40 | 267 | 2.12 | | 62228 | 27.39 | 16.45 | 1.66 |
| 80539 | NKTR | 112.26 | 52.92 | 491 | 2.12 | | 80539 | 125.76 | 49.45 | 2.54 |
| 91111 | ET | 44.08 | 20.78 | 226 | 2.12 | | 91111 | 44.12 | 20.68 | 2.13 |
| 71079 | RVT | 19.86 | 9.35 | 52 | 2.12 | | 71079 | 22.50 | 6.53 | 3.44 |
| 47626 | CNA | 36.55 | 17.20 | 169 | 2.13 | | 47626 | 23.82 | 12.56 | 1.90 |
| 45306 | IDCC | 40.78 | 19.17 | 142 | 2.13 | | 45306 | 42.14 | 15.30 | 2.75 |
| 83879 | LAMR | 51.10 | 24.02 | 451 | 2.13 | | 83879 | 47.81 | 25.06 | 1.91 |
| 56822 | NWLI | 25.75 | 12.10 | 78 | 2.13 | | 56822 | 18.12 | 7.26 | 2.49 |
| 91910 | USAT | 77.63 | 36.44 | 84 | 2.13 | | 91910 | 58.62 | 19.50 | 3.01 |
| 65365 | MYE | 29.58 | 13.88 | 89 | 2.13 | | 65365 | 23.49 | 8.99 | 2.61 |
| 92136 | PIO | 16.31 | 7.65 | 74 | 2.13 | | 92136 | 19.24 | 7.29 | 2.64 |
| 88905 | ATRS | 74.16 | 34.77 | 635 | 2.13 | | 88905 | 75.47 | 39.64 | 1.90 |
| 85604 | EPR | 29.48 | 13.82 | 115 | 2.13 | | 85604 | 13.17 | 8.66 | 1.52 |
| 91065 | CPA | 46.33 | 21.72 | 160 | 2.13 | | 91065 | 53.55 | 22.37 | 2.39 |
| 26710 | L | 15.97 | 7.49 | 61 | 2.13 | | 26710 | 14.18 | 5.74 | 2.47 |
| 86876 | WCC | 32.89 | 15.41 | 150 | 2.14 | | 86876 | 41.79 | 16.48 | 2.54 |
| 91683 | SBH | 36.48 | 17.07 | 200 | 2.14 | | 91683 | 34.63 | 15.86 | 2.18 |
| 92210 | | 45.06 | 21.08 | 168 | 2.14 | | 92210 | 40.66 | 16.32 | 2.49 |
| 59504 | BTI | 16.75 | 7.84 | 20 | 2.14 | | 59504 | 21.95 | 4.81 | 4.56 |
| 75233 | VGR | 17.86 | 8.35 | 16 | 2.14 | | 75233 | 25.19 | 5.21 | 4.83 |
| 53225 | GHC | 27.50 | 12.86 | 159 | 2.14 | | 53225 | 27.59 | 12.49 | 2.21 |
| 92690 | UMH | 20.48 | 9.58 | 99 | 2.14 | | 92690 | 22.37 | 9.08 | 2.46 |
| 89927 | TCOM | 70.46 | 32.92 | 355 | 2.14 | | 89927 | 80.15 | 34.15 | 2.35 |
| 44768 | TILE | 51.53 | 24.08 | 207 | 2.14 | | 44768 | 44.52 | 19.77 | 2.25 |
| 75854 | LSCC | 52.19 | 24.38 | 358 | 2.14 | | 75854 | 47.47 | 23.75 | 2.00 |
| 77857 | OHI | 20.45 | 9.55 | 120 | 2.14 | | 77857 | 18.89 | 9.73 | 1.94 |
| 12226 | MGPI | 199.02 | 92.91 | 8415 | 2.14 | | 12226 | 323.99 | 144.12 | 2.25 |
| 76119 | JOF | 14.47 | 6.75 | 18 | 2.14 | | 76119 | 18.31 | 3.24 | 5.65 |
| 92778 | DISCK | 32.01 | 14.93 | 245 | 2.14 | | 92778 | 39.57 | 18.72 | 2.11 |
| 92499 | PIZ | 17.79 | 8.29 | 68 | 2.15 | | 92499 | 21.35 | 7.30 | 2.93 |
| 77462 | M | 38.75 | 18.05 | 188 | 2.15 | | 77462 | 33.64 | 16.55 | 2.03 |
| 72980 | STC | 38.04 | 17.72 | 76 | 2.15 | | 72980 | 44.89 | 12.83 | 3.50 |
| 85449 | FDP | 21.68 | 10.09 | 26 | 2.15 | | 85449 | 25.41 | 5.28 | 4.82 |
| 91937 | TMUS | 43.12 | 20.07 | 114 | 2.15 | | 91937 | 40.93 | 15.23 | 2.69 |
| 47379 | KELYA | 29.52 | 13.68 | 57 | 2.16 | | 47379 | 34.39 | 9.47 | 3.63 |
| 49015 | LNC | 52.05 | 24.11 | 172 | 2.16 | | 49015 | 38.26 | 15.98 | 2.39 |
| 77643 | FCF | 23.69 | 10.98 | -2 | 2.16 | | 77643 | 36.54 | 6.77 | 5.40 |
| 82272 | MD | 26.25 | 12.16 | 108 | 2.16 | | 82272 | 34.32 | 12.05 | 2.85 |
| 60709 | | 31.87 | 14.75 | 190 | 2.16 | | 60709 | 38.96 | 16.76 | 2.32 |
| 86349 | IX | 38.16 | 17.66 | 152 | 2.16 | | 86349 | 27.90 | 12.90 | 2.16 |
| 91349 | | 76.20 | 35.21 | 101 | 2.16 | | 91349 | 69.36 | 31.95 | 2.17 |
| 84210 | FORR | 18.44 | 8.51 | 58 | 2.17 | | 84210 | 26.73 | 7.60 | 3.52 |
| 35263 | TRMK | 14.04 | 6.47 | 32 | 2.17 | | 35263 | 20.32 | 4.30 | 4.72 |
| 91979 | CAI | 96.65 | 44.55 | 633 | 2.17 | | 91979 | 96.29 | 46.67 | 2.06 |
| 18649 | BCO | 36.63 | 16.87 | 141 | 2.17 | | 18649 | 34.02 | 13.24 | 2.57 |
| 91333 | DIM | 14.88 | 6.85 | 53 | 2.17 | | 91333 | 16.61 | 5.56 | 2.99 |
| 91122 | XSD | 30.02 | 13.82 | 168 | 2.17 | | 91122 | 38.16 | 16.11 | 2.37 |
| 79363 | AZN | 17.55 | 8.08 | 85 | 2.17 | | 79363 | 14.52 | 7.27 | 2.00 |
| 90512 | PGJ | 26.90 | 12.37 | 112 | 2.17 | | 90512 | 35.13 | 12.59 | 2.79 |
| 91277 | QRTEA | 63.09 | 28.89 | 745 | 2.18 | | 91277 | 78.54 | 37.01 | 2.12 |
| 90957 | | 86.16 | 39.45 | 972 | 2.18 | | 90957 | 89.79 | 46.60 | 1.93 |
| 22293 | GLW | 27.36 | 12.53 | 217 | 2.18 | | 22293 | 38.84 | 17.57 | 2.21 |
| 77920 | BLX | 19.65 | 8.99 | 20 | 2.19 | | 77920 | 23.38 | 4.36 | 5.36 |
| 85459 | CHRW | 16.92 | 7.74 | 53 | 2.19 | | 85459 | 19.09 | 5.85 | 3.27 |
| 62341 | PDCE | 44.67 | 20.43 | 24 | 2.19 | | 62341 | 53.62 | 11.72 | 4.57 |
| 88280 | BHLB | 14.96 | 6.83 | -13 | 2.19 | | 88280 | 17.91 | 0.27 | 67.35 |
| 91626 | RNP | 32.80 | 14.97 | 187 | 2.19 | | 91626 | 24.59 | 13.29 | 1.85 |
| 86965 | FISI | 34.19 | 15.61 | 79 | 2.19 | | 86965 | 25.50 | 8.51 | 3.00 |
| 87356 | WWE | 73.32 | 33.46 | 574 | 2.19 | | 87356 | 60.04 | 32.42 | 1.85 |
| 75316 | | 60.84 | 27.76 | 347 | 2.19 | | 75316 | 63.38 | 29.46 | 2.15 |
| 82575 | BANR | 29.74 | 13.57 | -19 | 2.19 | | 82575 | 39.17 | 6.58 | 5.96 |
| 91131 | VPL | 12.59 | 5.74 | 38 | 2.19 | | 91131 | 14.41 | 4.23 | 3.41 |
| 89960 | KRO | 100.83 | 45.96 | 98 | 2.19 | | 89960 | 77.10 | 27.56 | 2.80 |
| 21573 | IP | 45.61 | 20.78 | 247 | 2.20 | | 21573 | 44.41 | 19.57 | 2.27 |
| 82656 | CLB | 33.28 | 15.15 | 99 | 2.20 | | 82656 | 46.79 | 15.62 | 3.00 |
| 89298 | RQI | 35.34 | 16.08 | 173 | 2.20 | | 89298 | 24.38 | 12.77 | 1.91 |
| 88290 | EWT | 18.94 | 8.61 | 108 | 2.20 | | 88290 | 26.86 | 10.34 | 2.60 |
| 80341 | MPAA | 89.68 | 40.72 | 327 | 2.20 | | 80341 | 81.07 | 33.47 | 2.42 |
| 80210 | RYN | 21.69 | 9.84 | 81 | 2.20 | | 80210 | 20.91 | 8.01 | 2.61 |
| 79033 | SGA | 81.39 | 36.94 | 571 | 2.20 | | 79033 | 45.13 | 27.80 | 1.62 |
| 86051 | AXTI | 97.28 | 44.14 | 222 | 2.20 | | 86051 | 92.41 | 38.28 | 2.41 |
| 88988 | IDT | 174.83 | 79.27 | 1533 | 2.21 | | 88988 | 164.91 | 81.57 | 2.02 |
| 20618 | CRS | 35.86 | 16.25 | 73 | 2.21 | | 20618 | 30.71 | 10.02 | 3.07 |
| 92318 | TOWN | 23.58 | 10.67 | 0 | 2.21 | | 92318 | 34.56 | 5.59 | 6.19 |
| 85926 | SEE | 30.61 | 13.82 | 133 | 2.22 | | 85926 | 36.63 | 13.74 | 2.67 |
| 14541 | CVX | 16.48 | 7.44 | 47 | 2.22 | | 14541 | 15.93 | 5.05 | 3.16 |
| 81696 | DISH | 32.77 | 14.77 | 125 | 2.22 | | 81696 | 41.20 | 15.41 | 2.67 |
| 84621 | LOGI | 47.04 | 21.19 | 101 | 2.22 | | 84621 | 39.17 | 14.21 | 2.76 |
| 80329 | FLEX | 41.00 | 18.46 | 197 | 2.22 | | 80329 | 64.05 | 23.97 | 2.67 |
| 88924 | FLO | 21.86 | 9.84 | 71 | 2.22 | | 88924 | 15.36 | 6.42 | 2.39 |
| 87067 | FNLC | 15.49 | 6.97 | 32 | 2.22 | | 87067 | 22.84 | 4.77 | 4.78 |
| 11809 | CLH | 24.32 | 10.94 | 56 | 2.22 | | 11809 | 25.69 | 7.08 | 3.63 |
| 58318 | TEX | 44.55 | 20.03 | 59 | 2.22 | | 58318 | 56.94 | 18.54 | 3.07 |
| 78418 | STKL | 54.56 | 24.53 | 146 | 2.22 | | 78418 | 64.38 | 24.20 | 2.66 |
| 34673 | NR | 53.19 | 23.85 | 86 | 2.23 | | 34673 | 36.66 | 12.29 | 2.98 |
| 59396 | ORI | 23.07 | 10.34 | 73 | 2.23 | | 59396 | 27.93 | 8.78 | 3.18 |
| 90600 | KRNY | 16.58 | 7.43 | 38 | 2.23 | | 90600 | 17.35 | 4.63 | 3.75 |
| 90839 | MGU | 17.33 | 7.76 | 56 | 2.23 | | 90839 | 19.53 | 6.31 | 3.09 |
| 88196 | SINA | 61.74 | 27.61 | 167 | 2.24 | | 88196 | 52.97 | 22.67 | 2.34 |
| 11369 | UBSI | 17.43 | 7.79 | -6 | 2.24 | | 11369 | 26.74 | 2.60 | 10.28 |
| 11891 | MGM | 54.09 | 24.18 | 76 | 2.24 | | 11891 | 43.19 | 12.55 | 3.44 |
| 76672 | RDNT | 84.51 | 37.74 | 204 | 2.24 | | 76672 | 134.13 | 40.48 | 3.31 |
| 11845 | WPP | 25.99 | 11.61 | 85 | 2.24 | | 11845 | 34.71 | 11.29 | 3.08 |
| 85741 | HOPE | 37.24 | 16.62 | 21 | 2.24 | | 85741 | 24.51 | 4.59 | 5.34 |
| 90436 | HALO | 50.20 | 22.39 | 161 | 2.24 | | 90436 | 61.12 | 23.20 | 2.63 |
| 64282 | LB | 45.87 | 20.45 | 156 | 2.24 | | 64282 | 43.29 | 18.42 | 2.35 |
| 79094 | JBL | 42.50 | 18.95 | 267 | 2.24 | | 79094 | 49.44 | 19.82 | 2.49 |
| 85939 | RLH | 34.91 | 15.54 | 245 | 2.25 | | 85939 | 39.51 | 18.22 | 2.17 |
| 26650 | CTO | 19.29 | 8.58 | 37 | 2.25 | | 26650 | 21.70 | 5.06 | 4.29 |
| 75342 | MTH | 33.90 | 15.07 | 202 | 2.25 | | 75342 | 32.35 | 15.84 | 2.04 |
| 76082 | COG | 46.49 | 20.66 | 244 | 2.25 | | 76082 | 45.57 | 21.07 | 2.16 |
| 83916 | TTEC | 30.64 | 13.62 | 242 | 2.25 | | 83916 | 46.90 | 19.47 | 2.41 |
| 21135 | FOE | 100.91 | 44.83 | 122 | 2.25 | | 21135 | 76.73 | 26.71 | 2.87 |
| 89195 | PFG | 39.16 | 17.38 | 96 | 2.25 | | 89195 | 31.85 | 11.15 | 2.86 |
| 77868 | PRCP | 42.27 | 18.76 | 140 | 2.25 | | 77868 | 49.86 | 16.75 | 2.98 |
| 83304 | PRGX | 33.13 | 14.70 | 132 | 2.25 | | 83304 | 28.14 | 12.10 | 2.33 |
| 88659 | ASX | 33.34 | 14.76 | 198 | 2.26 | | 88659 | 52.67 | 19.86 | 2.65 |
| 63125 | AGYS | 41.04 | 18.16 | 234 | 2.26 | | 63125 | 44.14 | 19.58 | 2.25 |
| 80691 | | 16.23 | 7.18 | 83 | 2.26 | | 80691 | 17.04 | 7.43 | 2.29 |
| 64742 | PCYO | 53.21 | 23.53 | 286 | 2.26 | | 64742 | 47.67 | 23.11 | 2.06 |
| 78971 | GILT | 33.50 | 14.78 | 258 | 2.27 | | 78971 | 34.35 | 18.06 | 1.90 |
| 88590 | | 16.20 | 7.14 | 85 | 2.27 | | 88590 | 20.10 | 7.83 | 2.57 |
| 82924 | JW | 19.22 | 8.48 | 32 | 2.27 | | 82924 | 21.97 | 4.98 | 4.41 |
| 91606 | CPRX | 107.43 | 47.38 | 7 | 2.27 | | 91606 | 145.99 | 50.36 | 2.90 |
| 89971 | TPX | 71.65 | 31.59 | 484 | 2.27 | | 89971 | 79.37 | 35.01 | 2.27 |
| 90505 | LVS | 100.22 | 44.17 | 778 | 2.27 | | 90505 | 80.29 | 40.73 | 1.97 |
| 55212 | SPXC | 39.00 | 17.18 | 166 | 2.27 | | 55212 | 56.22 | 21.22 | 2.65 |
| 76932 | ENB | 20.98 | 9.23 | 91 | 2.27 | | 76932 | 24.62 | 9.51 | 2.59 |
| 69586 | SBCF | 28.00 | 12.30 | -21 | 2.28 | | 69586 | 34.63 | 6.23 | 5.56 |
| 88266 | SWIR | 54.83 | 24.07 | 130 | 2.28 | | 88266 | 81.08 | 31.21 | 2.60 |
| 27633 | R | 26.71 | 11.72 | 24 | 2.28 | | 27633 | 29.53 | 6.50 | 4.54 |
| 65947 | HCN | 14.98 | 6.58 | 64 | 2.28 | | 65947 | 15.44 | 6.04 | 2.55 |
| 75819 | VTR | 22.73 | 9.97 | 100 | 2.28 | | 75819 | 14.84 | 8.12 | 1.83 |
| 79122 | AMRN | 169.94 | 74.49 | 1817 | 2.28 | | 79122 | 162.40 | 87.38 | 1.86 |
| 49154 | TGT | 22.01 | 9.64 | 91 | 2.28 | | 49154 | 16.99 | 7.88 | 2.16 |
| 31500 | EV | 23.71 | 10.38 | 67 | 2.28 | | 31500 | 29.67 | 9.25 | 3.21 |
| 89881 | NRO | 29.48 | 12.89 | 129 | 2.29 | | 89881 | 26.13 | 11.21 | 2.33 |
| 89258 | PRU | 38.55 | 16.85 | 169 | 2.29 | | 89258 | 32.99 | 14.45 | 2.28 |
| 91002 | KCE | 20.45 | 8.92 | 82 | 2.29 | | 91002 | 25.01 | 8.76 | 2.85 |
| 87077 | ECPG | 59.29 | 25.87 | 226 | 2.29 | | 87077 | 55.37 | 23.04 | 2.40 |
| 26614 | SJI | 17.40 | 7.58 | 40 | 2.30 | | 26614 | 21.12 | 5.16 | 4.09 |
| 76118 | IRL | 22.71 | 9.89 | 85 | 2.30 | | 76118 | 26.26 | 9.23 | 2.84 |
| 92587 | BEAT | 97.11 | 42.24 | 142 | 2.30 | | 92587 | 92.69 | 36.56 | 2.54 |
| 92096 | JAZZ | 253.34 | ##### | 6323 | 2.30 | | 92096 | 102.64 | 73.82 | 1.39 |
| 89234 | IXG | 19.31 | 8.39 | 62 | 2.30 | | 89234 | 18.67 | 6.46 | 2.89 |
| 90805 | DISCA | 34.89 | 15.15 | 244 | 2.30 | | 90805 | 44.05 | 19.57 | 2.25 |
| 88181 | NVMI | 170.46 | 73.94 | 3828 | 2.31 | | 88181 | 316.31 | 118.17 | 2.68 |
| 82775 | HIG | 46.95 | 20.36 | 171 | 2.31 | | 82775 | 29.30 | 14.25 | 2.06 |
| 92492 | IGF | 11.31 | 4.88 | 32 | 2.32 | | 92492 | 10.73 | 3.32 | 3.24 |
| 89453 | RRGB | 40.59 | 17.52 | 59 | 2.32 | | 89453 | 41.79 | 11.47 | 3.64 |
| 76722 | STFC | 17.40 | 7.51 | 13 | 2.32 | | 76722 | 23.73 | 3.88 | 6.11 |
| 89618 | GLNG | 72.29 | 31.19 | 233 | 2.32 | | 89618 | 71.71 | 28.04 | 2.56 |
| 38156 | WMB | 34.93 | 15.07 | 87 | 2.32 | | 38156 | 27.06 | 10.08 | 2.68 |
| 79864 | VSEC | 41.27 | 17.80 | 52 | 2.32 | | 79864 | 39.32 | 10.15 | 3.87 |
| 87179 | NTCT | 40.12 | 17.28 | 174 | 2.32 | | 87179 | 33.75 | 14.94 | 2.26 |
| 90253 | HEP | 21.37 | 9.19 | 168 | 2.32 | | 90253 | 28.59 | 12.95 | 2.21 |
| 88214 | EWY | 22.95 | 9.85 | 110 | 2.33 | | 88214 | 29.28 | 10.93 | 2.68 |
| 60986 | NWL | 33.66 | 14.43 | 90 | 2.33 | | 60986 | 31.28 | 11.08 | 2.82 |
| 83950 | SPPI | 101.56 | 43.53 | 487 | 2.33 | | 83950 | 124.32 | 53.40 | 2.33 |
| 91090 | CROX | 119.86 | 51.35 | 1995 | 2.33 | | 91090 | 127.59 | 67.30 | 1.90 |
| 58771 | RBC | 22.75 | 9.73 | 84 | 2.34 | | 58771 | 22.38 | 8.47 | 2.64 |
| 88159 | EXEL | 108.61 | 46.41 | 292 | 2.34 | | 88159 | 110.96 | 49.46 | 2.24 |
| 70033 | HOG | 31.42 | 13.42 | 101 | 2.34 | | 70033 | 30.01 | 11.25 | 2.67 |
| 18411 | SO | 10.91 | 4.66 | 19 | 2.34 | | 18411 | 11.95 | 2.33 | 5.12 |
| 78044 | UFI | 85.80 | 36.61 | 170 | 2.34 | | 78044 | 48.60 | 20.18 | 2.41 |
| 83720 | FFG | 81.64 | 34.83 | 325 | 2.34 | | 83720 | 19.56 | 17.02 | 1.15 |
| 77182 | PRGO | 41.89 | 17.87 | 20 | 2.34 | | 77182 | 38.76 | 9.25 | 4.19 |
| 91878 | GXC | 20.75 | 8.84 | 87 | 2.35 | | 91878 | 26.38 | 9.11 | 2.90 |
| 90300 | ACC | 19.96 | 8.50 | 102 | 2.35 | | 90300 | 21.70 | 9.42 | 2.30 |
| 78089 | TCO | 30.04 | 12.79 | 79 | 2.35 | | 78089 | 25.01 | 8.69 | 2.88 |
| 80795 | AMED | 52.43 | 22.26 | 183 | 2.36 | | 80795 | 55.51 | 24.09 | 2.30 |
| 79879 | VIAV | 55.32 | 23.47 | 362 | 2.36 | | 79879 | 46.94 | 23.54 | 1.99 |
| 87066 | EGOV | 32.18 | 13.64 | 171 | 2.36 | | 87066 | 40.20 | 16.48 | 2.44 |
| 47730 | ABM | 20.45 | 8.63 | 69 | 2.37 | | 47730 | 23.04 | 7.47 | 3.08 |
| 90718 | BFIN | 17.66 | 7.44 | 47 | 2.37 | | 90718 | 22.22 | 6.44 | 3.45 |
| 92769 | HOLI | 66.93 | 28.16 | 525 | 2.38 | | 92769 | 106.05 | 41.48 | 2.56 |
| 59600 | TWIN | 70.97 | 29.85 | 114 | 2.38 | | 59600 | 73.71 | 26.17 | 2.82 |
| 82654 | RY | 23.61 | 9.92 | 131 | 2.38 | | 82654 | 28.86 | 11.67 | 2.47 |
| 61241 | AMD | 117.67 | 49.41 | 755 | 2.38 | | 61241 | 142.29 | 64.62 | 2.20 |
| 92782 | MYRG | 26.61 | 11.15 | 182 | 2.39 | | 92782 | 37.74 | 15.80 | 2.39 |
| 89509 | XEC | 52.91 | 22.15 | 130 | 2.39 | | 89509 | 51.31 | 18.76 | 2.74 |
| 91877 | GMF | 19.98 | 8.34 | 99 | 2.39 | | 91877 | 27.32 | 9.92 | 2.75 |
| 77354 | SCHL | 28.54 | 11.89 | 196 | 2.40 | | 77354 | 38.03 | 15.48 | 2.46 |
| 86211 | MSTR | 41.22 | 17.17 | 244 | 2.40 | | 86211 | 51.24 | 20.51 | 2.50 |
| 81132 | STM | 53.51 | 22.29 | 109 | 2.40 | | 81132 | 43.34 | 15.03 | 2.88 |
| 13928 | COP | 24.66 | 10.27 | 55 | 2.40 | | 13928 | 17.25 | 5.92 | 2.91 |
| 90900 | | 36.34 | 15.10 | 231 | 2.41 | | 90900 | 27.50 | 15.07 | 1.82 |
| 87379 | CLCT | 62.85 | 26.11 | 288 | 2.41 | | 87379 | 77.12 | 31.99 | 2.41 |
| 66384 | WDC | 55.40 | 23.00 | 223 | 2.41 | | 66384 | 98.41 | 35.13 | 2.80 |
| 80236 | DSPG | 21.74 | 9.02 | 40 | 2.41 | | 80236 | 33.74 | 8.03 | 4.20 |
| 87394 | IMMR | 48.37 | 20.07 | 52 | 2.41 | | 87394 | 31.07 | 8.40 | 3.70 |
| 67467 | BIG | 27.31 | 11.32 | 100 | 2.41 | | 67467 | 38.94 | 13.18 | 2.96 |
| 82567 | OPK | 50.91 | 21.10 | 86 | 2.41 | | 82567 | 46.04 | 14.63 | 3.15 |
| 91614 | | 38.52 | 15.96 | 24 | 2.41 | | 91614 | 41.52 | 10.00 | 4.15 |
| 85636 | CM | 20.41 | 8.45 | 79 | 2.42 | | 85636 | 23.66 | 8.28 | 2.86 |
| 75938 | GF | 27.45 | 11.36 | 46 | 2.42 | | 75938 | 32.79 | 8.69 | 3.78 |
| 69999 | EEA | 17.76 | 7.34 | 42 | 2.42 | | 69999 | 19.72 | 5.27 | 3.75 |
| 75285 | MAN | 36.15 | 14.94 | 91 | 2.42 | | 75285 | 46.29 | 15.53 | 2.98 |
| 58334 | NWN | 11.33 | 4.68 | 37 | 2.42 | | 58334 | 8.04 | 3.45 | 2.33 |
| 75183 | PCF | 13.54 | 5.57 | 78 | 2.43 | | 75183 | 15.61 | 6.87 | 2.27 |
| 81035 | GGT | 27.41 | 11.29 | 69 | 2.43 | | 81035 | 31.23 | 9.37 | 3.33 |
| 32299 | GV | 186.88 | 76.93 | 495 | 2.43 | | 32299 | 221.33 | 79.70 | 2.78 |
| 78829 | UMBF | 19.19 | 7.90 | 24 | 2.43 | | 78829 | 29.41 | 5.36 | 5.48 |
| 90097 | UCTT | 135.65 | 55.73 | 321 | 2.43 | | 90097 | 99.12 | 44.31 | 2.24 |
| 12076 | SIG | 56.49 | 23.21 | 266 | 2.43 | | 12076 | 74.48 | 29.43 | 2.53 |
| 83148 | LNDC | 30.12 | 12.36 | 80 | 2.44 | | 83148 | 26.16 | 8.42 | 3.11 |
| 59628 | OLP | 57.93 | 23.76 | 175 | 2.44 | | 59628 | 29.14 | 13.29 | 2.19 |
| 89244 | WW | 113.72 | 46.59 | 31 | 2.44 | | 89244 | 97.14 | 22.37 | 4.34 |
| 81284 | BMO | 26.30 | 10.76 | 154 | 2.44 | | 81284 | 35.69 | 13.71 | 2.60 |
| 91945 | SPDW | 14.41 | 5.88 | 41 | 2.45 | | 91945 | 15.98 | 4.60 | 3.47 |
| 76788 | TVTY | 60.19 | 24.57 | 116 | 2.45 | | 76788 | 48.36 | 18.11 | 2.67 |
| 91234 | MWA | 42.60 | 17.36 | 8 | 2.45 | | 91234 | 55.58 | 11.24 | 4.95 |
| 79558 | BFS | 18.21 | 7.42 | 20 | 2.45 | | 79558 | 24.25 | 4.32 | 5.61 |
| 91575 | HRI | 48.28 | 19.66 | 70 | 2.46 | | 91575 | 59.09 | 18.76 | 3.15 |
| 72776 | TSI | 11.96 | 4.87 | 72 | 2.46 | | 72776 | 15.38 | 6.46 | 2.38 |
| 91944 | GWX | 17.04 | 6.93 | 50 | 2.46 | | 91944 | 19.93 | 5.89 | 3.38 |
| 57817 | JWN | 38.51 | 15.66 | 250 | 2.46 | | 57817 | 59.05 | 22.02 | 2.68 |
| 10382 | ASTE | 22.51 | 9.15 | -4 | 2.46 | | 10382 | 29.08 | 3.44 | 8.45 |
| 71810 | DNP | 10.17 | 4.13 | 71 | 2.46 | | 71810 | 17.57 | 6.73 | 2.61 |
| 51131 | SNE | 37.53 | 15.24 | 121 | 2.46 | | 51131 | 35.02 | 14.30 | 2.45 |
| 66800 | AIG | 36.41 | 14.78 | 50 | 2.46 | | 66800 | 42.22 | 11.57 | 3.65 |
| 25419 | WHR | 52.15 | 21.15 | 158 | 2.47 | | 25419 | 53.93 | 20.64 | 2.61 |
| 76661 | IONS | 75.93 | 30.64 | 281 | 2.48 | | 76661 | 92.23 | 31.22 | 2.95 |
| 30648 | COHU | 30.91 | 12.48 | 32 | 2.48 | | 30648 | 25.33 | 5.55 | 4.56 |
| 17144 | GIS | 15.37 | 6.20 | 28 | 2.48 | | 17144 | 15.72 | 3.79 | 4.15 |
| 82800 | SCCO | 30.25 | 12.19 | 94 | 2.48 | | 82800 | 44.90 | 14.41 | 3.12 |
| 86776 | INFY | 31.02 | 12.49 | 210 | 2.48 | | 86776 | 43.48 | 17.92 | 2.43 |
| 89803 | AHT | 80.96 | 32.60 | 494 | 2.48 | | 89803 | 101.75 | 41.20 | 2.47 |
| 75578 | MERC | 161.59 | 65.03 | 444 | 2.48 | | 75578 | 52.50 | 26.87 | 1.95 |
| 92766 | AAXJ | 19.57 | 7.87 | 89 | 2.49 | | 92766 | 27.04 | 9.34 | 2.89 |
| 81564 | | 18.86 | 7.59 | 34 | 2.49 | | 81564 | 22.02 | 4.86 | 4.53 |
| 27909 | CULP | 94.66 | 37.99 | 854 | 2.49 | | 27909 | 128.24 | 51.81 | 2.48 |
| 20053 | SNV | 30.40 | 12.20 | -45 | 2.49 | | 20053 | 45.26 | 6.49 | 6.97 |
| 42439 | HNI | 33.03 | 13.25 | 124 | 2.49 | | 42439 | 35.27 | 13.34 | 2.64 |
| 68187 | WRI | 20.89 | 8.37 | 20 | 2.50 | | 68187 | 16.15 | 2.99 | 5.41 |
| 76655 | TM | 19.17 | 7.68 | 77 | 2.50 | | 76655 | 19.37 | 7.39 | 2.62 |
| 88487 | PHG | 29.02 | 11.62 | 77 | 2.50 | | 88487 | 26.73 | 8.95 | 2.99 |
| 44943 | DIOD | 39.99 | 16.01 | 432 | 2.50 | | 44943 | 75.34 | 30.15 | 2.50 |
| 85972 | CRAI | 31.75 | 12.71 | 58 | 2.50 | | 85972 | 38.74 | 9.97 | 3.88 |
| 13610 | OLN | 27.21 | 10.86 | 11 | 2.51 | | 13610 | 29.92 | 5.18 | 5.77 |
| 92808 | DMRC | 37.92 | 15.13 | 45 | 2.51 | | 92808 | 45.60 | 12.73 | 3.58 |
| 24272 | ASH | 88.95 | 35.49 | 1251 | 2.51 | | 24272 | 83.95 | 42.47 | 1.98 |
| 90177 | ACAD | 246.53 | 98.18 | 1697 | 2.51 | | 90177 | 165.49 | 77.39 | 2.14 |
| 92500 | PIE | 20.83 | 8.30 | 65 | 2.51 | | 92500 | 25.89 | 7.79 | 3.32 |
| 16432 | GT | 41.77 | 16.63 | 242 | 2.51 | | 16432 | 49.69 | 20.64 | 2.41 |
| 85686 | ASRT | 67.42 | 26.78 | 119 | 2.52 | | 85686 | 56.43 | 21.85 | 2.58 |
| 28388 | VNO | 24.15 | 9.59 | 41 | 2.52 | | 28388 | 15.65 | 4.54 | 3.45 |
| 70228 | TOL | 27.84 | 11.05 | 54 | 2.52 | | 70228 | 28.60 | 7.54 | 3.79 |
| 89540 | DKS | 31.33 | 12.42 | 121 | 2.52 | | 89540 | 38.47 | 14.60 | 2.63 |
| 75032 | | 20.97 | 8.31 | 78 | 2.52 | | 75032 | 27.73 | 8.85 | 3.13 |
| 32707 | HP | 35.77 | 14.17 | 111 | 2.52 | | 32707 | 35.36 | 12.52 | 2.82 |
| 85072 | RL | 35.36 | 14.00 | 128 | 2.53 | | 85072 | 31.75 | 12.68 | 2.50 |
| 88865 | HAFC | 35.65 | 14.11 | 20 | 2.53 | | 88865 | 41.92 | 8.66 | 4.84 |
| 90715 | ZUMZ | 47.22 | 18.68 | 157 | 2.53 | | 90715 | 50.79 | 20.88 | 2.43 |
| 91262 | AAWW | 46.89 | 18.54 | 123 | 2.53 | | 91262 | 38.79 | 13.82 | 2.81 |
| 91733 | AVAV | 40.67 | 16.08 | 85 | 2.53 | | 91733 | 39.69 | 11.47 | 3.46 |
| 78038 | RDN | 138.10 | 54.59 | 345 | 2.53 | | 78038 | 72.90 | 35.70 | 2.04 |
| 89625 | | 84.79 | 33.50 | 152 | 2.53 | | 89625 | 49.31 | 18.47 | 2.67 |
| 83112 | ANDE | 38.79 | 15.31 | 172 | 2.53 | | 83112 | 44.83 | 17.96 | 2.50 |
| 47619 | KTCC | 95.07 | 37.37 | 482 | 2.54 | | 47619 | 99.84 | 38.90 | 2.57 |
| 59408 | BAC | 50.79 | 19.96 | 75 | 2.54 | | 59408 | 43.88 | 13.75 | 3.19 |
| 88379 | | 49.07 | 19.27 | 201 | 2.55 | | 88379 | 56.49 | 23.19 | 2.44 |
| 90790 | MPW | 34.87 | 13.69 | 155 | 2.55 | | 90790 | 20.12 | 11.31 | 1.78 |
| 83213 | EWG | 18.05 | 7.08 | 32 | 2.55 | | 83213 | 19.71 | 4.52 | 4.36 |
| 77309 | PFO | 24.33 | 9.53 | 129 | 2.55 | | 77309 | 26.38 | 10.91 | 2.42 |
| 87390 | | 93.02 | 36.43 | 611 | 2.55 | | 87390 | 98.86 | 46.44 | 2.13 |
| 81912 | | 19.88 | 7.78 | 92 | 2.55 | | 81912 | 25.43 | 9.26 | 2.75 |
| 79561 | PCM | 19.83 | 7.75 | 66 | 2.56 | | 79561 | 17.05 | 6.36 | 2.68 |
| 90120 | AGO | 52.71 | 20.56 | 236 | 2.56 | | 90120 | 37.60 | 17.80 | 2.11 |
| 79770 | LXP | 33.29 | 12.98 | 80 | 2.56 | | 79770 | 24.23 | 8.53 | 2.84 |
| 81566 | | 60.68 | 23.64 | 339 | 2.57 | | 81566 | 53.36 | 26.79 | 1.99 |
| 89394 | OSTK | 90.07 | 35.08 | 26 | 2.57 | | 89394 | 101.56 | 35.09 | 2.89 |
| 80774 | GEO | 32.54 | 12.67 | 64 | 2.57 | | 80774 | 29.93 | 8.78 | 3.41 |
| 85265 | SLG | 64.95 | 25.24 | 205 | 2.57 | | 85265 | 32.80 | 15.38 | 2.13 |
| 27975 | MSB | 82.24 | 31.95 | 173 | 2.57 | | 27975 | 90.05 | 37.22 | 2.42 |
| 89824 | PRSC | 112.08 | 43.52 | 4039 | 2.58 | | 89824 | 308.08 | 116.05 | 2.65 |
| 83244 | CENX | 100.83 | 39.15 | -27 | 2.58 | | 83244 | 77.55 | 24.67 | 3.14 |
| 10375 | | 23.06 | 8.95 | 43 | 2.58 | | 10375 | 20.62 | 5.47 | 3.77 |
| 89708 | JPC | 20.96 | 8.14 | 81 | 2.58 | | 89708 | 22.41 | 7.96 | 2.82 |
| 89968 | PIPR | 36.03 | 13.98 | 66 | 2.58 | | 89968 | 42.48 | 12.75 | 3.33 |
| 76961 | NTP | 46.99 | 18.24 | 58 | 2.58 | | 76961 | 61.65 | 16.73 | 3.68 |
| 81134 | CYD | 69.02 | 26.75 | 224 | 2.58 | | 81134 | 101.93 | 39.07 | 2.61 |
| 92187 | VEA | 14.25 | 5.52 | 35 | 2.58 | | 92187 | 15.52 | 4.11 | 3.78 |
| 90507 | MIC | 142.97 | 55.37 | 870 | 2.58 | | 90507 | 74.16 | 39.34 | 1.89 |
| 90377 | ONTO | 145.49 | 56.33 | 2297 | 2.58 | | 90377 | 279.75 | 101.40 | 2.76 |
| 46703 | HST | 40.82 | 15.80 | 120 | 2.58 | | 46703 | 29.47 | 11.91 | 2.48 |
| 79851 | OSBC | 88.58 | 34.27 | 12 | 2.58 | | 79851 | 96.44 | 25.50 | 3.78 |
| 71159 | TWN | 20.76 | 8.03 | 66 | 2.59 | | 71159 | 27.57 | 8.36 | 3.30 |
| 80635 | IMAX | 60.69 | 23.45 | 322 | 2.59 | | 80635 | 72.48 | 29.16 | 2.49 |
| 76888 | AXAS | 75.24 | 29.06 | 51 | 2.59 | | 76888 | 87.98 | 30.09 | 2.92 |
| 21055 | GCO | 38.98 | 15.06 | 162 | 2.59 | | 21055 | 37.01 | 16.24 | 2.28 |
| 88195 | SGMO | 106.06 | 40.97 | 230 | 2.59 | | 88195 | 150.24 | 57.84 | 2.60 |
| 91542 | PXI | 25.78 | 9.95 | 48 | 2.59 | | 91542 | 26.73 | 7.12 | 3.75 |
| 80069 | ARWR | 207.64 | 80.10 | 35 | 2.59 | | 80069 | 158.34 | 58.28 | 2.72 |
| 11499 | CECE | 41.98 | 16.18 | 179 | 2.59 | | 11499 | 52.77 | 24.45 | 2.16 |
| 88820 | PGC | 26.84 | 10.35 | -1 | 2.59 | | 88820 | 30.59 | 4.52 | 6.76 |
| 90012 | PETS | 36.96 | 14.23 | 32 | 2.60 | | 90012 | 43.17 | 10.47 | 4.12 |
| 89984 | ETG | 15.95 | 6.14 | 27 | 2.60 | | 89984 | 17.56 | 3.79 | 4.63 |
| 75405 | MPV | 10.93 | 4.20 | 66 | 2.60 | | 75405 | 13.07 | 5.91 | 2.21 |
| 65008 | ASYS | 90.83 | 34.88 | 23 | 2.60 | | 65008 | 100.27 | 37.28 | 2.69 |
| 46077 | | 28.93 | 11.10 | 80 | 2.61 | | 46077 | 29.21 | 9.69 | 3.01 |
| 34746 | FITB | 74.20 | 28.44 | 185 | 2.61 | | 34746 | 23.54 | 13.29 | 1.77 |
| 88287 | DLA | 55.39 | 21.21 | 364 | 2.61 | | 88287 | 65.18 | 28.14 | 2.32 |
| 26825 | K | 13.28 | 5.08 | 30 | 2.61 | | 26825 | 10.81 | 3.18 | 3.40 |
| 77117 | WNC | 169.39 | 64.81 | 191 | 2.61 | | 77117 | 170.46 | 51.42 | 3.32 |
| 85889 | SRDX | 43.97 | 16.82 | 87 | 2.61 | | 85889 | 33.51 | 11.38 | 2.94 |
| 91875 | JCE | 14.07 | 5.37 | 25 | 2.62 | | 91875 | 16.14 | 3.40 | 4.74 |
| 81893 | WSTG | 22.63 | 8.63 | 43 | 2.62 | | 81893 | 22.83 | 6.14 | 3.72 |
| 90756 | OSK | 93.77 | 35.68 | 590 | 2.63 | | 90756 | 103.79 | 43.14 | 2.41 |
| 90394 | SHO | 49.92 | 18.97 | 110 | 2.63 | | 90394 | 24.26 | 10.33 | 2.35 |
| 88351 | INSM | 88.61 | 33.67 | 181 | 2.63 | | 88351 | 79.82 | 32.74 | 2.44 |
| 89649 | FFC | 25.75 | 9.77 | 116 | 2.64 | | 89649 | 26.83 | 10.39 | 2.58 |
| 83219 | EWD | 21.64 | 8.21 | 81 | 2.64 | | 83219 | 23.39 | 8.35 | 2.80 |
| 90105 | ETO | 18.17 | 6.89 | 32 | 2.64 | | 90105 | 19.98 | 4.59 | 4.35 |
| 44134 | KMT | 32.80 | 12.44 | 50 | 2.64 | | 44134 | 39.32 | 11.01 | 3.57 |
| 46463 | GFF | 26.89 | 10.19 | 12 | 2.64 | | 46463 | 30.84 | 5.84 | 5.28 |
| 90490 | TBBK | 42.53 | 16.12 | 112 | 2.64 | | 90490 | 45.71 | 16.62 | 2.75 |
| 82646 | TLK | 23.32 | 8.84 | 110 | 2.64 | | 82646 | 24.54 | 9.99 | 2.46 |
| 41241 | HWKN | 29.64 | 11.20 | 168 | 2.65 | | 41241 | 40.64 | 16.17 | 2.51 |
| 81917 | SVC | 23.40 | 8.83 | 62 | 2.65 | | 81917 | 22.71 | 6.89 | 3.30 |
| 80362 | SCSC | 18.56 | 7.01 | 78 | 2.65 | | 80362 | 20.80 | 7.76 | 2.68 |
| 87053 | GNSS | 57.66 | 21.72 | 410 | 2.65 | | 87053 | 67.75 | 32.17 | 2.11 |
| 90698 | | 24.98 | 9.40 | 205 | 2.66 | | 90698 | 41.06 | 17.29 | 2.37 |
| 92249 | WPS | 14.66 | 5.50 | 46 | 2.66 | | 92249 | 17.22 | 5.08 | 3.39 |
| 80237 | | 89.81 | 33.71 | 539 | 2.66 | | 80237 | 123.89 | 51.07 | 2.43 |
| 88660 | SNP | 17.56 | 6.59 | 49 | 2.67 | | 88660 | 17.85 | 5.39 | 3.31 |
| 81084 | HDSN | 66.06 | 24.78 | -34 | 2.67 | | 81084 | 79.02 | 20.24 | 3.90 |
| 27430 | CTB | 64.17 | 24.07 | 425 | 2.67 | | 27430 | 75.68 | 31.68 | 2.39 |
| 11581 | EPAC | 26.50 | 9.93 | 10 | 2.67 | | 11581 | 22.87 | 3.17 | 7.22 |
| 88546 | MNDO | 46.17 | 17.30 | 171 | 2.67 | | 88546 | 62.93 | 21.42 | 2.94 |
| 76224 | BHE | 27.14 | 10.16 | 66 | 2.67 | | 76224 | 29.37 | 8.78 | 3.35 |
| 89808 | FLC | 27.19 | 10.17 | 114 | 2.67 | | 89808 | 26.88 | 10.30 | 2.61 |
| 82859 | REV | 76.69 | 28.68 | 278 | 2.67 | | 82859 | 57.86 | 24.70 | 2.34 |
| 81010 | SSYS | 67.04 | 25.06 | 68 | 2.68 | | 81010 | 70.73 | 24.58 | 2.88 |
| 70519 | C | 31.09 | 11.60 | -22 | 2.68 | | 70519 | 35.83 | 3.94 | 9.09 |
| 36397 | FHN | 21.44 | 8.00 | 42 | 2.68 | | 36397 | 25.28 | 6.62 | 3.82 |
| 88912 | IBCP | 81.33 | 30.33 | -3 | 2.68 | | 88912 | 99.07 | 32.46 | 3.05 |
| 10516 | ADM | 18.84 | 7.02 | 42 | 2.68 | | 10516 | 24.07 | 5.88 | 4.09 |
| 11701 | IMKTA | 35.84 | 13.35 | 55 | 2.69 | | 11701 | 28.43 | 7.84 | 3.63 |
| 91901 | GSIT | 43.64 | 16.25 | 88 | 2.69 | | 91901 | 45.97 | 15.13 | 3.04 |
| 85419 | CSU | 53.94 | 20.09 | 128 | 2.69 | | 85419 | 53.18 | 18.32 | 2.90 |
| 86128 | HZO | 89.97 | 33.44 | 440 | 2.69 | | 86128 | 58.75 | 27.57 | 2.13 |
| 88485 | EVC | 166.45 | 61.85 | 87 | 2.69 | | 88485 | 96.10 | 28.65 | 3.35 |
| 76558 | PFD | 23.61 | 8.77 | 96 | 2.69 | | 76558 | 27.41 | 9.57 | 2.86 |
| 71985 | SHYF | 51.38 | 19.08 | 53 | 2.69 | | 71985 | 72.51 | 19.59 | 3.70 |
| 72996 | SF | 25.52 | 9.47 | 36 | 2.69 | | 72996 | 24.70 | 5.73 | 4.31 |
| 67847 | RGLD | 32.71 | 12.14 | 74 | 2.69 | | 67847 | 35.29 | 11.43 | 3.09 |
| 77066 | | 19.63 | 7.29 | 96 | 2.69 | | 77066 | 27.23 | 9.67 | 2.82 |
| 88466 | SMTX | 120.04 | 44.52 | 656 | 2.70 | | 88466 | 89.85 | 39.86 | 2.25 |
| 81133 | TEO | 45.56 | 16.90 | 105 | 2.70 | | 81133 | 59.32 | 20.87 | 2.84 |
| 88957 | RDY | 45.03 | 16.68 | 285 | 2.70 | | 88957 | 49.37 | 21.16 | 2.33 |
| 78915 | UEIC | 46.26 | 17.13 | 56 | 2.70 | | 78915 | 47.92 | 13.85 | 3.46 |
| 16555 | UVV | 20.91 | 7.74 | 81 | 2.70 | | 16555 | 21.66 | 8.01 | 2.71 |
| 49744 | PCH | 22.57 | 8.36 | 22 | 2.70 | | 49744 | 23.47 | 4.62 | 5.08 |
| 12068 | ONB | 15.42 | 5.71 | -15 | 2.70 | | 12068 | 19.09 | -0.04 | -537.34 |
| 76273 | MXE | 27.68 | 10.24 | 61 | 2.70 | | 76273 | 23.92 | 7.11 | 3.36 |
| 76274 | CRD | 44.00 | 16.26 | 32 | 2.71 | | 76274 | 45.64 | 11.66 | 3.92 |
| 85768 | IESC | 51.72 | 19.11 | 78 | 2.71 | | 85768 | 59.44 | 18.24 | 3.26 |
| 38033 | WEYS | 10.57 | 3.90 | -12 | 2.71 | | 38033 | 14.66 | -0.24 | -60.61 |
| 89606 | SWBI | 67.13 | 24.79 | 467 | 2.71 | | 89606 | 57.68 | 30.09 | 1.92 |
| 92589 | KNDI | 120.11 | 44.34 | 363 | 2.71 | | 92589 | 158.20 | 58.83 | 2.69 |
| 79037 | ETH | 25.09 | 9.25 | 22 | 2.71 | | 79037 | 26.01 | 5.10 | 5.10 |
| 69032 | MS | 34.63 | 12.76 | 147 | 2.71 | | 69032 | 40.19 | 16.07 | 2.50 |
| 92170 | GOF | 20.95 | 7.72 | 76 | 2.71 | | 92170 | 22.69 | 7.75 | 2.93 |
| 89129 | EFA | 14.15 | 5.21 | 31 | 2.72 | | 89129 | 15.06 | 3.73 | 4.03 |
| 91533 | PSP | 19.83 | 7.30 | 40 | 2.72 | | 91533 | 18.85 | 5.01 | 3.76 |
| 89492 | HPI | 14.21 | 5.23 | 36 | 2.72 | | 89492 | 12.66 | 3.84 | 3.30 |
| 16505 | BPOP | 35.71 | 13.11 | -8 | 2.72 | | 16505 | 42.21 | 8.52 | 4.96 |
| 90716 | ATSG | 207.00 | 76.01 | 12572 | 2.72 | | 90716 | 424.72 | 173.78 | 2.44 |
| 57154 | GLT | 30.91 | 11.34 | 5 | 2.72 | | 57154 | 32.58 | 5.81 | 5.61 |
| 80951 | TESS | 54.06 | 19.84 | 107 | 2.73 | | 80951 | 52.11 | 18.29 | 2.85 |
| 30940 | FLS | 30.76 | 11.27 | 121 | 2.73 | | 30940 | 39.30 | 14.02 | 2.80 |
| 79909 | MHO | 47.56 | 17.42 | 99 | 2.73 | | 79909 | 62.09 | 17.94 | 3.46 |
| 23819 | HAL | 36.66 | 13.42 | 46 | 2.73 | | 23819 | 38.40 | 10.02 | 3.83 |
| 11600 | XRAY | 16.03 | 5.86 | 32 | 2.74 | | 11600 | 19.62 | 4.92 | 3.98 |
| 91854 | UFS | 81.32 | 29.66 | 251 | 2.74 | | 91854 | 57.65 | 21.68 | 2.66 |
| 77090 | ATNI | 33.78 | 12.31 | 169 | 2.75 | | 77090 | 41.01 | 16.27 | 2.52 |
| 91723 | CWI | 15.09 | 5.50 | 40 | 2.75 | | 91723 | 16.60 | 4.62 | 3.59 |
| 10302 | | 41.40 | 15.07 | 185 | 2.75 | | 10302 | 53.28 | 20.21 | 2.64 |
| 54148 | PHM | 57.23 | 20.83 | 138 | 2.75 | | 54148 | 66.39 | 20.12 | 3.30 |
| 78156 | HMSY | 39.22 | 14.26 | 168 | 2.75 | | 78156 | 37.50 | 16.22 | 2.31 |
| 78018 | CPSS | 150.31 | 54.63 | 662 | 2.75 | | 78018 | 169.01 | 65.13 | 2.60 |
| 84386 | BSAC | 30.35 | 11.03 | 122 | 2.75 | | 84386 | 35.06 | 12.64 | 2.77 |
| 89823 | GOOD | 20.86 | 7.58 | 111 | 2.75 | | 89823 | 25.60 | 10.19 | 2.51 |
| 84176 | SSP | 102.09 | 37.09 | 612 | 2.75 | | 84176 | 73.58 | 34.67 | 2.12 |
| 90050 | SCD | 17.30 | 6.27 | 38 | 2.76 | | 90050 | 18.43 | 4.86 | 3.79 |
| 86313 | UBA | 13.29 | 4.81 | 21 | 2.76 | | 86313 | 15.67 | 2.92 | 5.37 |
| 89598 | HPF | 14.22 | 5.15 | 42 | 2.76 | | 89598 | 13.38 | 4.35 | 3.07 |
| 90550 | DLB | 30.87 | 11.15 | 89 | 2.77 | | 90550 | 32.25 | 11.99 | 2.69 |
| 91271 | CODI | 21.81 | 7.87 | 11 | 2.77 | | 91271 | 23.82 | 3.56 | 6.69 |
| 91491 | | 102.11 | 36.85 | 134 | 2.77 | | 91491 | 145.01 | 54.13 | 2.68 |
| 75346 | FT | 14.53 | 5.24 | 69 | 2.77 | | 75346 | 19.72 | 6.88 | 2.87 |
| 60468 | GHM | 34.80 | 12.54 | 111 | 2.77 | | 60468 | 43.74 | 14.62 | 2.99 |
| 38295 | SPB | 32.19 | 11.60 | 13 | 2.78 | | 38295 | 42.66 | 9.44 | 4.52 |
| 63132 | DCO | 43.66 | 15.73 | 117 | 2.78 | | 63132 | 38.91 | 14.37 | 2.71 |
| 89990 | JTA | 17.79 | 6.41 | 34 | 2.78 | | 89990 | 22.08 | 5.03 | 4.39 |
| 88839 | HBIO | 35.57 | 12.80 | 55 | 2.78 | | 88839 | 23.06 | 7.02 | 3.29 |
| 33823 | BOOM | 64.09 | 23.05 | 82 | 2.78 | | 33823 | 54.08 | 17.30 | 3.13 |
| 89766 | HPS | 15.84 | 5.69 | 44 | 2.78 | | 89766 | 15.29 | 4.68 | 3.26 |
| 92040 | FOLD | 108.96 | 39.10 | 20 | 2.79 | | 92040 | 105.59 | 28.49 | 3.71 |
| 92301 | FANH | 81.45 | 29.22 | 150 | 2.79 | | 92301 | 68.51 | 24.30 | 2.82 |
| 76804 | MTG | 117.54 | 42.15 | 201 | 2.79 | | 76804 | 78.69 | 30.08 | 2.62 |
| 11154 | PLAB | 65.28 | 23.40 | 396 | 2.79 | | 11154 | 44.81 | 23.50 | 1.91 |
| 66835 | | 31.07 | 11.13 | 52 | 2.79 | | 66835 | 29.06 | 7.72 | 3.77 |
| 91886 | VEU | 15.29 | 5.48 | 41 | 2.79 | | 91886 | 17.28 | 4.77 | 3.63 |
| 85925 | RBA | 18.76 | 6.71 | 53 | 2.80 | | 85925 | 15.67 | 5.30 | 2.96 |
| 76279 | MTRX | 43.75 | 15.64 | 134 | 2.80 | | 76279 | 39.65 | 13.81 | 2.87 |
| 90523 | EOS | 12.59 | 4.50 | 37 | 2.80 | | 90523 | 16.72 | 4.40 | 3.80 |
| 89428 | BNS | 24.12 | 8.62 | 83 | 2.80 | | 89428 | 30.37 | 9.85 | 3.08 |
| 77236 | PHX | 30.95 | 11.03 | 72 | 2.80 | | 77236 | 28.66 | 9.10 | 3.15 |
| 90037 | HTH | 31.83 | 11.32 | 83 | 2.81 | | 90037 | 37.26 | 11.40 | 3.27 |
| 90346 | TTM | 70.45 | 24.99 | 177 | 2.82 | | 90346 | 96.97 | 34.51 | 2.81 |
| 92293 | TDC | 36.32 | 12.86 | 159 | 2.82 | | 92293 | 42.09 | 16.25 | 2.59 |
| 45225 | VHI | 86.18 | 30.49 | -46 | 2.83 | | 45225 | 91.46 | 29.02 | 3.15 |
| 79108 | SANM | 149.76 | 52.85 | 753 | 2.83 | | 79108 | 94.03 | 41.55 | 2.26 |
| 41217 | HVT | 35.08 | 12.38 | 101 | 2.83 | | 41217 | 38.12 | 12.35 | 3.09 |
| 79338 | SGMS | 84.97 | 29.95 | 2 | 2.84 | | 79338 | 97.04 | 23.65 | 4.10 |
| 91044 | DCP | 42.95 | 15.13 | 182 | 2.84 | | 91044 | 73.11 | 24.42 | 2.99 |
| 89517 | JPS | 18.44 | 6.50 | 64 | 2.84 | | 89517 | 17.56 | 6.34 | 2.77 |
| 89009 | PCQ | 12.60 | 4.43 | 87 | 2.84 | | 89009 | 16.99 | 7.60 | 2.23 |
| 69980 | GAB | 17.95 | 6.31 | 43 | 2.84 | | 69980 | 21.08 | 5.47 | 3.86 |
| 82649 | SWM | 49.89 | 17.54 | 150 | 2.84 | | 82649 | 82.51 | 24.05 | 3.43 |
| 81665 | SYX | 74.06 | 26.02 | 122 | 2.85 | | 81665 | 93.11 | 26.42 | 3.52 |
| 91584 | NOA | 78.89 | 27.68 | 166 | 2.85 | | 91584 | 70.87 | 30.25 | 2.34 |
| 84603 | CRESY | 37.33 | 13.09 | 52 | 2.85 | | 84603 | 36.90 | 10.52 | 3.51 |
| 91604 | ALLT | 80.44 | 28.21 | 268 | 2.85 | | 91604 | 77.12 | 29.85 | 2.58 |
| 90934 | PXE | 24.75 | 8.67 | 25 | 2.86 | | 90934 | 21.94 | 4.35 | 5.05 |
| 89444 | TRI | 15.07 | 5.27 | 50 | 2.86 | | 89444 | 15.63 | 5.32 | 2.94 |
| 88897 | OIS | 49.58 | 17.31 | 34 | 2.86 | | 88897 | 51.40 | 13.46 | 3.82 |
| 92400 | LL | 84.33 | 29.44 | -10 | 2.86 | | 92400 | 96.79 | 32.26 | 3.00 |
| 76478 | ESE | 20.89 | 7.28 | 61 | 2.87 | | 76478 | 22.78 | 6.89 | 3.30 |
| 89437 | PML | 13.03 | 4.53 | 97 | 2.87 | | 89437 | 19.50 | 8.44 | 2.31 |
| 79233 | HR | 18.77 | 6.53 | 21 | 2.87 | | 79233 | 15.42 | 2.92 | 5.29 |
| 87662 | IBN | 44.34 | 15.42 | 194 | 2.88 | | 87662 | 46.06 | 20.01 | 2.30 |
| 34817 | NUE | 16.00 | 5.56 | 12 | 2.88 | | 34817 | 20.40 | 2.80 | 7.28 |
| 48961 | LAWS | 36.74 | 12.76 | 38 | 2.88 | | 48961 | 46.24 | 10.73 | 4.31 |
| 86868 | GS | 28.10 | 9.73 | 98 | 2.89 | | 86868 | 41.93 | 14.06 | 2.98 |
| 66852 | RELL | 51.30 | 17.76 | 197 | 2.89 | | 66852 | 46.49 | 18.88 | 2.46 |
| 52329 | J | 22.30 | 7.71 | 22 | 2.89 | | 52329 | 25.26 | 4.65 | 5.43 |
| 80008 | RMT | 21.89 | 7.56 | 32 | 2.89 | | 80008 | 23.81 | 5.34 | 4.46 |
| 54704 | MOD | 94.16 | 32.50 | 122 | 2.90 | | 54704 | 58.73 | 20.50 | 2.86 |
| 92586 | ACWX | 15.07 | 5.20 | 37 | 2.90 | | 92586 | 16.73 | 4.36 | 3.84 |
| 81013 | TSEM | 148.03 | 51.04 | 634 | 2.90 | | 81013 | 199.53 | 75.82 | 2.63 |
| 89257 | PCN | 18.33 | 6.32 | 29 | 2.90 | | 89257 | 12.40 | 3.30 | 3.76 |
| 90796 | WPM | 59.07 | 20.36 | 201 | 2.90 | | 90796 | 69.86 | 26.62 | 2.62 |
| 90233 | | 37.10 | 12.79 | -5 | 2.90 | | 90233 | 26.25 | 2.60 | 10.12 |
| 59248 | TAP | 20.82 | 7.17 | 15 | 2.90 | | 59248 | 21.65 | 3.48 | 6.22 |
| 59010 | GPS | 33.92 | 11.68 | 92 | 2.91 | | 59010 | 36.94 | 12.38 | 2.98 |
| 18973 | BRC | 22.20 | 7.64 | 81 | 2.91 | | 18973 | 23.07 | 8.06 | 2.86 |
| 90527 | | 63.92 | 21.98 | 508 | 2.91 | | 90527 | 100.78 | 43.51 | 2.32 |
| 75113 | IEP | 43.10 | 14.81 | 126 | 2.91 | | 75113 | 51.13 | 16.06 | 3.18 |
| 92457 | CATM | 165.09 | 56.70 | 1916 | 2.91 | | 92457 | 237.18 | 95.35 | 2.49 |
| 90680 | DRH | 32.84 | 11.27 | 79 | 2.91 | | 90680 | 32.06 | 10.22 | 3.14 |
| 52936 | MCY | 14.54 | 4.98 | 12 | 2.92 | | 52936 | 16.99 | 2.42 | 7.02 |
| 89875 | ANIP | 107.77 | 36.85 | 25 | 2.92 | | 89875 | 86.86 | 27.30 | 3.18 |
| 88661 | TPR | 35.48 | 12.13 | 62 | 2.92 | | 88661 | 33.80 | 9.34 | 3.62 |
| 87666 | SLF | 22.06 | 7.53 | 43 | 2.93 | | 87666 | 25.32 | 6.64 | 3.81 |
| 91229 | IEO | 22.67 | 7.73 | 34 | 2.93 | | 91229 | 22.16 | 5.10 | 4.34 |
| 91128 | | 88.46 | 30.07 | 66 | 2.94 | | 91128 | 76.04 | 23.34 | 3.26 |
| 92263 | FLY | 32.12 | 10.89 | 56 | 2.95 | | 92263 | 23.60 | 6.72 | 3.51 |
| 92390 | FGD | 17.94 | 6.08 | 53 | 2.95 | | 92390 | 20.18 | 5.88 | 3.43 |
| 47511 | EBF | 23.64 | 8.00 | 59 | 2.96 | | 47511 | 23.33 | 7.04 | 3.31 |
| 67774 | TRP | 18.22 | 6.16 | 32 | 2.96 | | 67774 | 21.88 | 5.05 | 4.33 |
| 92259 | CZZ | 61.70 | 20.86 | 154 | 2.96 | | 92259 | 66.82 | 25.32 | 2.64 |
| 91725 | GII | 9.49 | 3.21 | 10 | 2.96 | | 91725 | 9.92 | 1.42 | 7.01 |
| 87084 | VHC | 144.63 | 48.80 | 62 | 2.96 | | 87084 | 138.97 | 44.92 | 3.09 |
| 81184 | USAP | 50.22 | 16.90 | 12 | 2.97 | | 81184 | 41.69 | 9.83 | 4.24 |
| 91498 | EHTH | 56.11 | 18.86 | 189 | 2.97 | | 91498 | 58.77 | 25.44 | 2.31 |
| 84203 | BBQ | 58.60 | 19.70 | 58 | 2.97 | | 84203 | 61.69 | 21.73 | 2.84 |
| 79646 | RFI | 22.59 | 7.59 | 46 | 2.98 | | 79646 | 24.10 | 6.18 | 3.90 |
| 84516 | CCBG | 25.37 | 8.51 | -15 | 2.98 | | 84516 | 25.21 | 1.76 | 14.30 |
| 11144 | AVD | 57.07 | 19.15 | 30 | 2.98 | | 11144 | 53.31 | 12.61 | 4.23 |
| 86047 | AMKR | 58.19 | 19.52 | 201 | 2.98 | | 86047 | 78.73 | 26.81 | 2.94 |
| 80297 | | 131.31 | 44.03 | 128 | 2.98 | | 80297 | 100.25 | 35.06 | 2.86 |
| 90534 | CZNC | 20.30 | 6.80 | 34 | 2.98 | | 90534 | 27.46 | 6.81 | 4.03 |
| 87842 | MET | 24.39 | 8.17 | 31 | 2.99 | | 87842 | 25.79 | 5.37 | 4.80 |
| 80399 | MAC | 60.95 | 20.37 | 138 | 2.99 | | 80399 | 36.67 | 13.76 | 2.67 |
| 46922 | RAD | 117.39 | 39.22 | 128 | 2.99 | | 46922 | 151.08 | 58.61 | 2.58 |
| 85914 | BBY | 48.17 | 16.09 | 88 | 2.99 | | 85914 | 82.91 | 23.65 | 3.51 |
| 90444 | ORA | 25.03 | 8.35 | 64 | 3.00 | | 90444 | 28.67 | 8.81 | 3.26 |
| 89546 | NGS | 34.51 | 11.51 | 62 | 3.00 | | 89546 | 41.42 | 11.31 | 3.66 |
| 90314 | CVGI | 202.48 | 67.51 | 513 | 3.00 | | 90314 | 182.17 | 73.04 | 2.49 |
| 47941 | TGNA | 88.04 | 29.35 | 165 | 3.00 | | 47941 | 35.79 | 14.63 | 2.45 |
| 82555 | BLDP | 84.98 | 28.30 | 112 | 3.00 | | 82555 | 77.67 | 26.00 | 2.99 |
| 90401 | ARCC | 42.10 | 14.01 | 146 | 3.00 | | 90401 | 32.48 | 12.69 | 2.56 |
| 84103 | | 26.82 | 8.92 | 52 | 3.01 | | 84103 | 36.50 | 9.50 | 3.84 |
| 88372 | UONEK | 236.58 | 78.66 | 632 | 3.01 | | 88372 | 400.69 | 149.32 | 2.68 |
| 52337 | THC | 86.91 | 28.88 | 273 | 3.01 | | 52337 | 120.19 | 40.17 | 2.99 |
| 80759 | MATW | 18.36 | 6.10 | 11 | 3.01 | | 80759 | 23.11 | 3.35 | 6.90 |
| 86382 | BUSE | 22.41 | 7.43 | -55 | 3.01 | | 86382 | 33.74 | 1.26 | 26.83 |
| 81527 | MDCA | 75.48 | 25.04 | 29 | 3.01 | | 81527 | 104.59 | 37.24 | 2.81 |
| 89756 | MARK | 84.07 | 27.85 | -68 | 3.02 | | 89756 | 81.89 | 16.57 | 4.94 |
| 92089 | GRBK | 185.81 | 61.48 | 150 | 3.02 | | 92089 | 286.46 | 112.10 | 2.56 |
| 92181 | MAG | 43.54 | 14.40 | 62 | 3.02 | | 92181 | 53.65 | 17.09 | 3.14 |
| 32054 | | 71.07 | 23.49 | 341 | 3.03 | | 32054 | 87.13 | 32.86 | 2.65 |
| 92406 | PKO | 15.83 | 5.23 | 49 | 3.03 | | 92406 | 14.15 | 4.92 | 2.88 |
| 89141 | PLPC | 39.22 | 12.96 | 18 | 3.03 | | 89141 | 24.01 | 4.18 | 5.74 |
| 10207 | FUND | 17.75 | 5.85 | 26 | 3.03 | | 10207 | 20.83 | 4.26 | 4.89 |
| 89852 | JDD | 19.57 | 6.43 | 46 | 3.04 | | 89852 | 20.49 | 5.51 | 3.72 |
| 90029 | DVAX | 88.33 | 29.02 | 9 | 3.04 | | 90029 | 131.27 | 42.08 | 3.12 |
| 87789 | LMNX | 19.65 | 6.45 | 8 | 3.05 | | 87789 | 17.75 | 2.31 | 7.69 |
| 91024 | LORL | 49.07 | 16.09 | 156 | 3.05 | | 91024 | 62.04 | 21.84 | 2.84 |
| 70121 | BDN | 39.19 | 12.83 | 67 | 3.05 | | 70121 | 23.30 | 7.62 | 3.06 |
| 90885 | EVRI | 110.11 | 36.05 | 132 | 3.05 | | 90885 | 114.08 | 42.12 | 2.71 |
| 91432 | | 35.21 | 11.52 | -43 | 3.05 | | 91432 | 35.12 | 1.45 | 24.22 |
| 75416 | RPT | 24.09 | 7.88 | 93 | 3.06 | | 75416 | 25.79 | 9.50 | 2.72 |
| 75175 | MBI | 45.06 | 14.73 | 119 | 3.06 | | 75175 | 71.99 | 21.88 | 3.29 |
| 54228 | INSI | 7.87 | 2.57 | 19 | 3.06 | | 54228 | 7.58 | 1.98 | 3.84 |
| 91163 | HIMX | 112.19 | 36.63 | 113 | 3.06 | | 91163 | 172.98 | 58.81 | 2.94 |
| 88203 | OSPN | 65.74 | 21.46 | 25 | 3.06 | | 88203 | 89.47 | 19.12 | 4.68 |
| 56565 | MCI | 13.12 | 4.28 | 53 | 3.06 | | 56565 | 14.55 | 5.22 | 2.79 |
| 89443 | WHG | 17.25 | 5.62 | 20 | 3.07 | | 89443 | 26.66 | 5.36 | 4.97 |
| 83604 | SKM | 23.75 | 7.75 | 47 | 3.07 | | 83604 | 25.50 | 6.65 | 3.83 |
| 75860 | IMGN | 74.17 | 24.16 | 12 | 3.07 | | 75860 | 88.28 | 31.94 | 2.76 |
| 59627 | VIRC | 29.58 | 9.63 | 105 | 3.07 | | 59627 | 40.50 | 13.86 | 2.92 |
| 90731 | HTGC | 25.09 | 8.17 | 40 | 3.07 | | 90731 | 21.84 | 5.29 | 4.13 |
| 88648 | TTMI | 47.34 | 15.33 | 87 | 3.09 | | 88648 | 55.70 | 16.20 | 3.44 |
| 91879 | SPEM | 21.04 | 6.81 | 72 | 3.09 | | 91879 | 27.97 | 8.49 | 3.29 |
| 89467 | LQD | 4.85 | 1.57 | 11 | 3.09 | | 89467 | 4.95 | 1.16 | 4.27 |
| 90605 | ULH | 38.05 | 12.30 | 28 | 3.09 | | 90605 | 34.20 | 7.81 | 4.38 |
| 88260 | PXLW | 131.39 | 42.44 | 303 | 3.10 | | 88260 | 115.21 | 45.03 | 2.56 |
| 86827 | HSII | 33.41 | 10.77 | 45 | 3.10 | | 86827 | 24.86 | 6.47 | 3.84 |
| 91161 | GPRE | 155.07 | 49.78 | 612 | 3.12 | | 91161 | 227.70 | 77.68 | 2.93 |
| 83422 | SYKE | 21.23 | 6.81 | 29 | 3.12 | | 83422 | 24.19 | 5.07 | 4.77 |
| 91689 | EXK | 87.52 | 28.09 | 111 | 3.12 | | 91689 | 102.33 | 34.91 | 2.93 |
| 90114 | LGI | 17.22 | 5.52 | 15 | 3.12 | | 90114 | 18.65 | 2.96 | 6.31 |
| 85792 | PWR | 24.98 | 8.00 | 52 | 3.12 | | 85792 | 28.53 | 7.39 | 3.86 |
| 84167 | GEL | 34.99 | 11.20 | 113 | 3.12 | | 84167 | 45.35 | 14.80 | 3.06 |
| 92696 | | 116.44 | 37.25 | -32 | 3.13 | | 92696 | 82.74 | 21.67 | 3.82 |
| 83212 | EWQ | 17.92 | 5.71 | 27 | 3.14 | | 83212 | 17.63 | 3.75 | 4.70 |
| 86763 | AUTO | 83.77 | 26.71 | 36 | 3.14 | | 86763 | 107.00 | 32.30 | 3.31 |
| 90179 | ANGO | 22.52 | 7.18 | 47 | 3.14 | | 90179 | 27.89 | 7.29 | 3.82 |
| 77379 | THFF | 14.68 | 4.68 | -2 | 3.14 | | 77379 | 23.16 | 1.87 | 12.37 |
| 88495 | | 40.74 | 12.98 | 227 | 3.14 | | 88495 | 43.71 | 19.90 | 2.20 |
| 86563 | HT | 37.60 | 11.96 | 46 | 3.14 | | 86563 | 39.14 | 8.68 | 4.51 |
| 84275 | UNFI | 30.17 | 9.60 | -41 | 3.14 | | 84275 | 41.26 | 7.04 | 5.86 |
| 88405 | IEV | 15.78 | 5.02 | 26 | 3.15 | | 88405 | 16.11 | 3.43 | 4.70 |
| 88360 | MRVL | 45.44 | 14.43 | 143 | 3.15 | | 88360 | 80.32 | 27.48 | 2.92 |
| 76135 | GVA | 20.78 | 6.60 | -8 | 3.15 | | 76135 | 24.49 | 1.89 | 12.94 |
| 13507 | ANAT | 28.27 | 8.97 | 73 | 3.15 | | 13507 | 32.00 | 9.35 | 3.42 |
| 90783 | FTK | 148.15 | 47.00 | -57 | 3.15 | | 90783 | 111.73 | 23.91 | 4.67 |
| 85567 | GEOS | 76.23 | 24.17 | 18 | 3.15 | | 85567 | 81.72 | 25.99 | 3.14 |
| 91607 | CSIQ | 224.62 | 71.21 | 122 | 3.15 | | 91607 | 267.34 | 98.18 | 2.72 |
| 89403 | SBS | 43.12 | 13.67 | 100 | 3.16 | | 89403 | 43.74 | 14.85 | 2.95 |
| 92581 | EPI | 28.94 | 9.17 | 119 | 3.16 | | 92581 | 36.90 | 13.25 | 2.79 |
| 90775 | THS | 26.15 | 8.27 | 86 | 3.16 | | 90775 | 25.58 | 9.29 | 2.75 |
| 89240 | JRS | 25.58 | 8.08 | 67 | 3.16 | | 89240 | 23.13 | 7.29 | 3.17 |
| 89790 | TELL | 358.63 | ##### | 34 | 3.17 | | 89790 | 619.25 | 197.46 | 3.14 |
| 87016 | PWFL | 38.89 | 12.27 | 38 | 3.17 | | 87016 | 23.74 | 5.82 | 4.08 |
| 84661 | AKO | 28.38 | 8.95 | 69 | 3.17 | | 84661 | 34.32 | 10.55 | 3.25 |
| 90517 | SVA | 80.27 | 25.32 | 403 | 3.17 | | 90517 | 118.09 | 43.60 | 2.71 |
| 92506 | GTS | 29.23 | 9.21 | 51 | 3.17 | | 92506 | 22.62 | 6.35 | 3.56 |
| 79881 | URBN | 42.64 | 13.42 | 122 | 3.18 | | 79881 | 47.60 | 15.24 | 3.12 |
| 86454 | XLE | 18.05 | 5.67 | 20 | 3.18 | | 86454 | 18.18 | 3.34 | 5.44 |
| 90701 | LAZ | 26.08 | 8.20 | 24 | 3.18 | | 90701 | 26.82 | 5.37 | 5.00 |
| 77516 | | 48.36 | 15.17 | 1 | 3.19 | | 77516 | 46.00 | 10.56 | 4.36 |
| 19583 | IVZ | 25.13 | 7.88 | 16 | 3.19 | | 19583 | 33.24 | 6.81 | 4.88 |
| 76686 | LCUT | 110.97 | 34.80 | 183 | 3.19 | | 76686 | 49.35 | 19.45 | 2.54 |
| 62156 | BXMT | 34.08 | 10.68 | -11 | 3.19 | | 62156 | 29.35 | 4.26 | 6.88 |
| 80725 | | 17.75 | 5.54 | 54 | 3.20 | | 80725 | 20.18 | 6.09 | 3.32 |
| 92496 | ITM | 4.48 | 1.40 | 19 | 3.21 | | 92496 | 5.25 | 1.88 | 2.80 |
| 90389 | BXMX | 9.10 | 2.83 | 12 | 3.21 | | 90389 | 11.44 | 1.76 | 6.52 |
| 92377 | DGS | 22.63 | 7.05 | 72 | 3.21 | | 92377 | 31.34 | 9.25 | 3.39 |
| 91832 | MDGL | 324.96 | ##### | -47 | 3.21 | | 91832 | 172.65 | 45.48 | 3.80 |
| 89290 | SYNA | 36.03 | 11.18 | 125 | 3.22 | | 89290 | 39.13 | 13.98 | 2.80 |
| 92354 | PZN | 49.81 | 15.45 | 105 | 3.22 | | 92354 | 51.59 | 16.22 | 3.18 |
| 30680 | CMC | 19.29 | 5.97 | 35 | 3.23 | | 30680 | 28.13 | 6.18 | 4.55 |
| 91331 | DWM | 13.88 | 4.29 | 21 | 3.23 | | 91331 | 14.24 | 2.79 | 5.10 |
| 86489 | SNBR | 428.67 | ##### | 12592 | 3.24 | | 86489 | 786.04 | 275.05 | 2.86 |
| 48347 | LZB | 185.35 | 57.20 | 1177 | 3.24 | | 48347 | 109.02 | 49.77 | 2.19 |
| 90371 | BBW | 54.88 | 16.91 | -19 | 3.24 | | 90371 | 72.21 | 15.97 | 4.52 |
| 89187 | EPP | 19.01 | 5.86 | 54 | 3.25 | | 89187 | 22.66 | 6.41 | 3.54 |
| 50286 | LEE | 220.19 | 67.85 | 415 | 3.25 | | 50286 | 243.81 | 90.72 | 2.69 |
| 87268 | CIR | 38.50 | 11.86 | -23 | 3.25 | | 87268 | 51.19 | 7.64 | 6.70 |
| 92603 | PIN | 23.82 | 7.33 | 96 | 3.25 | | 92603 | 30.91 | 10.58 | 2.92 |
| 77178 | QCOM | 21.19 | 6.51 | 59 | 3.25 | | 77178 | 19.11 | 6.46 | 2.96 |
| 91132 | VGK | 15.83 | 4.87 | 27 | 3.25 | | 91132 | 16.88 | 3.63 | 4.65 |
| 86547 | AXL | 165.90 | 50.91 | 284 | 3.26 | | 86547 | 64.67 | 26.02 | 2.49 |
| 91651 | RWX | 14.40 | 4.41 | 31 | 3.26 | | 91651 | 16.51 | 3.85 | 4.29 |
| 80306 | BCRX | 173.16 | 53.07 | 489 | 3.26 | | 80306 | 178.75 | 74.02 | 2.41 |
| 90890 | PID | 15.95 | 4.89 | 34 | 3.26 | | 90890 | 16.27 | 4.17 | 3.90 |
| 70703 | EQC | 30.52 | 9.32 | 123 | 3.27 | | 70703 | 34.35 | 12.46 | 2.76 |
| 80206 | IIF | 30.68 | 9.35 | 60 | 3.28 | | 80206 | 39.28 | 11.45 | 3.43 |
| 85349 | MTOR | 259.21 | 78.95 | 493 | 3.28 | | 85349 | 105.67 | 52.12 | 2.03 |
| 70092 | KBH | 50.81 | 15.46 | 40 | 3.29 | | 70092 | 59.82 | 15.48 | 3.86 |
| 10044 | RMCF | 18.99 | 5.77 | 60 | 3.29 | | 10044 | 23.42 | 7.14 | 3.28 |
| 89635 | PTY | 23.10 | 7.00 | 47 | 3.30 | | 89635 | 16.33 | 5.01 | 3.26 |
| 82234 | BTN | 70.77 | 21.42 | -7 | 3.30 | | 82234 | 84.64 | 24.24 | 3.49 |
| 92140 | PXF | 17.54 | 5.31 | 32 | 3.30 | | 92140 | 17.69 | 4.17 | 4.25 |
| 42059 | WMK | 17.00 | 5.14 | 42 | 3.31 | | 42059 | 24.46 | 6.22 | 3.93 |
| 80432 | AEO | 36.45 | 11.02 | 107 | 3.31 | | 80432 | 30.62 | 10.91 | 2.81 |
| 89925 | DZSI | 199.51 | 60.29 | 570 | 3.31 | | 89925 | 341.77 | 136.59 | 2.50 |
| 83532 | | 65.84 | 19.87 | 78 | 3.31 | | 83532 | 88.69 | 28.51 | 3.11 |
| 65330 | | 31.17 | 9.34 | 16 | 3.34 | | 65330 | 36.80 | 7.38 | 4.98 |
| 27511 | CUZ | 18.69 | 5.60 | -23 | 3.34 | | 27511 | 24.18 | 0.37 | 65.77 |
| 87121 | SALM | 150.29 | 45.00 | 179 | 3.34 | | 87121 | 226.89 | 70.20 | 3.23 |
| 90390 | PFN | 20.76 | 6.20 | 60 | 3.35 | | 90390 | 21.63 | 6.50 | 3.33 |
| 75985 | HALL | 18.14 | 5.39 | 22 | 3.36 | | 75985 | 19.25 | 3.53 | 5.46 |
| 83844 | AWRE | 39.59 | 11.76 | 93 | 3.37 | | 83844 | 44.33 | 13.82 | 3.21 |
| 49330 | ALX | 23.35 | 6.93 | 20 | 3.37 | | 49330 | 20.10 | 3.50 | 5.75 |
| 84375 | TEN | 218.14 | 64.76 | 828 | 3.37 | | 84375 | 162.13 | 64.23 | 2.52 |
| 79663 | SQM | 44.63 | 13.25 | 57 | 3.37 | | 79663 | 50.29 | 14.80 | 3.40 |
| 90825 | ACCO | 102.05 | 30.27 | 97 | 3.37 | | 90825 | 48.95 | 15.42 | 3.17 |
| 92854 | CLW | 96.22 | 28.47 | 481 | 3.38 | | 92854 | 177.29 | 59.65 | 2.97 |
| 92152 | DRW | 15.31 | 4.52 | 34 | 3.38 | | 92152 | 18.61 | 4.45 | 4.18 |
| 91727 | TA | 54.24 | 16.00 | 57 | 3.39 | | 91727 | 48.75 | 12.91 | 3.78 |
| 27618 | CRD | 56.70 | 16.70 | -38 | 3.39 | | 27618 | 61.20 | 10.95 | 5.59 |
| 87214 | | 118.69 | 34.92 | 978 | 3.40 | | 87214 | 253.45 | 87.62 | 2.89 |
| 91321 | TECK | 175.33 | 51.48 | 338 | 3.41 | | 91321 | 230.68 | 93.20 | 2.48 |
| 76052 | PPIH | 42.83 | 12.56 | 102 | 3.41 | | 76052 | 58.37 | 18.50 | 3.16 |
| 91447 | MXI | 20.87 | 6.11 | 47 | 3.42 | | 91447 | 25.81 | 6.55 | 3.94 |
| 78002 | TRIB | 61.45 | 17.99 | 43 | 3.42 | | 78002 | 72.60 | 21.64 | 3.36 |
| 91114 | KOP | 41.29 | 12.08 | -21 | 3.42 | | 91114 | 51.76 | 9.30 | 5.57 |
| 77595 | ARCB | 76.74 | 22.44 | 14 | 3.42 | | 77595 | 87.76 | 20.16 | 4.35 |
| 92528 | AHC | 123.10 | 35.92 | 55 | 3.43 | | 92528 | 65.14 | 18.83 | 3.46 |
| 77606 | KSS | 27.51 | 8.02 | 83 | 3.43 | | 77606 | 21.52 | 8.11 | 2.65 |
| 50702 | PEI | 52.08 | 15.19 | -20 | 3.43 | | 50702 | 41.32 | 4.99 | 8.28 |
| 83221 | EWS | 20.84 | 6.08 | 57 | 3.43 | | 83221 | 27.11 | 7.45 | 3.64 |
| 32548 | | 119.62 | 34.86 | 18 | 3.43 | | 32548 | 150.40 | 46.53 | 3.23 |
| 89730 | EEM | 20.26 | 5.90 | 56 | 3.43 | | 89730 | 27.32 | 7.40 | 3.69 |
| 78451 | OPY | 45.98 | 13.39 | 98 | 3.43 | | 78451 | 56.58 | 16.40 | 3.45 |
| 91125 | TX | 84.44 | 24.49 | 216 | 3.45 | | 91125 | 106.65 | 37.51 | 2.84 |
| 88421 | ARNA | 149.59 | 43.30 | -7 | 3.46 | | 88421 | 134.70 | 33.22 | 4.05 |
| 91902 | HAYN | 32.17 | 9.29 | 7 | 3.46 | | 91902 | 23.03 | 3.04 | 7.59 |
| 92754 | | 15.94 | 4.60 | 29 | 3.46 | | 92754 | 14.75 | 3.50 | 4.22 |
| 90188 | NUVA | 38.26 | 11.03 | 43 | 3.47 | | 90188 | 44.20 | 11.02 | 4.01 |
| 91066 | | 33.96 | 9.78 | 38 | 3.47 | | 91066 | 35.34 | 8.41 | 4.20 |
| 87832 | APT | 62.23 | 17.92 | 275 | 3.47 | | 87832 | 119.33 | 42.75 | 2.79 |
| 85627 | SAH | 90.98 | 26.15 | 246 | 3.48 | | 85627 | 53.57 | 20.90 | 2.56 |
| 78875 | CREE | 67.96 | 19.52 | 170 | 3.48 | | 78875 | 90.24 | 33.29 | 2.71 |
| 91642 | PGF | 16.67 | 4.79 | 32 | 3.48 | | 91642 | 10.59 | 3.27 | 3.24 |
| 82779 | NVAX | 69.10 | 19.82 | -3 | 3.49 | | 82779 | 68.54 | 22.40 | 3.06 |
| 92456 | ATHX | 113.83 | 32.65 | 220 | 3.49 | | 92456 | 264.46 | 81.93 | 3.23 |
| 86115 | CLS | 37.80 | 10.83 | 90 | 3.49 | | 86115 | 36.45 | 10.83 | 3.37 |
| 34497 | NPK | 25.41 | 7.28 | 52 | 3.49 | | 34497 | 27.93 | 7.75 | 3.60 |
| 75650 | CRH | 16.62 | 4.76 | 1 | 3.49 | | 75650 | 17.70 | 1.59 | 11.12 |
| 38850 | SKY | 112.90 | 32.28 | -27 | 3.50 | | 38850 | 113.00 | 26.87 | 4.21 |
| 91827 | QCLN | 32.78 | 9.36 | 56 | 3.50 | | 91827 | 37.25 | 9.65 | 3.86 |
| 92347 | PWZ | 5.42 | 1.54 | 20 | 3.51 | | 92347 | 6.26 | 2.05 | 3.06 |
| 90470 | ATLO | 13.46 | 3.82 | -4 | 3.52 | | 90470 | 16.91 | 0.77 | 21.83 |
| 83209 | EWO | 25.30 | 7.18 | 41 | 3.52 | | 83209 | 29.82 | 7.38 | 4.04 |
| 92379 | TGH | 61.31 | 17.36 | -6 | 3.53 | | 92379 | 74.68 | 18.03 | 4.14 |
| 86444 | INO | 113.92 | 32.24 | 92 | 3.53 | | 86444 | 160.01 | 46.66 | 3.43 |
| 85789 | BXS | 27.35 | 7.73 | 12 | 3.54 | | 85789 | 32.73 | 5.32 | 6.15 |
| 91130 | VWO | 21.03 | 5.94 | 62 | 3.54 | | 91130 | 28.71 | 7.94 | 3.62 |
| 82762 | | 95.48 | 26.95 | 78 | 3.54 | | 82762 | 133.03 | 46.27 | 2.87 |
| 92337 | CMF | 4.77 | 1.35 | 15 | 3.54 | | 92337 | 5.84 | 1.58 | 3.69 |
| 78763 | PRMW | 148.94 | 42.00 | 989 | 3.55 | | 78763 | 170.29 | 62.75 | 2.71 |
| 88953 | ABB | 21.16 | 5.97 | 27 | 3.55 | | 88953 | 22.58 | 4.73 | 4.77 |
| 92262 | AXU | 122.58 | 34.49 | -36 | 3.55 | | 92262 | 120.06 | 32.74 | 3.67 |
| 90161 | GLU | 11.74 | 3.30 | 14 | 3.56 | | 90161 | 14.60 | 2.22 | 6.57 |
| 66093 | T | 12.10 | 3.40 | 0 | 3.56 | | 66093 | 13.09 | 0.83 | 15.83 |
| 86866 | | 91.63 | 25.67 | 4 | 3.57 | | 86866 | 75.64 | 22.65 | 3.34 |
| 86560 | ETM | 180.55 | 50.53 | 364 | 3.57 | | 86560 | 153.38 | 52.35 | 2.93 |
| 79089 | FOSL | 78.19 | 21.85 | -6 | 3.58 | | 79089 | 67.04 | 19.83 | 3.38 |
| 87508 | NTWK | 67.12 | 18.75 | -4 | 3.58 | | 87508 | 53.92 | 15.16 | 3.56 |
| 92471 | TITN | 39.63 | 11.07 | -6 | 3.58 | | 92471 | 35.43 | 4.51 | 7.86 |
| 25590 | NFG | 26.56 | 7.41 | 63 | 3.58 | | 25590 | 30.40 | 8.91 | 3.41 |
| 91701 | USIG | 4.22 | 1.18 | 11 | 3.58 | | 91701 | 4.49 | 1.16 | 3.88 |
| 90042 | IGR | 25.40 | 7.08 | 55 | 3.59 | | 90042 | 25.19 | 6.92 | 3.64 |
| 85081 | ING | 34.17 | 9.51 | -4 | 3.59 | | 85081 | 27.70 | 3.12 | 8.88 |
| 90539 | PTE | 186.88 | 52.00 | -47 | 3.59 | | 90539 | 214.20 | 61.10 | 3.51 |
| 92635 | NX | 28.34 | 7.89 | 45 | 3.59 | | 92635 | 33.22 | 8.16 | 4.07 |
| 92498 | PBP | 7.09 | 1.97 | 12 | 3.60 | | 92498 | 8.26 | 1.41 | 5.84 |
| 90103 | CII | 13.12 | 3.65 | 26 | 3.60 | | 90103 | 16.42 | 3.42 | 4.80 |
| 90347 | VDE | 18.84 | 5.23 | 14 | 3.60 | | 90347 | 19.42 | 3.07 | 6.33 |
| 92859 | | 46.93 | 13.03 | -21 | 3.60 | | 92859 | 46.59 | 6.98 | 6.68 |
| 79857 | PTEN | 43.87 | 12.13 | -10 | 3.62 | | 79857 | 40.05 | 6.10 | 6.56 |
| 23835 | MDU | 23.62 | 6.52 | 10 | 3.62 | | 23835 | 26.96 | 3.80 | 7.10 |
| 90071 | NRG | 43.57 | 12.03 | 70 | 3.62 | | 90071 | 49.38 | 14.15 | 3.49 |
| 83390 | KVHI | 43.80 | 12.09 | 99 | 3.62 | | 83390 | 66.64 | 18.76 | 3.55 |
| 89854 | SHG | 27.99 | 7.72 | 51 | 3.63 | | 89854 | 26.98 | 7.20 | 3.74 |
| 25487 | CAR | 344.30 | 94.37 | 3111 | 3.65 | | 25487 | 557.72 | 194.13 | 2.87 |
| 92570 | DAN | 444.01 | ##### | 1742 | 3.65 | | 92570 | 429.86 | 147.14 | 2.92 |
| 90753 | ETV | 10.20 | 2.79 | 32 | 3.66 | | 90753 | 17.70 | 4.02 | 4.40 |
| 91706 | SHV | 0.05 | 0.01 | 0 | 3.66 | | 91706 | 0.07 | -0.01 | -8.89 |
| 34367 | FARM | 38.24 | 10.45 | -6 | 3.66 | | 34367 | 43.21 | 7.24 | 5.97 |
| 75831 | DGII | 23.45 | 6.41 | 24 | 3.66 | | 75831 | 20.84 | 4.26 | 4.89 |
| 79644 | KOF | 32.38 | 8.83 | 40 | 3.67 | | 79644 | 29.93 | 6.98 | 4.29 |
| 78137 | CKH | 23.10 | 6.30 | 16 | 3.67 | | 78137 | 25.61 | 4.31 | 5.94 |
| 91186 | VNDA | 297.20 | 80.88 | 5126 | 3.67 | | 91186 | 675.68 | 241.65 | 2.80 |
| 92542 | PGX | 9.41 | 2.55 | 8 | 3.70 | | 92542 | 7.01 | 1.03 | 6.81 |
| 49373 | HRB | 29.24 | 7.90 | 12 | 3.70 | | 49373 | 30.03 | 5.41 | 5.55 |
| 91063 | | 35.03 | 9.44 | 35 | 3.71 | | 91063 | 35.30 | 8.29 | 4.26 |
| 75607 | NLOK | 26.43 | 7.11 | 40 | 3.72 | | 75607 | 20.69 | 5.41 | 3.82 |
| 88298 | IYE | 17.68 | 4.76 | 10 | 3.72 | | 88298 | 17.72 | 2.40 | 7.39 |
| 90160 | FCT | 15.89 | 4.27 | 60 | 3.72 | | 90160 | 24.71 | 6.91 | 3.57 |
| 11208 | OFG | 66.14 | 17.78 | 172 | 3.72 | | 11208 | 46.38 | 19.71 | 2.35 |
| 87430 | KTOS | 51.67 | 13.88 | 1 | 3.72 | | 87430 | 44.04 | 8.61 | 5.11 |
| 89921 | OXSQ | 37.68 | 10.10 | 70 | 3.73 | | 89921 | 36.61 | 10.27 | 3.57 |
| 89236 | KB | 32.34 | 8.63 | 65 | 3.75 | | 89236 | 43.38 | 12.03 | 3.61 |
| 77519 | TRST | 16.27 | 4.34 | -28 | 3.75 | | 77519 | 24.31 | -0.66 | -36.68 |
| 92097 | LLNW | 53.04 | 14.15 | -4 | 3.75 | | 92097 | 52.32 | 11.66 | 4.49 |
| 91702 | IGIB | 2.98 | 0.79 | 10 | 3.75 | | 91702 | 3.34 | 0.96 | 3.47 |
| 75278 | AB | 32.34 | 8.61 | 31 | 3.76 | | 75278 | 24.42 | 5.75 | 4.24 |
| 86369 | EVF | 20.80 | 5.54 | 76 | 3.76 | | 86369 | 28.55 | 8.44 | 3.38 |
| 79238 | SKT | 19.23 | 5.12 | 7 | 3.76 | | 79238 | 18.74 | 2.34 | 7.99 |
| 90423 | INVA | 38.95 | 10.34 | 82 | 3.77 | | 90423 | 41.08 | 12.92 | 3.18 |
| 92125 | IDV | 18.90 | 5.02 | 41 | 3.77 | | 92125 | 21.06 | 5.20 | 4.05 |
| 83211 | EWC | 17.90 | 4.75 | 37 | 3.77 | | 83211 | 22.50 | 5.33 | 4.22 |
| 88397 | EZU | 18.27 | 4.85 | 15 | 3.77 | | 88397 | 18.33 | 2.84 | 6.45 |
| 89773 | JQC | 21.95 | 5.81 | 52 | 3.78 | | 89773 | 21.60 | 5.98 | 3.61 |
| 22323 | ISR | 138.84 | 36.64 | 50 | 3.79 | | 22323 | 129.19 | 37.47 | 3.45 |
| 92110 | CHW | 15.34 | 4.05 | 13 | 3.79 | | 92110 | 20.74 | 3.17 | 6.54 |
| 90793 | RDS | 19.25 | 5.08 | 17 | 3.79 | | 90793 | 17.95 | 3.12 | 5.75 |
| 70704 | MUX | 101.57 | 26.77 | 111 | 3.79 | | 70704 | 107.68 | 39.57 | 2.72 |
| 76221 | RIO | 34.66 | 9.13 | 118 | 3.80 | | 76221 | 51.81 | 16.56 | 3.13 |
| 88507 | CGEN | 157.61 | 41.52 | 405 | 3.80 | | 88507 | 327.91 | 99.67 | 3.29 |
| 91704 | TLH | 7.24 | 1.91 | 10 | 3.80 | | 91704 | 9.14 | 1.31 | 7.00 |
| 91705 | IEI | 2.69 | 0.71 | 5 | 3.81 | | 91705 | 3.14 | 0.54 | 5.84 |
| 12166 | FRBK | 46.02 | 12.08 | -32 | 3.81 | | 12166 | 47.08 | 5.78 | 8.14 |
| 89814 | PFL | 22.10 | 5.79 | 56 | 3.81 | | 89814 | 21.82 | 6.35 | 3.44 |
| 78170 | SM | 57.81 | 15.14 | -24 | 3.82 | | 78170 | 54.67 | 10.23 | 5.35 |
| 12503 | NAV | 65.55 | 17.14 | 21 | 3.83 | | 12503 | 95.70 | 29.51 | 3.24 |
| 89475 | SNY | 17.71 | 4.62 | 35 | 3.83 | | 89475 | 15.66 | 4.11 | 3.81 |
| 25320 | CPB | 18.02 | 4.70 | 10 | 3.84 | | 25320 | 17.46 | 2.44 | 7.15 |
| 92667 | EWX | 25.07 | 6.52 | 64 | 3.84 | | 92667 | 35.20 | 9.58 | 3.67 |
| 91968 | TFSL | 16.63 | 4.32 | 25 | 3.85 | | 91968 | 18.35 | 3.80 | 4.84 |
| 92408 | PHB | 5.31 | 1.38 | 10 | 3.86 | | 92408 | 6.00 | 1.14 | 5.27 |
| 88888 | CEO | 27.93 | 7.23 | 60 | 3.86 | | 88888 | 32.58 | 9.11 | 3.58 |
| 89965 | NG | 64.01 | 16.54 | 193 | 3.87 | | 89965 | 111.32 | 37.75 | 2.95 |
| 91106 | BTA | 13.14 | 3.40 | 57 | 3.87 | | 91106 | 18.66 | 5.95 | 3.13 |
| 26607 | BH | 41.07 | 10.60 | 11 | 3.88 | | 26607 | 64.65 | 16.13 | 4.01 |
| 88683 | WIT | 49.68 | 12.81 | 136 | 3.88 | | 88683 | 60.32 | 18.52 | 3.26 |
| 91792 | DIG | 36.56 | 9.42 | -19 | 3.88 | | 91792 | 34.11 | 3.29 | 10.37 |
| 86098 | WYY | 77.57 | 19.96 | 99 | 3.89 | | 86098 | 139.96 | 47.96 | 2.92 |
| 90202 | EFT | 20.24 | 5.21 | 58 | 3.89 | | 90202 | 27.41 | 7.22 | 3.80 |
| 89941 | TZOO | 97.89 | 25.17 | 77 | 3.89 | | 89941 | 89.82 | 26.55 | 3.38 |
| 12490 | IBM | 17.96 | 4.61 | 35 | 3.89 | | 12490 | 23.97 | 5.35 | 4.48 |
| 92449 | REED | 88.75 | 22.77 | 89 | 3.90 | | 92449 | 136.11 | 39.15 | 3.48 |
| 87364 | QUIK | 77.73 | 19.93 | 17 | 3.90 | | 87364 | 107.40 | 35.02 | 3.07 |
| 89690 | EZA | 19.85 | 5.09 | 31 | 3.90 | | 89690 | 26.65 | 5.82 | 4.58 |
| 91629 | BTZ | 17.24 | 4.42 | 34 | 3.90 | | 91629 | 13.84 | 3.73 | 3.71 |
| 92491 | EMB | 7.18 | 1.84 | 11 | 3.90 | | 92491 | 7.77 | 1.31 | 5.94 |
| 77129 | KIM | 23.58 | 6.03 | -20 | 3.91 | | 77129 | 21.59 | -0.10 | -226.13 |
| 89856 | WF | 41.86 | 10.69 | 148 | 3.92 | | 89856 | 55.79 | 19.59 | 2.85 |
| 85280 | | 54.77 | 13.98 | -30 | 3.92 | | 85280 | 51.93 | 6.49 | 8.00 |
| 25785 | F | 85.41 | 21.73 | 234 | 3.93 | | 25785 | 110.34 | 34.98 | 3.15 |
| 46340 | SFE | 53.58 | 13.63 | 108 | 3.93 | | 46340 | 54.68 | 16.15 | 3.38 |
| 62367 | GTY | 22.13 | 5.63 | 40 | 3.93 | | 62367 | 28.04 | 7.72 | 3.63 |
| 91038 | AAU | 114.71 | 29.16 | 3 | 3.93 | | 91038 | 119.73 | 29.47 | 4.06 |
| 88568 | SPTN | 26.72 | 6.79 | -26 | 3.93 | | 88568 | 40.79 | 3.54 | 11.54 |
| 89839 | BKK | 7.10 | 1.80 | 40 | 3.94 | | 89839 | 12.76 | 4.01 | 3.18 |
| 91567 | FOF | 12.89 | 3.27 | 21 | 3.94 | | 91567 | 14.61 | 2.85 | 5.13 |
| 90073 | JFR | 21.58 | 5.47 | 53 | 3.94 | | 90073 | 25.20 | 6.52 | 3.87 |
| 79198 | TWI | 70.22 | 17.76 | -44 | 3.95 | | 79198 | 82.70 | 16.50 | 5.01 |
| 76613 | IO | 138.14 | 34.93 | -90 | 3.95 | | 76613 | 91.32 | 8.15 | 11.20 |
| 27983 | XRX | 28.41 | 7.18 | -13 | 3.96 | | 27983 | 34.84 | 3.28 | 10.64 |
| 83143 | IRM | 21.73 | 5.48 | 31 | 3.97 | | 83143 | 18.39 | 4.32 | 4.26 |
| 83264 | GEF | 27.10 | 6.82 | 31 | 3.97 | | 83264 | 30.90 | 6.73 | 4.59 |
| 77570 | TGB | 128.87 | 32.43 | -17 | 3.97 | | 77570 | 218.03 | 72.33 | 3.01 |
| 89012 | PNF | 13.43 | 3.38 | 69 | 3.98 | | 89012 | 18.59 | 6.71 | 2.77 |
| 87344 | SIFY | 84.91 | 21.30 | -2 | 3.99 | | 87344 | 60.21 | 11.92 | 5.05 |
| 83462 | TUP | 42.61 | 10.68 | 39 | 3.99 | | 83462 | 43.23 | 10.60 | 4.08 |
| 26084 | CTG | 46.00 | 11.48 | 27 | 4.01 | | 26084 | 56.74 | 13.16 | 4.31 |
| 92024 | SPIP | 5.07 | 1.26 | 14 | 4.01 | | 92024 | 5.91 | 1.47 | 4.02 |
| 89959 | TIP | 4.59 | 1.14 | 10 | 4.02 | | 89959 | 5.12 | 1.11 | 4.61 |
| 91379 | HUSA | 150.82 | 37.47 | -94 | 4.03 | | 91379 | 83.34 | 17.63 | 4.73 |
| 85570 | CDZI | 59.85 | 14.85 | -18 | 4.03 | | 85570 | 54.59 | 7.90 | 6.91 |
| 90464 | SPOK | 19.98 | 4.94 | 15 | 4.04 | | 90464 | 26.57 | 4.16 | 6.39 |
| 85285 | EDAP | 59.85 | 14.80 | 28 | 4.04 | | 85285 | 59.34 | 17.90 | 3.32 |
| 90525 | IAU | 16.18 | 4.00 | 42 | 4.04 | | 90525 | 16.66 | 4.84 | 3.44 |
| 91991 | FGB | 29.30 | 7.25 | 33 | 4.04 | | 91991 | 25.79 | 5.63 | 4.58 |
| 87236 | | 63.15 | 15.62 | -58 | 4.04 | | 87236 | 69.00 | 13.20 | 5.23 |
| 91335 | DOL | 13.68 | 3.38 | 10 | 4.05 | | 91335 | 13.66 | 1.80 | 7.57 |
| 37584 | BEN | 29.07 | 7.18 | 40 | 4.05 | | 37584 | 30.88 | 7.36 | 4.20 |
| 90269 | JRO | 26.56 | 6.54 | 68 | 4.06 | | 90269 | 34.90 | 9.13 | 3.82 |
| 83651 | CASI | 143.30 | 35.25 | 126 | 4.07 | | 83651 | 139.45 | 51.45 | 2.71 |
| 87198 | ARLP | 39.41 | 9.67 | 29 | 4.07 | | 87198 | 42.57 | 12.27 | 3.47 |
| 66616 | KF | 21.22 | 5.21 | -19 | 4.07 | | 66616 | 19.44 | -0.30 | -65.12 |
| 85571 | | 44.35 | 10.85 | -59 | 4.09 | | 85571 | 50.41 | 3.95 | 12.75 |
| 90706 | CTRN | 43.81 | 10.63 | 39 | 4.12 | | 90706 | 45.79 | 13.14 | 3.48 |
| 92351 | PCY | 8.27 | 2.01 | 31 | 4.13 | | 92351 | 11.66 | 3.33 | 3.50 |
| 90604 | MTEM | 113.81 | 27.56 | -36 | 4.13 | | 90604 | 114.40 | 35.91 | 3.19 |
| 88392 | ERJ | 26.14 | 6.30 | 37 | 4.15 | | 88392 | 23.77 | 5.77 | 4.12 |
| 86143 | VVR | 21.23 | 5.11 | 53 | 4.16 | | 86143 | 23.68 | 6.32 | 3.75 |
| 91948 | BLV | 8.33 | 2.00 | 9 | 4.17 | | 91948 | 9.89 | 1.30 | 7.61 |
| 77007 | IMBI | 263.96 | 63.22 | 21 | 4.17 | | 77007 | 438.62 | 142.41 | 3.08 |
| 90865 | | 55.29 | 13.22 | -52 | 4.18 | | 90865 | 56.90 | 12.39 | 4.59 |
| 90448 | GLD | 16.20 | 3.87 | 40 | 4.18 | | 90448 | 16.67 | 4.72 | 3.53 |
| 83633 | | 62.87 | 15.03 | -40 | 4.18 | | 83633 | 117.07 | 29.96 | 3.91 |
| 85616 | | 51.56 | 12.32 | 29 | 4.18 | | 85616 | 57.52 | 15.27 | 3.77 |
| 80070 | SU | 22.33 | 5.33 | 43 | 4.19 | | 80070 | 31.36 | 7.25 | 4.32 |
| 85653 | AAIC | 86.62 | 20.67 | 113 | 4.19 | | 85653 | 115.19 | 32.23 | 3.57 |
| 10812 | HTLD | 23.14 | 5.52 | 16 | 4.19 | | 10812 | 26.62 | 4.59 | 5.80 |
| 70332 | | 32.85 | 7.83 | 14 | 4.20 | | 70332 | 30.83 | 5.33 | 5.78 |
| 89167 | CCRN | 34.90 | 8.32 | -17 | 4.20 | | 89167 | 43.37 | 4.89 | 8.86 |
| 84084 | HIBB | 35.01 | 8.34 | -9 | 4.20 | | 84084 | 38.13 | 5.70 | 6.69 |
| 92599 | EIS | 20.23 | 4.81 | 59 | 4.20 | | 92599 | 28.59 | 7.75 | 3.69 |
| 83223 | EWW | 22.45 | 5.34 | 28 | 4.20 | | 83223 | 24.19 | 4.77 | 5.07 |
| 33209 | ENZ | 54.74 | 13.01 | -43 | 4.21 | | 33209 | 39.52 | 3.84 | 10.28 |
| 77467 | MYF | 13.67 | 3.25 | 41 | 4.21 | | 77467 | 17.22 | 4.74 | 3.64 |
| 90794 | RDS | 17.95 | 4.27 | 10 | 4.21 | | 90794 | 16.49 | 2.29 | 7.20 |
| 44274 | AEGN | 24.98 | 5.94 | -17 | 4.21 | | 44274 | 26.57 | 1.62 | 16.37 |
| 92793 | EC | 54.77 | 13.01 | -14 | 4.21 | | 92793 | 47.34 | 9.72 | 4.87 |
| 90383 | FXI | 18.61 | 4.40 | 34 | 4.23 | | 90383 | 21.34 | 4.83 | 4.42 |
| 82747 | ASTC | 150.91 | 35.68 | 267 | 4.23 | | 82747 | 213.73 | 75.60 | 2.83 |
| 90376 | EOI | 11.81 | 2.78 | 8 | 4.24 | | 90376 | 14.20 | 1.66 | 8.56 |
| 91959 | ESSA | 10.52 | 2.47 | 10 | 4.26 | | 91959 | 12.27 | 1.72 | 7.14 |
| 89445 | BDSI | 103.74 | 24.37 | 28 | 4.26 | | 89445 | 148.14 | 49.50 | 2.99 |
| 89745 | HYT | 16.34 | 3.84 | 48 | 4.26 | | 89745 | 23.99 | 5.98 | 4.01 |
| 83208 | EWA | 20.97 | 4.92 | 37 | 4.26 | | 83208 | 24.04 | 5.37 | 4.48 |
| 57074 | PAI | 10.05 | 2.36 | 26 | 4.26 | | 57074 | 11.12 | 2.88 | 3.85 |
| 76712 | AES | 22.48 | 5.27 | 75 | 4.27 | | 76712 | 27.93 | 8.88 | 3.15 |
| 86979 | JNPR | 28.09 | 6.58 | 54 | 4.27 | | 86979 | 26.76 | 7.74 | 3.46 |
| 85234 | CXW | 29.16 | 6.83 | 9 | 4.27 | | 85234 | 32.80 | 4.77 | 6.87 |
| 75261 | | 187.29 | 43.78 | -4 | 4.28 | | 75261 | 115.73 | 26.28 | 4.40 |
| 92820 | ERX | 57.06 | 13.33 | -61 | 4.28 | | 92820 | 50.56 | 2.54 | 19.92 |
| 89892 | EFR | 21.04 | 4.90 | 61 | 4.29 | | 89892 | 29.93 | 7.74 | 3.87 |
| 32651 | HL | 72.05 | 16.79 | -16 | 4.29 | | 32651 | 80.23 | 18.44 | 4.35 |
| 91914 | AWP | 25.53 | 5.94 | 26 | 4.29 | | 91914 | 28.73 | 5.69 | 5.05 |
| 27828 | HPQ | 37.43 | 8.71 | 15 | 4.30 | | 27828 | 46.68 | 10.47 | 4.46 |
| 89828 | UVSP | 18.11 | 4.21 | -33 | 4.30 | | 89828 | 26.62 | -0.53 | -50.69 |
| 92753 | ERII | 101.67 | 23.61 | -11 | 4.31 | | 92753 | 36.60 | 4.64 | 7.88 |
| 91458 | DBV | 7.23 | 1.68 | 21 | 4.31 | | 91458 | 8.46 | 2.24 | 3.78 |
| 84032 | NOV | 35.79 | 8.30 | 18 | 4.31 | | 84032 | 37.08 | 7.36 | 5.04 |
| 81292 | | 13.79 | 3.19 | 1 | 4.33 | | 81292 | 11.34 | 0.68 | 16.61 |

| | | | | | | | | | |
|---|---|---|---|---|---|---|---|---|---|
| 58421 | FRD | 31.09 | 7.17 | 6 | 4.33 | 58421 | 24.14 | 2.90 | 8.32 |
| 90409 | HURN | 36.54 | 8.43 | -10 | 4.34 | 90409 | 40.45 | 6.24 | 6.48 |
| 84607 | EMKR | 50.17 | 11.57 | -19 | 4.34 | 84607 | 24.95 | 0.73 | 34.30 |
| 13303 | NL | 73.39 | 16.91 | -74 | 4.34 | 13303 | 74.37 | 9.46 | 7.86 |
| 79448 | MUA | 11.31 | 2.60 | 48 | 4.35 | 79448 | 16.99 | 5.21 | 3.26 |
| 92279 | PXH | 22.21 | 5.09 | 42 | 4.37 | 92279 | 27.94 | 6.62 | 4.22 |
| 88564 | SRGA | 33.69 | 7.71 | 34 | 4.37 | 88564 | 34.09 | 7.55 | 4.51 |
| 91754 | III | 79.21 | 18.11 | 25 | 4.37 | 91754 | 89.99 | 19.07 | 4.72 |
| 89550 | PMX | 12.29 | 2.80 | 73 | 4.38 | 89550 | 20.34 | 7.10 | 2.87 |
| 10886 | DXYN | 85.31 | 19.46 | -54 | 4.38 | 10886 | 107.82 | 22.81 | 4.73 |
| 56936 | JHS | 11.18 | 2.54 | 30 | 4.39 | 56936 | 15.08 | 3.59 | 4.19 |
| 92397 | BKF | 24.13 | 5.49 | 50 | 4.40 | 92397 | 32.55 | 7.88 | 4.13 |
| 86242 | BELFB | 32.74 | 7.45 | -13 | 4.40 | 86242 | 33.36 | 2.92 | 11.43 |
| 92517 | XIN | 47.10 | 10.71 | 56 | 4.40 | 92517 | 58.03 | 19.06 | 3.05 |
| 92149 | VTA | 18.85 | 4.28 | 45 | 4.40 | 92149 | 24.29 | 5.85 | 4.15 |
| 75336 | CIF | 17.89 | 4.06 | 34 | 4.40 | 75336 | 28.48 | 5.75 | 4.95 |
| 88391 | CNQ | 30.44 | 6.91 | 22 | 4.41 | 88391 | 36.41 | 7.16 | 5.09 |
| 82759 | NUAN | 23.22 | 5.27 | 28 | 4.41 | 82759 | 29.26 | 6.10 | 4.80 |
| 91801 | CORR | 33.13 | 7.51 | 49 | 4.41 | 91801 | 50.79 | 12.98 | 3.91 |
| 82508 | MYGN | 39.10 | 8.86 | -9 | 4.41 | 82508 | 48.39 | 8.68 | 5.57 |
| 90835 | EFV | 15.71 | 3.56 | 12 | 4.42 | 90835 | 15.28 | 2.13 | 7.17 |
| 91448 | JXI | 7.93 | 1.79 | 5 | 4.42 | 91448 | 7.44 | 0.70 | 10.59 |
| 11379 | DXLG | 134.71 | 30.34 | 317 | 4.44 | 11379 | 116.54 | 41.42 | 2.81 |
| 85706 | NYMX | 113.05 | 25.34 | -61 | 4.46 | 85706 | 234.38 | 64.85 | 3.61 |
| 83509 | FCN | 31.03 | 6.95 | 49 | 4.46 | 83509 | 24.60 | 6.50 | 3.79 |
| 87285 | NSL | 27.08 | 6.06 | 68 | 4.47 | 87285 | 39.16 | 9.65 | 4.06 |
| 85905 | DSU | 17.85 | 3.99 | 28 | 4.47 | 85905 | 16.28 | 3.57 | 4.56 |
| 65752 | TISI | 33.09 | 7.37 | -47 | 4.49 | 65752 | 30.02 | -0.77 | -39.03 |
| 92349 | PZA | 5.85 | 1.30 | 18 | 4.50 | 92349 | 7.40 | 1.91 | 3.88 |
| 89879 | FRA | 16.86 | 3.73 | 36 | 4.52 | 89879 | 22.13 | 4.83 | 4.58 |
| 75461 | PMM | 11.27 | 2.49 | 33 | 4.52 | 75461 | 13.01 | 3.60 | 3.62 |
| 92539 | MLN | 7.27 | 1.61 | 25 | 4.52 | 92539 | 9.18 | 2.63 | 3.49 |
| 92130 | SYN | 136.16 | 30.10 | -91 | 4.52 | 92130 | 91.68 | 21.88 | 4.19 |
| 90387 | | 59.23 | 13.09 | 22 | 4.52 | 90387 | 37.58 | 7.73 | 4.86 |
| 23085 | | 19.67 | 4.34 | 34 | 4.53 | 23085 | 20.30 | 5.31 | 3.82 |
| 87337 | PLUG | 524.26 | ##### | -88 | 4.53 | 87337 | 92.77 | 8.13 | 11.41 |
| 89468 | TLT | 12.23 | 2.70 | 2 | 4.53 | 89468 | 16.11 | 1.33 | 12.12 |
| 92245 | AROC | 58.25 | 12.83 | -43 | 4.54 | 92245 | 63.91 | 10.61 | 6.02 |
| 89016 | EQNR | 23.53 | 5.18 | 27 | 4.55 | 89016 | 22.96 | 4.60 | 4.99 |
| 76684 | HWC | 27.86 | 6.12 | -24 | 4.55 | 76684 | 28.92 | 0.43 | 67.74 |
| 91874 | PFF | 14.17 | 3.11 | 17 | 4.56 | 91874 | 10.89 | 2.09 | 5.20 |
| 85724 | PRDO | 70.19 | 15.41 | -36 | 4.56 | 85724 | 70.70 | 13.01 | 5.43 |
| 76644 | X | 77.84 | 17.07 | -51 | 4.56 | 76644 | 109.30 | 20.59 | 5.31 |

| | | | | | | | | | | |
|---|---|---|---|---|---|---|---|---|---|---|
| 83450 | | 23.92 | 5.24 | -1 | 4.56 | | 83450 | 18.70 | 1.89 | 9.92 |
| 75466 | WBK | 26.80 | 5.87 | 45 | 4.56 | | 75466 | 32.42 | 7.34 | 4.42 |
| 75649 | TG | 30.44 | 6.67 | -13 | 4.57 | | 75649 | 35.07 | 3.46 | 10.14 |
| 24328 | EQT | 29.30 | 6.40 | 7 | 4.58 | | 24328 | 28.75 | 4.38 | 6.56 |
| 80912 | DWSN | 60.16 | 13.13 | -29 | 4.58 | | 80912 | 69.26 | 15.48 | 4.47 |
| 89449 | KIRK | 98.84 | 21.58 | 260 | 4.58 | | 89449 | 181.30 | 55.91 | 3.24 |
| 89469 | IEF | 5.33 | 1.16 | 6 | 4.59 | | 89469 | 6.66 | 0.76 | 8.76 |
| 78927 | SITC | 62.36 | 13.57 | 39 | 4.60 | | 78927 | 39.11 | 9.16 | 4.27 |
| 81116 | VECO | 126.23 | 27.24 | 17 | 4.63 | | 81116 | 140.71 | 36.05 | 3.90 |
| 92120 | DEX | 20.20 | 4.35 | 22 | 4.64 | | 92120 | 25.81 | 4.57 | 5.65 |
| 63335 | | 79.53 | 17.12 | -78 | 4.64 | | 63335 | 81.77 | 5.74 | 14.24 |
| 89307 | MANT | 24.18 | 5.20 | -3 | 4.65 | | 89307 | 19.38 | 1.24 | 15.57 |
| 85168 | HLX | 51.06 | 10.98 | -25 | 4.65 | | 85168 | 42.94 | 8.15 | 5.27 |
| 82107 | WABC | 13.09 | 2.81 | 9 | 4.66 | | 82107 | 18.22 | 2.21 | 8.25 |
| 89744 | CHY | 15.15 | 3.25 | 11 | 4.67 | | 89744 | 17.29 | 2.36 | 7.32 |
| 51692 | | 339.73 | 72.76 | -17 | 4.67 | | 51692 | 407.57 | 129.70 | 3.14 |
| 89562 | ADRE | 19.36 | 4.14 | 32 | 4.68 | | 89562 | 25.80 | 5.35 | 4.82 |
| 90219 | SFL | 33.60 | 7.18 | -2 | 4.68 | | 90219 | 40.16 | 6.99 | 5.74 |
| 92439 | IFGL | 14.63 | 3.12 | 24 | 4.69 | | 92439 | 17.44 | 3.40 | 5.13 |
| 85424 | DRQ | 37.88 | 8.08 | 46 | 4.69 | | 85424 | 63.30 | 14.99 | 4.22 |
| 87061 | CYRN | 36.35 | 7.75 | 96 | 4.69 | | 87061 | 47.66 | 13.97 | 3.41 |
| 81040 | KEP | 28.08 | 5.99 | 27 | 4.69 | | 81040 | 17.68 | 3.79 | 4.67 |
| 75455 | CXE | 14.36 | 3.06 | 64 | 4.69 | | 75455 | 24.42 | 7.12 | 3.43 |
| 89571 | | 69.84 | 14.87 | -59 | 4.70 | | 89571 | 116.23 | 29.63 | 3.92 |
| 89611 | ACTG | 86.85 | 18.47 | -2 | 4.70 | | 89611 | 94.94 | 28.13 | 3.37 |
| 75263 | HYB | 20.84 | 4.42 | 68 | 4.71 | | 75263 | 34.44 | 8.95 | 3.85 |
| 90648 | ETB | 11.44 | 2.43 | 7 | 4.71 | | 90648 | 15.16 | 1.69 | 8.97 |
| 86416 | | 17.89 | 3.79 | 47 | 4.72 | | 86416 | 23.67 | 5.93 | 3.99 |
| 90533 | WTI | 82.63 | 17.48 | -71 | 4.73 | | 90533 | 37.96 | -3.03 | -12.55 |
| 90223 | DRAD | 47.11 | 9.94 | -2 | 4.74 | | 90223 | 93.64 | 25.11 | 3.73 |
| 91461 | WU | 15.43 | 3.25 | 19 | 4.75 | | 91461 | 18.21 | 3.19 | 5.70 |
| 91547 | LDOS | 20.61 | 4.34 | -4 | 4.75 | | 91547 | 23.44 | 1.84 | 12.72 |
| 90302 | BGT | 17.08 | 3.59 | 46 | 4.76 | | 90302 | 27.59 | 6.49 | 4.25 |
| 89323 | UCBI | 37.50 | 7.88 | -68 | 4.76 | | 89323 | 47.68 | 1.07 | 44.62 |
| 75070 | INT | 25.39 | 5.33 | 16 | 4.76 | | 75070 | 26.14 | 4.62 | 5.66 |
| 83883 | OPCH | 81.11 | 16.98 | 61 | 4.78 | | 83883 | 112.16 | 39.09 | 2.87 |
| 92350 | PLW | 7.10 | 1.49 | 3 | 4.78 | | 92350 | 9.16 | 0.70 | 13.03 |
| 89551 | PZC | 13.44 | 2.81 | 54 | 4.79 | | 89551 | 18.56 | 5.80 | 3.20 |
| 89231 | IXP | 9.18 | 1.92 | -2 | 4.79 | | 89231 | 10.28 | 0.25 | 40.79 |
| 84781 | MHD | 13.85 | 2.89 | 50 | 4.79 | | 84781 | 18.48 | 5.47 | 3.38 |
| 86593 | GBL | 24.38 | 5.06 | 29 | 4.82 | | 86593 | 37.71 | 8.22 | 4.59 |
| 89188 | IGE | 19.06 | 3.95 | 7 | 4.82 | | 89188 | 21.28 | 2.71 | 7.86 |
| 90510 | PHD | 16.03 | 3.32 | 40 | 4.83 | | 90510 | 22.26 | 5.14 | 4.33 |
| 87133 | BSD | 13.46 | 2.79 | 45 | 4.83 | | 87133 | 17.53 | 5.04 | 3.48 |
| 91997 | FXN | 26.53 | 5.48 | 1 | 4.84 | | 91997 | 26.01 | 3.06 | 8.49 |
| 92273 | MUB | 4.09 | 0.84 | 10 | 4.85 | | 92273 | 4.64 | 1.01 | 4.59 |
| 81521 | | 15.51 | 3.19 | -10 | 4.86 | | 81521 | 28.11 | 2.75 | 10.23 |
| 71722 | CMU | 13.35 | 2.74 | 51 | 4.87 | | 71722 | 20.33 | 5.63 | 3.61 |
| 92824 | AGZ | 1.66 | 0.34 | 4 | 4.87 | | 92824 | 1.64 | 0.38 | 4.36 |
| 91334 | DOO | 14.80 | 3.03 | 11 | 4.88 | | 91334 | 13.74 | 1.83 | 7.50 |
| 90827 | BDJ | 10.99 | 2.25 | -7 | 4.88 | | 90827 | 10.63 | -0.17 | -62.55 |
| 83233 | GEF | 31.62 | 6.43 | 11 | 4.91 | | 83233 | 36.71 | 6.64 | 5.53 |
| 79229 | PMO | 10.61 | 2.15 | 29 | 4.94 | | 79229 | 12.42 | 3.22 | 3.86 |
| 62068 | SOR | 23.90 | 4.83 | 15 | 4.94 | | 62068 | 27.54 | 5.03 | 5.48 |
| 89726 | AVK | 18.33 | 3.70 | 15 | 4.95 | | 89726 | 20.10 | 3.09 | 6.51 |
| 90646 | GLQ | 17.68 | 3.56 | 14 | 4.96 | | 90646 | 22.07 | 3.28 | 6.73 |
| 78854 | MNP | 12.82 | 2.58 | 43 | 4.96 | | 78854 | 15.56 | 4.69 | 3.32 |
| 80193 | BZH | 95.62 | 19.26 | 20 | 4.97 | | 80193 | 76.06 | 19.27 | 3.95 |
| 77729 | TGA | 86.78 | 17.46 | -24 | 4.97 | | 77729 | 128.12 | 27.37 | 4.68 |
| 90162 | GNW | 170.56 | 34.29 | 65 | 4.97 | | 90162 | 106.87 | 32.04 | 3.34 |
| 86222 | DHY | 21.64 | 4.34 | 43 | 4.98 | | 86222 | 31.45 | 6.90 | 4.56 |
| 92682 | TAOP | 132.48 | 26.60 | -83 | 4.98 | | 92682 | 232.11 | 56.05 | 4.14 |
| 91336 | DTH | 15.03 | 3.01 | 9 | 4.99 | | 91336 | 15.15 | 1.84 | 8.24 |
| 92291 | TFI | 4.39 | 0.88 | 12 | 4.99 | | 92291 | 5.37 | 1.30 | 4.14 |
| 91530 | SLX | 40.62 | 8.11 | 19 | 5.01 | | 91530 | 52.56 | 11.49 | 4.57 |
| 84052 | | 97.33 | 19.44 | -34 | 5.01 | | 84052 | 49.03 | 6.06 | 8.09 |
| 75033 | VOXX | 57.28 | 11.43 | -21 | 5.01 | | 75033 | 56.87 | 8.07 | 7.05 |
| 43123 | ATI | 44.89 | 8.92 | -15 | 5.04 | | 43123 | 42.77 | 7.94 | 5.39 |
| 91665 | GLDD | 30.78 | 6.09 | 60 | 5.05 | | 91665 | 34.66 | 10.55 | 3.29 |
| 91318 | XOP | 28.10 | 5.55 | -10 | 5.06 | | 91318 | 28.91 | 2.74 | 10.53 |
| 89637 | TS | 32.34 | 6.39 | 2 | 5.06 | | 89637 | 42.61 | 6.51 | 6.54 |
| 91095 | LQDT | 63.20 | 12.47 | -26 | 5.07 | | 91095 | 65.90 | 13.18 | 5.00 |
| 90762 | | 20.72 | 4.09 | -2 | 5.07 | | 90762 | 20.62 | 1.82 | 11.35 |
| 89782 | | 11.64 | 2.29 | 42 | 5.08 | | 89782 | 15.33 | 4.49 | 3.41 |
| 82279 | SENEA | 17.72 | 3.48 | 36 | 5.09 | | 82279 | 17.97 | 4.54 | 3.96 |
| 27422 | CLF | 104.39 | 20.49 | -70 | 5.09 | | 27422 | 149.61 | 33.30 | 4.49 |
| 86218 | SJR | 14.03 | 2.75 | 2 | 5.10 | | 86218 | 17.99 | 1.91 | 9.41 |
| 90919 | CGO | 17.50 | 3.43 | 7 | 5.10 | | 90919 | 25.26 | 3.24 | 7.80 |
| 50550 | TPC | 43.04 | 8.44 | -32 | 5.10 | | 50550 | 44.80 | 3.82 | 11.71 |
| 92468 | PSTI | 53.47 | 10.44 | 92 | 5.12 | | 92468 | 68.45 | 21.52 | 3.18 |
| 90886 | GLP | 35.95 | 6.99 | 44 | 5.15 | | 90886 | 40.58 | 9.84 | 4.12 |
| 89901 | WLL | 69.33 | 13.47 | -66 | 5.15 | | 89901 | 57.09 | 4.28 | 13.33 |
| 86872 | NAC | 12.45 | 2.42 | 33 | 5.15 | | 86872 | 15.04 | 3.89 | 3.87 |
| 89138 | BG | 19.39 | 3.75 | 3 | 5.17 | | 89138 | 17.68 | 1.74 | 10.14 |
| 52425 | WRE | 16.31 | 3.15 | -19 | 5.17 | | 52425 | 14.54 | -1.08 | -13.44 |
| 78936 | BKN | 14.04 | 2.71 | 53 | 5.17 | | 78936 | 20.09 | 5.85 | 3.43 |
| 65285 | HOV | 85.73 | 16.55 | -60 | 5.18 | | 65285 | 136.70 | 34.27 | 3.99 |
| 76798 | EZPW | 54.99 | 10.59 | -49 | 5.19 | | 76798 | 51.00 | 3.65 | 13.96 |
| 78056 | MCA | 12.14 | 2.34 | 34 | 5.20 | | 78056 | 14.91 | 3.94 | 3.79 |
| 92679 | SB | 74.04 | 14.24 | -73 | 5.20 | | 92679 | 100.25 | 20.18 | 4.97 |
| 15069 | MRO | 35.35 | 6.77 | -14 | 5.22 | | 15069 | 27.33 | 2.69 | 10.16 |
| 92601 | NOG | 99.77 | 19.11 | -13 | 5.22 | | 92601 | 125.65 | 29.49 | 4.26 |
| 89858 | PAA | 29.59 | 5.66 | 16 | 5.23 | | 89858 | 32.57 | 7.07 | 4.61 |
| 77818 | MMU | 11.63 | 2.22 | 36 | 5.24 | | 77818 | 12.72 | 3.81 | 3.34 |
| 83906 | SRCL | 24.24 | 4.62 | -30 | 5.24 | | 83906 | 27.90 | 0.41 | 68.25 |
| 89221 | BHK | 9.88 | 1.88 | 11 | 5.25 | | 89221 | 8.82 | 1.45 | 6.08 |
| 89780 | NHS | 21.01 | 3.96 | 52 | 5.30 | | 89780 | 30.25 | 7.21 | 4.20 |
| 91739 | | 85.27 | 16.07 | -83 | 5.31 | | 91739 | 54.60 | 0.69 | 79.69 |
| 89162 | NZF | 10.84 | 2.04 | 34 | 5.31 | | 89162 | 14.25 | 3.81 | 3.74 |
| 92857 | EDC | 61.21 | 11.52 | -27 | 5.31 | | 92857 | 77.42 | 17.58 | 4.40 |
| 90092 | SBRA | 32.91 | 6.19 | -1 | 5.31 | | 90092 | 33.81 | 4.42 | 7.65 |
| 89262 | WNEB | 12.36 | 2.32 | -3 | 5.32 | | 89262 | 13.55 | 0.59 | 23.06 |
| 92619 | GTE | 44.47 | 8.33 | -22 | 5.34 | | 92619 | 48.52 | 7.02 | 6.91 |
| 88173 | LXRX | 42.10 | 7.87 | -32 | 5.35 | | 88173 | 49.67 | 5.05 | 9.83 |
| 84827 | RMBS | 47.56 | 8.86 | -52 | 5.37 | | 84827 | 47.10 | 3.07 | 15.37 |
| 85917 | MUH | 14.06 | 2.62 | 38 | 5.38 | | 85917 | 17.87 | 4.54 | 3.93 |
| 83656 | AVDL | 60.10 | 11.15 | -34 | 5.39 | | 83656 | 77.48 | 16.08 | 4.82 |
| 14277 | SLB | 24.59 | 4.56 | -15 | 5.39 | | 14277 | 30.16 | 2.60 | 11.62 |
| 72813 | EMF | 31.18 | 5.77 | 45 | 5.40 | | 72813 | 43.72 | 10.13 | 4.31 |
| 89348 | WEA | 13.69 | 2.53 | 36 | 5.40 | | 89348 | 18.21 | 4.35 | 4.19 |
| 37402 | FSTR | 43.52 | 8.05 | -49 | 5.41 | | 37402 | 49.28 | 5.27 | 9.34 |
| 75475 | VLT | 17.47 | 3.23 | 41 | 5.41 | | 75475 | 22.48 | 5.29 | 4.25 |
| 89897 | NMZ | 11.76 | 2.17 | 29 | 5.41 | | 89897 | 16.53 | 3.67 | 4.51 |
| 76708 | ATGE | 40.78 | 7.51 | -18 | 5.43 | | 76708 | 32.79 | 3.25 | 10.10 |
| 34833 | OXY | 19.68 | 3.62 | 6 | 5.43 | | 34833 | 19.16 | 2.15 | 8.90 |
| 80375 | ZEUS | 48.21 | 8.86 | -30 | 5.44 | | 80375 | 48.02 | 4.59 | 10.45 |
| 92747 | URG | 62.84 | 11.53 | 12 | 5.45 | | 92747 | 100.41 | 25.56 | 3.93 |
| 77152 | MYD | 12.66 | 2.32 | 27 | 5.45 | | 77152 | 14.67 | 3.36 | 4.36 |
| 84007 | OCN | 66.42 | 12.16 | -80 | 5.46 | | 84007 | 67.69 | 4.84 | 14.00 |
| 87659 | EPC | 21.64 | 3.96 | -8 | 5.47 | | 87659 | 21.80 | 1.46 | 14.88 |
| 92095 | INFN | 46.63 | 8.52 | -55 | 5.47 | | 92095 | 40.08 | -0.48 | -82.66 |
| 91975 | ACM | 23.90 | 4.36 | -14 | 5.48 | | 91975 | 17.96 | 0.08 | 236.96 |
| 76934 | ISIG | 87.83 | 16.03 | 52 | 5.48 | | 76934 | 129.72 | 37.07 | 3.50 |
| 75573 | ODP | 92.18 | 16.82 | -13 | 5.48 | | 75573 | 57.07 | 11.30 | 5.05 |
| 20415 | JEF | 28.86 | 5.26 | -10 | 5.48 | | 20415 | 24.42 | 1.79 | 13.63 |
| 88729 | | 67.79 | 12.36 | -51 | 5.48 | | 88729 | 86.62 | 16.55 | 5.23 |
| 75459 | MEN | 12.42 | 2.26 | 36 | 5.49 | | 75459 | 16.25 | 4.15 | 3.92 |
| 10890 | UIS | 114.96 | 20.91 | 37 | 5.50 | | 10890 | 121.83 | 33.65 | 3.62 |
| 91453 | GSL | 76.29 | 13.86 | -78 | 5.50 | | 91453 | 98.12 | 11.95 | 8.21 |
| 91966 | PNNT | 42.15 | 7.66 | 76 | 5.51 | | 91966 | 51.56 | 13.46 | 3.83 |
| 81165 | MIND | 53.67 | 9.72 | -36 | 5.52 | | 81165 | 56.67 | 10.30 | 5.50 |
| 92396 | ECH | 26.02 | 4.71 | 39 | 5.53 | | 92396 | 37.59 | 8.62 | 4.36 |
| 88853 | FLR | 24.53 | 4.44 | -28 | 5.53 | | 88853 | 28.03 | 0.22 | 126.31 |
| 77460 | PPR | 16.99 | 3.07 | 37 | 5.53 | | 77460 | 21.98 | 4.94 | 4.45 |
| 85916 | MUC | 12.66 | 2.28 | 35 | 5.55 | | 85916 | 15.82 | 4.13 | 3.83 |
| 90853 | ATCO | 34.20 | 6.16 | -12 | 5.56 | | 90853 | 27.41 | 2.28 | 12.00 |
| 89330 | VALE | 63.90 | 11.48 | 9 | 5.57 | | 89330 | 71.03 | 18.51 | 3.84 |
| 79139 | MVT | 14.46 | 2.59 | 44 | 5.58 | | 79139 | 20.49 | 5.30 | 3.86 |
| 92294 | UEC | 146.46 | 26.24 | 303 | 5.58 | | 92294 | 357.39 | 107.36 | 3.33 |
| 63781 | | 60.81 | 10.89 | -47 | 5.59 | | 63781 | 52.55 | 4.88 | 10.77 |
| 91184 | CBIO | 192.99 | 34.49 | -96 | 5.60 | | 91184 | 168.20 | 22.62 | 7.43 |
| 89235 | IXC | 16.30 | 2.91 | 0 | 5.61 | | 89235 | 16.41 | 1.26 | 13.00 |
| 92776 | WPRT | 79.84 | 14.23 | -74 | 5.61 | | 92776 | 100.76 | 21.74 | 4.63 |
| 90279 | | 104.05 | 18.49 | -89 | 5.63 | | 90279 | 90.65 | 10.81 | 8.38 |
| 79790 | | 75.91 | 13.48 | -37 | 5.63 | | 79790 | 66.63 | 14.87 | 4.48 |
| 80520 | CUBA | 18.63 | 3.31 | 17 | 5.63 | | 80520 | 23.73 | 3.98 | 5.96 |
| 89777 | BLW | 11.34 | 2.01 | 25 | 5.65 | | 89777 | 14.93 | 3.13 | 4.77 |
| 75623 | CXH | 12.73 | 2.25 | 41 | 5.66 | | 75623 | 17.99 | 4.75 | 3.79 |
| 86102 | FHI | 19.83 | 3.50 | 57 | 5.66 | | 86102 | 32.59 | 9.23 | 3.53 |
| 89459 | BLE | 13.54 | 2.39 | 42 | 5.66 | | 89459 | 19.73 | 5.06 | 3.90 |
| 77466 | MYC | 12.94 | 2.28 | 42 | 5.67 | | 77466 | 16.73 | 4.69 | 3.57 |
| 70826 | MFM | 13.58 | 2.40 | 52 | 5.67 | | 70826 | 23.38 | 6.16 | 3.79 |
| 90205 | FEN | 18.08 | 3.19 | 37 | 5.68 | | 90205 | 28.96 | 6.66 | 4.35 |
| 91707 | MVO | 52.88 | 9.21 | -11 | 5.74 | | 91707 | 68.02 | 15.49 | 4.39 |
| 92189 | DEM | 20.22 | 3.52 | 23 | 5.74 | | 92189 | 24.20 | 4.52 | 5.36 |
| 37381 | FORD | 51.24 | 8.92 | -42 | 5.74 | | 37381 | 30.71 | -0.93 | -32.93 |
| 77925 | MQT | 13.12 | 2.28 | 33 | 5.74 | | 77925 | 16.31 | 4.02 | 4.06 |
| 90989 | CCO | 49.76 | 8.65 | -16 | 5.75 | | 90989 | 37.11 | 4.53 | 8.19 |
| 92493 | LRN | 38.49 | 6.68 | 32 | 5.76 | | 92493 | 43.78 | 10.47 | 4.18 |
| 89281 | | 11.98 | 2.08 | 43 | 5.77 | | 89281 | 13.17 | 4.37 | 3.02 |
| 78990 | | 74.59 | 12.93 | -28 | 5.77 | | 78990 | 65.74 | 11.78 | 5.58 |
| 78057 | MFT | 13.41 | 2.32 | 35 | 5.77 | | 78057 | 17.13 | 4.24 | 4.04 |
| 76858 | | 132.38 | 22.74 | -68 | 5.82 | | 76858 | 132.46 | 35.33 | 3.75 |
| 90088 | MCHX | 58.86 | 10.11 | -55 | 5.82 | | 90088 | 54.39 | 2.65 | 20.50 |
| 90433 | NGD | 65.80 | 11.29 | -47 | 5.83 | | 90433 | 83.34 | 18.68 | 4.46 |
| 90978 | SPWR | 140.54 | 24.10 | -87 | 5.83 | | 90978 | 147.73 | 19.87 | 7.43 |
| 89543 | SPEU | 14.71 | 2.52 | 3 | 5.83 | | 89543 | 15.25 | 1.28 | 11.91 |
| 77486 | | 91.32 | 15.65 | -32 | 5.84 | | 77486 | 172.70 | 54.91 | 3.15 |
| 92580 | UGA | 27.18 | 4.64 | 18 | 5.86 | | 92580 | 35.13 | 6.47 | 5.43 |
| 25304 | HSC | 51.79 | 8.83 | -28 | 5.87 | | 25304 | 37.65 | 3.58 | 10.53 |
| 87204 | CX | 43.07 | 7.34 | -25 | 5.87 | | 87204 | 44.40 | 5.39 | 8.23 |
| 89361 | BBK | 14.09 | 2.39 | 47 | 5.90 | | 89361 | 19.85 | 5.43 | 3.66 |
| 92348 | PZT | 5.73 | 0.97 | 15 | 5.91 | | 92348 | 7.24 | 1.63 | 4.44 |
| 91714 | DGL | 16.12 | 2.73 | 23 | 5.91 | | 91714 | 16.57 | 3.38 | 4.90 |
| 84041 | TGI | 36.51 | 6.17 | -46 | 5.92 | | 84041 | 40.88 | 1.70 | 24.07 |
| 92429 | | 87.57 | 14.77 | -88 | 5.93 | | 92429 | 83.19 | 5.61 | 14.82 |
| 89103 | GLAD | 24.49 | 4.12 | -10 | 5.95 | | 89103 | 24.79 | 1.59 | 15.56 |
| 78034 | PDCO | 22.23 | 3.74 | 5 | 5.95 | | 78034 | 25.15 | 3.54 | 7.11 |
| 89538 | BAF | 13.28 | 2.23 | 35 | 5.96 | | 89538 | 16.33 | 4.16 | 3.92 |
| 91301 | NHF | 28.36 | 4.76 | -13 | 5.96 | | 91301 | 27.15 | 2.77 | 9.82 |
| 79839 | ITRI | 30.00 | 4.99 | -26 | 6.01 | | 79839 | 31.48 | 0.93 | 33.79 |
| 83976 | ANF | 48.03 | 7.98 | -13 | 6.02 | | 83976 | 38.54 | 5.40 | 7.13 |
| 91715 | DBB | 23.79 | 3.95 | 29 | 6.02 | | 91715 | 34.80 | 6.87 | 5.07 |
| 88970 | SHBI | 30.80 | 5.11 | -39 | 6.03 | | 88970 | 36.71 | 1.22 | 29.99 |
| 91933 | HYG | 6.91 | 1.15 | 7 | 6.03 | | 91933 | 7.32 | 0.89 | 8.27 |
| 89545 | FEZ | 18.30 | 3.02 | -3 | 6.05 | | 89545 | 17.43 | 1.02 | 17.07 |
| 67598 | PEAK | 20.30 | 3.35 | 10 | 6.06 | | 67598 | 15.41 | 2.07 | 7.44 |
| 89336 | NVG | 10.20 | 1.68 | 22 | 6.06 | | 89336 | 12.15 | 2.66 | 4.57 |
| 78059 | MIY | 12.69 | 2.09 | 31 | 6.07 | | 78059 | 15.46 | 3.71 | 4.17 |
| 91320 | XME | 40.14 | 6.61 | -6 | 6.07 | | 91320 | 51.43 | 9.26 | 5.55 |
| 85340 | MFL | 13.42 | 2.19 | 39 | 6.13 | | 85340 | 17.37 | 4.62 | 3.76 |
| 75840 | MFV | 22.01 | 3.58 | 27 | 6.14 | | 75840 | 28.82 | 5.30 | 5.44 |
| 47677 | CBB | 30.16 | 4.90 | -19 | 6.16 | | 47677 | 45.60 | 7.12 | 6.41 |
| 17961 | | 19.92 | 3.23 | -26 | 6.16 | | 17961 | 24.65 | 0.23 | 104.93 |
| 79636 | GSS | 101.55 | 16.46 | -37 | 6.17 | | 79636 | 136.38 | 39.61 | 3.44 |
| 85931 | WDR | 41.84 | 6.75 | 17 | 6.20 | | 85931 | 50.15 | 10.84 | 4.63 |
| 91611 | FSLR | 70.69 | 11.38 | -69 | 6.21 | | 91611 | 57.93 | 4.02 | 14.42 |
| 11267 | CATO | 28.68 | 4.61 | -5 | 6.22 | | 11267 | 27.44 | 3.12 | 8.78 |
| 90384 | MTL | 128.49 | 20.66 | -75 | 6.22 | | 90384 | 151.33 | 39.94 | 3.79 |
| 52396 | SUP | 28.32 | 4.55 | -54 | 6.22 | | 52396 | 37.98 | 0.73 | 51.71 |
| 85172 | AEG | 30.54 | 4.91 | -23 | 6.22 | | 85172 | 31.60 | 1.45 | 21.73 |
| 59555 | HMC | 17.23 | 2.76 | 24 | 6.24 | | 59555 | 26.48 | 5.09 | 5.20 |
| 76711 | RGS | 21.34 | 3.42 | 17 | 6.24 | | 76711 | 9.96 | 2.01 | 4.94 |
| 90307 | KRG | 22.77 | 3.65 | -37 | 6.24 | | 90307 | 21.71 | -2.35 | -9.22 |
| 64805 | SBR | 22.72 | 3.64 | -9 | 6.25 | | 64805 | 29.42 | 2.99 | 9.84 |
| 90056 | TYG | 22.71 | 3.63 | 17 | 6.26 | | 90056 | 33.79 | 6.01 | 5.62 |
| 89011 | PMF | 15.04 | 2.40 | 64 | 6.26 | | 89011 | 22.78 | 7.02 | 3.24 |
| 90988 | BKD | 65.41 | 10.44 | 20 | 6.27 | | 90988 | 79.52 | 17.76 | 4.48 |
| 88434 | INFI | 96.30 | 15.31 | -85 | 6.29 | | 88434 | 108.71 | 15.22 | 7.14 |
| 92862 | UGL | 32.25 | 5.12 | 18 | 6.30 | | 92862 | 31.32 | 6.44 | 4.86 |
| 85744 | RMTI | 59.41 | 9.42 | -46 | 6.31 | | 85744 | 38.16 | 0.79 | 48.06 |
| 89050 | BFK | 13.49 | 2.14 | 47 | 6.32 | | 89050 | 20.09 | 5.48 | 3.67 |
| 46392 | | 36.41 | 5.74 | 33 | 6.35 | | 46392 | 48.56 | 10.11 | 4.80 |
| 91934 | RSX | 37.54 | 5.91 | 43 | 6.35 | | 91934 | 51.41 | 12.32 | 4.17 |
| 88930 | NXJ | 12.29 | 1.93 | 41 | 6.35 | | 88930 | 15.42 | 4.42 | 3.49 |
| 23799 | CIA | 21.98 | 3.46 | -22 | 6.36 | | 23799 | 22.73 | -0.14 | -167.58 |
| 79702 | PTN | 74.62 | 11.74 | -21 | 6.36 | | 79702 | 109.19 | 26.79 | 4.08 |
| 39917 | WY | 28.28 | 4.43 | -29 | 6.38 | | 39917 | 32.59 | 2.27 | 14.36 |
| 11790 | ALCO | 30.96 | 4.84 | -28 | 6.40 | | 11790 | 36.22 | 1.45 | 24.91 |
| 83565 | MFIN | 53.99 | 8.39 | -39 | 6.44 | | 83565 | 30.69 | 0.44 | 70.52 |
| 10333 | XOMA | 141.61 | 21.98 | -93 | 6.44 | | 10333 | 253.88 | 65.86 | 3.86 |
| 61815 | | 27.42 | 4.24 | -24 | 6.47 | | 61815 | 29.13 | 1.25 | 23.34 |
| 89760 | BBL | 31.52 | 4.87 | 16 | 6.47 | | 89760 | 34.91 | 6.95 | 5.03 |
| 11850 | XOM | 13.50 | 2.09 | -15 | 6.47 | | 11850 | 14.02 | -0.67 | -20.81 |
| 90689 | PGP | 28.61 | 4.42 | 22 | 6.47 | | 90689 | 33.89 | 5.97 | 5.68 |
| 91202 | SLV | 38.03 | 5.86 | 30 | 6.49 | | 91202 | 34.87 | 7.14 | 4.88 |
| 86556 | EVN | 14.55 | 2.24 | 45 | 6.49 | | 86556 | 20.53 | 5.50 | 3.73 |
| 24441 | CDE | 92.83 | 14.27 | -49 | 6.50 | | 24441 | 100.88 | 19.51 | 5.17 |
| 76083 | CEE | 33.05 | 5.07 | 34 | 6.52 | | 76083 | 39.58 | 8.59 | 4.61 |
| 87825 | UTSI | 58.60 | 8.95 | -51 | 6.55 | | 87825 | 65.12 | 4.39 | 14.82 |
| 86921 | TCP | 25.03 | 3.82 | 38 | 6.55 | | 86921 | 33.76 | 8.19 | 4.12 |
| 91579 | KBR | 29.43 | 4.49 | 0 | 6.56 | | 91579 | 28.67 | 3.78 | 7.59 |
| 75060 | CIK | 14.72 | 2.24 | 20 | 6.58 | | 75060 | 18.50 | 3.23 | 5.72 |
| 88436 | ENDP | 57.29 | 8.70 | -72 | 6.58 | | 88436 | 66.23 | 4.31 | 15.36 |
| 10239 | PTVCB | 11.48 | 1.74 | -8 | 6.59 | | 10239 | 17.24 | 0.45 | 38.12 |
| 49488 | LXU | 57.87 | 8.75 | -34 | 6.61 | | 49488 | 45.51 | 8.21 | 5.54 |
| 79461 | AWF | 17.68 | 2.67 | 38 | 6.62 | | 79461 | 26.69 | 5.74 | 4.65 |
| 77806 | MQY | 12.70 | 1.92 | 29 | 6.62 | | 77806 | 16.25 | 3.66 | 4.44 |
| 90068 | MFD | 18.03 | 2.71 | -4 | 6.65 | | 90068 | 17.53 | 1.05 | 16.66 |
| 89600 | NKX | 12.84 | 1.93 | 23 | 6.66 | | 89600 | 14.92 | 3.05 | 4.90 |
| 75811 | MUFG | 20.64 | 3.08 | -22 | 6.69 | | 75811 | 21.88 | -0.15 | -150.14 |
| 28484 | HES | 29.71 | 4.44 | -24 | 6.70 | | 28484 | 29.22 | 0.81 | 36.08 |
| 77469 | MYN | 11.70 | 1.75 | 34 | 6.70 | | 77469 | 15.33 | 3.95 | 3.88 |
| 79131 | | 29.67 | 4.42 | -17 | 6.72 | | 79131 | 26.47 | 1.58 | 16.76 |
| 82587 | IVAC | 62.94 | 9.36 | 3 | 6.72 | | 82587 | 59.20 | 12.87 | 4.60 |
| 85935 | MFA | 13.69 | 2.04 | 13 | 6.73 | | 85935 | 16.86 | 2.54 | 6.63 |
| 92844 | YCS | 22.72 | 3.37 | 13 | 6.73 | | 92844 | 20.36 | 2.94 | 6.92 |
| 24643 | HWM | 38.81 | 5.76 | -33 | 6.74 | | 24643 | 36.45 | 2.10 | 17.33 |
| 58975 | OII | 35.54 | 5.26 | -17 | 6.76 | | 58975 | 45.44 | 6.09 | 7.46 |
| 89429 | CHI | 15.51 | 2.29 | 7 | 6.76 | | 89429 | 20.59 | 2.40 | 8.56 |
| 81734 | PAAS | 53.34 | 7.88 | -14 | 6.77 | | 81734 | 56.40 | 9.20 | 6.13 |
| 75064 | GSK | 12.11 | 1.79 | 3 | 6.77 | | 75064 | 13.34 | 1.04 | 12.89 |
| 91710 | DBP | 19.37 | 2.85 | 22 | 6.78 | | 91710 | 19.46 | 3.73 | 5.22 |
| 75320 | USM | 15.80 | 2.32 | 20 | 6.82 | | 75320 | 17.71 | 3.17 | 5.59 |
| 89922 | XPER | 35.41 | 5.18 | 55 | 6.83 | | 89922 | 47.50 | 12.79 | 3.71 |
| 68866 | SSL | 23.67 | 3.43 | -3 | 6.90 | | 68866 | 21.82 | 1.79 | 12.18 |
| 91700 | GVI | 2.02 | 0.29 | 1 | 6.90 | | 91700 | 1.83 | 0.15 | 11.97 |
| 78061 | MPA | 12.03 | 1.74 | 35 | 6.91 | | 78061 | 17.51 | 4.27 | 4.10 |
| 66245 | MDC | 31.65 | 4.55 | 5 | 6.95 | | 66245 | 40.56 | 5.91 | 6.87 |
| 80233 | DAKT | 28.03 | 4.02 | -21 | 6.97 | | 80233 | 35.30 | 2.72 | 12.96 |
| 92847 | EUO | 18.79 | 2.69 | 14 | 6.98 | | 92847 | 15.16 | 2.35 | 6.46 |
| 88595 | PSO | 25.10 | 3.58 | 25 | 7.01 | | 88595 | 24.81 | 5.15 | 4.82 |
| 87043 | QMCO | 112.35 | 16.02 | -31 | 7.01 | | 87043 | 233.62 | 56.29 | 4.15 |
| 78963 | CHS | 47.74 | 6.80 | 34 | 7.02 | | 78963 | 82.98 | 19.39 | 4.28 |
| 91076 | CLMT | 51.07 | 7.25 | -75 | 7.04 | | 91076 | 62.20 | 7.27 | 8.55 |
| 89911 | TRQ | 81.35 | 11.48 | -6 | 7.09 | | 89911 | 147.57 | 34.68 | 4.26 |
| 87212 | MFC | 26.82 | 3.78 | -17 | 7.10 | | 87212 | 26.88 | 1.56 | 17.23 |
| 69446 | VGZ | 98.54 | 13.87 | -53 | 7.11 | | 69446 | 94.82 | 21.37 | 4.44 |
| 88779 | ICAD | 69.07 | 9.72 | -35 | 7.11 | | 88779 | 60.49 | 8.85 | 6.83 |
| 91086 | ACOR | 29.35 | 4.12 | -24 | 7.12 | | 91086 | 27.40 | 1.55 | 17.62 |
| 91834 | BBDC | 28.70 | 4.03 | -12 | 7.12 | | 91834 | 29.49 | 2.83 | 10.41 |
| 75837 | DSM | 12.69 | 1.78 | 26 | 7.13 | | 75837 | 16.62 | 3.48 | 4.78 |
| 26201 | CMTL | 41.11 | 5.75 | -47 | 7.15 | | 26201 | 36.98 | -0.89 | -41.48 |
| 77078 | TOT | 17.04 | 2.38 | -6 | 7.16 | | 77078 | 13.18 | 0.20 | 64.37 |
| 92608 | WIP | 8.19 | 1.14 | 8 | 7.19 | | 92608 | 9.62 | 1.21 | 7.96 |
| 90608 | INUV | 81.18 | 11.24 | 78 | 7.22 | | 90608 | 155.14 | 55.24 | 2.81 |
| 92684 | MHLD | 31.67 | 4.37 | -47 | 7.24 | | 92684 | 55.80 | 9.00 | 6.20 |
| 80368 | | 321.87 | 44.44 | -99 | 7.24 | | 80368 | 156.34 | 0.05 | 2845.08 |
| 82213 | RWT | 26.11 | 3.60 | 1 | 7.25 | | 82213 | 27.53 | 3.21 | 8.58 |
| 75039 | BHP | 32.16 | 4.43 | 20 | 7.25 | | 75039 | 36.50 | 7.38 | 4.95 |
| 92536 | GAU | 92.49 | 12.69 | -31 | 7.29 | | 92536 | 195.24 | 52.62 | 3.71 |
| 90803 | CNSL | 28.55 | 3.91 | -17 | 7.31 | | 90803 | 32.52 | 3.48 | 9.34 |
| 90843 | GPM | 10.09 | 1.38 | -15 | 7.31 | | 90843 | 11.49 | -0.97 | -11.81 |
| 89848 | AGG | 2.72 | 0.37 | 2 | 7.34 | | 89848 | 2.72 | 0.25 | 10.84 |
| 63773 | TDS | 14.50 | 1.97 | 11 | 7.36 | | 63773 | 13.68 | 2.02 | 6.79 |
| 77548 | MYI | 12.74 | 1.73 | 33 | 7.36 | | 77548 | 14.95 | 3.83 | 3.90 |
| 90443 | NAK | 145.04 | 19.67 | -85 | 7.37 | | 90443 | 198.20 | 43.38 | 4.57 |
| 77069 | NUM | 11.23 | 1.52 | 28 | 7.38 | | 77069 | 12.78 | 3.18 | 4.02 |
| 90129 | | 75.82 | 10.27 | -29 | 7.38 | | 90129 | 51.27 | 6.39 | 8.02 |
| 89051 | BBF | 12.68 | 1.72 | 23 | 7.39 | | 89051 | 15.26 | 3.13 | 4.88 |
| 86924 | | 49.97 | 6.76 | -30 | 7.39 | | 86924 | 37.75 | 2.73 | 13.85 |
| 79915 | | 44.61 | 6.03 | -26 | 7.40 | | 79915 | 57.84 | 8.79 | 6.58 |
| 84734 | FBC | 59.63 | 8.06 | -63 | 7.40 | | 84734 | 101.67 | 18.11 | 5.62 |
| 86023 | MUS | 14.12 | 1.89 | 24 | 7.46 | | 86023 | 17.32 | 3.44 | 5.04 |
| 89413 | BGFV | 64.40 | 8.60 | -50 | 7.49 | | 89413 | 87.77 | 15.74 | 5.58 |
| 86122 | ENG | 95.56 | 12.72 | -82 | 7.51 | | 86122 | 86.71 | 10.56 | 8.21 |
| 89783 | MUI | 11.33 | 1.50 | 25 | 7.53 | | 89783 | 12.50 | 2.91 | 4.30 |
| 90703 | TGP | 34.14 | 4.52 | -27 | 7.54 | | 90703 | 42.53 | 6.93 | 6.14 |
| 86754 | MUE | 13.08 | 1.73 | 22 | 7.55 | | 86754 | 17.16 | 3.21 | 5.35 |
| 12073 | PBCT | 12.19 | 1.61 | -19 | 7.56 | | 12073 | 14.86 | -1.12 | -13.32 |
| 89581 | CHA | 17.81 | 2.35 | 33 | 7.59 | | 89581 | 13.28 | 3.71 | 3.58 |
| 84511 | VCEL | 94.61 | 12.42 | -78 | 7.62 | | 84511 | 84.88 | 11.49 | 7.39 |
| 77530 | | 122.73 | 16.08 | -5 | 7.63 | | 77530 | 273.43 | 74.24 | 3.68 |
| 91228 | IEZ | 28.64 | 3.73 | -23 | 7.67 | | 91228 | 33.23 | 2.23 | 14.94 |
| 92538 | KOL | 49.87 | 6.46 | -16 | 7.72 | | 92538 | 63.59 | 12.02 | 5.29 |
| 39538 | MAT | 31.23 | 4.04 | -38 | 7.73 | | 39538 | 29.65 | -0.18 | -162.72 |
| 24010 | ETR | 12.39 | 1.60 | 4 | 7.74 | | 24010 | 16.60 | 1.51 | 11.02 |
| 89438 | PNI | 13.61 | 1.76 | 49 | 7.75 | | 89438 | 21.17 | 5.74 | 3.69 |
| 79356 | VCV | 13.16 | 1.70 | 46 | 7.76 | | 79356 | 19.84 | 5.32 | 3.73 |
| 77607 | MYJ | 13.31 | 1.71 | 29 | 7.77 | | 77607 | 15.70 | 3.59 | 4.37 |
| 85339 | MHN | 13.77 | 1.77 | 36 | 7.78 | | 85339 | 19.52 | 4.63 | 4.22 |
| 10656 | | 47.49 | 6.09 | -92 | 7.79 | | 10656 | 70.62 | 4.51 | 15.67 |
| 81299 | GCV | 13.82 | 1.77 | -19 | 7.81 | | 81299 | 15.88 | -0.95 | -16.72 |
| 92428 | APEI | 31.26 | 3.97 | -23 | 7.88 | | 92428 | 23.50 | 0.37 | 64.27 |
| 89707 | NCV | 27.82 | 3.52 | 12 | 7.89 | | 89707 | 37.79 | 6.23 | 6.06 |
| 78107 | NAZ | 13.76 | 1.74 | 29 | 7.89 | | 78107 | 17.47 | 3.83 | 4.56 |
| 78071 | BIOL | 91.10 | 11.52 | -86 | 7.91 | | 78071 | 51.86 | -2.92 | -17.79 |
| 91306 | PLM | 60.15 | 7.58 | 42 | 7.93 | | 91306 | 117.75 | 27.74 | 4.24 |
| 92171 | | 18.57 | 2.34 | -5 | 7.94 | | 92171 | 20.38 | 1.23 | 16.64 |
| 75157 | LEO | 13.57 | 1.71 | 27 | 7.94 | | 75157 | 17.60 | 3.71 | 4.74 |
| 79447 | MSD | 14.50 | 1.82 | 16 | 7.95 | | 79447 | 17.85 | 2.84 | 6.29 |
| 81038 | HNP | 31.48 | 3.96 | -14 | 7.95 | | 81038 | 34.57 | 2.96 | 11.69 |
| 86870 | NAD | 11.33 | 1.42 | 26 | 7.97 | | 86870 | 15.25 | 3.34 | 4.57 |
| 88526 | CRNT | 62.79 | 7.88 | -25 | 7.97 | | 88526 | 72.11 | 16.46 | 4.38 |
| 89630 | IAG | 64.28 | 8.06 | -40 | 7.97 | | 89630 | 83.74 | 17.96 | 4.66 |
| 89553 | PYN | 11.85 | 1.48 | 42 | 8.00 | | 89553 | 17.86 | 4.84 | 3.69 |
| 81774 | FCX | 54.89 | 6.86 | -16 | 8.01 | | 81774 | 88.62 | 22.65 | 3.91 |
| 88307 | MVC | 21.86 | 2.73 | -25 | 8.02 | | 88307 | 20.68 | -0.83 | -24.77 |
| 77776 | CHN | 25.74 | 3.21 | 2 | 8.03 | | 77776 | 32.00 | 4.21 | 7.61 |
| 91909 | SUMR | 73.46 | 9.13 | -51 | 8.04 | | 91909 | 58.39 | 8.25 | 7.08 |
| 10258 | CLDX | 106.82 | 13.27 | -98 | 8.05 | | 10258 | 111.07 | 10.02 | 11.08 |
| 75239 | NCA | 9.04 | 1.12 | 10 | 8.06 | | 75239 | 8.55 | 1.25 | 6.85 |
| 89513 | NBH | 10.12 | 1.26 | 12 | 8.06 | | 89513 | 11.11 | 1.65 | 6.72 |
| 88818 | ERF | 51.80 | 6.43 | -60 | 8.06 | | 88818 | 69.50 | 7.33 | 9.48 |
| 89952 | CMP | 18.03 | 2.23 | -29 | 8.07 | | 89952 | 21.19 | -1.06 | -19.91 |
| 88282 | CHU | 24.74 | 3.06 | -13 | 8.08 | | 88282 | 21.15 | 0.47 | 45.42 |
| 75401 | KTF | 14.06 | 1.72 | 26 | 8.15 | | 75401 | 18.03 | 3.60 | 5.01 |
| 12060 | GE | 26.57 | 3.26 | -53 | 8.16 | | 12060 | 29.25 | -2.36 | -12.40 |
| 91703 | IGSB | 1.15 | 0.14 | 2 | 8.18 | | 91703 | 1.19 | 0.25 | 4.83 |
| 76927 | STRL | 46.10 | 5.64 | -41 | 8.18 | | 76927 | 40.78 | 1.24 | 32.86 |
| 89698 | HSON | 59.51 | 7.27 | -60 | 8.18 | | 89698 | 36.45 | -2.67 | -13.65 |
| 87361 | PCTI | 24.88 | 3.04 | -35 | 8.19 | | 87361 | 28.28 | 0.01 | 2598.98 |
| 91471 | ACER | 146.48 | 17.86 | -68 | 8.20 | | 91471 | 294.40 | 74.27 | 3.96 |
| 75415 | JMM | 11.42 | 1.39 | 17 | 8.22 | | 75415 | 12.34 | 2.26 | 5.46 |
| 80089 | TV | 22.59 | 2.74 | -16 | 8.23 | | 80089 | 25.02 | 1.11 | 22.53 |
| 86062 | EVOL | 53.84 | 6.51 | -26 | 8.27 | | 86062 | 101.74 | 22.60 | 4.50 |
| 79777 | | 68.62 | 8.24 | -70 | 8.33 | | 79777 | 45.79 | -1.28 | -35.66 |
| 92248 | THM | 96.61 | 11.60 | -59 | 8.33 | | 92248 | 162.32 | 44.02 | 3.69 |
| 87033 | HSBC | 22.48 | 2.69 | -16 | 8.34 | | 87033 | 21.91 | 0.37 | 59.33 |
| 90907 | SNSS | 97.97 | 11.73 | -96 | 8.35 | | 90907 | 125.34 | 16.17 | 7.75 |
| 89049 | BFZ | 12.64 | 1.51 | 32 | 8.38 | | 89049 | 17.87 | 4.04 | 4.42 |
| 92726 | ATAX | 12.98 | 1.54 | 14 | 8.40 | | 92726 | 16.18 | 2.37 | 6.81 |
| 77501 | CNR | 30.76 | 3.66 | -91 | 8.41 | | 77501 | 45.83 | -4.30 | -10.67 |
| 80303 | | 86.72 | 10.29 | -76 | 8.43 | | 80303 | 132.62 | 25.13 | 5.28 |
| 88511 | CRIS | 72.13 | 8.50 | -82 | 8.48 | | 88511 | 128.15 | 28.16 | 4.55 |
| 80200 | IFN | 25.56 | 3.01 | 11 | 8.49 | | 80200 | 31.18 | 5.26 | 5.92 |
| 75354 | MVF | 12.04 | 1.42 | 24 | 8.49 | | 75354 | 15.81 | 3.20 | 4.94 |
| 90338 | | 36.59 | 4.28 | 20 | 8.56 | | 90338 | 47.83 | 10.06 | 4.75 |
| 63715 | POWL | 29.57 | 3.45 | -14 | 8.57 | | 63715 | 35.06 | 3.90 | 8.99 |
| 89784 | NCZ | 29.12 | 3.38 | 5 | 8.61 | | 89784 | 37.36 | 5.59 | 6.69 |
| 91414 | FEO | 21.68 | 2.52 | 19 | 8.61 | | 91414 | 30.71 | 5.22 | 5.88 |
| 77515 | KOPN | 42.71 | 4.95 | -51 | 8.63 | | 77515 | 44.30 | 1.95 | 22.75 |
| 84624 | | 49.86 | 5.77 | -26 | 8.65 | | 84624 | 66.68 | 12.54 | 5.32 |
| 90108 | ERH | 12.20 | 1.41 | -3 | 8.65 | | 90108 | 14.47 | 0.62 | 23.29 |
| 76572 | NQP | 12.19 | 1.41 | 33 | 8.66 | | 76572 | 16.77 | 3.99 | 4.20 |
| 79767 | ENIA | 20.53 | 2.36 | 28 | 8.69 | | 79767 | 32.72 | 6.42 | 5.10 |
| 92511 | EDV | 20.76 | 2.38 | -22 | 8.74 | | 92511 | 27.37 | 0.99 | 27.54 |
| 75168 | SWZ | 19.95 | 2.27 | -44 | 8.79 | | 75168 | 23.21 | -2.91 | -7.98 |
| 87280 | NGG | 14.38 | 1.63 | -13 | 8.80 | | 87280 | 13.89 | -0.47 | -29.52 |
| 13688 | PCG | 16.96 | 1.92 | -39 | 8.83 | | 13688 | 22.56 | -2.05 | -10.98 |
| 89190 | ILF | 25.62 | 2.90 | 21 | 8.84 | | 89190 | 34.99 | 6.36 | 5.50 |
| 89456 | LYTS | 28.65 | 3.21 | -54 | 8.92 | | 89456 | 37.88 | -1.20 | -31.45 |
| 89788 | EHI | 18.07 | 2.01 | 14 | 8.97 | | 89788 | 21.65 | 3.15 | 6.87 |
| 83756 | GERN | 68.03 | 7.52 | -77 | 9.04 | | 83756 | 87.71 | 7.91 | 11.09 |
| 91949 | BIV | 4.10 | 0.45 | 2 | 9.08 | | 91949 | 3.83 | 0.28 | 13.64 |
| 88626 | DRRX | 52.27 | 5.75 | -86 | 9.09 | | 88626 | 78.17 | 2.02 | 38.64 |
| 52679 | JHI | 16.20 | 1.78 | 7 | 9.13 | | 52679 | 18.16 | 2.09 | 8.69 |
| 89695 | DMLP | 27.42 | 3.00 | -8 | 9.14 | | 89695 | 37.67 | 6.06 | 6.22 |
| 86165 | PEIX | 122.01 | 13.35 | -98 | 9.14 | | 86165 | 72.35 | -6.41 | -11.29 |
| 87270 | | 14.09 | 1.53 | 41 | 9.19 | | 87270 | 20.64 | 5.09 | 4.06 |
| 90213 | MGI | 65.72 | 7.13 | -75 | 9.22 | | 90213 | 78.39 | 11.74 | 6.68 |
| 92675 | CYB | 3.82 | 0.41 | 1 | 9.26 | | 92675 | 4.65 | 0.17 | 26.94 |
| 87078 | MDRX | 30.88 | 3.33 | -3 | 9.28 | | 87078 | 48.83 | 8.95 | 5.46 |
| 88264 | RBBN | 38.93 | 4.18 | -39 | 9.32 | | 88264 | 43.12 | 4.15 | 10.40 |
| 90267 | MCN | 14.27 | 1.53 | -1 | 9.35 | | 90267 | 17.45 | 1.15 | 15.24 |
| 89809 | FTF | 13.23 | 1.41 | 9 | 9.36 | | 89809 | 17.67 | 2.14 | 8.25 |
| 86552 | CEV | 13.81 | 1.47 | 45 | 9.37 | | 86552 | 20.78 | 5.39 | 3.85 |
| 75306 | NMI | 10.60 | 1.13 | 7 | 9.40 | | 75306 | 11.62 | 1.26 | 9.22 |
| 85942 | DHF | 18.37 | 1.95 | 22 | 9.41 | | 85942 | 25.13 | 4.19 | 6.00 |
| 75034 | BKR | 33.90 | 3.58 | -33 | 9.47 | | 75034 | 32.88 | 1.30 | 25.29 |
| 89460 | BFY | 14.64 | 1.54 | 32 | 9.49 | | 89460 | 20.07 | 4.41 | 4.55 |
| 92385 | CIM | 17.74 | 1.86 | 3 | 9.53 | | 92385 | 18.08 | 2.04 | 8.87 |
| 83445 | BVN | 52.76 | 5.52 | -19 | 9.56 | | 83445 | 67.16 | 15.11 | 4.44 |
| 83822 | GES | 41.87 | 4.36 | 35 | 9.60 | | 83822 | 63.09 | 14.34 | 4.40 |
| 90266 | LPL | 33.03 | 3.43 | -1 | 9.62 | | 90266 | 44.13 | 7.34 | 6.01 |
| 64629 | | 50.06 | 5.20 | -58 | 9.63 | | 64629 | 74.01 | 13.29 | 5.57 |
| 82824 | IDRA | 135.32 | 14.05 | -95 | 9.63 | | 82824 | 146.92 | 13.55 | 10.84 |
| 79360 | VPV | 13.49 | 1.40 | 38 | 9.65 | | 79360 | 19.38 | 4.72 | 4.11 |
| 75418 | VOD | 18.56 | 1.91 | -21 | 9.69 | | 75418 | 28.84 | 1.36 | 21.16 |
| 92600 | TUR | 35.98 | 3.71 | -9 | 9.70 | | 92600 | 47.42 | 7.80 | 6.08 |
| 10363 | | 64.44 | 6.64 | -58 | 9.71 | | 10363 | 45.15 | 1.54 | 29.40 |
| 80072 | VLY | 13.44 | 1.36 | -47 | 9.86 | | 80072 | 14.52 | -5.08 | -2.86 |
| 79859 | NYCB | 20.93 | 2.11 | -21 | 9.91 | | 79859 | 22.41 | -0.01 | -3402.18 |
| 90999 | PDS | 51.61 | 5.18 | -78 | 9.96 | | 90999 | 30.96 | -9.29 | -3.33 |
| 91570 | ETY | 12.33 | 1.24 | -6 | 9.96 | | 91570 | 15.15 | 0.46 | 32.75 |
| 92695 | FAN | 26.85 | 2.69 | -8 | 9.99 | | 92695 | 27.25 | 2.07 | 13.17 |
| 75268 | PPT | 14.66 | 1.46 | 21 | 10.03 | | 75268 | 20.72 | 3.48 | 5.96 |
| 80401 | RCS | 11.14 | 1.11 | 14 | 10.08 | | 80401 | 9.32 | 1.69 | 5.53 |
| 90126 | CYTK | 56.34 | 5.57 | -63 | 10.12 | | 90126 | 36.56 | -3.27 | -11.20 |
| 86871 | NAN | 10.78 | 1.07 | 30 | 10.12 | | 86871 | 14.65 | 3.51 | 4.17 |
| 90563 | NFJ | 15.77 | 1.55 | -18 | 10.16 | | 90563 | 13.92 | -1.01 | -13.77 |
| 89733 | PHK | 24.99 | 2.46 | 45 | 10.17 | | 89733 | 34.30 | 7.64 | 4.49 |
| 79756 | VKI | 13.35 | 1.31 | 32 | 10.17 | | 79756 | 19.70 | 4.30 | 4.59 |
| 89629 | FMN | 14.28 | 1.38 | 35 | 10.36 | | 89629 | 22.99 | 4.94 | 4.65 |
| 89252 | FVE | 59.48 | 5.68 | -69 | 10.48 | | 89252 | 68.44 | 7.42 | 9.22 |
| 81610 | CWCO | 25.75 | 2.46 | -7 | 10.49 | | 81610 | 35.01 | 3.65 | 9.59 |
| 89882 | MAV | 15.57 | 1.48 | 34 | 10.53 | | 89882 | 24.35 | 5.05 | 4.82 |
| 84072 | DRD | 90.46 | 8.50 | -63 | 10.65 | | 84072 | 85.06 | 10.14 | 8.39 |
| 21207 | NEM | 37.46 | 3.52 | -15 | 10.65 | | 21207 | 37.34 | 3.99 | 9.37 |
| 77591 | VTN | 13.66 | 1.28 | 35 | 10.71 | | 77591 | 22.51 | 4.93 | 4.56 |
| 91677 | | 86.68 | 8.07 | -80 | 10.75 | | 91677 | 80.17 | 15.00 | 5.34 |
| 85918 | MUJ | 12.04 | 1.12 | 25 | 10.77 | | 85918 | 15.69 | 3.32 | 4.73 |
| 91293 | FXF | 8.60 | 0.80 | 1 | 10.81 | | 91293 | 5.43 | 0.22 | 24.41 |
| 89753 | FORM | 34.54 | 3.19 | -3 | 10.82 | | 89753 | 37.31 | 7.08 | 5.27 |
| 81044 | PKX | 31.59 | 2.87 | -27 | 11.02 | | 81044 | 39.78 | 3.24 | 12.26 |
| 92865 | AGQ | 85.51 | 7.75 | -58 | 11.04 | | 92865 | 73.27 | 8.83 | 8.30 |
| 76127 | TTI | 57.20 | 5.18 | -65 | 11.05 | | 76127 | 56.21 | 1.57 | 35.83 |
| 88807 | INSG | 65.33 | 5.89 | -11 | 11.09 | | 88807 | 72.38 | 20.26 | 3.57 |
| 45728 | IVC | 34.78 | 3.13 | -72 | 11.10 | | 45728 | 42.40 | -1.30 | -32.54 |
| 22517 | PPL | 12.42 | 1.11 | -1 | 11.18 | | 22517 | 11.19 | 0.47 | 23.65 |

| | | | | | | | | | |
|---|---|---|---|---|---|---|---|---|---|
| 86558 | EVY | 14.89 | 1.32 | 56 | 11.28 | 86558 | 28.10 | 7.15 | 3.93 |
| 78944 | VMM | 10.52 | 0.92 | 14 | 11.39 | 78944 | 10.99 | 1.90 | 5.78 |
| 81621 | HLIT | 36.69 | 3.22 | -16 | 11.40 | 81621 | 29.54 | 2.55 | 11.58 |
| 92922 | JNK | 8.32 | 0.72 | 4 | 11.48 | 92922 | 9.06 | 0.72 | 12.58 |
| 21152 | CAJ | 19.36 | 1.67 | -12 | 11.60 | 21152 | 21.49 | 0.67 | 31.98 |
| 89688 | EAD | 14.48 | 1.24 | 31 | 11.65 | 89688 | 25.38 | 4.92 | 5.15 |
| 89282 | AFB | 11.94 | 1.02 | 28 | 11.67 | 89282 | 16.40 | 3.60 | 4.56 |
| 92011 | SMOG | 29.09 | 2.48 | -22 | 11.74 | 92011 | 29.62 | 1.23 | 24.06 |
| 90950 | GASS | 40.36 | 3.43 | -43 | 11.76 | 90950 | 49.11 | 4.65 | 10.57 |
| 89534 | BYM | 12.00 | 1.01 | 18 | 11.86 | 89534 | 14.62 | 2.60 | 5.62 |
| 89746 | EVV | 13.78 | 1.16 | 21 | 11.90 | 89746 | 18.49 | 3.15 | 5.87 |
| 75392 | DMF | 12.44 | 1.03 | 17 | 12.11 | 75392 | 14.38 | 2.49 | 5.78 |
| 75465 | KSM | 15.09 | 1.25 | 29 | 12.11 | 75465 | 21.01 | 4.19 | 5.01 |
| 88534 | ISSC | 42.47 | 3.49 | -43 | 12.17 | 88534 | 46.26 | 2.85 | 16.21 |
| 83224 | EWM | 22.01 | 1.80 | 2 | 12.21 | 83224 | 25.79 | 3.28 | 7.87 |
| 89764 | ERC | 12.93 | 1.06 | 9 | 12.22 | 89764 | 16.75 | 1.99 | 8.40 |
| 86799 | CNX | 45.27 | 3.70 | -51 | 12.22 | 86799 | 57.44 | 7.20 | 7.98 |
| 89537 | BSE | 13.20 | 1.08 | 31 | 12.23 | 89537 | 18.13 | 4.06 | 4.47 |
| 88885 | AMX | 26.45 | 2.12 | -8 | 12.45 | 88885 | 27.17 | 2.31 | 11.77 |
| 79444 | TALO | 87.54 | 6.99 | -97 | 12.53 | 79444 | 54.28 | -12.71 | -4.27 |
| 87583 | AGEN | 51.91 | 4.14 | -17 | 12.55 | 87583 | 51.86 | 10.12 | 5.13 |
| 92572 | SMB | 1.71 | 0.14 | 5 | 12.55 | 92572 | 2.15 | 0.53 | 4.09 |
| 79362 | YPF | 49.29 | 3.91 | -71 | 12.62 | 79362 | 53.63 | -1.05 | -51.22 |
| 92360 | BWX | 6.54 | 0.51 | 2 | 12.82 | 92360 | 5.06 | 0.26 | 19.17 |
| 89786 | MHI | 14.95 | 1.17 | 31 | 12.83 | 89786 | 22.54 | 4.49 | 5.02 |
| 91194 | GLO | 15.99 | 1.23 | -9 | 12.97 | 91194 | 21.40 | 1.02 | 21.00 |
| 80866 | TDF | 19.51 | 1.49 | -2 | 13.11 | 80866 | 26.18 | 2.55 | 10.26 |
| 88673 | BRFS | 36.20 | 2.73 | -14 | 13.25 | 88673 | 41.41 | 5.47 | 7.57 |
| 89256 | NMR | 40.85 | 3.08 | -55 | 13.27 | 89256 | 41.74 | -0.94 | -44.50 |
| 78841 | EGY | 62.78 | 4.72 | -80 | 13.30 | 78841 | 54.97 | -3.01 | -18.27 |
| 62930 | MXF | 26.46 | 1.99 | -13 | 13.31 | 62930 | 25.78 | 1.45 | 17.75 |
| 69761 | BCS | 44.70 | 3.36 | -17 | 13.32 | 69761 | 37.33 | 3.33 | 11.21 |
| 88841 | RGP | 17.21 | 1.29 | -13 | 13.35 | 88841 | 22.17 | 1.10 | 20.16 |
| 91758 | UUP | 7.31 | 0.53 | 3 | 13.69 | 91758 | 6.34 | 0.48 | 13.14 |
| 77316 | VGM | 12.90 | 0.94 | 30 | 13.73 | 77316 | 20.79 | 4.25 | 4.89 |
| 77377 | SBI | 9.15 | 0.66 | 11 | 13.78 | 77377 | 9.98 | 1.49 | 6.68 |
| 92733 | AFK | 19.82 | 1.43 | -7 | 13.86 | 92733 | 24.36 | 1.98 | 12.33 |
| 92135 | PBD | 25.85 | 1.86 | -18 | 13.90 | 92135 | 27.85 | 1.46 | 19.09 |
| 57007 | TRC | 20.39 | 1.46 | -28 | 13.93 | 57007 | 24.20 | -0.57 | -42.82 |
| 90244 | GLV | 15.53 | 1.11 | -6 | 13.95 | 90244 | 19.09 | 0.89 | 21.54 |
| 87439 | IDN | 69.34 | 4.97 | -84 | 13.96 | 87439 | 74.63 | 1.64 | 45.38 |
| 91670 | PTMN | 35.26 | 2.52 | -5 | 14.01 | 91670 | 29.31 | 3.12 | 9.39 |
| 87289 | DHC | 20.42 | 1.45 | -34 | 14.06 | 87289 | 20.61 | -1.88 | -10.97 |

| | | | | | | | | | | |
|---|---|---|---|---|---|---|---|---|---|---|
| 77808 | NXR | 8.23 | 0.58 | 5 | 14.15 | | 77808 | 8.21 | 0.84 | 9.77 |
| 89511 | NBW | 10.68 | 0.74 | 17 | 14.39 | | 89511 | 12.49 | 2.22 | 5.64 |
| 45495 | | 34.97 | 2.43 | -47 | 14.39 | | 45495 | 32.58 | -1.38 | -23.61 |
| 81784 | TK | 42.53 | 2.94 | -83 | 14.46 | | 81784 | 42.86 | -3.02 | -14.17 |
| 89591 | | 14.69 | 1.01 | 22 | 14.50 | | 89591 | 15.27 | 3.02 | 5.05 |
| 10952 | CDR | 42.44 | 2.91 | -56 | 14.59 | | 10952 | 22.13 | -5.12 | -4.32 |
| 90340 | KYN | 23.03 | 1.56 | -16 | 14.81 | | 90340 | 31.02 | 3.19 | 9.72 |
| 89898 | OMEX | 70.40 | 4.67 | -91 | 15.07 | | 89898 | 48.09 | -10.49 | -4.59 |
| 62359 | PBT | 34.81 | 2.30 | -57 | 15.13 | | 62359 | 36.36 | -2.15 | -16.90 |
| 77478 | | 95.06 | 6.24 | -92 | 15.23 | | 77478 | 66.94 | -2.71 | -24.71 |
| 35107 | PKE | 20.97 | 1.37 | -5 | 15.28 | | 35107 | 23.29 | 2.01 | 11.56 |
| 70420 | | 26.00 | 1.69 | -10 | 15.36 | | 70420 | 36.02 | 4.21 | 8.56 |
| 88468 | SPRT | 54.81 | 3.57 | -63 | 15.37 | | 88468 | 64.74 | 5.89 | 10.99 |
| 89603 | NEA | 10.94 | 0.71 | 18 | 15.43 | | 89603 | 14.44 | 2.57 | 5.61 |
| 89238 | LYG | 43.68 | 2.82 | -66 | 15.48 | | 89238 | 53.00 | 1.90 | 27.88 |
| 82599 | NWPX | 38.98 | 2.52 | -45 | 15.49 | | 82599 | 37.64 | 1.39 | 27.08 |
| 75411 | MHF | 9.50 | 0.61 | 8 | 15.49 | | 75411 | 9.73 | 1.16 | 8.39 |
| 80114 | STCN | 70.64 | 4.51 | -40 | 15.65 | | 80114 | 88.89 | 15.89 | 5.59 |
| 77071 | NUO | 12.84 | 0.81 | 8 | 15.87 | | 77071 | 15.85 | 1.92 | 8.27 |
| 92041 | GLRE | 28.85 | 1.78 | -34 | 16.24 | | 92041 | 40.42 | 3.42 | 11.83 |
| 78223 | AEM | 36.65 | 2.25 | -21 | 16.26 | | 78223 | 37.55 | 4.65 | 8.08 |
| 85421 | CHL | 13.79 | 0.85 | -6 | 16.29 | | 85421 | 10.29 | -0.13 | -82.23 |
| 86869 | KT | 20.25 | 1.24 | -3 | 16.35 | | 86869 | 16.19 | 0.89 | 18.26 |
| 89387 | CPSI | 27.41 | 1.67 | -6 | 16.37 | | 89387 | 32.83 | 4.24 | 7.74 |
| 90121 | AINV | 40.42 | 2.46 | -56 | 16.42 | | 90121 | 22.49 | -5.24 | -4.30 |
| 89596 | | 14.87 | 0.90 | 30 | 16.44 | | 89596 | 23.60 | 4.82 | 4.89 |
| 90882 | ETW | 12.30 | 0.73 | -6 | 16.73 | | 90882 | 17.95 | 0.75 | 23.89 |
| 91952 | BND | 2.76 | 0.16 | 0 | 16.75 | | 91952 | 2.70 | 0.06 | 42.65 |
| 89246 | ACH | 44.25 | 2.60 | -42 | 17.00 | | 89246 | 52.85 | 6.31 | 8.37 |
| 89489 | EIM | 12.18 | 0.72 | 27 | 17.01 | | 89489 | 16.38 | 3.55 | 4.61 |
| 92172 | DHX | 52.55 | 3.05 | -63 | 17.23 | | 92172 | 55.62 | 3.53 | 15.74 |
| 86557 | | 15.84 | 0.92 | 39 | 17.30 | | 86557 | 25.29 | 5.60 | 4.52 |
| 90976 | SEED | 90.13 | 5.15 | -75 | 17.49 | | 90976 | 161.80 | 31.48 | 5.14 |
| 30737 | | 49.59 | 2.82 | -53 | 17.60 | | 30737 | 49.12 | 2.62 | 18.78 |
| 92026 | SPTI | 2.10 | 0.12 | 0 | 17.71 | | 92026 | 2.35 | 0.06 | 36.98 |
| 88396 | EWZ | 32.90 | 1.83 | 9 | 17.96 | | 88396 | 46.74 | 8.38 | 5.57 |
| 92016 | HNW | 16.25 | 0.90 | 9 | 18.10 | | 92016 | 19.66 | 2.40 | 8.20 |
| 80791 | ADTN | 32.29 | 1.72 | -28 | 18.74 | | 80791 | 38.12 | 3.00 | 12.70 |
| 83217 | EWI | 22.82 | 1.22 | -26 | 18.78 | | 83217 | 18.58 | -1.41 | -13.19 |
| 92577 | DWX | 20.00 | 1.06 | 0 | 18.85 | | 92577 | 22.18 | 1.89 | 11.76 |
| 92357 | SHM | 1.14 | 0.06 | 2 | 18.96 | | 92357 | 1.18 | 0.22 | 5.34 |
| 76515 | CVA | 16.61 | 0.87 | -39 | 19.10 | | 76515 | 20.60 | -2.92 | -7.05 |
| 86942 | GUT | 15.74 | 0.82 | 6 | 19.31 | | 86942 | 24.55 | 3.13 | 7.86 |
| 51916 | VBF | 9.30 | 0.48 | 3 | 19.53 | | 51916 | 10.78 | 0.81 | 13.36 |
| 75082 | NUV | 7.23 | 0.36 | 8 | 19.82 | | 75082 | 8.15 | 1.04 | 7.81 |
| 83909 | SSRM | 45.32 | 2.28 | -24 | 19.88 | | 83909 | 40.64 | 5.34 | 7.61 |
| 92090 | BKCC | 28.29 | 1.39 | -46 | 20.33 | | 92090 | 19.84 | -4.30 | -4.62 |
| 77609 | NXQ | 7.82 | 0.38 | 4 | 20.42 | | 77609 | 7.35 | 0.68 | 10.77 |
| 77131 | VKQ | 12.49 | 0.61 | 26 | 20.52 | | 77131 | 17.83 | 3.54 | 5.04 |
| 91814 | ARAY | 33.35 | 1.62 | -34 | 20.58 | | 91814 | 29.07 | -0.39 | -75.22 |
| 82196 | | 77.04 | 3.70 | -84 | 20.84 | | 82196 | 46.39 | -5.37 | -8.63 |
| 33452 | ERIC | 23.73 | 1.11 | 14 | 21.35 | | 33452 | 22.78 | 3.82 | 5.97 |
| 88612 | UMC | 24.81 | 1.16 | -9 | 21.43 | | 88612 | 38.45 | 4.13 | 9.32 |
| 92453 | | 56.65 | 2.63 | -85 | 21.57 | | 92453 | 45.22 | -5.32 | -8.50 |
| 86117 | | 16.87 | 0.78 | -21 | 21.73 | | 86117 | 18.65 | -0.65 | -28.80 |
| 12266 | CDMO | 61.95 | 2.84 | -60 | 21.78 | | 12266 | 54.73 | 5.60 | 9.77 |
| 77239 | OFC | 22.19 | 1.01 | -31 | 22.04 | | 77239 | 25.50 | -0.68 | -37.28 |
| 83799 | STRA | 41.82 | 1.89 | -47 | 22.12 | | 83799 | 48.35 | 2.20 | 21.96 |
| 92023 | SPAB | 2.51 | 0.11 | 0 | 22.24 | | 92023 | 2.76 | 0.05 | 60.30 |
| 79303 | | 38.51 | 1.73 | -25 | 22.31 | | 79303 | 38.64 | 4.07 | 9.50 |
| 80857 | VLGEA | 19.05 | 0.84 | -7 | 22.61 | | 80857 | 16.01 | 0.47 | 34.01 |
| 75240 | NNY | 7.28 | 0.32 | 11 | 22.67 | | 75240 | 8.11 | 1.34 | 6.03 |
| 89970 | EMD | 15.97 | 0.68 | -2 | 23.40 | | 89970 | 17.96 | 1.15 | 15.68 |
| 84737 | GIFI | 40.00 | 1.71 | -50 | 23.43 | | 84737 | 30.64 | -1.97 | -15.56 |
| 79049 | NMT | 11.87 | 0.50 | 21 | 23.52 | | 79049 | 16.35 | 3.03 | 5.39 |
| 91119 | XAN | 40.67 | 1.71 | -35 | 23.83 | | 91119 | 27.68 | -0.68 | -40.59 |
| 77638 | VMO | 12.86 | 0.54 | 15 | 23.89 | | 77638 | 15.31 | 2.39 | 6.41 |
| 72231 | MMT | 10.11 | 0.42 | 6 | 23.94 | | 72231 | 14.86 | 1.54 | 9.65 |
| 89053 | BNY | 13.23 | 0.55 | 28 | 24.03 | | 89053 | 21.11 | 4.31 | 4.90 |
| 79918 | HIO | 13.64 | 0.56 | 10 | 24.35 | | 79918 | 18.38 | 2.21 | 8.30 |
| 92621 | IPI | 68.54 | 2.78 | -87 | 24.67 | | 92621 | 58.09 | -3.97 | -14.63 |
| 77124 | DTF | 11.37 | 0.46 | 4 | 24.73 | | 77124 | 11.09 | 0.90 | 12.32 |
| 79665 | TEI | 18.38 | 0.74 | 5 | 24.88 | | 79665 | 23.01 | 2.54 | 9.07 |
| 90781 | | 58.26 | 2.34 | -54 | 24.90 | | 90781 | 78.13 | 14.57 | 5.36 |
| 89436 | PCK | 11.20 | 0.44 | 27 | 25.18 | | 89436 | 17.19 | 3.63 | 4.74 |
| 89372 | PHT | 23.21 | 0.91 | 0 | 25.50 | | 89372 | 30.63 | 3.77 | 8.13 |
| 92843 | SCO | 62.55 | 2.41 | -63 | 25.98 | | 92843 | 61.72 | 3.75 | 16.48 |
| 85602 | SID | 75.00 | 2.86 | -66 | 26.25 | | 85602 | 96.91 | 15.66 | 6.19 |
| 77584 | BKE | 27.72 | 1.05 | -11 | 26.36 | | 77584 | 24.03 | 1.66 | 14.51 |
| 90031 | ONCT | 96.55 | 3.64 | -100 | 26.50 | | 90031 | 68.20 | -15.85 | -4.30 |
| 85261 | MT | 48.74 | 1.84 | -62 | 26.56 | | 85261 | 61.38 | 5.25 | 11.69 |
| 92665 | OCUP | 75.38 | 2.77 | -90 | 27.25 | | 92665 | 53.09 | -6.18 | -8.59 |
| 92543 | SOL | 80.95 | 2.89 | -95 | 28.00 | | 92543 | 66.86 | -3.93 | -17.01 |
| 77157 | OI | 32.77 | 1.13 | -37 | 28.97 | | 77157 | 32.77 | -0.02 | -1546.37 |
| 77470 | NXP | 8.59 | 0.30 | 2 | 28.98 | | 77470 | 7.76 | 0.48 | 16.24 |
| 89505 | ATHE | 79.47 | 2.63 | -89 | 30.17 | | 89505 | 91.59 | 8.76 | 10.46 |
| 83366 | | 63.58 | 2.10 | -90 | 30.27 | | 83366 | 56.11 | -5.52 | -10.16 |
| 90548 | ARC | 61.90 | 1.90 | -70 | 32.59 | | 90548 | 80.97 | 6.03 | 13.43 |
| 90499 | FMO | 24.37 | 0.73 | -28 | 33.24 | | 90499 | 27.14 | 0.56 | 48.89 |
| 86033 | HIX | 19.28 | 0.58 | 14 | 33.30 | | 86033 | 29.06 | 4.15 | 7.00 |
| 51960 | | 53.89 | 1.62 | -59 | 33.30 | | 51960 | 88.73 | 10.43 | 8.51 |
| 29890 | BP | 18.66 | 0.56 | -19 | 33.45 | | 29890 | 17.83 | -0.62 | -28.81 |
| 91456 | CAF | 25.81 | 0.74 | -7 | 34.74 | | 91456 | 30.45 | 3.20 | 9.52 |
| 79866 | SCHN | 33.08 | 0.94 | -43 | 35.27 | | 79866 | 40.94 | 1.53 | 26.79 |
| 21785 | | 27.15 | 0.74 | -38 | 36.69 | | 21785 | 26.86 | -1.65 | -16.24 |
| 91319 | XES | 31.06 | 0.84 | -47 | 36.93 | | 91319 | 36.85 | -0.06 | -612.83 |
| 92467 | ORN | 50.93 | 1.34 | -56 | 38.01 | | 92467 | 71.92 | 10.66 | 6.75 |
| 91082 | PERI | 72.59 | 1.87 | -63 | 38.80 | | 91082 | 120.89 | 23.32 | 5.18 |
| 89935 | MEIP | 73.99 | 1.90 | -94 | 38.94 | | 89935 | 41.87 | -10.00 | -4.19 |
| 90965 | NM | 65.30 | 1.63 | -91 | 39.99 | | 90965 | 92.33 | 3.98 | 23.19 |
| 67838 | | 20.96 | 0.52 | -49 | 40.35 | | 67838 | 25.41 | -1.74 | -14.63 |
| 88963 | DGICA | 10.81 | 0.27 | -19 | 40.35 | | 88963 | 12.63 | -1.34 | -9.45 |
| 89586 | | 11.72 | 0.29 | 24 | 40.67 | | 89586 | 17.96 | 3.40 | 5.28 |
| 89547 | NRP | 74.32 | 1.75 | -78 | 42.44 | | 89547 | 67.03 | 7.55 | 8.88 |
| 90672 | RAIL | 27.53 | 0.64 | -63 | 43.11 | | 90672 | 30.33 | -4.43 | -6.85 |
| 90118 | SA | 44.60 | 0.94 | 1 | 47.21 | | 90118 | 40.99 | 8.39 | 4.89 |
| 81222 | RCII | 29.76 | 0.60 | -8 | 49.46 | | 81222 | 37.95 | 5.69 | 6.67 |
| 91357 | HWCC | 28.22 | 0.56 | -46 | 50.36 | | 91357 | 25.55 | -2.09 | -12.21 |
| 59089 | | 41.64 | 0.79 | -91 | 52.44 | | 59089 | 41.55 | -9.02 | -4.60 |
| 75844 | NTR | 31.02 | 0.59 | -23 | 52.69 | | 75844 | 29.63 | 1.69 | 17.50 |
| 84083 | HMY | 89.73 | 1.66 | -84 | 53.99 | | 84083 | 56.40 | -4.39 | -12.86 |
| 79039 | USEG | 80.17 | 1.48 | -93 | 54.32 | | 79039 | 110.38 | 15.77 | 7.00 |
| 88309 | MBT | 33.23 | 0.59 | -34 | 56.11 | | 88309 | 42.96 | 5.06 | 8.50 |
| 18964 | PEO | 16.98 | 0.30 | -25 | 56.16 | | 18964 | 17.29 | -1.41 | -12.25 |
| 62376 | | 37.60 | 0.65 | -87 | 57.86 | | 62376 | 43.51 | -4.49 | -9.69 |
| 91292 | FXA | 12.32 | 0.21 | -1 | 58.55 | | 91292 | 12.63 | 0.55 | 22.81 |
| 91135 | BTE | 60.41 | 1.03 | -85 | 58.87 | | 91135 | 66.22 | 2.22 | 29.82 |
| 92252 | SNCA | 83.53 | 1.33 | -99 | 62.85 | | 92252 | 73.46 | -12.30 | -5.97 |
| 87128 | NOK | 50.97 | 0.79 | -63 | 64.37 | | 87128 | 42.98 | -2.68 | -16.03 |
| 84604 | CTIC | 179.49 | 2.76 | -98 | 64.93 | | 84604 | 241.59 | 37.05 | 6.52 |
| 75642 | MCR | 9.64 | 0.15 | 0 | 66.17 | | 75642 | 12.83 | 0.71 | 18.09 |
| 64961 | NXGN | 29.81 | 0.36 | -31 | 81.88 | | 64961 | 26.94 | 0.32 | 84.59 |
| 89949 | LFC | 30.51 | 0.36 | -32 | 85.60 | | 89949 | 32.06 | 0.63 | 50.49 |
| 89301 | GME | 43.74 | 0.48 | -42 | 91.39 | | 89301 | 37.17 | -0.56 | -66.13 |
| 76461 | VERU | 45.53 | 0.42 | -61 | 108.68 | | 76461 | 41.11 | 0.43 | 94.91 |
| 81624 | ISNS | 40.24 | 0.29 | -29 | 141.08 | | 81624 | 42.16 | 4.36 | 9.66 |
| 91729 | | 64.37 | 0.43 | -92 | 149.00 | | 91729 | 39.89 | -12.16 | -3.28 |
| 79048 | NMY | 11.16 | 0.06 | 19 | 181.14 | | 79048 | 15.85 | 2.81 | 5.63 |
| 84566 | GOGL | 60.47 | 0.26 | -92 | 236.22 | | 84566 | 57.52 | -5.71 | -10.07 |
| 91858 | SAR | 18.98 | 0.08 | -22 | 244.09 | | 91858 | 20.46 | -0.47 | -43.90 |
| 92355 | NWG | 33.48 | 0.06 | -63 | 565.16 | | 92355 | 39.27 | -2.70 | -14.54 |
| 89602 | NRK | 9.75 | 0.02 | 12 | 648.39 | | 89602 | 13.00 | 1.90 | 6.85 |
| 92470 | SBLK | 79.54 | 0.05 | -95 | 1499.43 | | 92470 | 75.40 | 2.93 | 25.76 |
| 92521 | DGLY | 78.68 | 0.05 | -89 | 1701.29 | | 92521 | 69.59 | -3.88 | -17.94 |
| 92856 | TECS | 14.38 | -47.54 | -100 | -0.30 | | 92856 | 18.46 | -46.27 | -0.40 |
| 91386 | QID | 12.06 | -35.77 | -99 | -0.34 | | 91386 | 16.28 | -34.55 | -0.47 |
| 92817 | SPXS | 14.29 | -41.69 | -100 | -0.34 | | 92817 | 20.70 | -38.93 | -0.53 |
| 91385 | DXD | 11.38 | -28.30 | -96 | -0.40 | | 91385 | 14.88 | -25.99 | -0.57 |
| 91312 | PSQ | 7.79 | -19.28 | -88 | -0.40 | | 91312 | 10.78 | -18.57 | -0.58 |
| 91387 | SDS | 11.66 | -28.55 | -96 | -0.41 | | 91387 | 16.05 | -26.37 | -0.61 |
| 92813 | FAZ | 19.15 | -44.84 | -100 | -0.43 | | 92813 | 29.92 | -42.80 | -0.70 |
| 91313 | DOG | 6.84 | -14.96 | -78 | -0.46 | | 91313 | 8.65 | -13.53 | -0.64 |
| 92814 | TZA | 21.75 | -46.24 | -100 | -0.47 | | 92814 | 31.37 | -41.44 | -0.76 |
| 91311 | SH | 7.11 | -15.03 | -78 | -0.47 | | 91311 | 9.34 | -13.70 | -0.68 |
| 91721 | TWM | 17.85 | -32.08 | -98 | -0.56 | | 91721 | 24.27 | -28.68 | -0.85 |
| 91782 | SKF | 17.32 | -30.64 | -99 | -0.57 | | 91782 | 25.66 | -29.39 | -0.87 |
| 91314 | MYY | 9.79 | -16.16 | -82 | -0.61 | | 91314 | 13.78 | -14.60 | -0.94 |
| 85293 | ABIO | 26.09 | -41.26 | -100 | -0.63 | | 85293 | 21.62 | -44.64 | -0.48 |
| 91720 | RWM | 10.79 | -16.80 | -82 | -0.64 | | 91720 | 13.94 | -14.90 | -0.94 |
| 92700 | SEF | 11.15 | -16.01 | -83 | -0.70 | | 92700 | 14.87 | -15.17 | -0.98 |
| 91776 | SRS | 21.18 | -29.45 | -99 | -0.72 | | 91776 | 25.47 | -30.03 | -0.85 |
| 92410 | EWV | 16.24 | -20.23 | -87 | -0.80 | | 92410 | 23.59 | -15.85 | -1.49 |
| 92860 | | 25.22 | -30.86 | -98 | -0.82 | | 92860 | 35.70 | -26.71 | -1.34 |
| 91615 | YTEN | 37.68 | -38.41 | -100 | -0.98 | | 91615 | 28.73 | -43.27 | -0.66 |
| 92409 | FXP | 24.71 | -24.08 | -97 | -1.03 | | 92409 | 26.69 | -24.67 | -1.08 |
| 92858 | EDZ | 33.24 | -30.89 | -100 | -1.08 | | 92858 | 41.21 | -27.98 | -1.47 |
| 77437 | MBOT | 45.92 | -42.66 | -100 | -1.08 | | 77437 | 58.28 | -37.85 | -1.54 |
| 92353 | EFZ | 11.48 | -10.22 | -65 | -1.12 | | 92353 | 14.22 | -8.92 | -1.59 |
| 90922 | EGF | 3.25 | -2.81 | -22 | -1.16 | | 90922 | 4.84 | -2.30 | -2.10 |
| 91947 | UNG | 27.00 | -22.70 | -97 | -1.19 | | 91947 | 24.35 | -25.43 | -0.96 |
| 92819 | ERY | 37.29 | -30.48 | -99 | -1.22 | | 92819 | 43.41 | -25.23 | -1.72 |
| 92411 | EEV | 26.17 | -20.59 | -96 | -1.27 | | 92411 | 33.65 | -19.38 | -1.74 |
| 75260 | MIN | 7.14 | -5.40 | -41 | -1.32 | | 75260 | 7.42 | -4.92 | -1.51 |
| 92664 | PST | 11.11 | -8.35 | -59 | -1.33 | | 92664 | 11.72 | -7.75 | -1.51 |
| 91691 | GDL | 5.36 | -3.96 | -30 | -1.35 | | 91691 | 5.62 | -3.39 | -1.66 |
| 75075 | MGF | 6.85 | -4.99 | -43 | -1.37 | | 75075 | 5.02 | -5.37 | -0.94 |
| 63546 | USAU | 43.34 | -30.15 | -99 | -1.44 | | 63546 | 77.76 | -16.22 | -4.79 |
| 92412 | EUM | 15.34 | -10.19 | -74 | -1.51 | | 92412 | 20.55 | -10.20 | -2.01 |
| 64822 | | 42.58 | -28.25 | -100 | -1.51 | | 64822 | 43.61 | -29.26 | -1.49 |
| 91779 | DUG | 28.33 | -18.71 | -90 | -1.51 | | 91779 | 32.85 | -15.15 | -2.17 |
| 92662 | TBT | 20.70 | -13.63 | -77 | -1.52 | | 92662 | 26.27 | -9.92 | -2.65 |
| 87608 | GIGM | 26.28 | -17.07 | -89 | -1.54 | | 87608 | 24.94 | -16.32 | -1.53 |
| 79006 | | 55.24 | -35.63 | -100 | -1.55 | | 79006 | 69.64 | -25.43 | -2.74 |
| 90724 | EGLE | 53.79 | -34.33 | -100 | -1.57 | | 90724 | 87.88 | -23.28 | -3.77 |
| 57913 | VXRT | 42.28 | -26.58 | -99 | -1.59 | | 57913 | 40.41 | -28.30 | -1.43 |
| 92173 | ETJ | 9.44 | -5.83 | -55 | -1.62 | | 92173 | 10.38 | -7.12 | -1.46 |
| 91964 | OPTT | 44.18 | -26.25 | -99 | -1.68 | | 91964 | 41.58 | -31.03 | -1.34 |
| 91742 | SNOA | 37.98 | -22.29 | -99 | -1.70 | | 91742 | 33.12 | -24.76 | -1.34 |
| 90483 | NCTY | 36.76 | -21.55 | -97 | -1.71 | | 90483 | 48.71 | -21.29 | -2.29 |
| 57293 | CRF | 15.53 | -9.07 | -63 | -1.71 | | 57293 | 19.45 | -7.93 | -2.45 |
| 91021 | PSTV | 55.20 | -31.22 | -100 | -1.77 | | 91021 | 52.04 | -34.38 | -1.51 |
| 91525 | IRR | 15.68 | -8.60 | -68 | -1.82 | | 91525 | 20.18 | -8.74 | -2.31 |
| 75047 | CLM | 17.01 | -9.23 | -63 | -1.84 | | 75047 | 24.67 | -7.08 | -3.49 |
| 90955 | SSKN | 55.06 | -29.19 | -98 | -1.89 | | 90955 | 93.71 | -4.72 | -19.87 |
| 90611 | GGN | 18.72 | -9.71 | -72 | -1.93 | | 90611 | 21.57 | -9.66 | -2.23 |
| 63706 | NRT | 21.04 | -10.87 | -74 | -1.94 | | 63706 | 24.79 | -9.53 | -2.60 |
| 78172 | | 34.09 | -17.58 | -96 | -1.94 | | 78172 | 46.03 | -14.01 | -3.29 |
| 50606 | TDW | 39.75 | -20.30 | -98 | -1.96 | | 50606 | 38.72 | -22.44 | -1.73 |
| 92131 | PLG | 45.99 | -23.11 | -99 | -1.99 | | 92131 | 48.10 | -22.41 | -2.15 |
| 18403 | | 33.63 | -16.64 | -95 | -2.02 | | 18403 | 39.05 | -16.29 | -2.40 |
| 40416 | | 35.26 | -17.45 | -94 | -2.02 | | 40416 | 33.62 | -17.90 | -1.88 |
| 85346 | NAT | 35.55 | -17.36 | -94 | -2.05 | | 85346 | 35.89 | -16.94 | -2.12 |
| 79586 | FTEK | 32.28 | -15.74 | -89 | -2.05 | | 79586 | 37.08 | -13.57 | -2.73 |
| 90456 | | 26.39 | -12.01 | -88 | -2.20 | | 90456 | 20.99 | -16.40 | -1.28 |
| 88240 | AEZS | 56.41 | -25.03 | -99 | -2.25 | | 88240 | 63.84 | -12.75 | -5.01 |
| 91856 | EOD | 11.74 | -5.15 | -51 | -2.28 | | 91856 | 11.58 | -6.30 | -1.84 |
| 23887 | | 38.18 | -16.72 | -98 | -2.28 | | 23887 | 39.81 | -20.70 | -1.92 |
| 63765 | SWN | 31.65 | -13.83 | -88 | -2.29 | | 63765 | 44.31 | -8.80 | -5.03 |
| 11394 | THMO | 48.41 | -20.71 | -99 | -2.34 | | 11394 | 51.90 | -14.50 | -3.58 |
| 79903 | | 37.98 | -16.03 | -96 | -2.37 | | 79903 | 46.21 | -16.44 | -2.81 |
| 82298 | | 28.32 | -11.92 | -84 | -2.38 | | 82298 | 35.09 | -11.40 | -3.08 |
| 85472 | CSR | 12.34 | -5.08 | -54 | -2.43 | | 85472 | 12.39 | -6.81 | -1.82 |
| 79022 | BBI | 44.02 | -18.10 | -92 | -2.43 | | 79022 | 70.81 | -5.22 | -13.56 |
| 79237 | RIG | 31.38 | -12.77 | -85 | -2.46 | | 79237 | 40.43 | -9.75 | -4.15 |
| 79315 | | 39.57 | -15.98 | -97 | -2.48 | | 79315 | 39.49 | -17.66 | -2.24 |
| 92345 | NBY | 41.98 | -16.55 | -97 | -2.54 | | 92345 | 59.16 | -7.94 | -7.45 |
| 78852 | DUC | 8.03 | -3.12 | -19 | -2.57 | | 78852 | 10.56 | -1.56 | -6.78 |
| 85576 | RNWK | 27.34 | -10.43 | -84 | -2.62 | | 85576 | 24.30 | -13.05 | -1.86 |
| 88871 | MCF | 35.40 | -13.40 | -94 | -2.64 | | 88871 | 36.81 | -15.95 | -2.31 |
| 89068 | TAC | 24.98 | -9.34 | -79 | -2.68 | | 89068 | 31.28 | -9.58 | -3.26 |
| 89648 | EGO | 39.51 | -14.57 | -93 | -2.71 | | 89648 | 45.49 | -12.99 | -3.50 |
| 88208 | CLSN | 61.09 | -22.17 | -99 | -2.76 | | 88208 | 137.46 | 6.03 | 22.79 |
| 77699 | FCEL | 50.57 | -18.08 | -99 | -2.80 | | 77699 | 42.86 | -24.41 | -1.76 |
| 91391 | | 38.46 | -13.70 | -97 | -2.81 | | 91391 | 37.68 | -18.41 | -2.05 |
| 75049 | TEF | 20.73 | -7.38 | -62 | -2.81 | | 75049 | 18.04 | -7.86 | -2.30 |
| 91734 | ESEA | 46.52 | -16.47 | -97 | -2.82 | | 91734 | 36.66 | -22.55 | -1.63 |
| 90537 | | 33.18 | -11.67 | -87 | -2.84 | | 90537 | 40.10 | -11.62 | -3.45 |
| 89139 | FRO | 46.76 | -16.41 | -96 | -2.85 | | 89139 | 33.55 | -18.84 | -1.78 |
| 69892 | | 23.40 | -8.17 | -69 | -2.86 | | 69892 | 22.83 | -8.40 | -2.72 |
| 83762 | AIM | 57.98 | -20.20 | -96 | -2.87 | | 83762 | 40.30 | -18.22 | -2.21 |
| 83520 | JAKK | 32.54 | -11.26 | -93 | -2.89 | | 83520 | 32.79 | -17.99 | -1.82 |
| 84820 | SRL | 29.89 | -10.22 | -84 | -2.92 | | 84820 | 27.39 | -11.17 | -2.45 |
| 89349 | TNP | 32.53 | -11.11 | -85 | -2.93 | | 89349 | 33.45 | -12.50 | -2.68 |
| 92863 | GLL | 28.20 | -9.55 | -81 | -2.95 | | 92863 | 33.36 | -10.42 | -3.20 |
| 32791 | | 36.65 | -12.15 | -95 | -3.02 | | 32791 | 44.23 | -12.45 | -3.55 |
| 44813 | CVM | 61.15 | -19.91 | -96 | -3.07 | | 44813 | 86.82 | 0.17 | 498.94 |
| 92155 | | 51.42 | -16.61 | -97 | -3.10 | | 92155 | 42.75 | -23.29 | -1.84 |
| 90943 | DHT | 40.93 | -12.92 | -94 | -3.17 | | 90943 | 44.62 | -11.31 | -3.95 |
| 92272 | IID | 15.64 | -4.83 | -38 | -3.24 | | 92272 | 21.47 | -2.71 | -7.93 |
| 90612 | IGD | 13.06 | -3.94 | -40 | -3.32 | | 90612 | 15.39 | -3.87 | -3.98 |
| 86233 | LEU | 81.00 | -24.17 | -100 | -3.35 | | 86233 | 141.66 | -1.75 | -80.88 |
| 34666 | CLBS | 51.98 | -15.44 | -92 | -3.37 | | 34666 | 88.54 | 2.43 | 36.47 |
| 86083 | IDXG | 54.95 | -16.30 | -98 | -3.37 | | 86083 | 58.68 | -15.91 | -3.69 |
| 77175 | AMSC | 57.17 | -16.86 | -93 | -3.39 | | 77175 | 95.40 | 7.04 | 13.54 |
| 80682 | | 35.64 | -10.31 | -96 | -3.46 | | 80682 | 41.44 | -14.78 | -2.80 |
| 91978 | BGY | 15.50 | -4.45 | -41 | -3.48 | | 91978 | 20.31 | -3.37 | -6.03 |
| 92842 | UCO | 45.20 | -12.91 | -98 | -3.50 | | 92842 | 29.53 | -23.88 | -1.24 |
| 92027 | BIL | 0.07 | -0.02 | 0 | -3.62 | | 92027 | 0.09 | -0.01 | -6.74 |
| 77420 | | 30.15 | -8.29 | -82 | -3.64 | | 77420 | 35.99 | -6.41 | -5.61 |
| 68021 | CMO | 12.67 | -3.48 | -38 | -3.64 | | 68021 | 17.05 | -3.33 | -5.12 |
| 92864 | ZSL | 47.86 | -12.69 | -98 | -3.77 | | 92864 | 51.08 | -16.88 | -3.03 |
| 85414 | NLY | 13.56 | -3.59 | -38 | -3.77 | | 85414 | 15.03 | -3.60 | -4.17 |
| 91395 | CBLI | 57.42 | -15.06 | -98 | -3.81 | | 91395 | 90.21 | -1.85 | -48.66 |
| 91895 | CPLP | 26.05 | -6.81 | -73 | -3.82 | | 91895 | 32.38 | -7.62 | -4.25 |
| 25452 | JCS | 28.86 | -7.36 | -74 | -3.92 | | 25452 | 30.63 | -8.47 | -3.62 |
| 77292 | PDLI | 25.64 | -6.53 | -53 | -3.92 | | 77292 | 26.19 | -3.30 | -7.94 |
| 81043 | PHI | 19.10 | -4.86 | -54 | -3.93 | | 81043 | 19.55 | -5.56 | -3.52 |
| 89973 | UTI | 38.34 | -9.68 | -79 | -3.96 | | 89973 | 35.69 | -8.31 | -4.30 |
| 40440 | DBD | 26.23 | -6.58 | -91 | -3.99 | | 40440 | 28.92 | -12.70 | -2.28 |
| 91760 | UDN | 7.42 | -1.85 | -20 | -4.01 | | 91760 | 6.54 | -1.99 | -3.29 |
| 92518 | TNK | 45.05 | -11.19 | -93 | -4.03 | | 92518 | 44.25 | -11.59 | -3.82 |
| 83885 | MVIS | 51.38 | -12.61 | -96 | -4.07 | | 83885 | 58.16 | -9.12 | -6.38 |
| 82526 | ACHV | 95.37 | -23.05 | -100 | -4.14 | | 82526 | 220.14 | 21.83 | 10.08 |
| 82860 | SPH | 17.97 | -4.33 | -46 | -4.15 | | 82860 | 25.14 | -2.87 | -8.77 |
| 11644 | CRK | 50.37 | -12.09 | -98 | -4.17 | | 11644 | 30.44 | -26.20 | -1.16 |
| 92594 | SLS | 77.95 | -18.44 | -100 | -4.23 | | 92594 | 124.06 | -3.19 | -38.84 |
| 92694 | OCSL | 20.39 | -4.81 | -44 | -4.24 | | 92694 | 19.87 | -4.03 | -4.94 |
| 90164 | GHL | 30.33 | -7.14 | -65 | -4.25 | | 90164 | 30.43 | -5.11 | -5.96 |
| 91291 | FXB | 7.48 | -1.74 | -15 | -4.30 | | 91291 | 8.10 | -1.33 | -6.07 |
| 88411 | TKC | 22.77 | -5.29 | -61 | -4.31 | | 88411 | 32.77 | -3.99 | -8.22 |
| 85897 | ANH | 13.65 | -3.10 | -37 | -4.40 | | 85897 | 17.77 | -3.06 | -5.80 |
| 90675 | BOE | 14.23 | -3.22 | -42 | -4.42 | | 90675 | 14.99 | -4.20 | -3.57 |
| 71862 | TURN | 24.58 | -5.50 | -56 | -4.47 | | 71862 | 22.35 | -5.59 | -4.00 |
| 78054 | QUMU | 43.63 | -9.70 | -86 | -4.50 | | 78054 | 45.10 | -6.41 | -7.04 |
| 92271 | HTY | 12.14 | -2.69 | -39 | -4.50 | | 92271 | 11.92 | -4.15 | -2.87 |
| 91208 | USO | 26.20 | -5.82 | -71 | -4.50 | | 91208 | 21.28 | -8.98 | -2.37 |
| 88177 | SEEL | 74.56 | -16.28 | -99 | -4.58 | | 88177 | 65.52 | -9.31 | -7.04 |
| 90685 | FMY | 9.06 | -1.95 | -21 | -4.65 | | 90685 | 6.71 | -2.13 | -3.15 |
| 75274 | GIM | 11.98 | -2.57 | -22 | -4.66 | | 75274 | 11.93 | -1.88 | -6.34 |
| 91047 | FXE | 8.79 | -1.86 | -22 | -4.71 | | 91047 | 7.55 | -2.16 | -3.49 |
| 91712 | DBA | 13.84 | -2.89 | -35 | -4.78 | | 91712 | 11.08 | -3.73 | -2.97 |
| 75672 | WWR | 92.76 | -19.31 | -100 | -4.80 | | 75672 | 127.32 | -9.57 | -13.31 |
| 91684 | AOD | 18.74 | -3.83 | -42 | -4.89 | | 91684 | 23.72 | -2.78 | -8.53 |
| 91269 | CBAT | 58.19 | -11.66 | -95 | -4.99 | | 91269 | 51.82 | -12.03 | -4.31 |
| 89199 | DB | 34.99 | -6.97 | -74 | -5.02 | | 89199 | 37.11 | -6.34 | -5.85 |
| 86745 | BB | 39.19 | -7.80 | -82 | -5.02 | | 86745 | 47.26 | -4.59 | -10.30 |
| 83188 | CCJ | 25.69 | -5.04 | -34 | -5.10 | | 83188 | 38.32 | 2.15 | 17.84 |
| 92720 | ICLN | 26.12 | -5.08 | -61 | -5.15 | | 92720 | 25.53 | -6.00 | -4.26 |
| 89692 | KGC | 46.32 | -8.97 | -82 | -5.16 | | 89692 | 38.96 | -9.29 | -4.19 |
| 91560 | DAC | 34.24 | -6.61 | -89 | -5.18 | | 91560 | 39.48 | -12.85 | -3.07 |
| 91381 | GSG | 18.77 | -3.62 | -51 | -5.19 | | 91381 | 16.51 | -5.42 | -3.05 |
| 91938 | EDD | 19.89 | -3.70 | -42 | -5.37 | | 91938 | 21.98 | -3.06 | -7.19 |
| 87471 | TGC | 44.87 | -8.33 | -85 | -5.39 | | 87471 | 31.36 | -12.10 | -2.59 |
| 88462 | SAVA | 52.27 | -9.66 | -98 | -5.41 | | 88462 | 47.53 | -19.46 | -2.44 |
| 18075 | AP | 31.07 | -5.71 | -86 | -5.44 | | 18075 | 40.79 | -8.16 | -5.00 |
| 87373 | ALSK | 35.60 | -6.54 | -85 | -5.44 | | 87373 | 39.71 | -8.19 | -4.85 |
| 76586 | PICO | 29.85 | -5.43 | -66 | -5.50 | | 76586 | 28.32 | -6.35 | -4.46 |
| 91656 | | 41.27 | -7.25 | -89 | -5.69 | | 91656 | 47.63 | -7.59 | -6.28 |
| 75367 | ZTR | 10.93 | -1.88 | -34 | -5.82 | | 75367 | 15.15 | -2.88 | -5.26 |
| 92687 | | 110.47 | -18.94 | -100 | -5.83 | | 92687 | 116.89 | -24.57 | -4.76 |
| 82281 | SMSI | 53.10 | -9.10 | -92 | -5.83 | | 82281 | 58.54 | 1.40 | 41.85 |
| 86080 | PLX | 56.63 | -9.71 | -83 | -5.83 | | 86080 | 97.59 | 8.21 | 11.88 |
| 89289 | | 41.90 | -7.13 | -91 | -5.87 | | 89289 | 50.20 | -9.41 | -5.34 |
| 91950 | BSV | 1.07 | -0.18 | -2 | -6.03 | | 91950 | 0.76 | -0.25 | -3.08 |
| 90101 | SVRA | 65.63 | -10.57 | -94 | -6.21 | | 90101 | 138.18 | 20.88 | 6.62 |
| 91747 | FXY | 10.42 | -1.67 | -21 | -6.22 | | 91747 | 9.64 | -1.88 | -5.13 |
| 92032 | ALBO | 58.85 | -9.35 | -96 | -6.29 | | 92032 | 47.52 | -14.43 | -3.29 |
| 30796 | GFI | 34.21 | -5.42 | -59 | -6.31 | | 30796 | 38.71 | 0.21 | 180.12 |
| 90755 | FSP | 19.00 | -3.01 | -58 | -6.31 | | 90755 | 21.28 | -5.97 | -3.57 |
| 91294 | FXC | 7.69 | -1.21 | -12 | -6.38 | | 91294 | 9.12 | -0.92 | -9.91 |
| 89817 | | 25.54 | -4.00 | -66 | -6.38 | | 89817 | 17.58 | -8.63 | -2.04 |
| 60599 | LUMN | 20.02 | -3.13 | -45 | -6.40 | | 60599 | 24.49 | -2.95 | -8.29 |
| 88960 | NS | 26.22 | -4.10 | -49 | -6.40 | | 88960 | 28.68 | -2.61 | -11.00 |
| 60186 | OMI | 22.17 | -3.45 | -75 | -6.42 | | 60186 | 28.08 | -7.43 | -3.78 |
| 75652 | TEVA | 28.13 | -4.37 | -64 | -6.44 | | 75652 | 30.09 | -5.15 | -5.85 |
| 89029 | MOSY | 64.40 | -9.95 | -99 | -6.47 | | 89029 | 60.74 | -17.62 | -3.45 |
| 92061 | REM | 12.11 | -1.85 | -31 | -6.53 | | 92061 | 11.79 | -2.91 | -4.05 |
| 78189 | PESI | 32.21 | -4.93 | -62 | -6.54 | | 78189 | 39.09 | -2.80 | -13.94 |
| 78877 | | 41.71 | -6.33 | -86 | -6.59 | | 78877 | 49.35 | -5.10 | -9.68 |
| 90735 | LINC | 60.79 | -9.19 | -76 | -6.62 | | 90735 | 39.41 | -6.67 | -5.90 |
| 89470 | SHY | 0.62 | -0.09 | -1 | -6.64 | | 89470 | 0.84 | -0.12 | -7.02 |
| 90291 | PSEC | 17.00 | -2.54 | -47 | -6.69 | | 90291 | 15.05 | -5.16 | -2.92 |
| 89154 | CS | 31.65 | -4.71 | -61 | -6.71 | | 89154 | 36.78 | -3.28 | -11.20 |
| 91279 | PRTK | 52.61 | -7.76 | -94 | -6.78 | | 91279 | 28.48 | -19.15 | -1.49 |
| 79667 | | 30.55 | -4.45 | -30 | -6.86 | | 79667 | 31.37 | 1.18 | 26.68 |
| 92777 | | 46.85 | -6.81 | -98 | -6.88 | | 92777 | 49.03 | -7.54 | -6.50 |
| 50017 | RRC | 33.93 | -4.89 | -72 | -6.95 | | 50017 | 40.63 | -3.67 | -11.08 |
| 91305 | | 59.52 | -8.56 | -96 | -6.95 | | 91305 | 60.95 | -11.19 | -5.45 |
| 90630 | DSX | 37.01 | -5.32 | -74 | -6.96 | | 90630 | 40.00 | -6.18 | -6.47 |
| 91283 | BATL | 75.51 | -10.80 | -100 | -6.99 | | 91283 | 70.25 | -19.21 | -3.66 |
| 86820 | CFFN | 11.92 | -1.70 | -37 | -7.02 | | 86820 | 16.51 | -3.20 | -5.16 |
| 91374 | AGD | 19.95 | -2.84 | -40 | -7.02 | | 91374 | 22.45 | -2.77 | -8.11 |
| 90926 | ALT | 62.08 | -8.85 | -100 | -7.02 | | 90926 | 71.61 | -3.18 | -22.54 |
| 24459 | PBI | 35.09 | -4.93 | -77 | -7.12 | | 24459 | 47.33 | -6.20 | -7.63 |
| 84255 | SEAC | 33.79 | -4.69 | -83 | -7.21 | | 84255 | 46.57 | -3.80 | -12.26 |
| 86341 | AU | 43.49 | -5.96 | -55 | -7.29 | | 86341 | 35.23 | -1.10 | -32.02 |
| 39490 | APA | 26.62 | -3.61 | -65 | -7.37 | | 39490 | 30.14 | -5.86 | -5.14 |
| 90067 | | 46.05 | -5.96 | -91 | -7.73 | | 90067 | 44.62 | -11.27 | -3.96 |
| 89796 | | 40.85 | -5.26 | -85 | -7.77 | | 89796 | 41.20 | -7.41 | -5.56 |
| 77895 | EQS | 19.99 | -2.47 | -53 | -8.09 | | 77895 | 15.38 | -6.16 | -2.50 |
| 92240 | SRV | 32.64 | -3.94 | -62 | -8.29 | | 92240 | 42.01 | -1.32 | -31.89 |
| 68427 | IAF | 21.58 | -2.59 | -34 | -8.33 | | 68427 | 28.98 | -0.95 | -30.66 |
| 62391 | SJT | 33.64 | -4.02 | -85 | -8.37 | | 62391 | 42.56 | -7.06 | -6.03 |
| 91040 | | 85.49 | -9.90 | -100 | -8.63 | | 91040 | 57.28 | -26.31 | -2.18 |
| 91836 | AVNW | 46.14 | -5.25 | -79 | -8.79 | | 91836 | 46.40 | -4.34 | -10.68 |
| 92035 | CLNE | 50.81 | -5.71 | -72 | -8.90 | | 92035 | 58.01 | -1.56 | -37.07 |
| 79499 | INOD | 42.92 | -4.79 | -40 | -8.97 | | 79499 | 49.78 | 4.03 | 12.35 |
| 91713 | DBO | 27.41 | -3.02 | -56 | -9.08 | | 91713 | 25.35 | -4.63 | -5.47 |
| 90366 | STON | 35.20 | -3.87 | -82 | -9.09 | | 90366 | 44.55 | -4.78 | -9.31 |
| 64020 | LUB | 29.28 | -3.19 | -71 | -9.19 | | 64020 | 40.00 | -4.61 | -8.68 |
| 88332 | CPST | 58.84 | -6.10 | -96 | -9.65 | | 88332 | 44.93 | -12.38 | -3.63 |
| 90494 | HPR | 46.59 | -4.78 | -88 | -9.75 | | 90494 | 52.78 | -5.81 | -9.08 |
| 26649 | ASA | 31.86 | -3.26 | -44 | -9.77 | | 26649 | 34.94 | -0.56 | -62.16 |
| 59483 | YRCW | 99.65 | -10.00 | -100 | -9.97 | | 59483 | 75.57 | -21.08 | -3.59 |
| 91872 | IAE | 20.14 | -2.00 | -25 | -10.07 | | 91872 | 28.15 | 0.18 | 157.64 |
| 91348 | ATEC | 62.83 | -6.22 | -92 | -10.11 | | 91348 | 55.08 | -9.12 | -6.04 |
| 91845 | SNMP | 41.55 | -4.06 | -94 | -10.24 | | 91845 | 51.06 | -10.31 | -4.95 |
| 92405 | SCU | 49.60 | -4.79 | -82 | -10.36 | | 92405 | 68.97 | 0.06 | 1087.76 |
| 76215 | ICON | 56.14 | -5.40 | -99 | -10.41 | | 76215 | 62.44 | -5.64 | -11.06 |
| 92427 | ANTE | 63.71 | -5.87 | -95 | -10.85 | | 92427 | 61.30 | -8.34 | -7.35 |
| 10530 | VIVO | 18.37 | -1.67 | -32 | -11.00 | | 10530 | 23.32 | -1.20 | -19.41 |
| 75331 | BKT | 6.54 | -0.59 | -4 | -11.07 | | 75331 | 6.64 | -0.21 | -31.07 |
| 75809 | AEF | 27.90 | -2.47 | -35 | -11.32 | | 75809 | 39.92 | 1.88 | 21.24 |
| 90704 | | 49.53 | -4.37 | -84 | -11.34 | | 90704 | 63.42 | -1.63 | -38.80 |
| 82637 | E | 15.83 | -1.35 | -34 | -11.70 | | 82637 | 13.38 | -3.22 | -4.15 |
| 38682 | RRD | 46.39 | -3.96 | -82 | -11.70 | | 38682 | 56.26 | -5.01 | -11.24 |
| 80485 | | 73.06 | -6.18 | -96 | -11.81 | | 80485 | 48.16 | -14.79 | -3.26 |
| 87339 | PTNR | 42.44 | -3.58 | -71 | -11.85 | | 87339 | 35.92 | -5.33 | -6.74 |
| 88810 | RIGL | 33.07 | -2.78 | -71 | -11.91 | | 88810 | 35.87 | -5.70 | -6.30 |
| 79052 | NPV | 11.25 | -0.93 | 4 | -12.16 | | 79052 | 14.56 | 1.36 | 10.73 |
| 91873 | MBB | 2.21 | -0.18 | 0 | -12.41 | | 91873 | 2.23 | 0.00 | -815.13 |
| 76123 | JOE | 22.12 | -1.78 | -46 | -12.44 | | 76123 | 26.45 | -3.08 | -8.59 |
| 83421 | | 33.86 | -2.64 | -78 | -12.84 | | 83421 | 38.16 | -5.23 | -7.29 |
| 23026 | FE | 13.92 | -1.08 | -23 | -12.91 | | 23026 | 16.58 | -1.31 | -12.69 |
| 77369 | CRT | 28.50 | -2.18 | -61 | -13.07 | | 77369 | 29.69 | -4.56 | -6.52 |
| 75622 | BPT | 34.20 | -2.62 | -70 | -13.08 | | 75622 | 31.79 | -5.86 | -5.43 |
| 92534 | GCC | 15.46 | -1.17 | -20 | -13.16 | | 92534 | 14.19 | -1.34 | -10.63 |
| 91822 | NCMI | 26.74 | -2.03 | -36 | -13.18 | | 91822 | 35.77 | 1.72 | 20.83 |
| 92010 | JMP | 20.57 | -1.53 | -30 | -13.42 | | 92010 | 31.31 | 0.21 | 150.73 |
| 90618 | PBW | 29.36 | -2.17 | -50 | -13.52 | | 90618 | 32.57 | -1.98 | -16.43 |
| 80962 | | 56.30 | -4.05 | -87 | -13.91 | | 80962 | 39.52 | -11.18 | -3.53 |
| 91749 | EXG | 12.84 | -0.87 | -24 | -14.85 | | 91749 | 16.28 | -1.55 | -10.50 |
| 12209 | LWAY | 36.03 | -2.42 | -79 | -14.87 | | 12209 | 43.30 | -4.03 | -10.74 |
| 10838 | | 82.40 | -5.44 | -96 | -15.15 | | 10838 | 128.57 | 17.86 | 7.20 |
| 79225 | | 9.49 | -0.63 | 7 | -15.15 | | 79225 | 12.61 | 1.34 | 9.42 |
| 90973 | OIIM | 39.20 | -2.58 | -23 | -15.20 | | 90973 | 59.96 | 7.34 | 8.17 |
| 89976 | AUY | 45.07 | -2.89 | -69 | -15.57 | | 89976 | 40.06 | -2.69 | -14.87 |
| 69366 | FAX | 15.67 | -0.99 | -10 | -15.83 | | 69366 | 20.00 | 0.65 | 30.92 |
| 90927 | IGA | 11.57 | -0.72 | -17 | -15.99 | | 90927 | 14.07 | -0.98 | -14.35 |
| 77458 | FCO | 15.82 | -0.99 | -16 | -16.00 | | 77458 | 21.35 | 0.21 | 103.34 |
| 89889 | CIDM | 55.69 | -3.47 | -85 | -16.03 | | 89889 | 88.28 | 9.07 | 9.74 |
| 91129 | DBC | 17.69 | -1.10 | -32 | -16.14 | | 91129 | 16.69 | -2.36 | -7.06 |
| 85425 | ORAN | 23.05 | -1.42 | -42 | -16.21 | | 85425 | 20.10 | -3.56 | -5.64 |
| 92213 | RIOT | 101.21 | -6.15 | -100 | -16.45 | | 92213 | 221.33 | 22.51 | 9.83 |
| 89398 | | 33.64 | -1.96 | -47 | -17.16 | | 89398 | 40.62 | 1.69 | 24.10 |
| 91287 | GROW | 52.13 | -3.02 | -78 | -17.26 | | 91287 | 88.63 | 11.28 | 7.86 |
| 75381 | BBVA | 31.59 | -1.79 | -57 | -17.64 | | 75381 | 30.73 | -3.71 | -8.27 |
| 75269 | DX | 15.57 | -0.87 | -13 | -17.93 | | 75269 | 18.87 | 0.22 | 85.72 |
| 91232 | GDX | 32.29 | -1.79 | -38 | -18.04 | | 91232 | 32.52 | 0.48 | 68.02 |
| 90434 | FAM | 17.65 | -0.95 | -10 | -18.52 | | 90434 | 22.77 | 1.01 | 22.52 |
| 90098 | CYCC | 116.12 | -6.02 | -98 | -19.30 | | 90098 | 71.22 | -11.32 | -6.29 |
| 92426 | NMM | 46.48 | -2.35 | -88 | -19.79 | | 92426 | 62.39 | -1.36 | -45.83 |
| 33099 | IMO | 15.60 | -0.74 | -25 | -21.06 | | 33099 | 12.57 | -2.06 | -6.11 |
| 89570 | MMLP | 29.57 | -1.40 | -30 | -21.07 | | 89570 | 45.33 | 3.25 | 13.95 |
| 90495 | BGR | 21.79 | -1.03 | -31 | -21.09 | | 90495 | 29.25 | -0.28 | -102.71 |
| 92822 | SUB | 0.85 | -0.04 | 3 | -21.39 | | 92822 | 1.29 | 0.31 | 4.15 |
| 11628 | CPF | 33.54 | -1.54 | -88 | -21.75 | | 11628 | 40.04 | -4.97 | -8.05 |
| 92787 | | 31.74 | -1.41 | -34 | -22.56 | | 92787 | 32.66 | 0.89 | 36.83 |
| 92509 | USL | 23.88 | -1.06 | -40 | -22.61 | | 92509 | 22.90 | -2.36 | -9.69 |
| 21776 | EXC | 15.14 | -0.65 | -19 | -23.25 | | 21776 | 22.16 | 0.15 | 151.32 |
| 91582 | MFG | 18.47 | -0.77 | -48 | -24.12 | | 91582 | 23.93 | -3.63 | -6.59 |
| 88284 | CYH | 52.48 | -2.15 | -77 | -24.45 | | 88284 | 67.63 | 5.41 | 12.50 |
| 71298 | GOLD | 48.28 | -1.96 | -63 | -24.60 | | 71298 | 47.93 | -1.48 | -32.47 |
| 77910 | NIM | 6.34 | -0.25 | 2 | -25.00 | | 77910 | 6.57 | 0.37 | 17.86 |
| 85237 | ELP | 36.96 | -1.46 | -26 | -25.25 | | 85237 | 43.69 | 4.48 | 9.74 |
| 87844 | PTR | 21.29 | -0.83 | -31 | -25.70 | | 87844 | 21.24 | -1.38 | -15.44 |
| 92020 | GEN | 55.41 | -2.11 | -86 | -26.30 | | 92020 | 50.32 | -2.45 | -20.58 |
| 89134 | OVV | 42.03 | -1.38 | -77 | -30.42 | | 89134 | 55.07 | -1.94 | -28.38 |
| 10463 | REFR | 46.10 | -1.50 | -28 | -30.81 | | 10463 | 46.72 | 7.56 | 6.18 |
| 87137 | DVN | 29.29 | -0.95 | -66 | -30.81 | | 87137 | 28.17 | -6.05 | -4.66 |
| 90875 | UEPS | 34.33 | -1.09 | -66 | -31.45 | | 90875 | 41.86 | -1.81 | -23.14 |
| 29867 | | 39.03 | -1.22 | -60 | -32.02 | | 29867 | 42.01 | -1.51 | -27.91 |
| 89855 | WIA | 6.34 | -0.19 | -2 | -33.36 | | 89855 | 8.56 | 0.14 | 62.37 |
| 89593 | | 12.82 | -0.37 | 20 | -34.42 | | 89593 | 19.14 | 3.39 | 5.64 |
| 90386 | MOS | 28.47 | -0.81 | -16 | -34.97 | | 90386 | 32.52 | 2.69 | 12.08 |
| 88490 | PBR | 49.44 | -1.37 | -47 | -36.21 | | 88490 | 61.77 | 6.41 | 9.64 |
| 29102 | NBR | 44.78 | -1.23 | -83 | -36.45 | | 29102 | 54.61 | -2.95 | -18.50 |
| 11018 | FBP | 52.34 | -1.42 | -95 | -36.81 | | 11018 | 62.25 | -4.04 | -15.41 |
| 90791 | | 43.72 | -1.16 | -89 | -37.79 | | 90791 | 63.12 | 4.79 | 13.17 |
| 91709 | DBE | 23.76 | -0.62 | -38 | -38.53 | | 91709 | 22.58 | -2.06 | -10.98 |
| 10253 | DSS | 57.71 | -1.45 | -90 | -39.86 | | 10253 | 82.20 | 4.99 | 16.48 |
| 77693 | NXN | 7.03 | -0.17 | 3 | -40.24 | | 77693 | 6.82 | 0.50 | 13.67 |
| 79226 | | 10.82 | -0.26 | 10 | -40.96 | | 79226 | 13.91 | 1.75 | 7.93 |
| 75313 | PIM | 12.27 | -0.29 | 5 | -41.93 | | 75313 | 17.37 | 1.58 | 11.01 |
| 90552 | EVG | 10.28 | -0.24 | 7 | -42.43 | | 90552 | 14.73 | 1.52 | 9.69 |
| 89490 | ENX | 12.27 | -0.28 | 18 | -43.31 | | 89490 | 18.65 | 3.04 | 6.13 |
| 77659 | BBBY | 32.08 | -0.74 | -55 | -43.55 | | 77659 | 36.11 | -1.48 | -24.43 |
| 90695 | TRX | 62.98 | -1.38 | -92 | -45.79 | | 90695 | 70.99 | -0.43 | -163.84 |
| 77763 | | 38.66 | -0.74 | -79 | -52.14 | | 77763 | 40.34 | -6.12 | -6.59 |
| 92466 | OESX | 67.95 | -1.22 | -89 | -55.91 | | 92466 | 109.30 | 3.72 | 29.35 |
| 91418 | | 65.67 | -0.98 | -81 | -67.21 | | 91418 | 86.77 | 10.13 | 8.57 |
| 75152 | SAN | 35.49 | -0.53 | -48 | -67.21 | | 75152 | 35.23 | -1.00 | -35.24 |
| 77073 | | 11.04 | -0.16 | 10 | -68.81 | | 77073 | 14.81 | 1.88 | 7.86 |
| 90057 | WIW | 6.86 | -0.09 | -3 | -74.80 | | 90057 | 9.12 | 0.05 | 176.06 |
| 91928 | DNN | 45.65 | -0.59 | -54 | -77.17 | | 91928 | 63.00 | 5.13 | 12.29 |
| 78688 | | 55.77 | -0.72 | -91 | -77.32 | | 78688 | 52.93 | -8.12 | -6.52 |
| 80779 | CLI | 24.24 | -0.30 | -20 | -82.07 | | 80779 | 22.03 | -0.18 | -125.21 |
| 28345 | MUR | 26.52 | -0.30 | -39 | -86.98 | | 28345 | 31.52 | 0.34 | 91.73 |
| 90930 | PXJ | 29.17 | -0.31 | -49 | -94.33 | | 90930 | 31.09 | -2.11 | -14.71 |
| 89512 | NBO | 10.87 | -0.11 | 6 | -102.31 | | 89512 | 11.85 | 1.22 | 9.72 |
| 89491 | EVM | 12.34 | -0.09 | 24 | -131.20 | | 89491 | 17.24 | 3.33 | 5.18 |
| 90285 | MNKD | 80.79 | -0.51 | -94 | -158.45 | | 90285 | 79.38 | -1.35 | -58.78 |
| 18438 | SCX | 29.03 | -0.13 | -68 | -220.77 | | 18438 | 36.08 | -4.26 | -8.47 |
| 89332 | | 49.18 | -0.18 | -13 | -267.80 | | 89332 | 61.25 | 9.16 | 6.69 |
| 70295 | | 13.15 | -0.04 | -25 | -332.03 | | 70295 | 16.08 | -1.54 | -10.46 |
| 92680 | AGNC | 18.29 | -0.02 | -18 | -1132.91 | | 92680 | 17.30 | -0.47 | -36.83 |
| 83220 | EWP | 23.71 | 0.00 | -29 | -5112.87 | | 83220 | 20.24 | -1.61 | -12.60 |
| 89043 | CEQP | 49.50 | -0.01 | -67 | -9043.15 | | 89043 | 49.55 | 1.55 | 31.99 |